\documentclass[12pt,a4paper,openright]{book}
\usepackage{graphicx}
\usepackage{background}
\usepackage{fancyhdr}   
\usepackage{multicol}
\usepackage{makeidx}
\usepackage{hyperref} 
\usepackage{amsmath}
\usepackage{amsthm}
\usepackage{amssymb}
\usepackage{array}
\usepackage{pstricks,pst-node}
\usepackage{setspace}
\usepackage{lineno}
\usepackage{cite}
\usepackage{latexsym}
\usepackage{setspace}
\usepackage[top=25mm, left=35mm, bottom=25mm, right=25mm]{geometry}
\usepackage{slashbox}
\usepackage{caption}
\usepackage{rotating}
\usepackage{enumerate}
\usepackage{afterpage}
\usepackage{mwe}
\usepackage{xcolor}
\usepackage{colortbl}
\usepackage{multirow}
\usepackage{mathtools}
\usepackage[b]{esvect}
\usepackage{gensymb}
\usepackage{algpseudocode}
\usepackage{csquotes}
\usepackage{subcaption}
\usepackage{float}
\usepackage{multirow}

\usepackage{tikz}
\usetikzlibrary{shapes.geometric, arrows, calc,automata,positioning,decorations,decorations.text}
\usepackage{authblk}
\usepackage{lipsum}
\usepackage{float}
\usepackage{environ}

\newtheorem{Theorem}{Theorem}

\NewEnviron{NORMAL}{%
    \scalebox{2}{$\BODY$}
}
 
\NewEnviron{HUGE}{%
    \scalebox{5}{$\BODY$}
}

\newcounter{relctr} 
\everydisplay\expandafter{\the\everydisplay\setcounter{relctr}{0}} 

\AtBeginDocument{} 

\makeatletter
\newsavebox\myboxA
\newsavebox\myboxB
\newlength\mylenA

\newcommand*\xoverline[2][0.75]{%
    \sbox{\myboxA}{$\m@th#2$}%
    \setbox\myboxB\null
    \ht\myboxB=\ht\myboxA%
    \dp\myboxB=\dp\myboxA%
    \wd\myboxB=#1\wd\myboxA
    \sbox\myboxB{$\m@th\overline{\copy\myboxB}$}
    \setlength\mylenA{\the\wd\myboxA}
    \addtolength\mylenA{-\the\wd\myboxB}%
    \ifdim\wd\myboxB<\wd\myboxA%
       \rlap{\hskip 0.5\mylenA\usebox\myboxB}{\usebox\myboxA}%
    \else
        \hskip -0.5\mylenA\rlap{\usebox\myboxA}{\hskip 0.5\mylenA\usebox\myboxB}%
    \fi}
\makeatother

\def\dept{Department of Electrical Engineering}
\def\IIT{Indian Institute of Technology Kanpur}
\def\instaddress{Kanpur, INDIA - 208016}
\title{Modeling and Control of Smart Standalone Microgrids within CPS framework}
\author{MEHER PREETAM KORUKONDA}

\def\thesisdate{February, 2021}

\backgroundsetup{contents={}}

\makeindex

\begin{document}
\makeatletter
\thispagestyle{empty}
\begin{titlepage}
\begin{center}
\LARGE
\textbf{Modeling and Control of Smart Standalone Microgrids within Cyber Physical System Framework} \\
\vspace{2cm}
\Large
{\em A Thesis Submitted}\\ in Partial Fulfillment of the Requirements\\
for the Degree of\\ \textbf{Doctor of Philosophy}\\
\vspace{2cm}
by\\
\textbf{Meher Preetam Korukonda} \\\textbf{(12104172)}
\vfill
\begin{figure}[!h]
\centering
\includegraphics[width=0.35\linewidth]{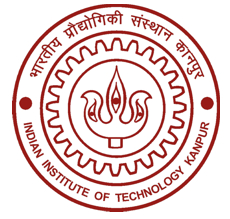}
\label{fig:spvdg0}
\end{figure}
{\em to the} \\
\dept \\
\MakeUppercase{\IIT} \\
\instaddress\\
\thesisdate 

\end{center}
\end{titlepage}
\makeatother
\doublespacing
\onehalfspacing
\newpage \null\thispagestyle{empty}  








 \begin{figure*}[!h]
\centering
\includegraphics[width=1.0\linewidth,trim={4cm 0 2cm 2cm},clip]{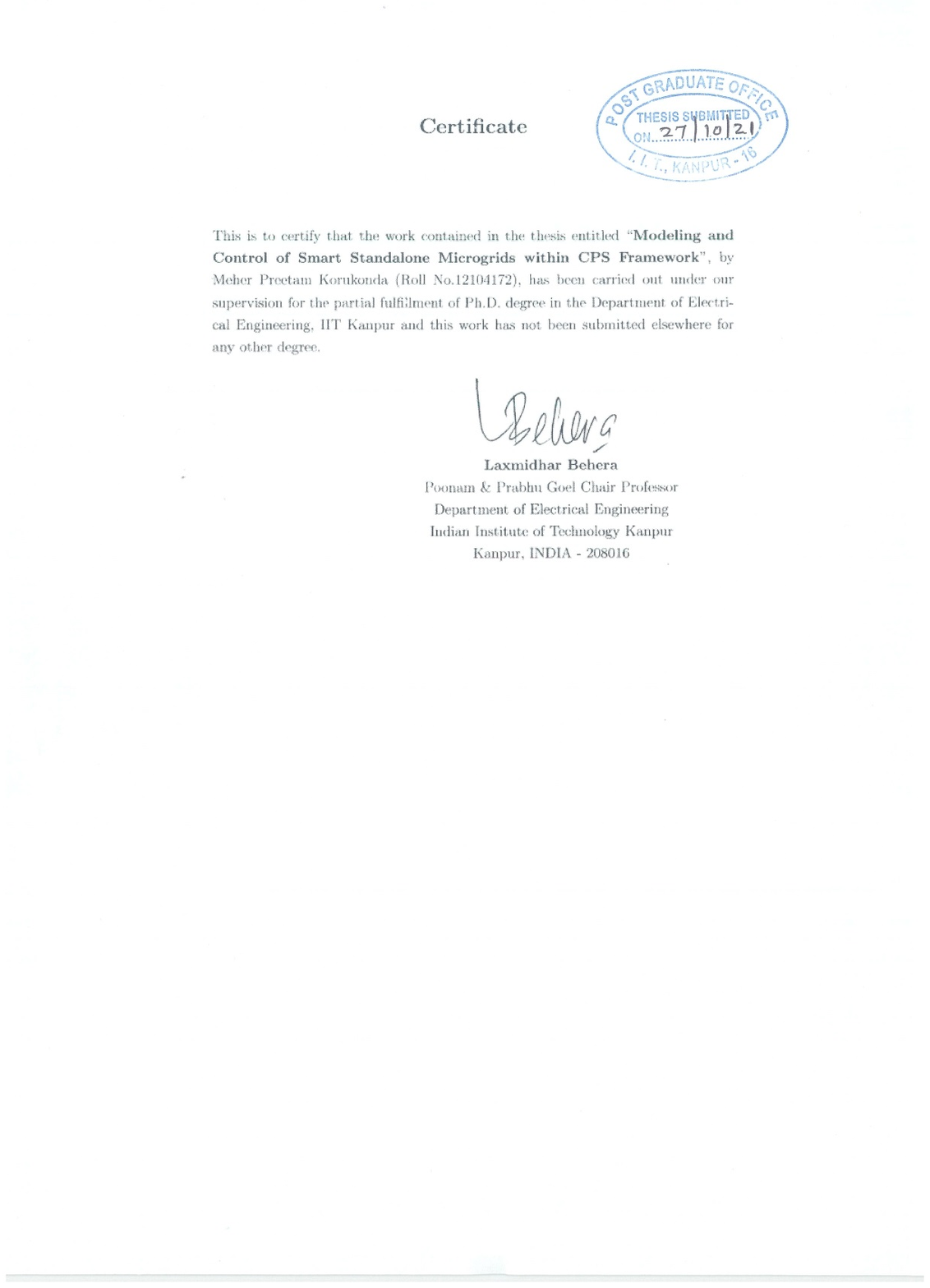}
\label{fig:certificate}
\end{figure*}
 \begin{figure*}[!h]
\centering
\includegraphics[width=1.0\linewidth]{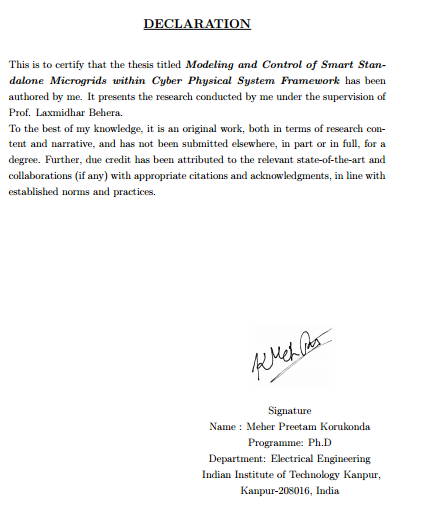}
\label{fig:certificate}
\end{figure*}
\begin{center}
\begin{large}
{\it{\bf ABSTRACT/SYNOPSIS} }
\end{large}
\end{center} 
\hrule
\vspace{1ex}
\noindent
Name of Student: Meher Preetam Korukonda \hspace{5ex}Roll no.: 12104172\\
Degree for which submitted: Ph.D \hspace{9ex}Department: Electrical Engineering\\
Title: Modeling and Control of Smart Standalone Microgrids within Cyber Physical System Framework.\\
Name of Thesis Supervisor: Prof. Laxmidhar Behera

\hspace{-0.6cm}Month and year of Thesis submission: October, 2021.
\vspace{1ex}
\hrule

\vspace{3ex}
\noindent
The ability of grid-connected microgrids to operate in islanded mode makes them an efficient solution for improving power quality and reliability. This property of microgrid is very much beneficial for remote and undeveloped areas in progressing countries. Moreover, the development of information and communications technology (ICT) has led to the development of Smart Standalone Microgrids (SSMG), which are inundated with a plethora of sensing devices. This allows multiple microgrids to be controlled in a coordinated way to achieve self-sufficiency in power.  In such systems, there is much interdependency between various power, control and communication parameters. Owing to these developments, the control of physical variables like voltage get affected by cyber parameters like, communication structure, delay and link loss. Moreover, due to isolation from main grid and abundance of distributed renewable power generation units like solar photovoltaics, the inertia of these standalone grids is reduced greatly and calls for deployment of advanced control algorithms which use an abundance of sensors. Hence, the stability of these systems is greatly affected by sensor failures apart from many physical parameters like load and environmental conditions. Hence, in this thesis a generic structure of an AC-DC hybrid microgrid is considered which is further subdivided into various AC and DC counterparts or SSMGs. The AC and DC SSMGs are separately modeled and control solutions are proposed to improve their stability. The first two contributions propose adaptive control schemes on the primary level of control in the hybrid microgrid. Their function is to provide fast and stable voltage and current regulation in the DCSSMG when subjected to change in atmospheric conditions along with faults in sensor readings. The third and fourth contributions cater to coordinated voltage control in the ACSSMG in the presence of simultaneous disturbances from both cyber and physical domains. 

In the first contribution of this thesis, a detailed model of a DCSSMG with Hybrid Energy Storage System (HESS) was derived consisting of solar photovoltaic panels, battery and supercapacitor. Nonlinear control techniques like backstepping are generally employed for extracting maximum power from the photovoltaic panels  and regulating the voltage of the DCSSMG in the presence of disturbances in load, irradiance and temperature. These techniques although effective, use a lot of sensors making the system expensive to implement and prone to sensor failure. In this work, a disturbance observer based back-stepping controller is proposed to obviate the necessity for measuring disturbance values. The effects of irradiation and temperature on PV arrays, the variations in loads and battery voltage are modeled in the form of disturbances. Instead of measuring these with sensors, the proposed observer update laws based on Lyapunov stability theory estimate their values. They are further utilized for effective control during intermittencies. It can be seen from the MATLAB simulation results that adoption of this technique contributes towards faster, cheaper and more reliable control of the DCSSMGs when subjected simultaneously to sensor faults and sudden changes in temperature, irradiance and load.

The second contribution deals with the development of an adaptive neural controller for carrying out voltage control and maximum power point tracking (MPPT) in the DCSSMG with unknown disturbances. This controller removes the necessity to keep track of system model parameters like resistances, inductance and capacitances apart from eliminating the need for expensive sensors for sensing load and environmental conditions. The neural network weight update laws of the controller are derived using the Lyapunov stability. It is shown that the proposed controller is able to ensure the uniformly ultimately boundedness (UUB) of all signals of the resulting closed-loop system. The performance of the proposed controller is evaluated in simulations against state-of-the-art controllers during disturbances and parameter intermittencies in the presence of sensor failures.

In the third contribution of this thesis, a generic, hybrid and customized cyber-physical framework is developed to jointly model the multi-disciplinary variables and their interactions present in a densely connected ACSSMG. This cyber-physical model is used to design adaptive controllers to ensure better control of  microgrid voltages irrespective of the changes in operating point brought about by changes in physical/cyber parameters.  The different operating conditions of the power system have been modeled as multiple subsystems of a hybrid switching system and controller design is carried out by solving the optimisation formulations developed for delay-free and delay-existent operation of the ACSSMG using the theory of common Lyapunov function (CLF). The optimisation is carried out using the block coordinate descent (BCD) methodology by converting the non-convex formulation into a series of convex problems to obtain a solution.

In the fourth and last contribution, the design of communication network between various sensors and controllers in a sparsely connected ACSSMG and its effect on improving voltage stability is explored. The design process involves automated examination of voltage stability in the presence of a number of topological combinations of the communication network, which increase exponentially with the number of nodes. The different characteristics and availability of various physical and communication resources in the network pose multiple constraints on this design. For this purpose, an integrated cyber physical controller design methodology is developed consisting of a connection finding algorithm and a generalised constraint-based sensor controller connection design (CBSCD) procedure. This effectively reduces the number of combinations, to design more stable cyber-physical controllers. To handle variations in multiple parameters in physical and communication domain, different controllers have been developed for different operating conditions that are deployed as per requirement. The methodology has been shown to effectively stabilise bus voltages in a smart grid scenario under variations in load, communication delays and loss of communication links.

\let\cleardoublepage\clearpage
\chapter*{\centering Acknowledgements}
\addcontentsline{toc}{chapter}{Acknowledgements}
 \begin{figure*}[!h]
\centering
\includegraphics[width=0.9\linewidth]{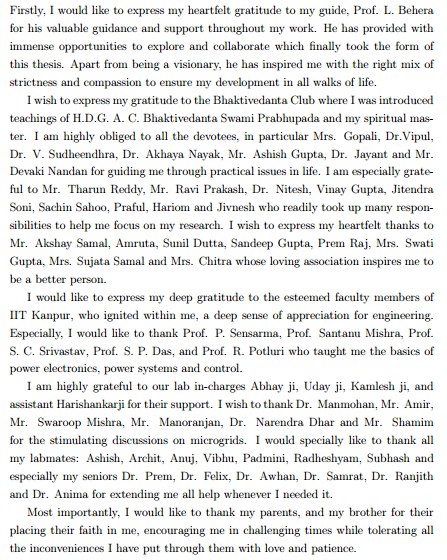}
\label{fig:ack}
\end{figure*}

\tableofcontents
\listoffigures \addcontentsline{toc}{chapter}{List of Figures}
\listoftables \addcontentsline{toc}{chapter}{List of Tables}

\mainmatter
\pagestyle{headings}
\pagenumbering{arabic}
\chapter{Introduction}\label{ch:introduction}
\pagenumbering{arabic}

Microgrids offer new approaches to repurpose, modify and enhance the existing power distribution systems and adapt to the growing demand of sustainable and renewable green power. A lot of power deficit exists in developing countries inspite of ongoing efforts to improve electrification. For instance, the Indian power sector declared that it has achieved a 100$\%$ electrification of its villages. This drive successfully connected all the Indian villages to the power grid. Still, more than 50$\%$ of them are victims of poor power management and do not have access to electricity for more than 12 hours a day \cite{powerindia19}. Distributed generators (DGs) can be installed locally which can harness the freely available renewable energies like solar and wind very easily to address the power deficit problem. The DGs can further be integrated on a modular basis into the existing grid infrastructure with the help of microgrid technologies for effective power management and reliable power supply to the end users.

From the grid point of view, the main advantage of a microgrid is that it is treated as a controlled entity within the power system which can operate as a single load or source. From customers’ point of view, microgrids are beneficial because they can meet their power requirement locally, supply uninterruptible power, improve power quality (PQ), reduce feeder loss, and provide voltage support. In their standalone form, microgrids can be used to provide power to remote and highly inaccessible areas such as mountainous regions \cite{mgmountain}, islands\cite{mgislands}, deserts\cite{mgdesert}, ships\cite{mgshipboard}, military\cite{mgmilitary}, etc. Furthermore microgrids reduce environmental pollution and global warming by utilizing low-carbon technology [4]. The choice of the distributed generators in microgrids mainly depends on the climate and topology of the region. Sustainability of a microgrid system depends on the energy scenario, strategy, and policy of that country and it varies from region to region.
 \begin{figure}[!h]
\centering
\includegraphics[width=1.0\linewidth]{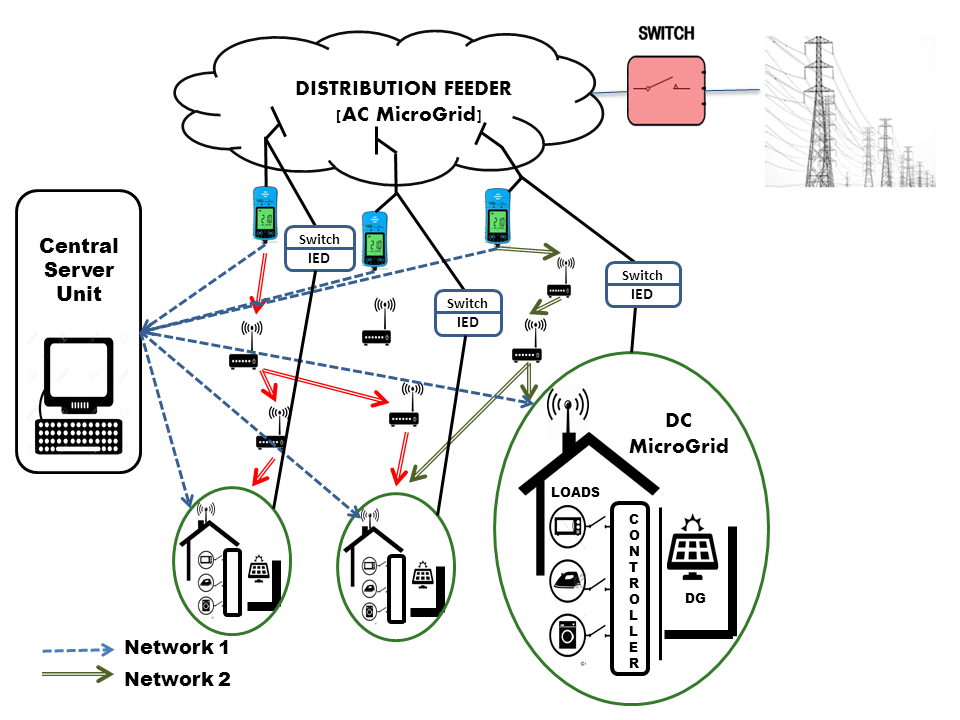}
\caption{Hybrid Microgrid Illustration}
\label{fig:hybridmg}
\end{figure}

Fig \ref{fig:hybridmg} illustrates the modular hybrid AC-DC microgrid structure that can be used for integrating the renewable energy sources to the existing distribution infrastructure. The local DC renewable energy sources like PV and energy storage devices like battery are connected together with DC loads to form a DC microgrid (DCMG). Each of this DCMG is further attached to the nearest point of common coupling (PCC) in a distribution feeder via an inverter (IED). It is followed by a switch which allows the DCMG either to stay connected to the distribution feeder or isolate itself to operate independently. The distribution feeder is interconnected to various DCMGs, AC sources and loads. It can thus be operated as an AC microgrid (ACMG) and is further connected to the transmission system through a switch which decides whether the ACMG functions in the standalone or the grid connected mode. The hybrid AC-DC microgrid can thus incorporate both AC and DC sources and loads with the help of modular ACMG and DCMG configurations.

The local DCMG operator may opt to operate it independently if the local generation effectively caters to its maximum and minimum load conditions. However, if the DCMG has insufficient generation to match its load or the DCMG operator wishes to utilize its additional generation capacity to make extra revenue, the DCMG may be connected to the ACMG and be operated in a co-ordinated fashion with other DCMGs to achieve power balance in the ACMG. Each DCMG can be envisioned to work effectively either as a load or a source when connected to the ACMG. To achieve coordination among various DCMGs, a communication network consisting of additional sensors, receivers and routing equipment is put in place. Moreover, inorder to maintain quality power delivery, the DCMG itself needs to coordinate between its existing renewable DC sources, energy storage systems and loads to counter the problem of low inertia \cite{overview3}. This  problem can be countered by having sufficient sensing and communication capabilities for facilitating distributed and adaptive control strategies within the microgrid. Thus, with the inundation of communication and sensors, both the DC and AC microgrids can be converted to smart microgrids. When these are operated in standalone mode, they are termed as Smart Standalone Microgrids (SSMG). Due to absence of external grid connection, the SSMGs become vulnerable to disturbances from both cyber and physical domains. The DCSSMG stability becomes highly dependent on various parameters of the cyber world like sensor failure, communication delay, communication failure and the structure of the communication network itself and factors from the physical world such as load, PV irradiance, PV temperature and wind, etc. 


This means that the SSMG needs to be conjointly modeled with both the cyber and physical domain parameters and the developed controllers need to be tested against vulnerabilities from both the cyber and physical domains. Cyber physical systems (CPS) \cite{lee2008cyber} research aims to provide integrated multi-disciplinary frameworks for understanding and manipulating large scale complex distributed systems. The advances in multiple technological domains such as sensing, communications, control systems and information technology and their massive deployment in everyday systems have provided sufficient motivation to envision and actualise the emergent properties resulting from their co-existence. CPS methodologies are dedicated to exploiting these properties to provide the existing systems with a fresh set of capabilities in areas \cite{cpsiet} such as safety, utility, resiliency, security, adaptability, scalability, and reliability. CPS have already begun to show many implementations in various fields like smart grid \cite{7017600}, robotics \cite{fink2012robust}, health care \cite{lee2012challenges}, data centers \cite{parolini2010cyber} and many more. Today's SSMGs are a result of convergence of many technologies including renewable distributed generation, power electronics, communications, advanced sensing, and embedded technologies which make them viable candidates for positioning CPS technologies  \cite{khaitan2015design}. A separate field called cyber physical energy systems (CPES) has advented for exploring the possibilities arising out of these overlaps. 

CPS frameworks are especially helpful in multi-domain modeling of the complex problems like coordinated control in SSMGs. This thesis studies the various challenges in carrying out coordinated voltage control in a typical hybrid AC-DC microgrid consisting of AC and DC SSMGs. It has contributed to developing models capturing various heterogenous parameters in the SSMGs and developing cyber physical control solutions for handling the effect of uncertainties in coordinated voltage control both from cyber and physical domains.


\section{Standalone Microgrids - A Brief Description}

\begin{figure}[!h]
\centering
\includegraphics[width=1.0\linewidth]{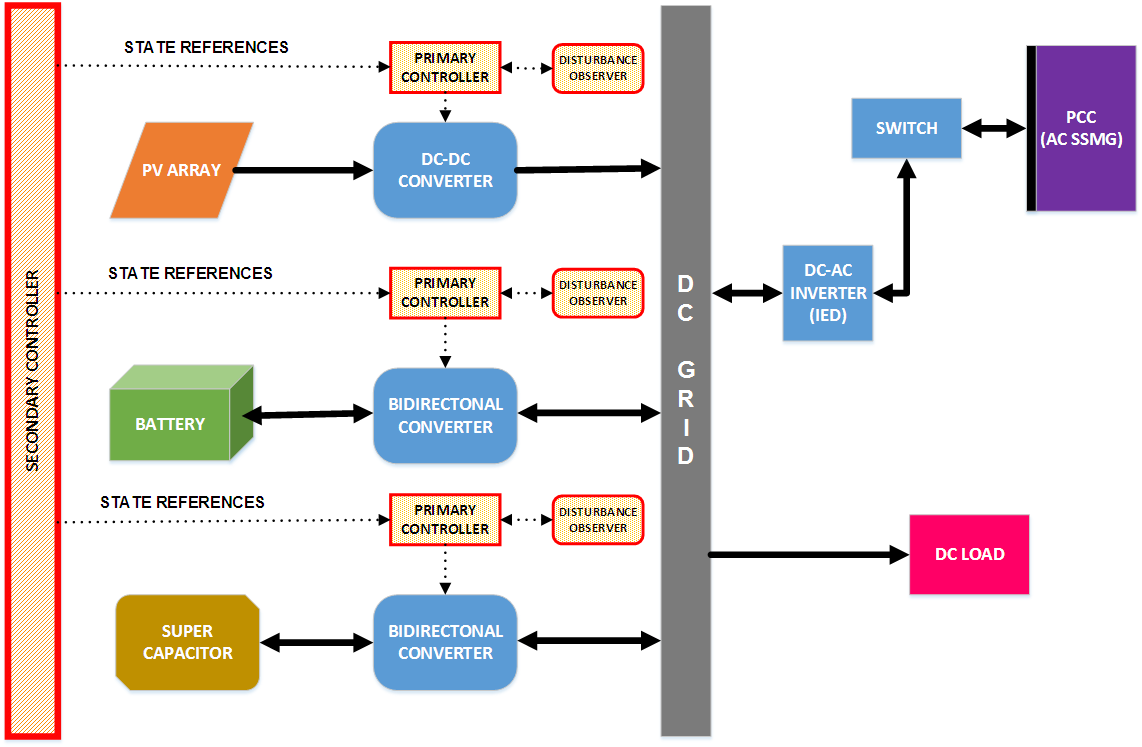}
\caption{DCSSMG Illustration}
\label{fig:dcssmg}
\end{figure}

The AC-DC hybrid microgrid shown in fig \ref{fig:hybridmg} can be clustered into two SSMGs- namely the DC SSMG and the AC SSMG as shown in fig \ref{fig:dcssmg} and fig \ref{fig:acssmg} respectively.
  
\subsection{DCSSMG}
A typical DCMG that can inhabit a general Indian household has been shown in fig.1.2. The solar photovoltaic array has been chosen as the main power source as solar power is available in abundance. A battery is provided to support the PV array in times of low ambient light. The supercapacitor is an auxiliary of the battery and functions to enhance the shelf life of the battery by smoothening the battery current when the DCMG is subjected to disturbances. The battery is also referred to as the battery energy storage system (BESS) and the battery-supercapacitor storage combination is popularly known as Hybrid Energy Storage System (HESS). The PV is connected with the DC bus via a DC/DC boost converter while storage devices like battery and supercapacitor are integrated with the DC grid through bidirectional converters (BDC). The power transfer from/to these sources are regulated by the duty cycles of the respective converters. The bidirectional converters enable power to flow from grid to battery and vice-versa. The DC bus is connected to the PCC of the ACSSMG through an inverter and a switch. When the switch is open, the DCMG functions as a DCSSMG.

One of the major control functions of the DCSSMG is to extract the maximum power from the PV array technically called Maximum Power Point Tracking (MPPT). This means that for a given PV array operating at a particular set of atmospheric conditions, the output voltage and current of the PV array need to be regulated to specific values so as to extract the maximum power from it. The secondary controller determines these references and transmit them to the primary controllers for regulation. The second control function of the DCSSMG is to maintain the power balance in the DCSSMG. This is accomplished by the battery through the BDC. The BDC charges the battery when PV array delivers excess power and the battery delivers power to load when the PV array has power deficit when compared to the load demand. This power balance also leads to DC grid voltage control in the steady state.  
However, the battery current needs to be steady with very few transients over time so as to increase battery life. To achieve this, the supercapacitor is installed which directly gets involved in the control of the DC grid responding instantaneously to perturbations in grid voltage and reducing transient load on the battery while the battery slowly adjusts to the new steady state power exchange conditions.  

The DCMG also flaunts an interlinking converter (IED) which works bidirectionally as an inverter when the DCMG needs to transfer power to the ACMG and as a rectifier when the ACMG needs to transfer power to the DCMG. The IED is followed by a switch which controlled by the DCMG operator. When the switch is open, the DCMG operates in isolated mode making it DCSSMG.

\subsection{ACSSMG}
 \begin{figure}[!h]
\centering
\includegraphics[width=0.95\linewidth]{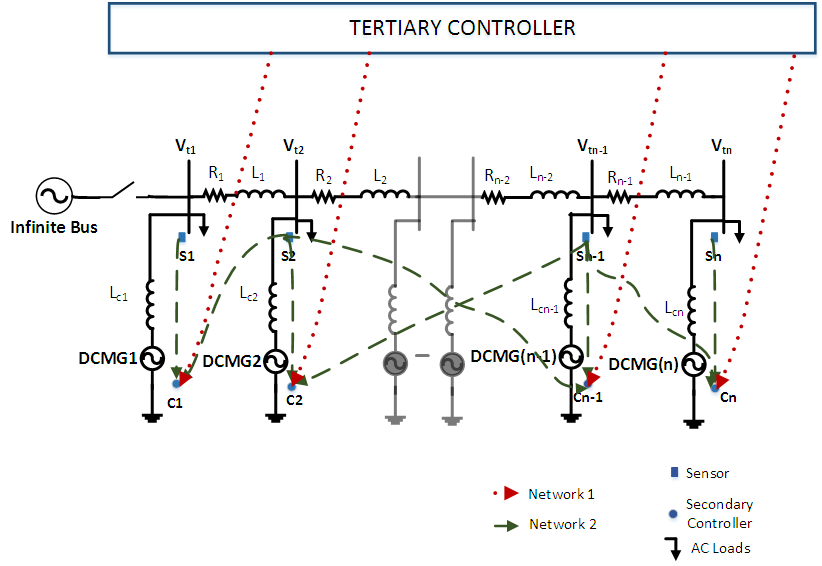}
\caption{ACSSMG Illustration}
\label{fig:acssmg}
\end{figure}

The ACMG provides the framework for coordinating between the various DCMGs and other sources to provide quality power to the AC loads. The ACMG does not require additional infrastructure. Rather, the existing distribution feeders can be used to integrate the advanced renewable power generation technologies like solar PV, electric vehicles, etc which are recently gaining a lot of popularity due to their low carbon footprint.

A depiction of the ACMG is given in fig \ref{fig:acssmg}. The transmission line is connected to the distribution feeder with a switch which allows the ACMG to work either in the standalone or the grid connected modes. The DCMGs are attached to the ACMG at the PCC and exchange AC power through IEDs. The ACMG operator can choose to operate independently as long as it has sufficient power to feed the AC loads and the connected DCMGs or during power failure from the transmission line. The power balance is achieved through effective coordination between connected DCMGs which may be separated by distances measuring upto kilometers. Hence, communication networks are established with the help of sensors, receivers and intermediary transmission nodes to facilitate this coordination. 

The major control function at the level of ACSSMG is to maintain the frequency and voltage at various PCCs.
 The voltage/frequency at each PCC in the ACMG is sensed by a sensor and sent to the DCMG wherein the IED references are controlled to maintain voltage/frequency stability at PCCs. An established procedure for this problem would be to let each inverter control the voltage of the respective PCC with the help of data obtained from sensors placed at that particular PCC. However, following a paradigm of distributed control between DCMGs, a set of distributed controllers are deployed with the help of communication networks which can provide enhanced control using the information acquired from sensors present at multiple PCCs. Since, the distances between PCCs can be large, communication parameters like delay, packet loss, communication failure and placement of communication nodes play an important role in the overall voltage/frequency stability of the ACSSMG. 


 \begin{figure}[!h]
\centering
\includegraphics[width=0.95\linewidth]{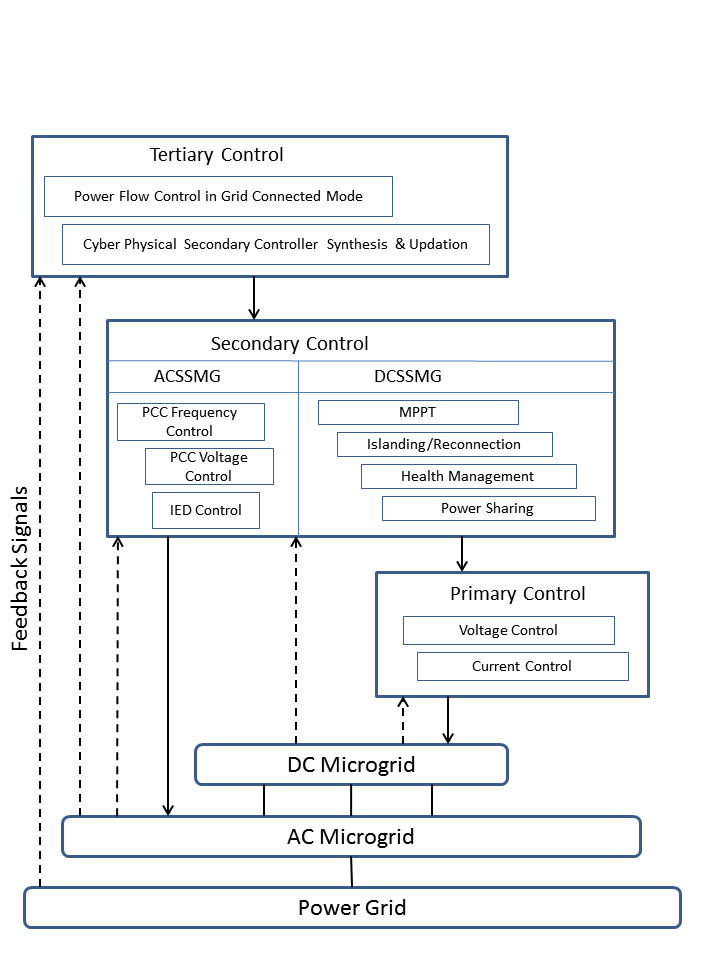}
\caption{Hierarchical Control Structure}
\label{fig:hcmg}
\end{figure}

\section{Cyber Physical Hierarchical Control}
\textcolor{black}{The hierarchical control paradigm in microgrids has become prominent in the microgrids around 2010 after being inspired from the hierarchical control used in power dispatching of the ac power systems. Over time this paradigm has seen a number of successful and practical implementations \cite{aalborg,hcmghil2}, eg: Aalborg University microgrid.} Fig \ref{fig:hcmg} sums up the entire control in the AC-DC microgrid. It contains three levels of control- primary, secondary and tertiary.  

\subsection{Primary Control}
{Primary control} is responsible for achieving fast and stable regulation of local voltage and current values related to individual power components in the DCSSMG as the PV, battery and supercapacitor. The set-points for the voltage and current values of different converters are given by the higher level controllers keeping in mind various control functions like maximum power point tracking (MPPT) and power balance. The control actions are carried out by deciding the value of duty cycles for the power converters attached to the respective devices. The individual primary controllers in the DCSSMG may need to exchange the values of voltages and currents between themselves for better coordination but since they are located quite close to each other, these values can be exchanged via the secondary controller and the effect of communication parameters is minimal on the stability of DCSSMG. 

Most literature related to DCSSMGs in the last decade employed PI controllers\cite{kollimalla2} for carrying out the primary level of control. These are designed using small signal system models and tuned to work effectively only around a particular operating point \cite{kabalan2016large}.

However, in the course of operation of the DCSSMG, the operating point may vary due to environmental and loading conditions which can compromise the stability of these controllers. Added to this, the lower inertia of the SPVDG due to renewable power resources make the DCSSMG vulnerable to sudden disturbances. Thus more advanced controllers are necessary which can be developed using large-signal modeling of the overall system.

In the existing literature, nonlinear controllers have been proposed for improving the dynamic performance in power electronics based systems. For instance, \cite{wang2019continuous} proposed a continuous non-singular sliding mode technique for DC-DC boost converter in the presence of time-varying disturbances. A modular back-stepping approach has been explored in \cite{iovine2017nonlinear} for designing controllers with complete stability analysis for an SPVDG system with batteries and supercapacitors to handle transients at multiple time scales. Partial feedback linearization is applied along with robust H-infinity mixed-sensitivity loop shaping in \cite{mahmud2020robust} for an SPVDG system to improve performance during sudden disturbances.
A model predictive control (MPC) technique is proposed in \cite{bambang2014energy} for computing power references and sending them to both battery and supercapacitor controllers. A modified backstepping based controller strategy has been used to satisfy diverse control objectives of an islanded microgrid with HESS in\cite{iovine2017nonlinear}. Similarly, a Lure-Lyapunov framework has been designed in \cite{mane2017improving} for handling various loading transients in an electric vehicle.
Even PI controllers at the primary level have been coupled with advanced higher-level control techniques like model predictive control to add control flexibility and reduce cost in an isolated wind/solar/battery system \cite{kong2019hierarchical}.

However, most of these advanced controllers heavily rely on precise system models that demand continuous inflow of many internal system states and parameters. Successful implementation of these techniques require installing a huge number of sensors. Sensing is an intricate part of the cyber system and if the number of sensors are increased, it is difficult to maintain them and the stability of the SSMG becomes highly affected by sensor failures. It also becomes difficult to locate the faulty sensor if the number of sensors is huge in the system. It also increases the cost of the system. Keeping these in mind, it is prudent to minimize the  overall sensor count in the SSMG especially at the level of primary control.

\subsection{Secondary Control}
Secondary control of the AC-DC Microgrid is the major interface between the AC and the DC microgrids. Secondary control computes the set points for the primary controllers. The primary control is locally implemented at each distributed generator(DG) in the DCSSMG while the secondary controller exploits a centralized control structure. Central controllers issue global commands based on information gathered from the entire system and require a  two-way communication network. The references of the secondary controllers are set by the tertiary controllers. Since the secondary control pertains to both the AC and DC microgrids, it needs to fulfill the control requirements specific to both the AC and DC microgrids.  

When the DCMG is isolated, the secondary controller references are generated so as to achieve various device level goals in the DCSSMG like MPPT, power balance, DC voltage control and battery life improvement. No references are generated for the IED connected to the DCMG since it is non-functional in standalone mode. The popular MPPT algorithms are mainly based on different techniques like perturb and observe (P\&O) \cite{a9}\cite{a10}, incremental conductance (INC) \cite{a11}\cite{a12}, hill climbing, fractional open-circuit voltage \cite{a13}, fractional short-circuit current \cite{a14}, ripple correlation control \cite{a15}, fuzzy logic control \cite{a16}, \cite{a17}, particle swarm optimization \cite{a18}, \cite{a19}, artificial neural network \cite{a20}, genetic algorithm \cite{a21}. These algorithms differ from one another based on their ease of implementation, tracking speed, number of sensors used, tracking efficiency, cost, etc. The primary references related to power balance is generally calculated through power balance equations and other functions like battery health maintenance etc are decided on various techniques presented in \cite{kollimalla2}\cite{gridhessref}\cite{microgridhessref}. The voltage of the DCSSMG is decided by the operator and fed to the secondary controllers.

However, when the DCMG is connected to the ACMG, then the voltage and current references of the interconnecting inverter IED are computed so as to to maintain voltage and frequency at PCCs. In such case, the various references of the other DCMG components are modified accordingly by the secondary controllers. Communication has been used for enhancing the performance of power systems in terms of voltage and frequency control \cite{olivares2014trends}. The centralized type of control  \cite{rezaei2015robust,mehrizi2012constrained} which has been the conventional norm in microgrid control is slowly fading out in practise. \textcolor{black}{
Decentralized control is a very well established area of research and can surely make the system less complicated in terms of control and management. However, decentralized control can be easily destabilized due to communication failures or cyber attacks due to lack of redundancy in the information. If the existing meagre communication resources gets targeted by the hackers,  it would compromise the system stability. Since the grid is interconnected, it is good to explore a more distributed \cite{zhou2016consensus,sun2015multiagent,bidram2013secondary} form of control where we have receive data from other channels whcih can improve reliable and resilient operation. This additional data can help us in reconstructing the compromised data or support the control strategy even in the absence of a local communication channel. To ensure the stability of grid in difficult situations, we have chosen a more communication intensive microgrid in this thesis.Moreover, communication technologies are advancing at such a fast pace that the communication resources themselves would become cheaper and having extensive communication in the microgrid would be a regular situation.}

The authors in \cite{xin2011cooperative} use a cooperative control strategy for multiple solar plants using minimal communications. A distributed secondary control technique is developed in \cite{shafiee2013distributed} and implemented using a densely connected communication framework to augment the working of traditional droop. Improving on these works, a hybrid strategy has been adopted in \cite{liang2013stability} which uses different strategies in the presence of centralized communication network and otherwise. Certain works such as \cite{majumder2012power} focus on innovating communication protocols and routing techniques to improve power sharing in different configurations of the microgrid. Most of the works found in the literature are either specific to power domain or communication domain or computation domain even though they portray application to the smart grid. However, they do not model the inter-dependencies between parameters of various domains.


This calls for hybrid CPS modeling frameworks which contain inputs and outputs belonging to multiple fields including both cyber and physical domains. Moreover, these frameworks need to be generic to provide a direct pathway for many problems to be modeled into its structure. The CPS frameworks developed need to support all of these problems using a unified framework. The framework should also be customisable to the needs of a particular problem. For example, there might be some localities suffering from power deficit and are highly sensitive to changes in load. There might be certain other places such as hilly areas, where communication is a problem and the ACSSMG becomes sensitive to delay. The erratic climatic conditions in some places may lead to damage in communication equipment. Similarly, the ACSSMG may also experience different levels of sensitivities to different parameters at different times of the day. The load is generally high during the day and less in the morning. If the communication network of the grid is shared jointly for Internet browsing and downloading, the ACSSMG may become sensitive to delay due to excessive data usage during evening times. Hence, to study these effects and provide relevant solutions, hybrid frameworks are necessary.
 
\subsection{Tertiary Control}
Tertiary control is the highest control level in the hybrid AC-DC microgrid. It is highly concerned with the optimal operation of the microgrid \cite{tertiary1}. It also manages the power flow between AC microgrid and the main grid. In the grid-connected mode, the power flow between ACSSMG and the transmission line can be managed by adjusting the DG voltage and frequency. 

For instance, the work in \cite{tertiary3} proposes to provide additional support to a microgrid with lower generation or couple the ones with higher generation to other suitable microgrids to incur less cost. 
Instead of using extra compensation equipment, which may
bring more cost, the authors in \cite{tertiary4} proposes a tertiary control to employ DGs as distributed compensators, and achieve optimal unbalance compensation. \cite{tertiary5} studies a distributed two-level tertiary control system to adjust the voltage set points of individual microgrids and
balance the loading among all the sources throughout the cluster. Traditionally both the secondary and tertiary controllers have been designed at two different levels with two different purposes and time-scales of operation. However, of late, there have been many works \cite{tertiary21}\cite{tertiary22}\cite{tertiary20} which have unified these two levels of operation in the microgrids through a distributed architecture.

Also, the tertiary controllers may be used for resource planning of the entire microgrid system \cite{tertiary20}. The resource planning is generally dedicated only to the cyber parameters or to the physical parameters. Moreover, have most resource planning has been carried out offline till now. There is a great scope to carry out resource planning online based on changing structure of the microgrid which in turn can influence the effective control of microgrid.

\section{Thesis Objectives}
Based on the discussions so far in the previous sections of this chapter, the control of AC and DC SSMGs face several challenges such as intensive sensor and communication requirement, high vulnerability to disturbances from cyber and physical domains and optimization of resources etc. In light of these challenges, the following objectives have been defined:

\begin{enumerate}
\item	To develop cyber-physical frameworks that can jointly model both the effect of both physical and communication parameters in SSMGs.
\item	To design controllers which can counter the effect of reduced inertia in DCSSMGs with renewable rich power sources in the presence of atmospheric and load changes. 
\item	To design distributed controllers which can work with least number of sensors in the DCSSMG so as to make it robust to sensor failures and reduce overall cost.
\item	To design controllers which can achieve higher voltage stability at points of common coupling  of the ACSSMG in the presence of disturbances from changes in atmospheric conditions, load and other communication constraints like bandwidth, etc.

\end{enumerate}

\section{Thesis Outline}
The major objective of this thesis is to harness the growing power of cheaply available computation to provide solutions for effective control of the SSMGs which is highly dependent on both physical conditions and communication network.

The different works carried out in this thesis are outlined chapter-wise as shown below:


\textbf{Chapter \ref{chapter:fastestmppt}} presents a detailed modeling of a DC SSMG with Hybrid Energy Storage System (HESS) consisting of PV, battery and supercapacitor. Nonlinear control techniques like backstepping are generally employed for effective MPPT  and DC voltage control in the presence of disturbances in load, irradiance and temperature. These techniques although effective, use a lot of sensors making the system expensive to implement and prone to sensor failure. In this work, a disturbance observer based back-stepping controller is proposed to obviate the necessity for measuring disturbance values. The effects of irradiation and temperature on PV arrays, the variations in loads and battery voltage are modeled in the form of disturbances. Instead of measuring these with sensors, the proposed observer update laws based on Lyapunov stability theory estimate their values. They are further utilized for effective control during intermittencies. It can be seen from the MATLAB simulation results that adoption of this technique contributes towards faster, cheaper and more reliable control of the DCSSMGs for an increased set of operating conditions.

In \textbf{Chapter \ref{chapter:DirectPerturbMPPT}}, an adaptive neural controller is proposed for the MPPT and grid voltage control of an unknown battery based DC SSMG with unknown disturbances. This controller removes the necessity to keep track of system model parameters like resistances, inductance and capacitances apart from eliminating the need for expensive sensors for sensing load and environmental conditions. The neural network weight update laws of the controller are derived using the Lyapunov stability. It is shown that the proposed controller is able to ensure the uniformly ultimately boundedness (UUB) of all signals of the resulting closed-loop system. The performance of the proposed controller is evaluated in simulations against state-of-the-art controllers during disturbances and parameter intermittencies in the presence of sensor failures.

In \textbf{Chapter \ref{chapter:nonlinear}}, a generic, hybrid and customized cyber-physical framework is developed to jointly model the multi-disciplinary variables and their interactions present in a densely connected ACSSMG. This cyber-physical model is used to design adaptive controllers to ensure better control of  microgrid voltages irrespective of the changes in operating point brought about by changes in physical/cyber parameters.  The different operating conditions of the power system have been modeled as multiple subsystems of a hybrid switching system and controller design is carried out by solving the optimisation formulations developed for delay-free and delay-existent operation of the ACSSMG using the theory of common Lyapunov function (CLF). The optimisation is carried out using the block coordinate descent (BCD) methodology by converting the non-convex formulation into a series of convex problems to obtain a solution.

\textbf{Chapter \ref{chapter:distobs}} deals with the design of communication network between various sensors and controllers in a sparsely connected ACSSMG and its effect on improving voltage stability is explored. The design process involves automated examination of voltage stability in the presence of a number of topological combinations of the communication network, which increase exponentially with the number of nodes. The different characteristics and availability of various physical and communication resources in the network pose multiple constraints on this design. For this purpose, a generalised constraint-based sensor controller connection design (CBSCD) methodology was designed, which effectively reduces the number of combinations, to design more stable cyber-physical controllers. To handle variations in multiple parameters in physical and communication domain, different controllers have been developed for different operating conditions that are deployed as per requirement. The methodology has been shown to effectively stabilise bus voltages in a smart grid scenario under variations in load, communication delays and loss of communication links.

\textbf{Chapter \ref{sec:conclusion}} extracts the major conclusions of the thesis along with some observations and proposes future directions to extend and strengthen the novel contributions of the thesis.

\section{Research Contributions}
The thesis provides advanced control solutions for the DC and AC SSMG configurations to handle issues emerging from the confluence of sensing, control, communication and power system. The notable contributions have been listed chapter-wise as follows.

 \begin{figure}[!h]
\centering
\includegraphics[width=1.0\linewidth]{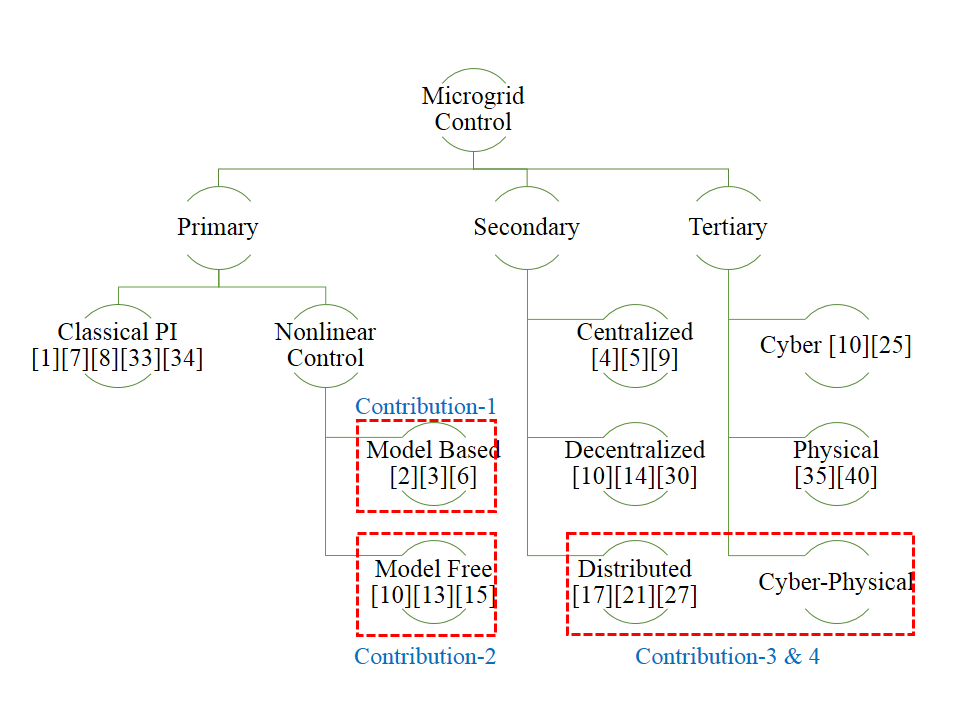}
\caption{Thesis Contributions}
\label{fig:contributions}
\end{figure}
\begin{itemize}

\item \textbf{Sensor-Free Control of DC SSMG with Hybrid Energy Storage using Disturbance Observers}
\begin{itemize}
\item An adaptive back-stepping based controller design for the DCSSMG system consisting of solar array, battery and supercapacitor.

\item An update law for computing the disturbance values in an online manner and reduce the need for additional sensors in the DCSSMG.

\item A comprehensive stability analysis of the proposed adaptive backstepping control using Lyapunov theory.
\end{itemize}

\item \textbf{Adaptive Neural Controller for Model-free Control of DC SSMG with Unknown Disturbances}
\begin{itemize}
 \item A  novel adaptive neural controller for model-free control of a DC sub-microgrid with solar PV and battery in presence of unknown disturbances.
 \item A comprehensive stability proof for the DC sub-microgrid including various subsystems and their interconnections using Lyapunov stability for achieving the uniformly ultimately boundedness (UUB) of all states when applied for real-time system estimation and control.
 \item A comparative analysis of the proposed adaptive neural controller with state of the art adaptive and nonlinear controllers when real-time information regarding system model parameters and disturbances is unknown.
\end{itemize}

\item \textbf{Hybrid Adaptive Cyber-Physical Framework for Densely Connected SSMG}
\begin{itemize}
\item A generic, hybrid and customizable framework to capture the dynamics of communication and control using the theory of hybrid switching systems.
\item Optimization formulations using Common Lyapunov Function (CLF) to design controllers acknowledging variations in both physical and communication parameters in delay-free and delay existent systems.   
\item A BCD based technique for finding a solution to the above formulations which are non-convex in nature. 
\end{itemize}
\item \textbf{Optimal Communication Design based Cyber Physical Control Framework for Sparsely Connected SSMG}
A generalized constraint based sensor controller connection design (CBSCD) methodology to design controllers for CPES considering:
\begin{itemize}
\item Connection constraints that can be specified to define the boundaries of the communication network.
\item Resource constraints like communication bandwidth that can be specified to describe the connecting capabilities of various sensors and controllers.
\item Utility based constraints like cost, can be specified, so as to accommodate required operational demands of power utilities.
\item Physical variable constraints like load requirement on the buses can be specified. 
\end{itemize}
\end{itemize}

       \chapter{Adaptive Observer based Sensor-Free Control }\label{chapter:fastestmppt}
\chaptermark{Sensor Free Controller}
\index{Estimation}\index{MPPT!Maximum Power Point Tracking} \index{Sensorless}
 
In this chapter, an observer based back-stepping controller has been designed using Lyapunov theory. Based on the structure of the system model, the overall model has been divided into different sub-systems. Furthermore, controllers have been designed for each subsystem to ensure different control functionalities in a DC microgrid equipped with a Hybrid Energy Storage System.  

\section{Introduction} \label{sec:introMPPT} \index{Hyrbid Energy Storage Systems}
Energy storage systems play an important role in today's evolving power systems as they provide flexibility in deploying renewable energy sources which are highly intermittent in nature. While battery based energy storage enjoys great popularity, they cannot be used for applications that demand high power surges due to low power density. There are other energy storage options with high power density like ultracapacitors and superconducting magnets but they discharge very quickly and cannot provide power for prolonged time intervals. Hybridizing these two different types of energy storage devices like batteries with high energy density and supercapacitors with high power density is an attractive storage alternative that can be used in many practical situations. 
\begin{figure}[!t]
\centering
\includegraphics[scale=0.45]{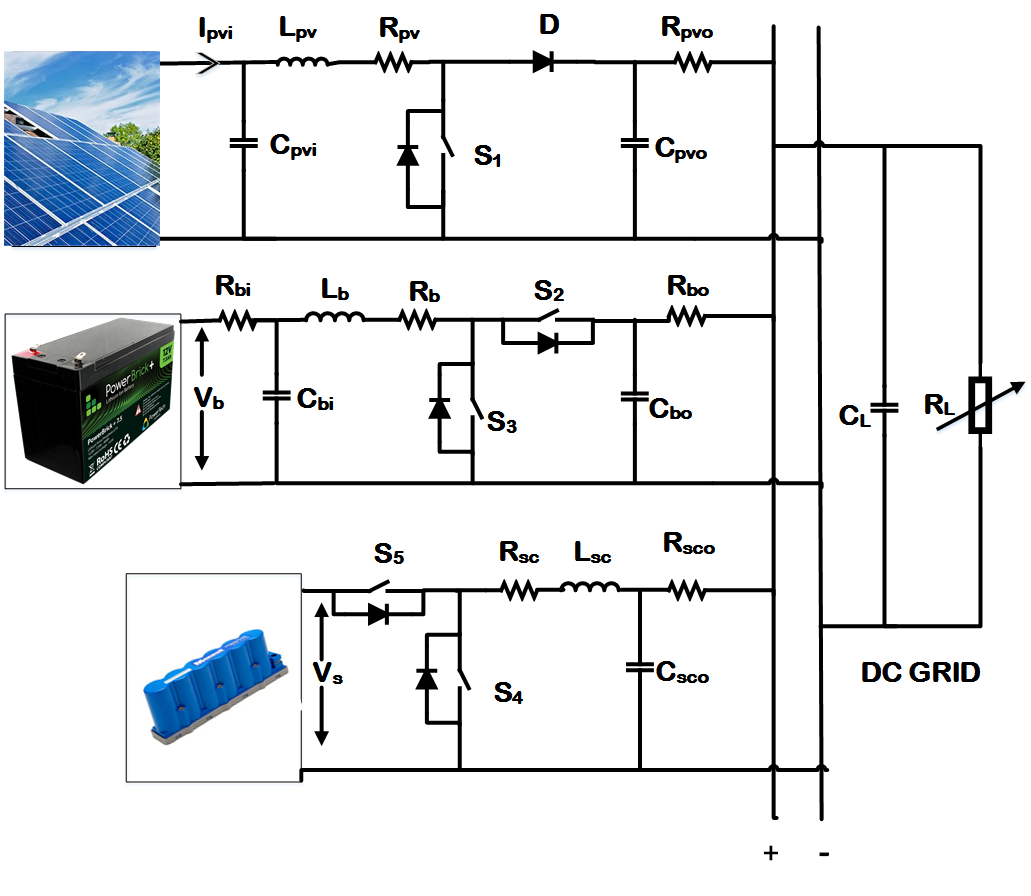}
\caption{Power Architecture of the DC Microgrid with HESS}
\label{fig:spvdg1}
\end{figure}
 Super-capacitor based HESS were initially conceived as a good power support for the electrical vehicle have found their way into trains\cite{trainsref}, ships\cite{shipsref} and aircrafts\cite{aircraftsref}.They have also diversified into a myriad of applications in the energy sector, from residential homes \cite{residenceref} to distribution systems \cite{gridhessref}, and other upcoming technological sectors like data centers\cite{datacenterref}. 
Especially, in the standalone microgrids which are based heavily on renewables like PV and wind, HESS have great potential to handle the low inertia and intermittent environment dependent power generation/consumption.\textcolor{black}{The DCSSMGs especially which are the main source of power for this grid are driven by solar PV cells or wind generators. Thus the power generation in the grid is highly dependent on irradiance and temperature. For instance, the daily variation of irradiance in Delhi around the month of January ranges between 5 to 800 W/m2. Moreover, during winters, the temperature in Delhi changes between 7 degrees to 21 degrees celsius. These parameters greatly affect PV generation which can vary from almost 5$\%$ to 80$\%$ of its generation capacity during the day drops to zero during the night. Moreover, the greatest factor that adds to the randomness of the solar energy is the presence of traveling clouds which brings sudden changes in the power generated by PV cells. The prediction of wind energy is much more difficult when compared to solar irradiance due to ever changing climatic conditions. For instance, the total wind energy produced by India in 2020 suddenly dropped 45$\%$ compared to previous year in spite of having added additional generation capacity. The large and random variations in power generation by the renewables contribute to reduction in stability and resiliency of a microgrid and the power mismatches created by their unpredictable nature reduces the reliability of the microgrid.} These microgrids can easily be destabilized in presence of sudden drop in solar generation due to rain or load curtailment due to fault. With proper sizing as suggested in \cite{systemsizingref1}\cite{systemsizingref2}, they can regulate critical microgrid voltages/currents without necessarily making the grid bulky or costly. 
For instance, in \cite{deadbeathesstse19}, a deadbeat control methodology is proposed after examining the optimal combination of batteries and supercapacitors for the microgrid. It is preferred that the HESS can be controlled under a decentralized mechanism where no communication will be incurred. In \cite{decentralizedhesstpec15}, a battery and a supercapacitor are separately assigned with a low-pass filter and a high-pass filter for different types of transient power allocation. Similarly, a distributed framework is demonstrated in \cite{distributedhesstse18} to minimize DC bus voltage deviation and ensure accurate power sharing among different energy sources.
Most of the works on microgrids are aimed at high level control \cite{self2}, \cite{self3}, \cite{selfnpsc16} development and deal only with the overall power management of the grid. They assume that storage interfaced converters are highly stable and can seamlessly carry out desired voltage/current regulation as per their generated references.They generally use PI controllers like the ones developed in \cite{kollimalla2} and \cite{battula2020analysis} for carrying out the primary level of control which are designed using small signal system models and  tuned to work effectively only around a particular operating point \cite{lyapunovmgtsg17}. 

Advanced nonlinear controllers like the one proposed in \cite{korukonda2021modular} have shown great promise in ensuring large signal stability in microgrids for an extended region of operation.
However, most of these advanced controllers heavily rely on precise system models that demand continuous inflow of many internal system states and parameters. Successful implementation of these techniques require installing a huge number of sensors. Sensing is an intricate part of the cyber system and if the number of sensors are increased, it is difficult to maintain them and the stability of the SSMG becomes highly affected by sensor failures. It also becomes difficult to locate the faulty sensor if the number of sensors is huge in the system. In the recent past, the IC technology has progressed at an enormous pace. This advancement has made computing power cheaper and more accessible. Hence, it may be reasoned that expensive sensors be replaced by estimating the sensor values \cite{selfdobsiecon20} to reduce the SSMG's dependence on sensors. However, to the best of our knowledge, there are not many works in the literature to apply any computation based estimation algorithms to a HESS based DC microgrid (DCMG). 

Hence, in this work, a DCSSMG consisting of PV, battery and a supercapacitor as shown in Fig.\ref{fig:spvdg1} is considered and an adaptive  observer based backstepping technique is developed to estimate PV output current, battery voltage, supercapacitor voltage and load impedance in real-time. The control technique is designed so as to provide high speed performance in the presence of disturbances to compensate for the problem of low inertia. The update law is designed for computing the disturbance values in an online manner so as to reduce the need for additional sensors in the DCSSMG. Furthermore, a comprehensive stability analysis of the proposed adaptive backstepping control is delineated using Lyapunov theory.

Section \ref{sec:aobsmodel} presents the overall system model of the DCSSMG system used and defines the exact control problem that is addressed in this chapter. Section \ref{sec:aobsdesign} describes the proposed control technique along with an extensive stability proof. Section \ref{sec:aobsresults} delineates the various comparative simulation studies performed on the system at hand in the presence of known and unknown disturbances. Finally, Section \ref{sec:aobssummary} summarizes the work done in this chapter.

\section{Mathematical Modeling}\label{sec:aobsmodel}
\index{PV Model}
\begin{figure}[!t]
\centering
\includegraphics[scale=0.45]{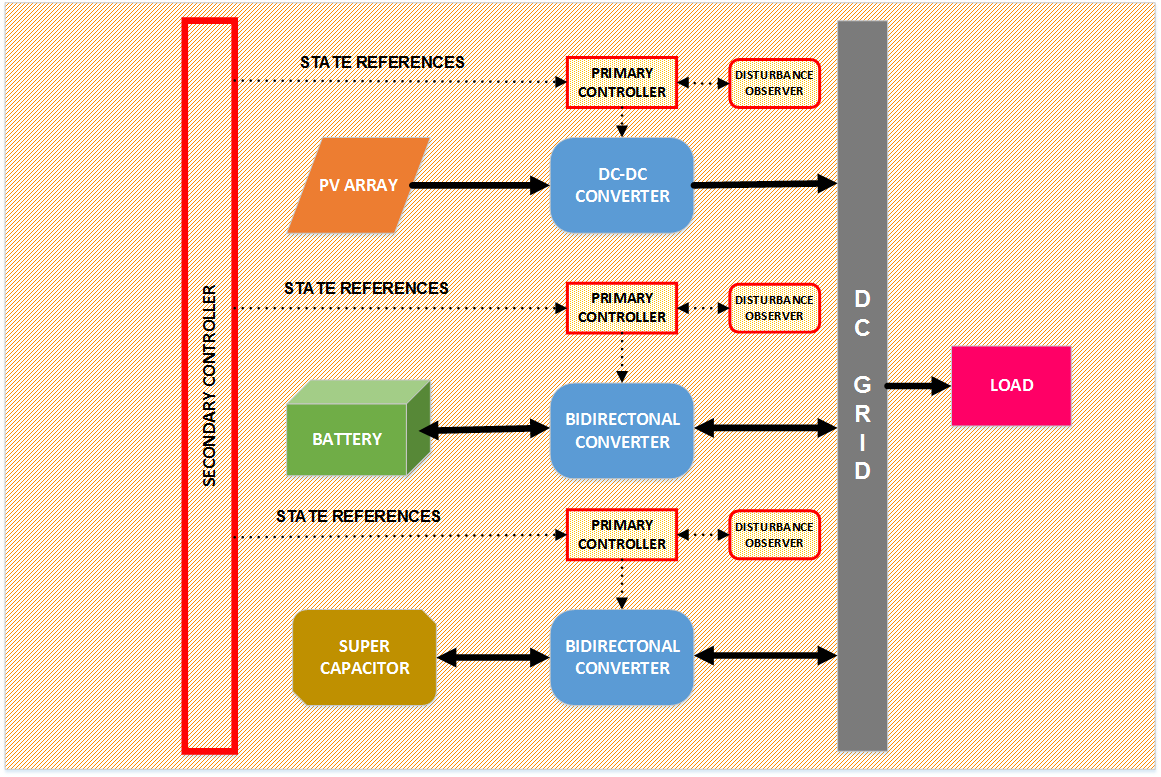}
\caption{Control Structure of the DC Microgrid with HESS}
\label{fig:dcmgoverall}
\end{figure}

A solar panel based DCSSMG is chosen which is coupled with a battery and supercapacitor for operating in the standalone mode. The complete circuit diagram of the DCSSMG is shown in Fig. \ref{fig:spvdg1}. The PV is connected with the DC bus via a DC/DC boost converter which feeds the load with the help of a Hybrid Energy Storage System consisting of battery and supercapacitor. The battery and the supercapacitor are integrated with the DC grid through bidirectional converters.The power transfer from/to these sources are regulated by the duty cycles of the respective converters. The bidirectional converters are used to control the flow of power both out of and into the storage devices so as to add power support to the PV in times of shortage and to charge the storage devices respectively .

The PV array generates an output current whose equation is described by \eqref{eq:ipv} as discussed in \cite{self1},  
\begin{align}\label{eq:ipv}
\scalebox{.9}{$
i_{pvi}=n_{p}I_{g}-n_{p}I_{s}\left(e^{\dfrac{q(v_{pv}+i_{pv}R_{s})}{n_{s}pKT}}-1\right)-\dfrac{v_{pv}+i_{pv}R_{s}}{R_{sh}}$}
\end{align}
The PV output current $i_{pv}$ and the maximum power point of the PV array are intricately dependent on temperature and irradiance. It can be observed in the  state-space model, the output current of the PV array is denoted by $d_1$. The voltage across the output capacitor $C_{pvi}$ of the PV array is described as $x_1$ while the input capacitor $C_{pvo}$ voltage is modeled as $x_2$. The PV inductor $L_{pv}$ current is further termed $x_3$. Change in temperature and irradiance affect $i_{pvi}$ and are thus modeled into disturbance $d_1$. 

The battery is assumed to be a continuous DC voltage source with voltage $V_b$ and is modeled into the state-space as disturbance $d_2$. The voltage across battery output capacitor $C_{bi}$ is chosen to be $x_4$. While the inductor $L_b$ sports a current designated as $x_6$, the battery output capacitor $C_{bo}$ voltage is termed as $x_5$.  

The supercapacitor voltage which is $d_3$ is modeled using the following equation as given in \cite{scref}
\begin{align}
\frac{d u_c}{dt}=-\frac{1}{C_{sc} R_p} + \frac{1}{C_{sc}}i_{sc} \nonumber\\
V_{sc}= u_c + i_{sc}R_s  
\end{align} 
where capacitance $C_{sc}$ and resistance $R_p$ are connected in parallel and $
R_s$ is connected in series with the previous combination. While the current through the supercapacitor $i_{sc}$ is designated as $x_7$, the  voltage across $C_{sco}$ is assumed to be $x_8$. The grid voltage  of the DCSSMG which is measured at capacitor $C_L$ is denoted as $x_9$ and the resistive load admittance $\frac{1}{R_L}$ is termed as $d_4$. The following subsections show a thorough modeling of the DCSSMG system with its various subsystems using the famed state space averaging technique,
\subsection{PV Modeling}
\textcolor{black}{The circuit of the PV subsystem is depicted in Fig.\ref{fig:pv}. Upon observation, it can be seen that the PV converter works in two different modes depending on the the status of switch $S_1$.
\vspace{-0.5cm}
\begin{figure}[H]
\centering
\includegraphics[width=3.2in]{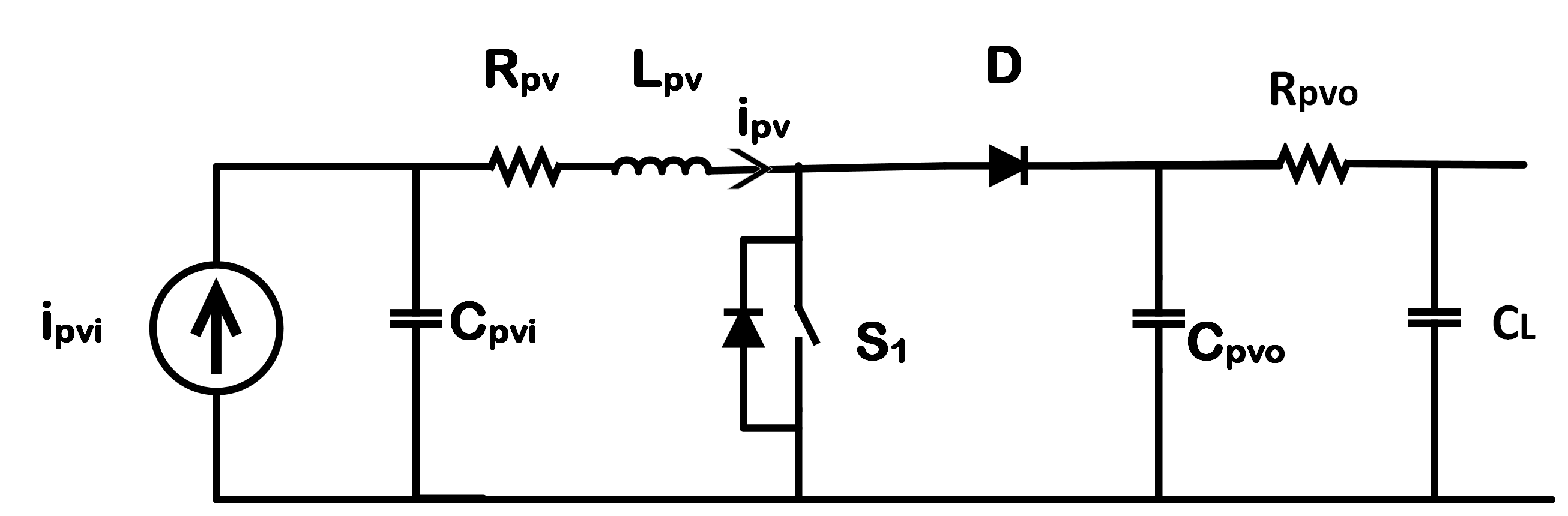}
\caption{PV Subsystem}
\label{fig:pv}
\end{figure}
\vspace{-0.8cm}
\subsection{Mode-1}
If switch $S_1$ is closed, then the circuit in Fig.\ref{fig:pv} can be divided into the following circuits in Fig.\ref{fig:pv1}.  Upon applying Kirchoff's laws on these circuits we get,
\begin{figure}[H]
\centering
\includegraphics[width=3.2in]{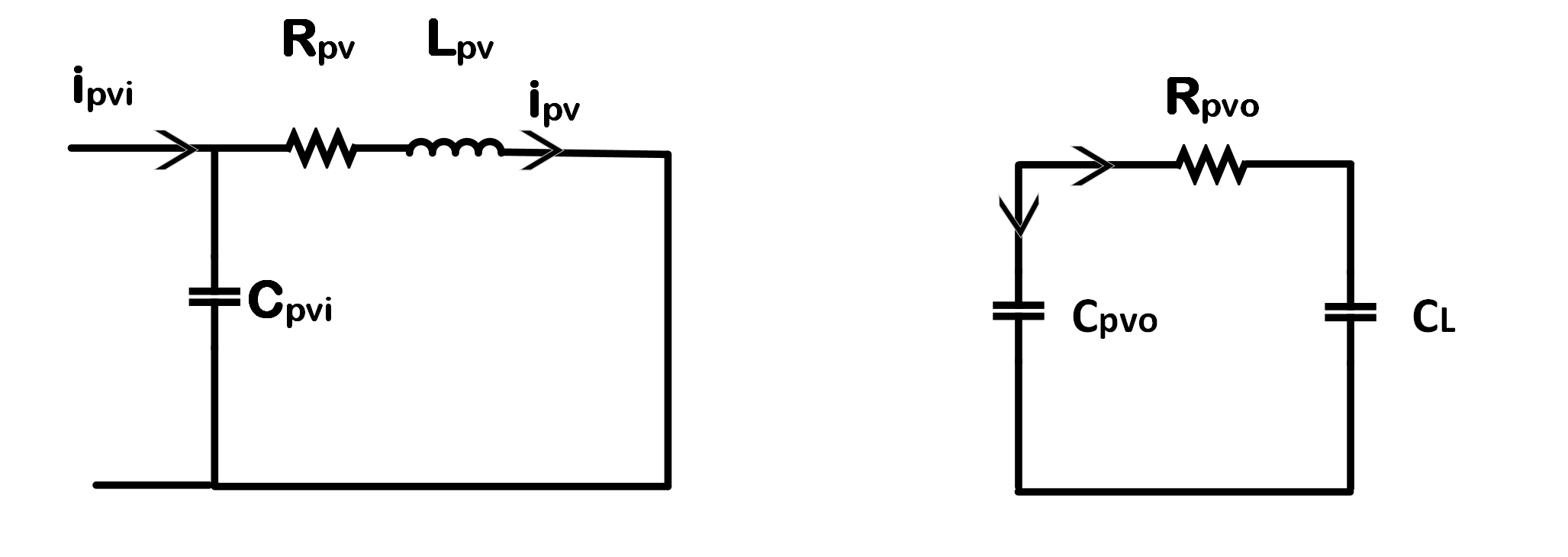}
\caption{PV Subsystem- Mode 1}
\label{fig:pv1}
\end{figure}
\vspace{-0.7cm}
\begin{align}
i_{pvi}&= i_{cpvi}+ i_{pv} \nonumber \\
i_{pvi}&= C_{pvi}\frac{dV_{pvi}}{dt} + i_{pv} \label{eq:pv11}\\
V_{pvi}&-i_{pv}R_{pv}-L_{pv}\frac{di_{pv}}{dt}=0  \label{eq:pv12}\\
   C_{pvo}\frac{dV_{pvo}}{dt} &+ \frac{V_{pvo}-V_{CL}}{R_{pvo}} =0 \label{eq:pv13}
\end{align}}
\vspace{-0.6cm}
\subsection{Mode-2}
\begin{figure}[H]
\centering
\includegraphics[width=2.8in]{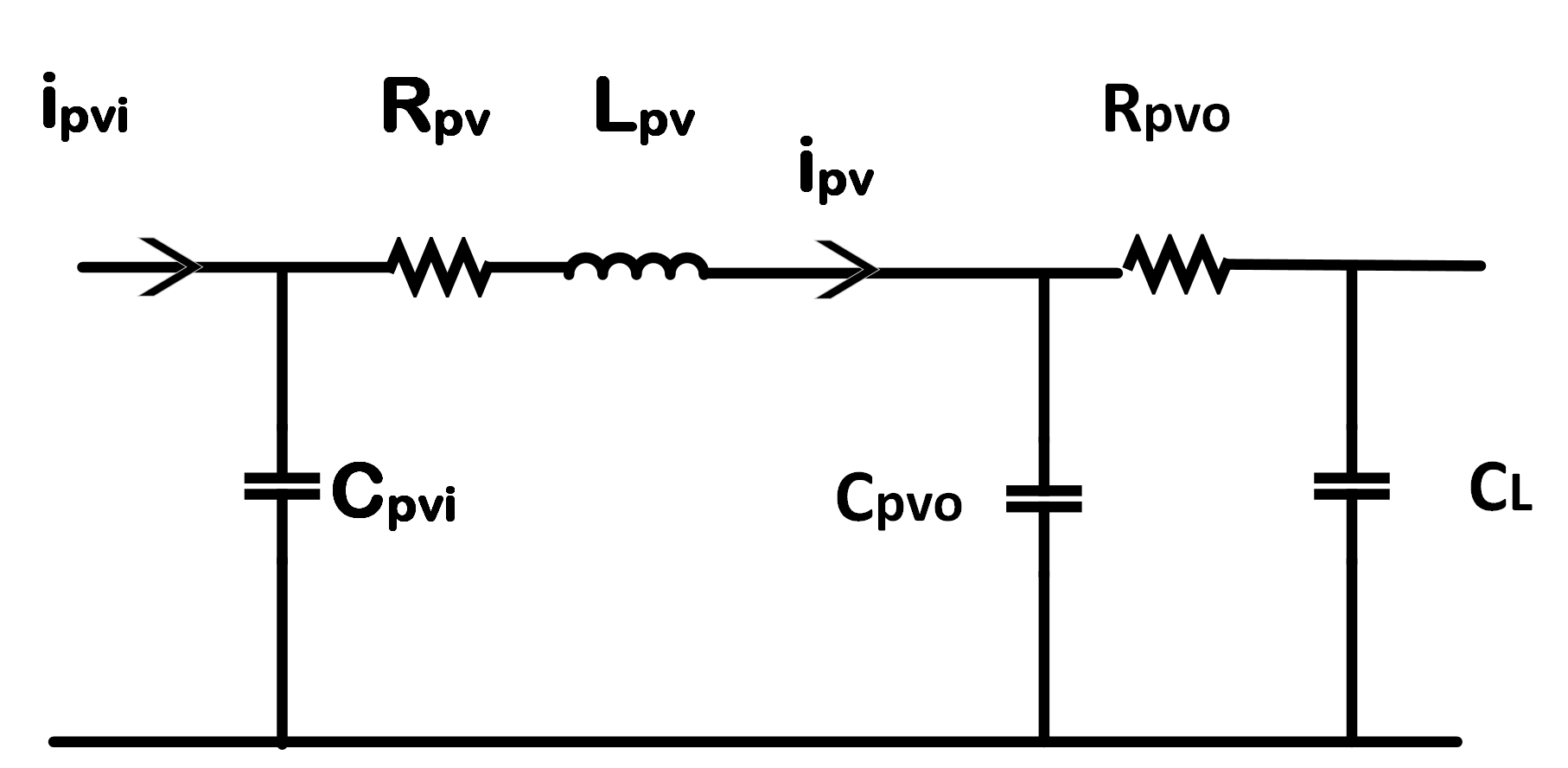}
\caption{PV Subsystem- Mode 2}
\label{fig:pv2}
\end{figure}

When switch $S_1$ is open, the circuit assumes the form which is given in Fig.\ref{fig:pv2}
\vspace{-0.3cm}
\vspace{-0.5cm}
\begin{align}
i_{pvi}&= C_{pvi}\frac{d_{V_{pvi}}}{dt}+ i_{pv} \label{eq:pv21}\\
V_{pvi}&-R_{pv}i_{pv}-L_{pv}\frac{di_{pv}}{dt}-V_{pvo}=0 \label{eq:pv22}\\
i_{pv}&=C_{pvo}\frac{dV_{pvo}}{dt}+ \frac{V_{pvo}-V_{CL}}{V_{pvo}} \label{eq:pv23}
\end{align}
Now, for a single time period, we combine the operation of both the modes and average them with respect to the time period to get the state space average model of the PV subsystem. Upon combining equations \eqref{eq:pv11} and \eqref{eq:pv21}, we get the following equations,
\begin{align}
    i_{pvi}&= C_{pvi}\frac{dV_{pvi}}{dt}+i_{pv} \\
    \frac{dV_{pvi}}{dt}&= \frac{i_{pvi}}{C_{pvi}}-\frac{i_{pv}}{C_{pvi}}
\end{align}
This means,
\begin{align}
    \dot{x}_1= \frac{d_1}{C_{pvi}}- \frac{x_3}{C_{pvi}}
\end{align}
Similarly, combining equations \eqref{eq:pv12} and \eqref{eq:pv22},
\begin{align}
V_{pvi}&-R_{pv}i_{pv}-L_{pv}\frac{di_{pv}}{dt}-V_{pvo}(1-u_1)=0  \nonumber \\
\frac{d i_{pv}}{dt}&= \frac{V_{pvi}}{L_{pv}}-\frac{R_{pv}}{L_{pv}}i_{pv} - \frac{V_{pvo}}{L_{pv}}+ \frac{V_{pvo}}{L_{pv}}u_1 
\end{align}
This means,
\begin{align}
\dot{x}_3= \frac{x_1-x_2}{L_{pv}}-\frac{R_{pv}}{L_{pv}}x_3 + \frac{x_2}{L_{pv}}u_1
\end{align}
Finally, the output capacitor dynamics is derived by combining equations  \eqref{eq:pv13} and \eqref{eq:pv23} as follows,
\begin{align}
    i_{pv}(1-u_1)=C_{pvo}\frac{d V_{pvo}}{dt} + \frac{V_{pvo-V_{CL}}}{R_{pvo}} \nonumber\\
    \frac{d V_{pvo}}{dt}= \frac{i_{pv}}{C_{pvo}}(1-u_1) + \frac{V_{CL}-V_{pvo}}{R_{pvo}C_{pvo}}
\end{align}
Noting the states, the equation can be rewritten as follows,
\begin{align}
    \dot{x}_2= \frac{x_3}{C_{pvo}}+ \frac{x_6}{R_{pvo}C_{pvo}}- \frac{x_2}{R_{pvo}C_{pvo}}-\frac{x_3}{C_{pvo}}u_1
\end{align}
\subsection{Battery Modeling}
\begin{figure}[H]
\centering
\includegraphics[width=3.5in]{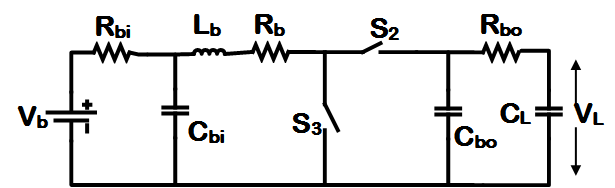}
\caption{Battery Subsystem Circuit}
\label{fig:bat0}
\end{figure}

The circuit of the battery subsystem can be found in Fig.\ref{fig:bat0}. The bidirectional converter attached to the battery contains switches $S_2$ and $S_3$ which are activated in a complementary manner. It means that if $S_3$ is switched ON, then $S_2$ is OFF and when $S_3$ is OFF, then $S_2$ is ON. These two modes of operation are analysed to derive the final state space model.
\vspace{-0.2cm}
\vspace{-0.5cm}
\subsubsection{Mode-1}
\vspace{-0.5cm}
\begin{figure}[H]
\centering
\includegraphics[width=3.5in]{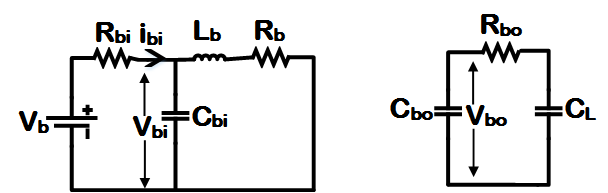}
\caption{Battery Subsystem - Mode 1}
\label{fig:bmode1}
\end{figure}
In mode-1, we consider the case where switch $S_3$ is ON and $S_2$ is OFF.  The circuit for this mode is given in Fig.\ref{fig:bmode1}. Applying Kirchoff's laws on the circuit in Fig.\ref{fig:bmode1}, we get,
\begin{align}
    V_b &- i_{bi} R_{bi} -V_{bi} =0 \label{eq:b11}\\
    i_{bi} &= \frac{V_b- V_{bi}}{R_{bi}} \\
    i_{bi} & = C_{bi} \frac{d V_{bi}}{dt} +i_b \label{eq:b12} 
    \end{align}
Upon rearranging \eqref{eq:b12}, the following dynamics are obtained, 
\begin{align}
\frac{d V_{bi}}{dt}  & = \frac{1}{C_{bi}} (i_{bi}-i_b) \label{4}
\end{align} 
\vspace{-0.3cm}
\begin{align}
V_{bi} -L_b \frac{di_b}{dt} - R_b i_b v= 0 
\end{align} 
\vspace{-0.3cm}
\begin{align}
\frac{d i_{b}}{dt}  & = \frac{1}{L_b} (V_{bi}-i_bR_{b}) 
\end{align} 
\vspace{-0.3cm}
\begin{align}
&= C_{bo} \frac{dV_{bo}}{dt} + \frac{V_{bo}-V_L}{R_{bo}} 
\end{align} 
\vspace{-0.3cm}
\begin{align}
\frac{dV_{bo}}{dt} &= -\frac{1}{C_{bo}}\frac{V_{bo}-V_L}{R_{bo}}
\end{align}
\subsubsection{Mode-2}
\begin{figure}[!ht]
\centering
\includegraphics[width=3.6in]{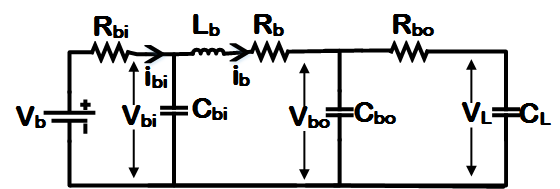}
\caption{Battery Subsystem- Mode 2}
\label{fig:bmode2}
\end{figure}

In mode-2, the switch $S_3$ is OFF and $S_2$ is ON. This circuit has been given in Fig.\ref{fig:bmode2},
Applying Kirchoff's laws on this circuit, we get the following equations,

\begin{align}
    i_{bi} &=C_{bi} \frac{d V_{bi}}{dt} +i_b \\
    \frac{d V_{bi}}{dt}  & = \frac{1}{C_{bi}} (i_{bi}-i_b) \label{10}\\
    V_{bi} &-L_b \frac{di_b}{dt} - R_b i_b -V_{bo} = 0 \\
    \frac{d i_{b}}{dt}  & = \frac{1}{L_b} (V_{bi}-i_bR_b -V_{bo}) \\
    i_b &= C_{bo} \frac{dV_{bo}}{dt} + \frac{V_{bo}-V_L}{R_{bo}} \\
    \frac{dV_{bo}}{dt} &= -\frac{1}{C_{bo}}(i_b-\frac{V_{bo}}{R_{bo}}+\frac{V_L}{R_{bo}})
\end{align}

Combining equations \eqref{4} and \eqref{10} and averaging for both and ON and OFF cycle,

\begin{align}
    \frac{d V_{bi}}{dt} u_2+  \frac{d V_{b_i}}{dt} (1-u_2) =  \frac{1}{C_{bi}} (i_{bi}-i_b) u_2 +  \frac{1}{C_{bi}} (i_{bi}-i_b) (1-u_2)
\end{align}

\begin{align}
    \frac{d V_{bi}}{dt}  & =  \frac{1}{C_{bi}} i_{bi}u_2 - \frac{1}{C_{bi}}i_b u_2 +  \frac{1}{C_{bi}} i_{bi} - \frac{1}{C_{bi}}i_{bi} u_2 - \frac{1}{C_{bi}} i_b + \frac{1}{C_{bi}} i_b u_2 \nonumber\\
      & =  \frac{1}{C_{bi}} - \frac{1}{C_{bi}}i_b \nonumber \\
      & =  \frac{1}{C_{bi}} (\frac{V_b- V_{bi}}{R_{bi}}) - \frac{1}{C_{bi}}i_b \nonumber 
\end{align}
This finally results in,
\begin{align}
    \frac{d V_{bi}}{dt}  & =  \frac{V_b}{R_b C_{bi}} -\frac{V_b}{R_{bi}C_{bi}} - \frac{i_b}{C_{bi}}  \label{eq:battt1}
\end{align}

Similarly,

\begin{align}
    \frac{d i_{b}}{dt} u_2+  \frac{d i_{b}}{dt} (1-u_2) =  \frac{1}{L_{b}} (V_{bi}-i_bR_b) u_2 +  \frac{1}{L_{b}} (V_{bi}-i_bR_b-V_{bo}) (1-u_2) \nonumber
\end{align}
This gets simplified to,
\begin{align}
    \frac{d i_{b}}{dt} =  \frac{V_{bi}}{L_{b}}- \frac{R_b i_b}{L_{b}} + \frac{V_{bo}}{L_{b}} -  \frac{V_{bo} u_2}{L_{b}}  \label{eq:battt2}
\end{align}

\begin{align}
    \frac{d V_{bo}}{dt} u_2+  \frac{d V_{bo}}{dt} (1-u_2) =  \left[ -\frac{1}{C_{bo} R_{bo}} (V_{b}-V_L) \right] u_2 +  \left[ \frac{1}{C_{bo}} (i_{b}-\frac{V_{bo}}{R_{bo}}+ \frac{V_L}{R_{bo}}) \right] (1-u_2) \nonumber
\end{align}
This finally gets reduced to,
\begin{align}
    \frac{d V_{bo}}{dt} =  \frac{i_{b}}{C_{bo}}- \frac{V_{bo}}{C_{bo}R_{bo}} +\frac{V_L}{R_{bo}C_{bo}} -  \frac{i_bu_2}{C_{bo}} \label{eq:battt3} 
\end{align}

Upon replacing the variables with their respective state notations in equations \eqref{eq:battt1} to \eqref{eq:battt3}, we get the following,
\begin{align}
    \dot{x}_4=&\frac{d_2}{R_{bi}C_{bi}}-\frac{x_4}{R_{bi}C_{bi}}-\frac{x_6}{C_{bi}}  \\ 
 \dot{x}_5=& \frac{x_6}{C_{bo}} - \frac{x_5}{R_{bo}C_{bo}}+\frac{x_9}{R_{bo}C_{bo}} -\frac{x_6}{C_{bo}}u_2  \\
 \dot{x}_6=& \frac{x_4}{L_b}-\frac{x_5}{L_{b}} - \frac{x_{6}R_{b}}{L_{b}} + \frac{x_5}{L_{b}}u_2
\end{align}

\subsection{Supercapacitor Modeling}
\begin{figure}[!ht]
\centering
\includegraphics[width=3.6in]{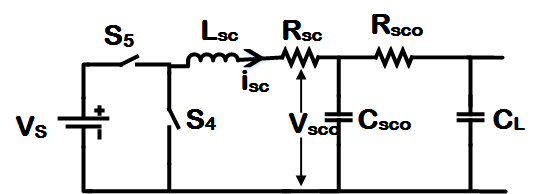}
\caption{Supercapacitor Subsystem}
\label{fig:sc0}
\end{figure}
The circuit of the battery subsystem can be found in Fig.\ref{fig:sc0}. The bidirectional converter attached to the supercapacitor contains switches $S_4$ and $S_5$ which are activated in a complementary manner. It means that if $S_4$ is switched ON, then $S_5$ is OFF and when $S_4$ is OFF, then $S_5$ is ON. These two modes of operation are analysed to derive the final state space model.
\subsubsection{Mode 1}
\begin{figure}[!ht]
\centering
\includegraphics[width=3.6in]{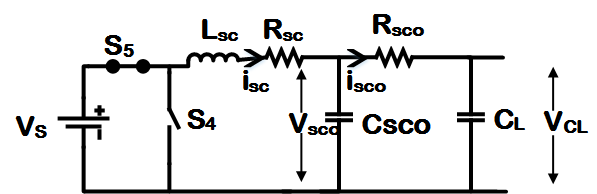}
\caption{Supercapacitor Subsystem- Mode 1}
\label{fig:sc1}
\end{figure}

If switch $S_5$ is closed and $S_4$ is open, then the circuit in Fig.\ref{fig:sc0} gets transformed to Fig.\ref{fig:sc1}.  Upon applying Kirchoff's laws on this circuit we get,

\begin{align}
    V_s - L_{sc}\frac{di_{sc}}{dt} - R_{sc}i_{sc} - V_{sco} &= 0 \nonumber \\
 L_{sc}\frac{di_{sc}}{dt} &= V_s  - R_{sc}i_{sc} - V_{sco} \nonumber \\
 \frac{di_{sc}}{dt} &= \frac{1}{L_{sc}}(V_s  - R_{sc}i_{sc} - V_{sco}) \label{eq:3} 
 \end{align}
 \begin{align}
 i_{sc} &= C_{sco}\frac{dV_{sco}}{dt} + i_{sco} \nonumber \\
 C_{sco}\frac{dV_{sco}}{dt} &= i_{sc} - i_{sco} \nonumber \\
 \frac{dV_{sco}}{dt} &= \frac{1}{C_{sco}}(i_{sc} - i_{sco}) \label{eq:sc12}
\end{align}
\subsubsection{Mode 2}
\begin{figure}[!ht]
\centering
\includegraphics[width=3.6in]{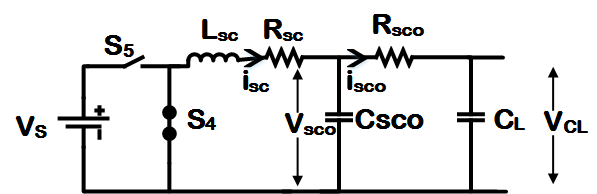}
\caption{Supercapacitor Subsystem- Mode 2}
\label{fig:sc2}
\end{figure}

When $S_5$ is OFF and $S_4$ is ON, the circuit in Fig.\ref{fig:sc0} changes to the one shown in Fig.\ref{fig:sc2} which is analysed as follows,
In the first loop,
\begin{align}
    - L_{sc}\frac{di_{sc}}{dt} - R_{sc}i_{sc} - V_{sco} &= 0 \nonumber\\
 L_{sc}\frac{di_{sc}}{dt} &=   - R_{sc}i_{sc} - V_{sco} \nonumber \\
 \frac{di_{sc}}{dt} &= \frac{1}{L_{sc}}(  - R_{sc}i_{sc} - V_{sco}) \label{eq:9} 
 \end{align}
 In the second loop,
 \begin{align}
 i_{sc} &= C_{sco}\frac{dV_{sco}}{dt} + i_{sco} \\\nonumber
 C_{sco}\frac{dV_{sco}}{dt} &= i_{sc} - i_{sco}\\
 i_{sco} &= \frac{v_{sco}-v_{cl}}{R_{sco}} \\
 \frac{dV_{sco}}{dt} &= \frac{1}{C_{sco}}(i_{sc} - i_{sco}) \label{eq:sc22}
\end{align}
By combining equations \eqref{eq:3} and \eqref{eq:9} and averaging them over a cycle, the following equations are obtained, 
\begin{align}
    \frac{di_{sc}}{dt}u_3 &+ \frac{di_{sc}}{dt}(1-u_3) = \frac{1}{L_{sc}}(V_s- R_{sc}i_{sc} - V_{sco})u_3 + \frac{1}{L_{sc}}(- R_{sc}i_{sc} - V_{sco})(1-u_3) \nonumber \\
 \frac{di_{sc}}{dt} &=   \frac{V_s}{L_{sc}}u_3- \frac{R_{sc}i_{sc}}{L_sc}u_3 - \frac{V_{sco}}{L_{sc}}u_3 - \frac{R_{sc}i_{sc}}{L_sc} + \frac{R_{sc}i_{sc}}{L_sc}u_3 -\frac{V_{sco}}{L_{sc}}+\frac{V_{sco}u_3}{L_{sc}} \nonumber \\
 \frac{di_{sc}}{dt} &= \frac{V_su_3}{L_{sc}} -  \frac{R_{sc}i_{sc}}{L_{sc}} - \frac{V_{sco}}{L_{sc}} \label{eq:sc1f} 
\end{align}

Similarly, combining equations \eqref{eq:sc12} and \eqref{eq:sc22}, the following dynamics are obtained
 \begin{align}
    \frac{dV_{sco}}{dt}u_3 + \frac{dV_{sco}}{dt}(1-u_3) &= \frac{1}{C_{sco}}(i_{sc} - i_{sco})u_3 + \frac{1}{C_{sco}}(i_{sc} - i_{sco})(1-u_3) \nonumber \\
 \frac{dV_{sco}}{dt} &=   \frac{i_{sc}}{C_{sco}}u_3+ \frac{i_{sc}}{C_{sco}} - \frac{i_{sc}}{C_{sco}}u_3  - \frac{i_{sco}}{C_{sco}} + \frac{i_{sco}}{C_{sco}}u_3 \nonumber \\
 \frac{dV_{sco}}{dt} & = \frac{i_{sc}}{C_{sco}} - \frac{i_{sco}}{C_{sco}} 
\end{align}

Here, substituting $i_{sco} = \frac{V_{sco}-V_{CL}}{R_{sco}}$

\begin{align}
\frac{dV_{sco}}{dt} & = \frac{i_{sc}}{C_{sco}} - \frac{1}{C_{sco}R_{sco}}(V_{sco}-V_{CL}) \\
  \frac{dV_{sco}}{dt} & = \frac{i_{sc}}{C_{sco}} - \frac{V_{sco}}{C_{sco}R_{sco}}+\frac{V_{CL}}{C_{sco}R_{sco}} \label{eq:sc2f}
\end{align}

When the state variables are substituted in equations \eqref{eq:sc1f} and \eqref{eq:sc2f}, the following are obtained,
\begin{align}
\dot{x}_7=& -\frac{x_8}{L_{sc}} - \frac{R_{sc}x_7}{L_{sc}}+\frac{d_3}{L_{sc}}u_3   \nonumber \\
 \dot{x}_8=& \frac{x_7}{C_{sco}} - \frac{x_8}{R_{sco}C_{sco}}+\frac{x_9}{R_{sco}C_{sco}} 
 \end{align}
\subsection{Load Modeling}
The load circuit diagram for both the ON and OFF modes remains same and hence, simple circuit analysis can yield the desired load dynamics. The load circuit diagram is given as shown in Fig.\ref{fig:load}. The following analysis is carried out to extract the load dynamics,
\textcolor{black}{\begin{figure}[!ht]
\centering
\includegraphics[width=2.5in]{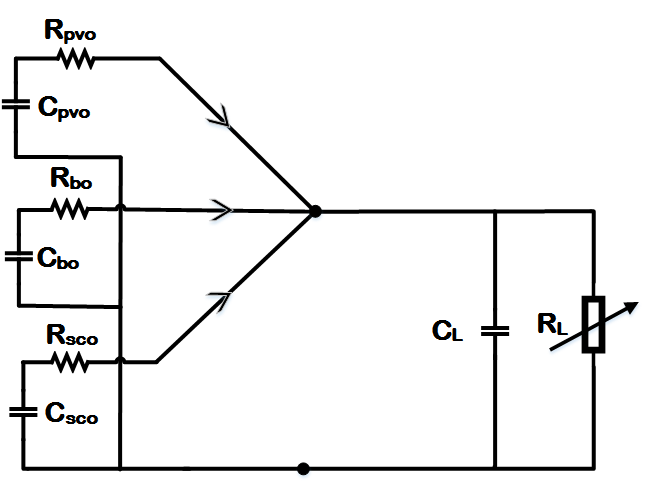}
\caption{Load Circuit Diagram}
\label{fig:load}
\end{figure}
\begin{align}
\frac{V_{pvo}-V_{CL}}{R_{pvo}} + \frac{V_{bo}-V_{CL}}{R_{bo}}+ \frac{V_{sco}-V_{CL}}{R_{sco}}= \frac{V_{CL}}{R_L} + C_L\frac{d V_{CL}}{dt}  \\
\frac{d V_{CL}}{dt}=\frac{V_{pvo}}{C_LR_{pvo}}+ \frac{V_{bo}}{C_LR_{bo}}+ \frac{V_{bo}}{C_LR_{bo}}- \frac{V_{CL}}{C_L}\Bigg[\frac{1}{R_{pvo}}+ \frac{1}{R_{bo}}+ \frac{1}{R_{sco}}+ \frac{1}{R_L}\Bigg] \\
\dot{x}_9= \frac{x_2}{C_LR_{pvo}} + \frac{x_5}{C_L R_{bo}}+ \frac{x_8}{C_L R_{sco}}-\frac{x_9}{C_L}\Bigg[\frac{1}{R_{pvo}}+ \frac{1}{R_{bo}}+ \frac{1}{R_{sco}}+ d_4\Bigg]
\end{align}}
\subsection{The DCSSMG State Space Model}
Combining all the dynamics derived in the previous subsections, the overall state space model for the DCSSMG is compiled as follows:
\begin{align}
 \dot{x}_1=& \frac{d_1}{C_{pvi}} - \frac{x_3}{C_{pvi}} \nonumber \\
 \dot{x}_2=& \frac{x_3}{C_{pvo}} - \frac{x_2}{R_{pvo}C_{pvo}} + \frac{x_9}{R_{pvo}C_{pvo}}- \frac{x_{3}}{C_{pvo}}u_{1} \nonumber \\
 \dot{x}_3=& \frac{x_1}{L_{pv}} - \frac{x_2}{L_{pv}} - \frac{x_{3}R_{pv}}{L_{pv}} + \frac{x_2}{L_{pv}}u_1  \nonumber \\  
 \dot{x}_4=&\frac{d_2}{R_{bi}C_{bi}}-\frac{x_4}{R_{bi}C_{bi}}-\frac{x_6}{C_{bi}} \nonumber \\ 
 \dot{x}_5=& \frac{x_6}{C_{bo}} - \frac{x_5}{R_{bo}C_{bo}}+\frac{x_9}{R_{bo}C_{bo}} -\frac{x_6}{C_{bo}}u_2  \label{eq:pvbatschess} \\
 \dot{x}_6=& \frac{x_4}{L_b}-\frac{x_5}{L_{b}} - \frac{x_{6}R_{b}}{L_{b}} + \frac{x_5}{L_{b}}u_2 \nonumber \\
 \dot{x}_7=& -\frac{x_8}{L_{sc}} - \frac{R_{sc}x_7}{L_{sc}}+\frac{d_3}{L_{sc}}u_3   \nonumber \\
 \dot{x}_8=& \frac{x_7}{C_{sco}} - \frac{x_8}{R_{sco}C_{sco}}+\frac{x_9}{R_{sco}C_{sco}}   \nonumber \\
 \dot{x}_9=& \frac{x_2}{C_{L}R_{pvo}} +\frac{x_5}{C_{L}R_{bo}}+\frac{x_8}{C_{L}R_{sco}}-\frac{x_9}{C_L}z-\frac{x_9}{C_L}d_4  
 \end{align}
 \begin{align}
 z=&\Bigg[\frac{1}{R_{pvo}}+\frac{1}{R_{bo}}+\frac{1}{R_{sco}}\Bigg] \nonumber \\
  x=& [V_{pvi}~~V_{pvo}~~I_{pv}~~V_{bi}~~V_{bo}~~I_b~~I_{sc}~~V_{sco}~~V_{dc} ] \nonumber\\ d =& \Bigg[I_{pvi} ~~V_b ~~V_s~~ \frac{1}{R_L}\Bigg], ~~
  u=[u_1~~u_2~~u_3]= [s_1~~s_3~~s_5] \nonumber
 \end{align}
The output of the DCMG system is $y=[x_1~~x_4~~x_9].$

\section{Controller Design }\label{sec:aobsdesign}
This section formulates the control problem  and develops the adaptive observer based controllers. The complete Lyapunov based stability is also proved. Due to this, the developed controllers can be assured of convergence and hence, can be deployed for online purposes.

\subsection{Problem Formulation}
As seen in \ref{fig:dcmgoverall}, there are two control levels for the DCSSMG system- primary and secondary. The references of the primary controllers are assumed to be known and made available from the secondary level. Thus, the current work is concerned only with designing the adaptive observer based controllers at the primary level.

 Given $y_d= [x_{1ref}~~x_{4ref}~~x_{9ref}]$, and the values of $x_{2ref},~x_{5ref}$ the control objective is to design the controllers $u_1$, $u_2$, $u_3$ such that the output $y=[x_1~~x_4~~x_9]$ converges to $y_d$ when disturbances $d = \Big[I_{pvi} ~~~V_b ~~~V_s~~  \frac{1}{R_L}\Big]$ and system model containing various system parameters including capacitances, inductances and resistances are unknown.


\subsection{Secondary Reference Generation:}
A conventional secondary reference generation is shown in this subsection. All the references are derived using power balance equations and are assumed to be deployed in a centralized manner for simplicity.

The DCSSMG operator fixes a grid operating voltage $V_{DC}$ which is assumed to be known. The maximum power point goals are desribed in the form of $V_{mpp}$ and $I_{mpp}$ which are also assumed to be coming from an MPPT algorithm such as any standard algorithm like perturb and observe, incremental conductance, etc. The rest of the secondary references of the DCSSMG are computed in the following manner:

\begin{align}
x_{9ref} &= V_{DC} \nonumber \\
x_{1ref}&= V_{mpp}\nonumber\\
x_{3ref}&= I_{mpp}\nonumber\\
& a_1=1 \hspace{1cm}  b_1=-x_{9ref} \nonumber \\
& c_1= x_{3ref}^2R_{pvo}R_{pv}-x_{1ref}x_{3ref}R_{pvo}\nonumber \\
x_{2ref}&=\frac{-b_1\pm \sqrt{{b^2_1-4a_1c_1}}}{2a_1} \nonumber \\
x_{5ref}&=-\frac{R_{bo}}{R_{pvo}}x_{2ref} + x_{9ref}\Bigg(1+\frac{R_{bo}}{R_{pvo}}+\frac{R_{bo}}{R_l}\Bigg) \nonumber \\ 
 &a_2=R_{bo}(R_{bi}+R_b)  \hspace{1cm}  b_2=-R_{bo}d_2 \nonumber \\
 &c_2= x_{5ref}^2-x_{5ref}x_{9ref}\nonumber \\
x_{6ref}&=\frac{-b_2\pm \sqrt{{b^2_2-4a_2c_2}}}{2a_2} \nonumber \\
x_{4ref} &= d_2-x_{6ref}{R_{bi}} \nonumber \\
x_{7ref}&=0 \nonumber \\
x_{8ref}&= V_{DC} \nonumber
\end{align}

\subsection{Controller Design Procedure:}
\begin{figure}[H]
\centering
\includegraphics[width=5.6in]{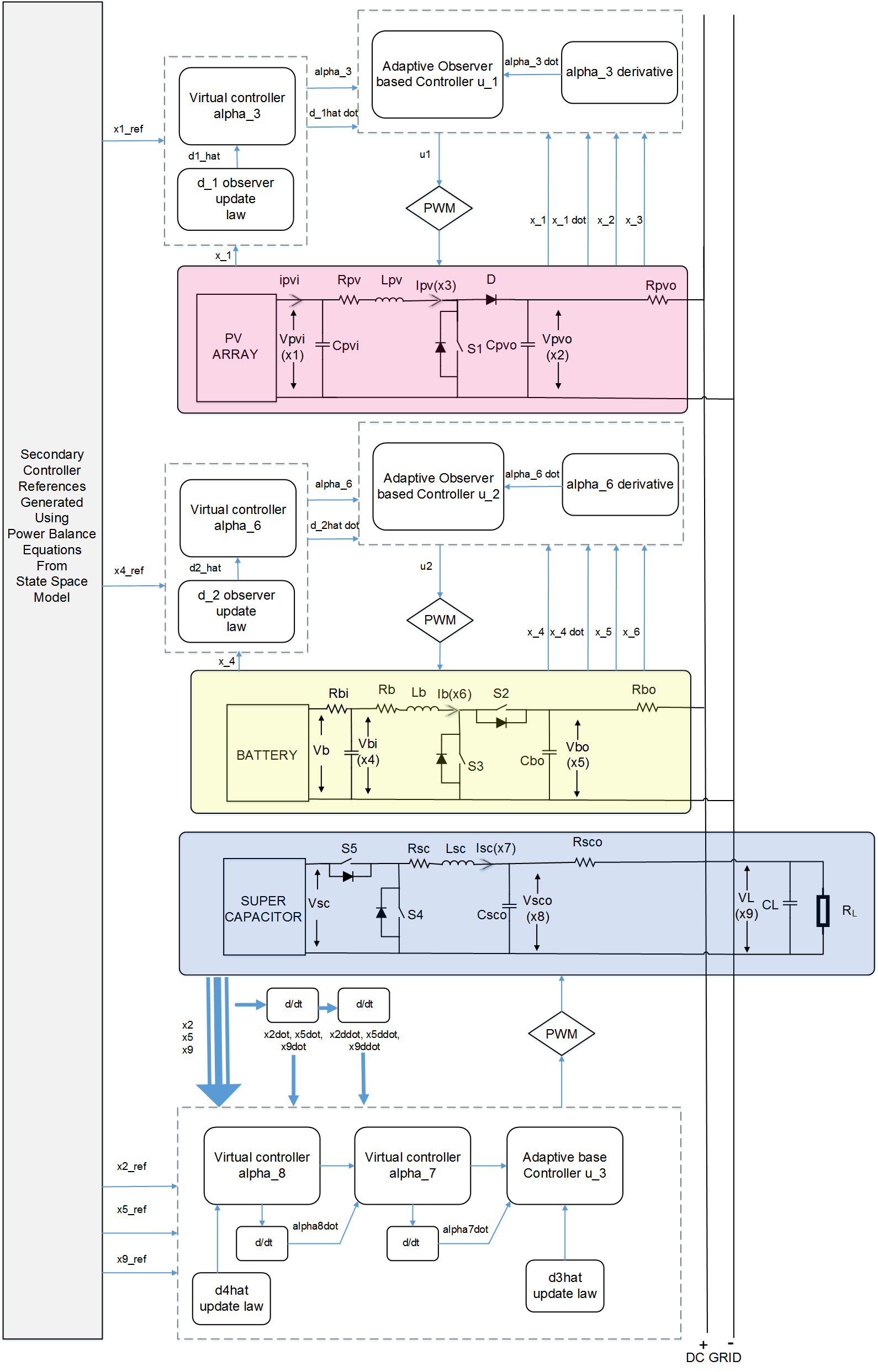}
\caption{Adaptive Observer Backstepping Controller Implementation Diagram}
\label{fig:dobs_implement}
\end{figure}

After analyzing the structure of the system dynamics as given in \eqref{eq:pvbatschess}, it can be partitioned into various subsystems and the controller can be developed for each subsystem. The controller for each subsystem will be developed using back-stepping type procedure. A detailed schematic has been presented in Fig.\ref{fig:dobs_implement}  illustrating the overall design procedure of the adaptive observer based controllers for various subsystems. The salient steps for controller development are also enumerated as follows:
\begin{enumerate}
\item The PV virtual controller $\alpha_3$ is computed so as to ensure regulation of state $x_1$ by reducing output error $e_1=x_1-x_1ref$.
\item With the help of PV virtual controller, the PV output $u_1$ and $\dot{\widehat{d}}_1$ are calculated to reduce the state errors $e_2=x_2-x_{2ref}$ and $e_3= x_3-\alpha_3$.
\item The battery virtual controller $alpha_6$ is found out so as to ensure regulation in battery input capacitor voltage by reducing $e_4=x_{4}-x_{4ref}$.
\item The battery duty cycle $u_2$ and $\dot{\widehat{d}}_2$ are computed to reduce the state errors $e_6= x_6-\alpha_6$ and $e_5= x_5-x_{5ref}$.
\item The virtual input $\alpha_8$ and $\dot{\widehat{d}}_4$ is computed such that error $e_9=x_9-x_{9ref}$ goes to zero.
\item The virtual input$\alpha_7$ is calculated such that $x_8$ follows the virtual input $\alpha_8$.
\item The supercapacitor controller $u_3$ and $\dot{\widehat{d}}_3$ are calculated such that $x_7$ follows $\alpha_7$.
\end{enumerate}
\subsection{Controller Design Proof:}
In this section, the proposed control design with stability proof is presented
\begin{Theorem}
Let the virtual controllers ${\alpha}_i, i=3,6,8,7 $ be chosen as $ \alpha_3 = \hat{d}_1 + K_1e_1$, $\alpha_6 = \frac{\widehat{d}_2-x_4}{R_{bi}} + K_4e_4$, $\alpha_8 = -R_{sco}K_9e_9 - \frac{R_{sco}}{R_{pvo}}x_2- \frac{R_{sco}}{R_{bo}}x_5+x_9\widehat{z}$, $\alpha_7= -K_8e_8 + \frac{x_8}{R_{sco}} - \frac{x_9}{R_{sco}} + C_{sco}\dot{\alpha}_8$ and controller inputs be chosen as 
\begin{align}
u_1 &= \frac{-K_3e_3-x_1+x_2+x_3R_{pv}-L_{pv}\dot{\alpha}_3}{x_2} , \nonumber\\
u_2 &= \frac{-K_6e_6 -x_4+ x_5+ x_6R_b+ L_b\dot{\alpha}_6}{x_5} ,\nonumber \\ 
 u_3 &= \frac{-K_7e_7+x_8+R_{sc}x_7+L_{sc}\dot{\alpha}_7}{\widehat{d}_3} \nonumber
 \end{align}
 and disturbance updation laws be chosen as $\dot{\widehat{d}}_1 = \gamma_1 e_1$, $\dot{\widehat{d}}_2 = \gamma_2e_4$,  $\dot{\widehat{d}}_3 = \gamma_3 e_7 u_3$, $\dot{\widehat{d}}_4= -\gamma_4x^2_9$ where $\gamma_i>0$, $e_j=x_j-x_{jref}, ~j=1,2,4,5,9$, $e_k=x_k-\alpha_k, k= 3,6,7,8$, and $\widehat{z}= \Big[\frac{R_{sco}}{R_{pvo}}+\frac{R_{sco}}{R_{bo}}+1+ R_{sco}\widehat{d}_4\Big]$ then all the states of the DCMG system will converge to the desired reference values.
\end{Theorem}

\noindent
\textit{Proof:}
In this proof, Lyapunov stability and backstepping theory \cite{purebstep1}  are employed for designing adaptive controllers with disturbance observers. A positive definite composite Lyapunov function is chosen consisting of Lyapunov functions from different subsystems like the PV subsystem, battery subsystem and the supercapacitor subsystem. \textcolor{black}{The main factor behind choice of the virtual controller lies in the selection of the Lyapunov function for the backstepping based controllers. The first derivative of the Lyapunov is computed and finally the controller or the virtual controller is chosen in such a way through mathematical manipulation that upon plugging in the virtual controller/controller expression, the Lyapunov derivative becomes negative definite. The existence of negative definite Lyapunov derivative ensures the convergence of errors thereby achieving the necessary control objectives.}

In the upcoming process, the controller for the PV boost converter is designed first followed by a controller for the bidirectional converter for the battery and finally that of the supercapacitor subsystem.
Consider the dynamics of the state $x_1$ and define a Lyapunov function with errors $e_1= x_1-x_{1ref}$ and $\widetilde{d}_1=d_1-\widehat{d}_1$,  as follows, $W_1=\frac{C_{pvi}}{2}e^2_1+\frac{1}{2\gamma_1}\widetilde{d}^2_1$. Upon differentiation, we get 
 \begin{align}
 \dot{W}_1 &= e_1(d_1-x_3)-\frac{1}{\gamma_1}\widetilde{d}_1\dot{\widehat{d}}_1 \nonumber \\
 &= e_1[\widehat{d}_1-x_3] + \widetilde{d}_1\Bigg[e_1- \frac{\dot{\widehat{d}}_1}{\gamma_1}\Bigg]
 \end{align}
 The virtual controller $\alpha_3$ and disturbance update law $\dot{\widehat{d}}_1$ are chosen as
\begin{align}
\alpha_3 &= d_1 + K_1e_1 \label{eq:nvu3} \\
\dot{\widehat{d}}_1 &= \gamma_1e_1
\end{align} 
This ensues the following result, $\dot{W}_1= -K_1e^2_1$ which means $\dot{W}_1 \leq 0$ when $K_1\geq0$. The next Lyapunov function is chosen to facilitate backstepping, $W_{3}=  \frac{L_{pv}}{2}e^2_3$ where, $e_3= x_3-\alpha_3$ .  On taking the derivative,
\begin{align}
\dot{W}_{3}&=  L_{pv}e_3(\dot{x}_3 - \dot{\alpha}_3) \nonumber \\
&= e_3(x_1-x_2-R_{pv}x_3+x_2u_1-L_{pv}\dot{\alpha}_3) 
\end{align}
The duty cycle $u_1$ is selected as follows,
\begin{align}
u_1 &= \frac{-K_3e_3-x_1+x_2+x_3R_{pv}-L_{pv}\dot{\alpha}_3}{x_2} \label{eq:nu1}
\end{align}
where
\begin{align}
\dot{\alpha}_3&= K_1 \dot{x}_1 + \dot{\widehat{d}}_1 
\end{align}
This ultimately results in $\dot{W}_{3}= -K_3e^2_3$ which means $\dot{W}_{3} \leq 0$ when $K_3 \geq 0$.
Hence, upon choosing $u_1$ as discussed above, it can be guaranteed that the states $x_1$, and $x_3$ will converge.

For the battery subsystem, a Lyapunov function using $e_4= x_4-x_{4ref}$ and $\widetilde{d}_2= d_2- \widehat{d}_2$ is chosen as $W_4=\frac{C_{bi}}{2}e^2_4 + \frac{1}{\gamma_2}\widetilde{d}^2_2$. The derivative of the Lyapunov is calculated as below,
\begin{align}
\dot{W}_4 &= e_4\Bigg[\frac{d_2-x_4}{R_{bi}}-x_6\Bigg]-\frac{1}{\gamma_2}\widetilde{d}_2\dot{\widehat{d}}_2 \nonumber \\
&= e_4\Bigg[\frac{\widehat{d}_2}{R_{bi}}-\frac{x_4}{R_{bi}}-x_6\Bigg] + \widetilde{d}_2\Bigg[e_4-\frac{\dot{\widehat{d}}_2}{\gamma_2}\Bigg]
\end{align}
 The virtual controller $\alpha_6$ is chosen as
\begin{align}
\alpha_6 &= \frac{\widehat{d}_2-x_4}{R_{bi}} + K_4e_4 \label{eq:nvu4} \\
\dot{\widehat{d}}_2 &= \gamma_2e_4 \\
\ddot{\widehat{d}}_2 &= \gamma_2\dot{x}_4 \\
\end{align} 
This leads to the follows expression of the Lyapunov derivative, $\dot{W}_4= -K_4e^2_4$ which means $\dot{W}_4\leq 0$ when $K_4\geq0$. 

The following Lyapunov function $W_{6}=  \frac{L_{b}}{2}e^2_6$ is chosen where, $e_6= x_6-\alpha_6$. Taking the Lyapunov derivative,
\begin{align}
\dot{W}_{6}&=  L_{b}e_6(\dot{x}_6 - \dot{\alpha}_6) \nonumber \\
&=  e_6(x_4-x_5-R_{b}x_6+ x_5u_2-L_{b}\dot{\alpha}_6) 
\end{align}
The duty cycle for the battery subsystem $u_2$ is chosen as follows:
\begin{align}
u_2 &= \frac{-K_6e_6 -x_4+ x_5+ x_6R_b+ L_b\dot{\alpha}_6}{x_5} \label{eq:nu2}
\end{align}
where
\begin{align}
\dot{\alpha}_6&= \Bigg(K_4-\frac{1}{R_{bi}}\Bigg) \dot{x}_4 + \frac{\dot{\widehat{d}}_2}{R_{bi}} 
\end{align}
then, it results in $\dot{W}_{6}= -K_6e^2_6$ which means $\dot{W}_{6} \leq 0$ when $K_6\geq0$.
It is seen that the selected virtual controller $\alpha_6$ and the duty cycle $u_2$ leads to convergence of states $x_4$ and $x_6$.

Now, for the supercapacitor subsystem, the following Lyapunov function is chosen $W_{2,5,9}= \frac{C_{pvo}}{2}e^2_2+\frac{C_{bo}}{2}e^2_5+\frac{C_{L}}{2}x^2_9 + \frac{1}{2\gamma_4}\widetilde{d}_4^2$ where $e_2= x_2-x_{2ref},~e_5= x_5-x_{5ref}$ and $\widetilde{d}_4=d_4-\widehat{d}_4$. Upon differentiation, we get 
\begin{align}
\dot{W}_{2,5,9}= C_{pvo}e_2\dot{x}_2+C_{bo}e_5\dot{x}_5+ C_Le_9\dot{x}_9 +\frac{1}{\gamma_4}\widetilde{d}_4(\dot{d}_4-\dot{\widehat{d}}_4)
\end{align}
 This translates to:
\begin{align}
\dot{W}_{2,5,9}=&e_2\Bigg[x_3+ \frac{x_9-x_2}{R_{pvo}} -x_3u_1\Bigg]+ e_5\Bigg[x_6+ \frac{x_9-x_5}{R_{bo}} -x_6u_2\Bigg] + x_9\Bigg[\frac{x_2}{R_{pvo}}+\frac{x_5}{R_{bo}}+\frac{x_8}{R_{sco}}\Bigg]\nonumber \\ & -x^2_9\Bigg[\frac{1}{R_{pvo}}+\frac{1}{R_{bo}}+\frac{1}{R_{sco}}+\widehat{d}_4\Bigg]+\widetilde{d}_4\Bigg(-x^2_9 + \frac{\dot{d}_4-\dot{\widehat{d}}_4}{\gamma_4}\Bigg)
\end{align}
The virtual controller $\alpha_8$ is chosen as,
\begin{align}
\alpha_8 =& -\frac{R_{sco}}{R_{pvo}}x_2-\frac{R_{sco}}{R_{bo}}x_5+ \frac{R_{sco}}{x_9}e_2\Bigg[-K_2e_2-x_3+ \frac{x_2-x_9}{R_{pvo}} + x_3u_1 \Bigg] -R_{sco}K_9\frac{e^2_9}{x_9}   \nonumber \\
 & +\frac{R_{sco}}{x_9}e_5\Bigg[-K_5e_5-x_6+ \frac{x_5-x_9}{R_{bo}} + x_6u_2 \Bigg] + R_{sco}x_9\Bigg[\frac{1}{R_{pvo}}+\frac{1}{R_{bo}}+\frac{1}{R_{sco}}+\widehat{d}_4\Bigg]\label{eq:nvu5}
\end{align} 
 and $\dot{\widehat{d}}_4$ as
 \begin{align}
 \dot{\widehat{d}}_4= -\gamma_4x^2_9
 \end{align}
 then, $\dot{W}_{2,5,9}= -K_2e^2_2-K_5e^2_5-K_9e^2_9$ which means $\dot{W}_9 \leq 0$ when $K_2,~K_5,~K_9\geq0$. In the next step, the Lyapunov function $W_{8}= \frac{C_{sco}}{2}e^2_8$ is chosen where, $e_8= x_8-\alpha_8$ .  Upon differentiation the following expression is obtained,
  \begin{align}
 \dot{W}_{8} =& C_{sco}e_8(\dot{x}_8 - \dot{\alpha}_8) \nonumber \\
 = & e_8\Bigg[x_7-\frac{x_8}{R_{sco}}+\frac{x_9}{R_{sco}}-C_{sco}\dot{\alpha}_8 \Bigg] 
 \end{align}
 \begin{align}
 \dot{\alpha}_8 = &  -\frac{R_{sco}}{R_{pvo}}\dot{x}_2-\frac{R_{sco}}{R_{bo}}\dot{x}_5 + R_{sco}\dot{x}_9c+ R_{sco}x_9\dot{\widehat{d}}_4 - \frac{2R_{sco}K_9e_9\dot{x}_9}{{x}_9} +\frac{\dot{x}_9R_{sco}K_9e^2_9}{x^2_9} \nonumber \\ 
 &  + \frac{R_{sco}(\dot{e}_2a+ e_2\dot{a})}{x_9}- \frac{R_{sco}\dot{x}_9e_2a}{x^2_9}  + \frac{R_{sco}(\dot{e}_5b+ e_5\dot{b})}{x_9}- \frac{R_{sco}\dot{x}_9e_5b}{x^2_9}
 \end{align}
\begin{align}
a= &-K_2e_2-x_3+ \frac{x_2-x_9}{R_{pvo}}+ x_3u_1 \\
\dot{a}= &-K_2\dot{e}_2-\dot{x}_3+ \frac{\dot{x}_2-\dot{x}_9}{R_{pvo}}+ \dot{x}_3u_1 + x_3\dot{u}_1 \end{align}
\begin{align}
b= &-K_5e_5-x_6+ \frac{x_5-x_9}{R_{bo}}+ x_6u_2 \\
\dot{b}= &-K_2\dot{e}_5-\dot{x}_6+ \frac{\dot{x}_5-\dot{x}_9}{R_{bo}}+ \dot{x}_6u_2 + x_6\dot{u}_2 \\
c= & \frac{1}{R_{pvo}}+\frac{1}{R_{bo}}+\frac{1}{R_{sco}}+\widehat{d}_4
 \end{align}
If $x_7$ is chosen as follows,
\begin{align}
\alpha_7= -K_8e_8 + \frac{x_8}{R_{sco}} - \frac{x_9}{R_{sco}} + C_{sco}\dot{\alpha}_8
\end{align}
it leads to $\dot{W}_8=-K_8e^2_8$ which means $\dot{W}_8\leq0$ provided $K_8\geq0$. \\
Finally, the following Lyapunov is chosen
\begin{align}
W_7= \frac{L_{sc}}{2}e^2_7 + \frac{1}{2\gamma_3}\widetilde{d}^2_3
\end{align} where $e_7=x_7-\alpha_7$
Upon differentiation, we get,
\begin{align}
\dot{W}_7=& L_{sc}e_7(\dot{x}_7-\dot{\alpha}_7) + \frac{1}{\gamma_3}\widetilde{d}_3(\dot{
d}_3-\dot{\widehat{d}}_3) \nonumber \\
=& e_7\Bigg[-x_8-R_{sc}x_7+d_3u_3-L_{sc}\dot{\alpha}_7\Bigg] + \frac{1}{\gamma_3}(\dot{d}_3-\dot{\widehat{d}}_3) \nonumber \\
=& e_7\Bigg[-x_8-R_{sc}x_7+d_3u_3-L_{sc}\dot{\alpha}_7\Bigg] + \widetilde{d}_3\Bigg[ e_7u_3 + \frac{\dot{d}_3-\dot{\widehat{d}}_3}{\gamma_3}\Bigg]
\end{align}
If the duty cycle $u_3$ and $\dot{\widehat{d}}_3$ are chosen as follows:
\begin{align}
u_3=& \frac{-K_7e_7+x_8+R_{sc}x_7+L_{sc}\dot{\alpha}_7}{\widehat{d}_3}  \\
\dot{\widehat{d}}_3 =& \gamma_3e_7u_3 
\end{align}
Applying this $u_3$, $\dot{W}_7$ becomes $\dot{W}_7=-K_7e^2_7$ which means $\dot{W}_7\leq0$ when $K_7\geq0$. \\ \\
Applying the designed controllers and virtual controllers, it is evident that the total Lyapunov function of the entire system $V= V_1+ W_{3}+ W_4 + W_{6}+ W_{7} + W_8 + W_{2,5,9} \geq 0$ and $\dot{W}\leq0$. This means that all the states in the microgrid are stabilized and the state errors will eventually converge to zero.
\section{Results}\label{sec:aobsresults}
\begin{figure*}[!ht]
\captionsetup[subfigure]{aboveskip=-1pt,belowskip=-1pt}
\centering
\begin{subfigure}{0.32\textwidth}
  \centering
\includegraphics[width =  5cm,height=3cm]{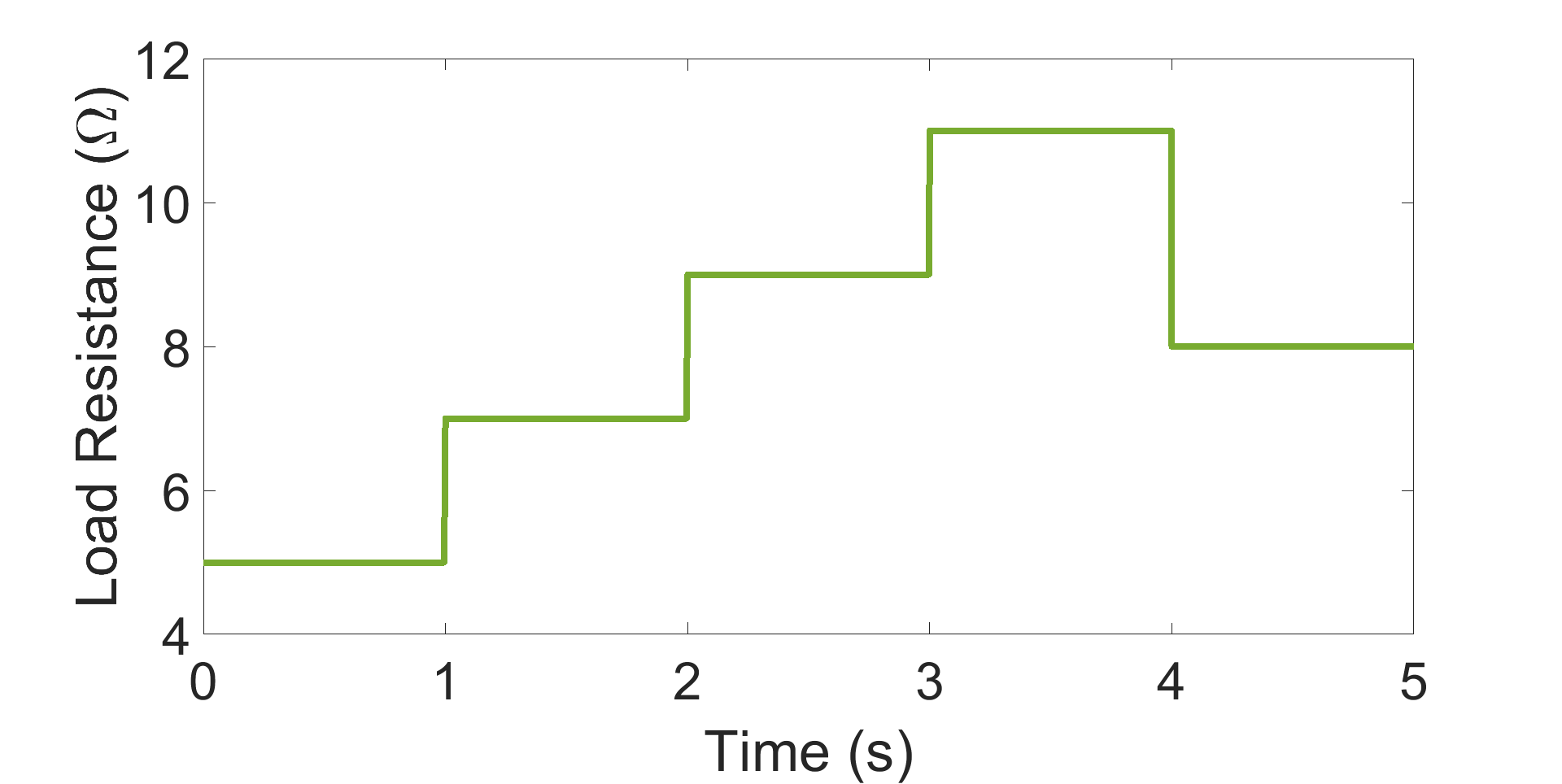}
\caption{\scriptsize{case 1: Change in Load}}
\label{fig:loadchange_loadresistance}
\end{subfigure}
\begin{subfigure}{.32\textwidth}
  \centering
\includegraphics[width =  5cm,height=3cm]{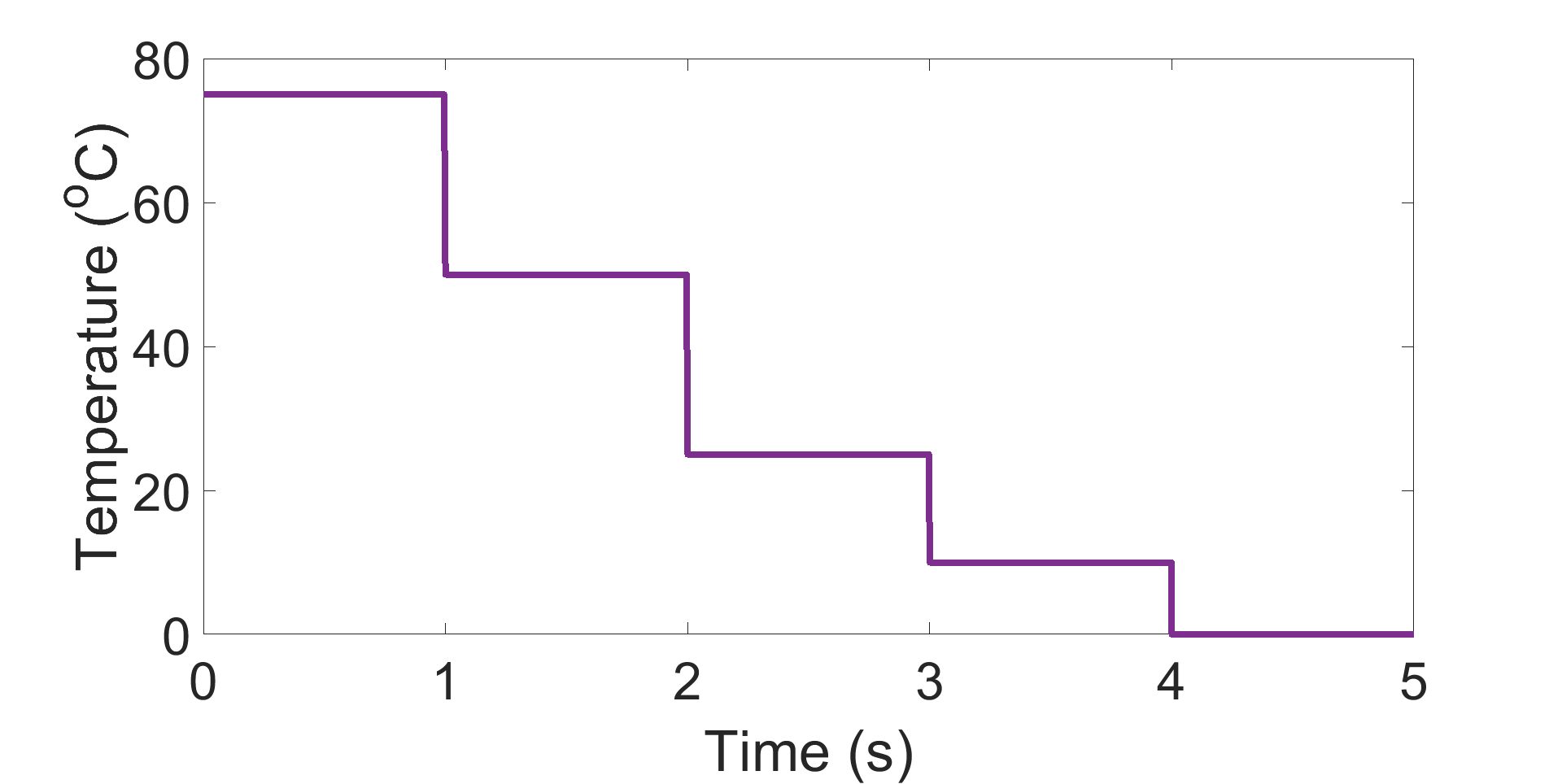}
\caption{\scriptsize{case 2: Change in Temperature}}
\label{fig:tempchange_tempvariation}
\end{subfigure}
\begin{subfigure}{.32\textwidth}
  \centering
\includegraphics[width =  5cm,height=3cm]{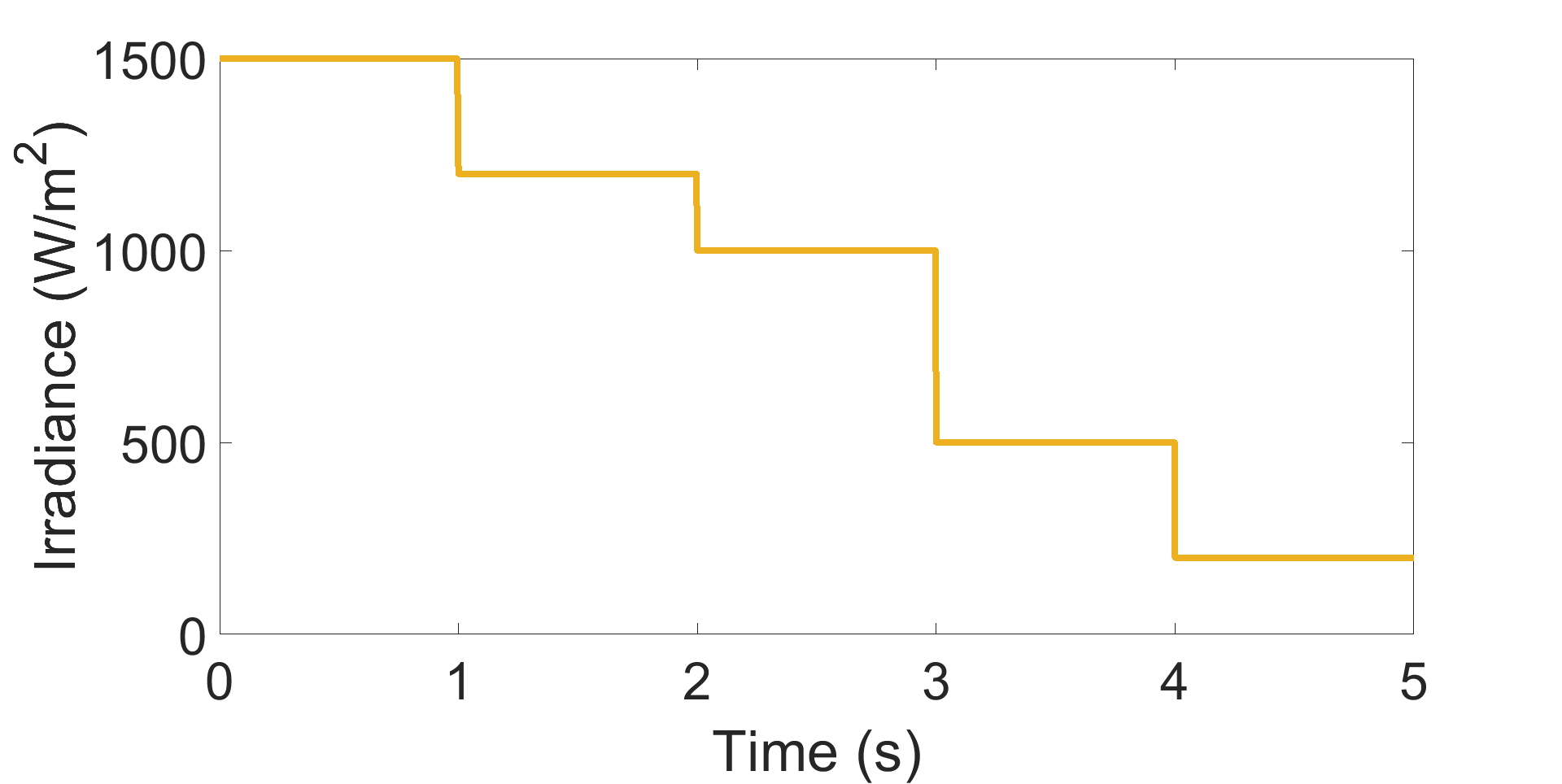}
\caption{\scriptsize{case 3: Change in Irradiance}}
\label{fig:irrchange_irrprofile}
\end{subfigure}
\caption{Test cases for validating the proposed algorithm}
\label{fig:testcases}
\end{figure*}

The reference output of the DCSSMG is $y_d=[x_{1ref} ~x_{4ref} ~x_{9ref}]$ where $x_{9ref}$ respresents the DCSSMG system voltage and $x_{1ref}$ is the PV array output voltage and $x_{4ref}$ is the reference battery output voltage. 
\begin{table}
\begin{tabular}{c}
\includegraphics[width=1.0\textwidth]{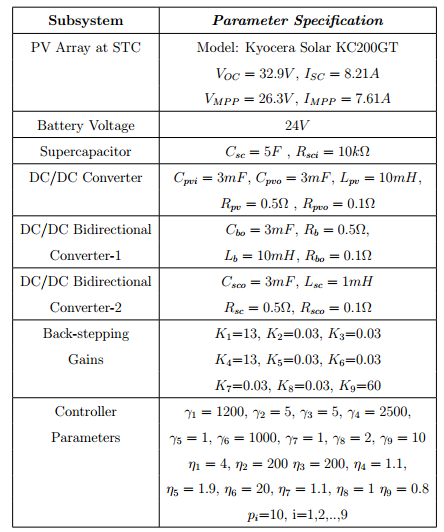}
\end{tabular}
\caption{DCSSMG Specifications}
\label{table:parameters_distest}
\end{table}
It is to be noted that $x_{9ref}$ always set to 40$V$, $x_{1ref}$ is the MPPT voltage set by the secondary controller depending on the PV characteristic of the chosen PV array, and $x_{4ref}$ is decided based on the power requirement from/to the battery. Table \ref{table:parameters_distest} shows the details of the various components used in the DCSSMG system. The backstepping controller parameters and the adaptive observer constants which are tuned to provide the best controller performance are also mentioned. A total of four observers are designed and the efficacy of the proposed controller is compared with that of a state of the art back-stepping algorithm in \cite{iovinetase17}. All simulations are carried out using MATLAB 2018b software. Fig.\ref{fig:testcases} shows the different test cases that have been selected for checking the efficiacy of the proposed algorithm.
\vspace{-0.5cm}
\subsection{Case-1: Change in Load}
\begin{figure}[H]
\captionsetup[subfigure]{aboveskip=-1pt,belowskip=-1pt}
\centering
\begin{subfigure}{0.32\textwidth}
  \centering
\includegraphics[width =  5cm,height=3cm]{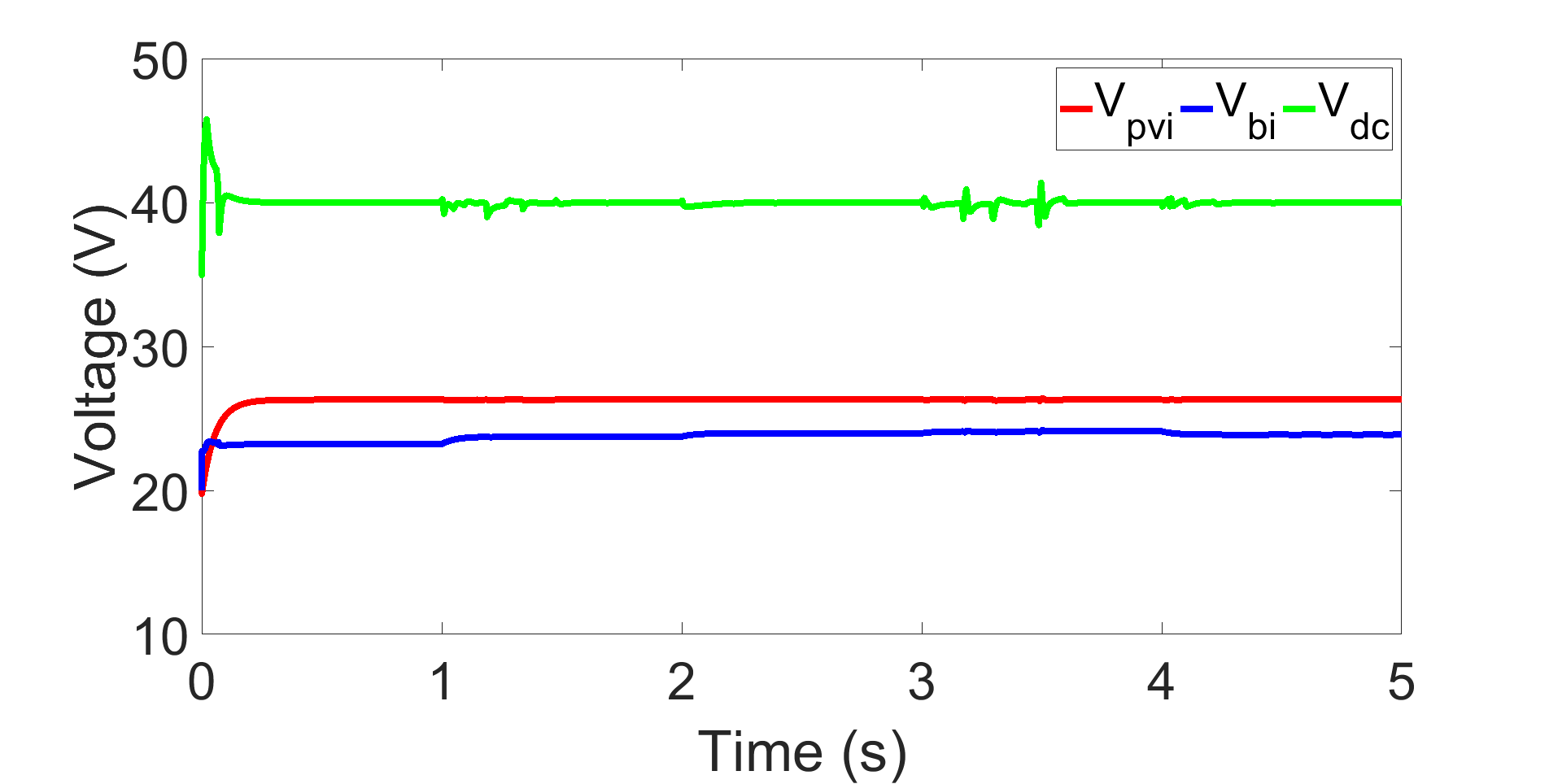}
\caption{\scriptsize{Backstepping and Known Disturbances: Outputs.}}
\label{fig:loadchange_outputs_bstep}
\end{subfigure}
\begin{subfigure}{.32\textwidth}
  \centering
\includegraphics[width =  5cm,height=3cm]{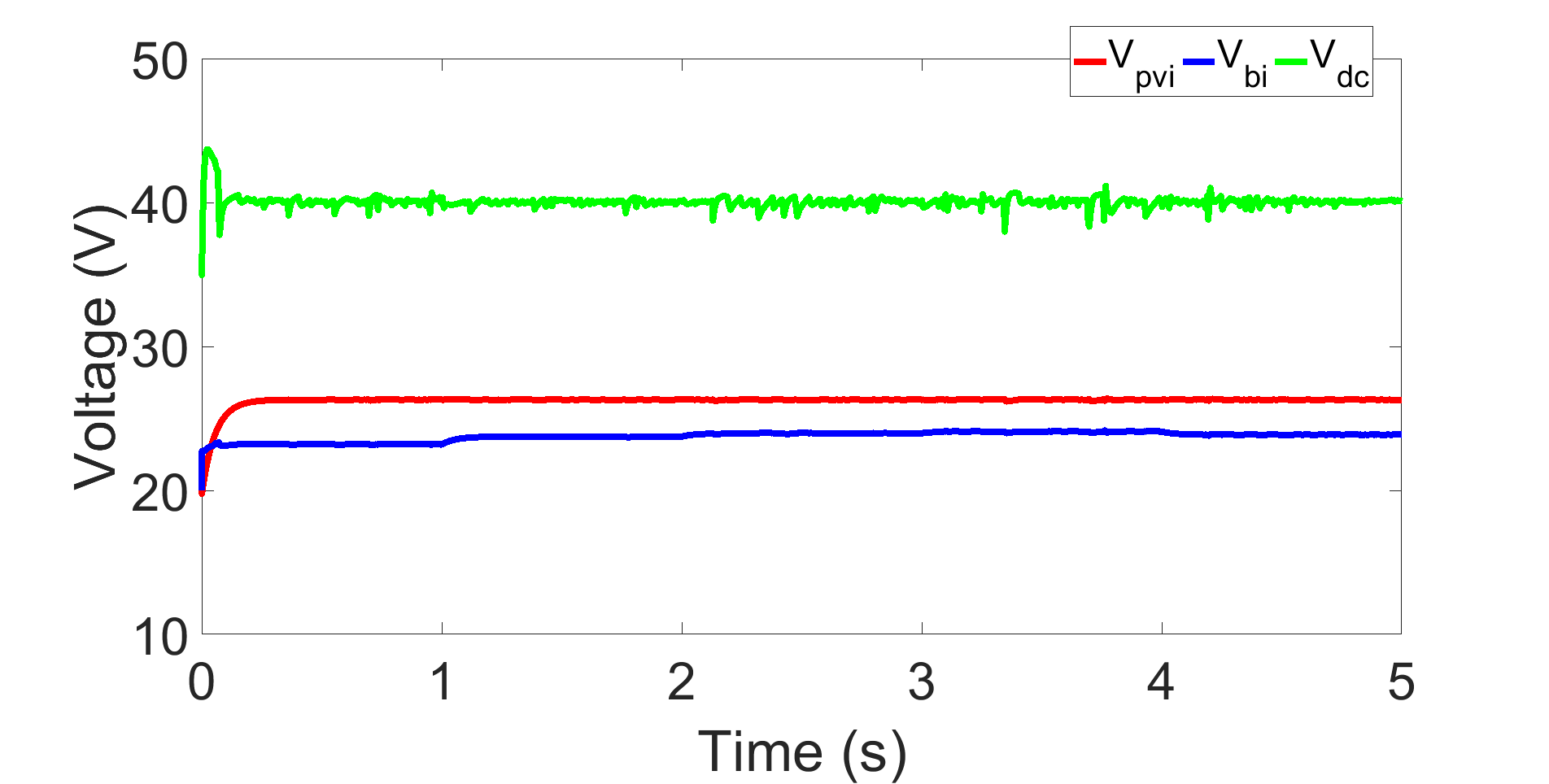}
\caption{\scriptsize{Backstepping and Unknown Disturbances:  Outputs.}}
\label{fig:loadchange_outputs_bstepunknown}
\end{subfigure}
\begin{subfigure}{.32\textwidth}
  \centering
\includegraphics[width =  5cm,height=3cm]{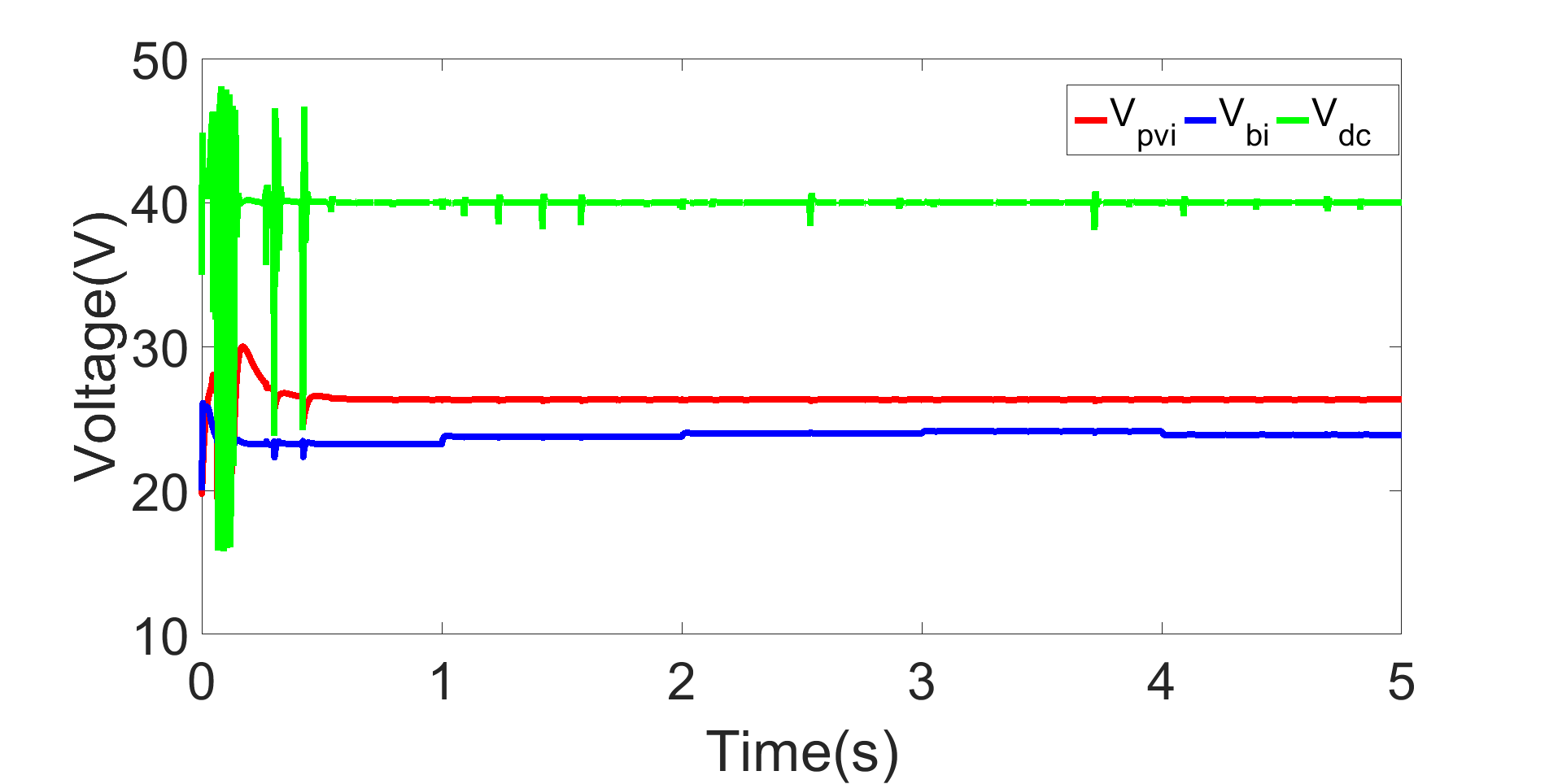}
\caption{\scriptsize{Proposed Control and Unknown Disturbances: Outputs.}}
\label{fig:loadchange_outputs_distest}
\end{subfigure}\\
\begin{subfigure}{.32\textwidth}
  \centering
\includegraphics[width =  5cm,height=3cm]{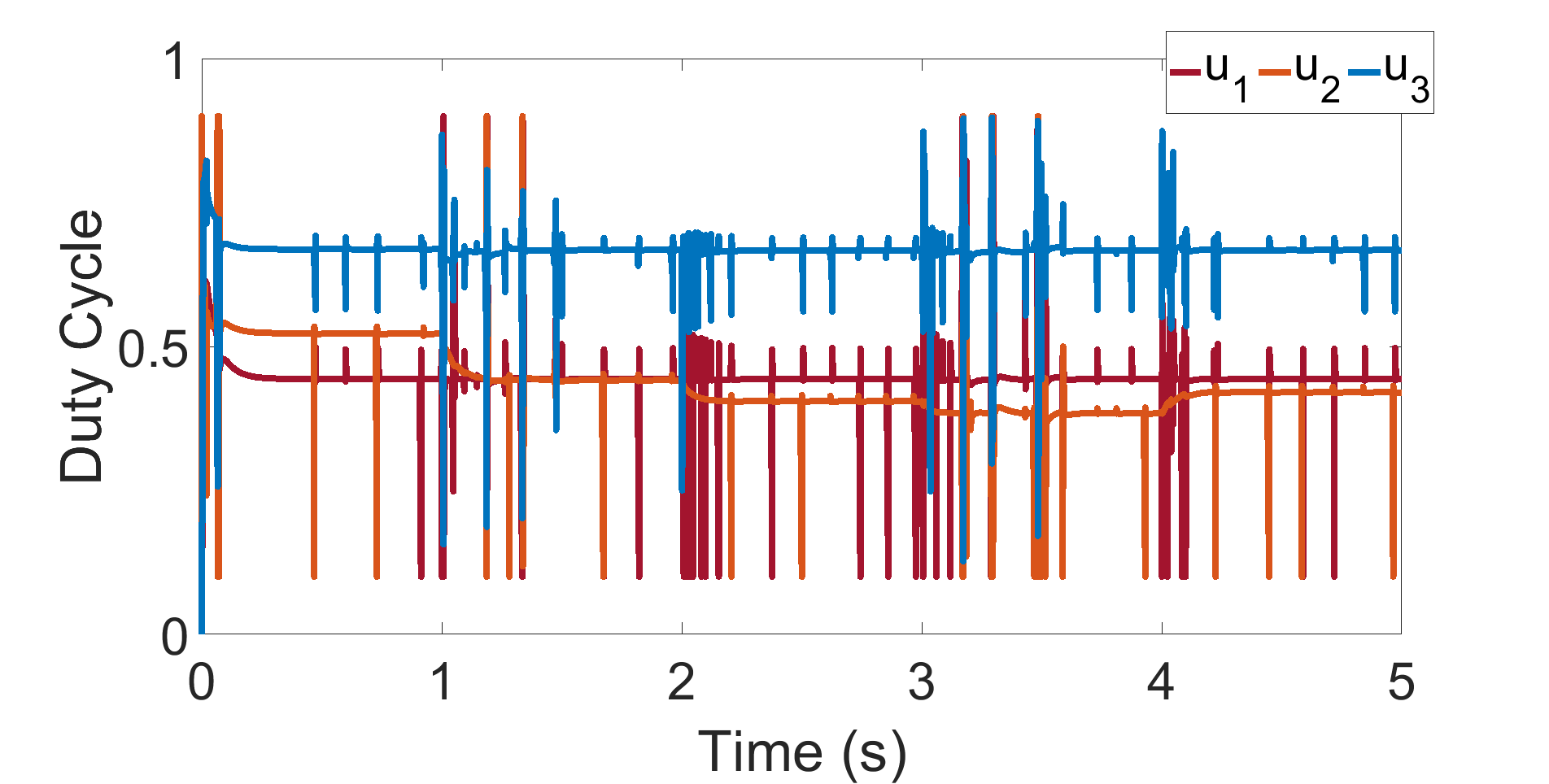}
\caption{\scriptsize{Backstepping and Known Disturbances: Inputs.}}
\label{fig:loadchange_inputs_bstep}
\end{subfigure}
\begin{subfigure}{.32\textwidth}
  \centering
\includegraphics[width =  5cm,height=3cm]{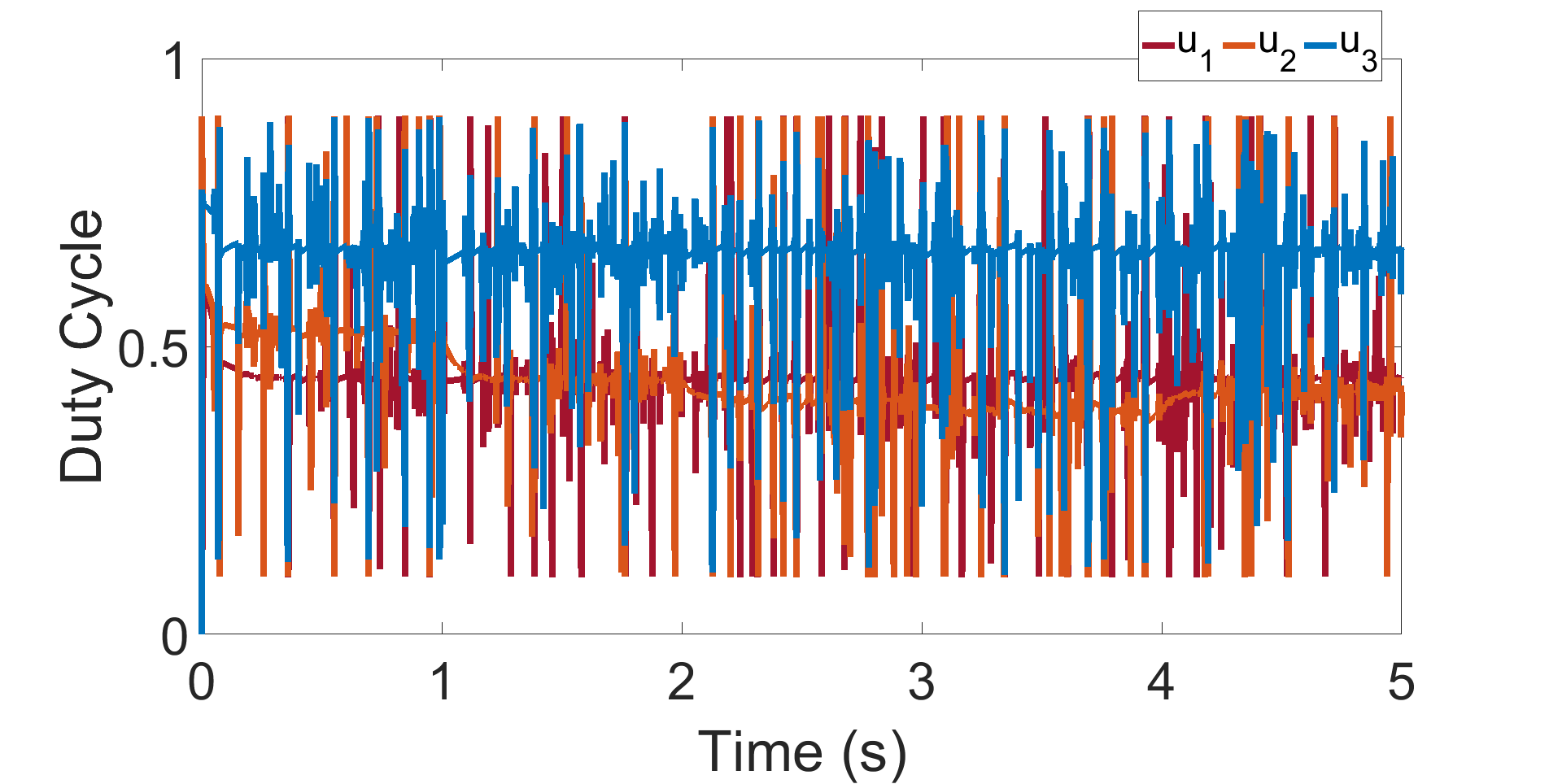}
\caption{\scriptsize{Backstepping and Unknown Disturbances: Inputs.}}
\label{fig:loadchange_inputs_bstepunkown}
\end{subfigure}
\begin{subfigure}{.32\textwidth}
  \centering
\includegraphics[width =  5cm,height=3cm]{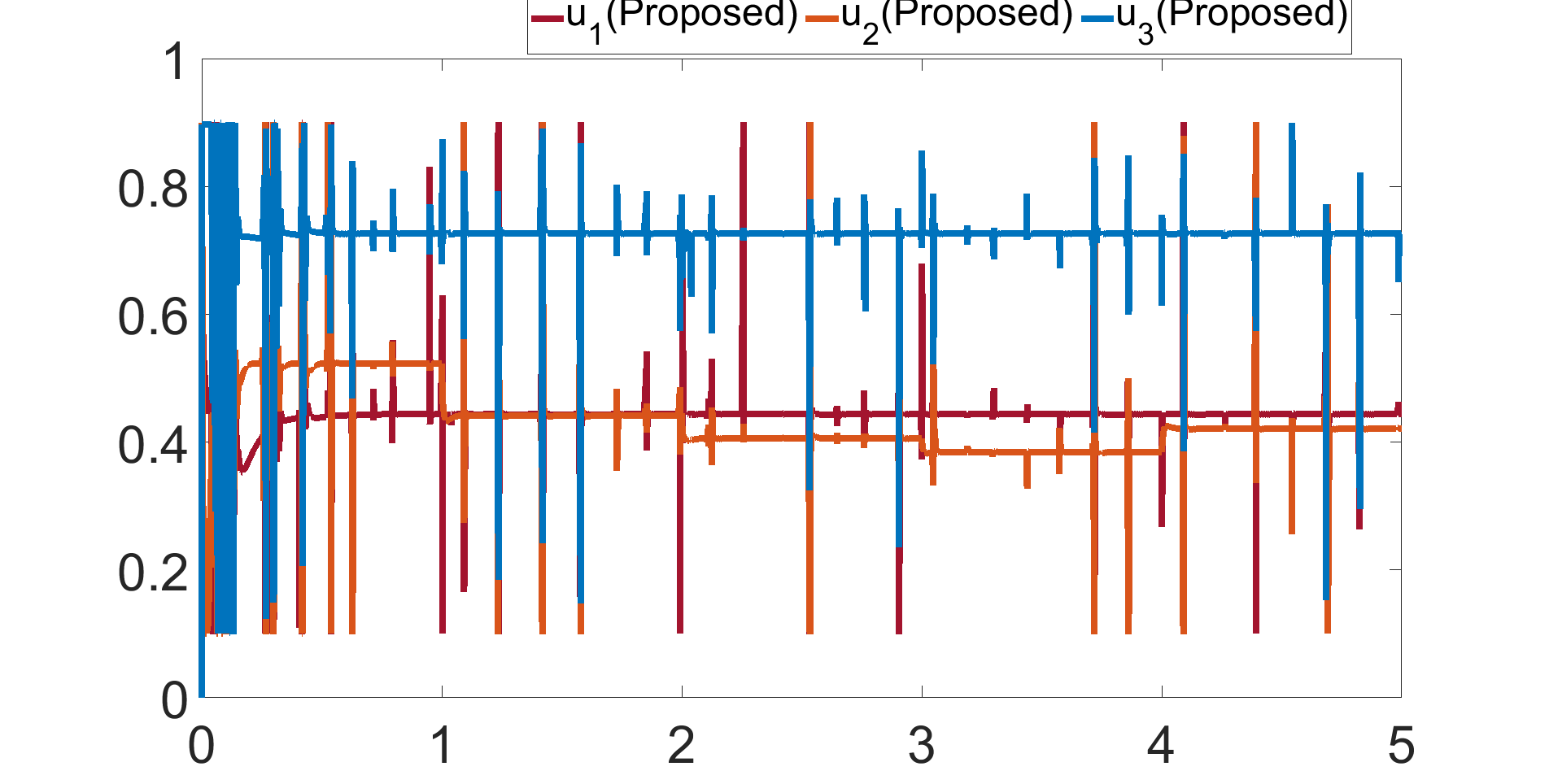}
\caption{\scriptsize{Proposed Control and Unknown Disturbances: Inputs.}}
\label{fig:loadchange_inputs_distest}
\end{subfigure}\\
\begin{subfigure}{.32\textwidth}
  \centering
\includegraphics[width =  5cm,height=3cm]{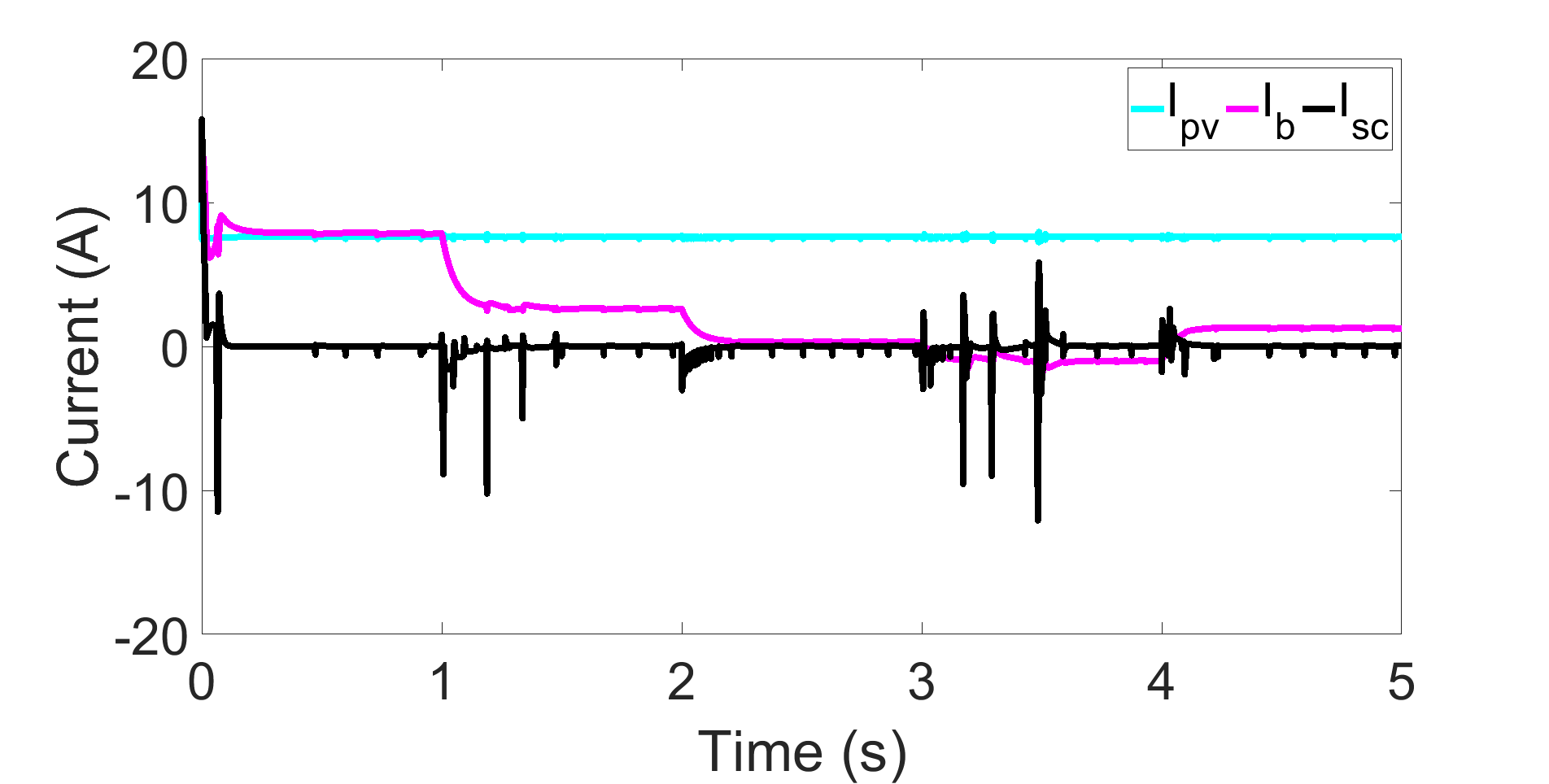}
\caption{\scriptsize{Backstepping and Known Disturbances: Currents.}}
\label{fig:loadchange_currents_bstep}
\end{subfigure}
\begin{subfigure}{.32\textwidth}
  \centering
\includegraphics[width =  5cm,height=3cm]{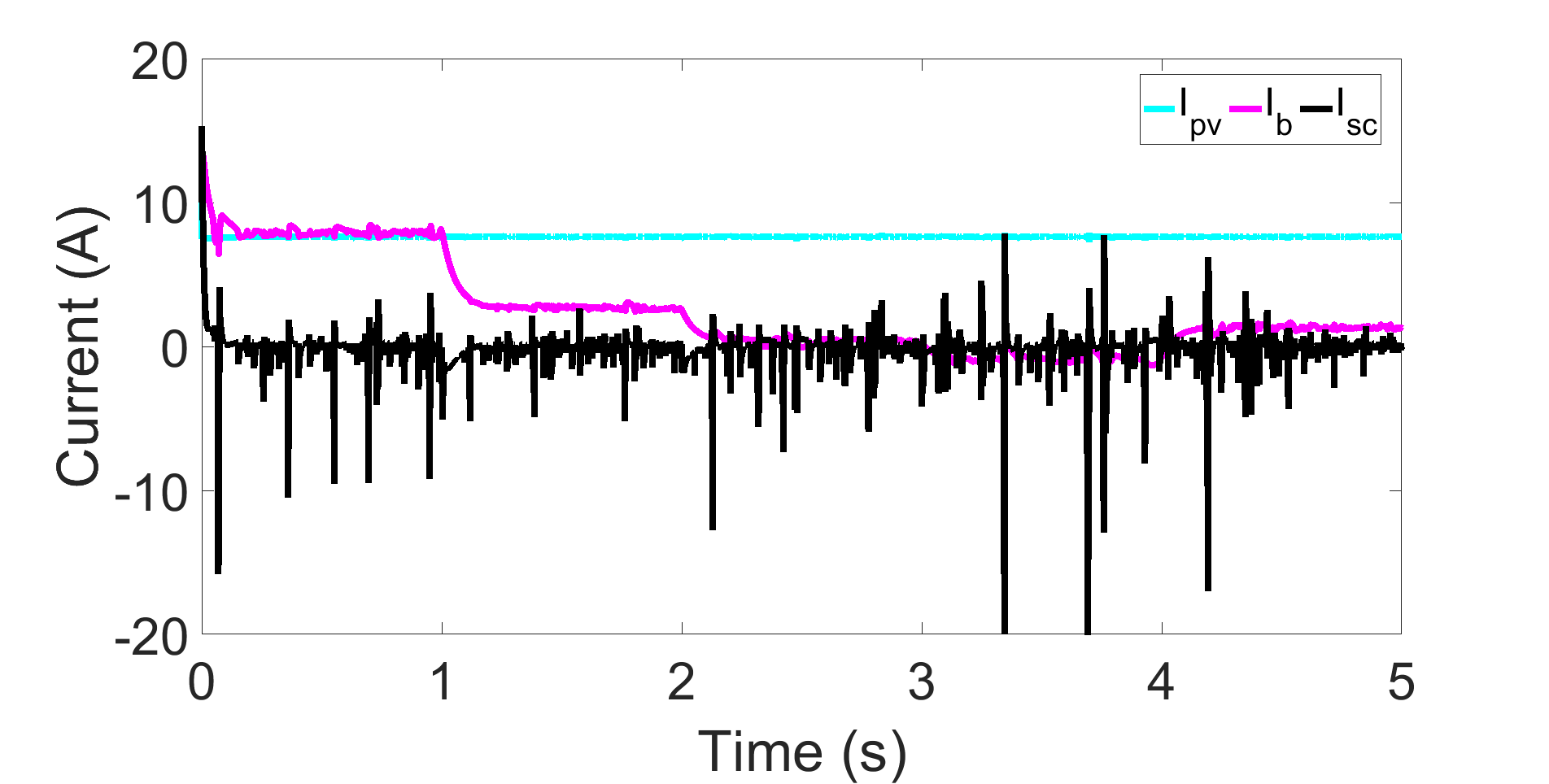}
\caption{\scriptsize{ Backstepping and Unknown Disturbances: Currents.}}
\label{fig:loadchange_currents_bstepunkown}
\end{subfigure}
\begin{subfigure}{.32\textwidth}
  \centering
\includegraphics[width =  5cm,height=3cm]{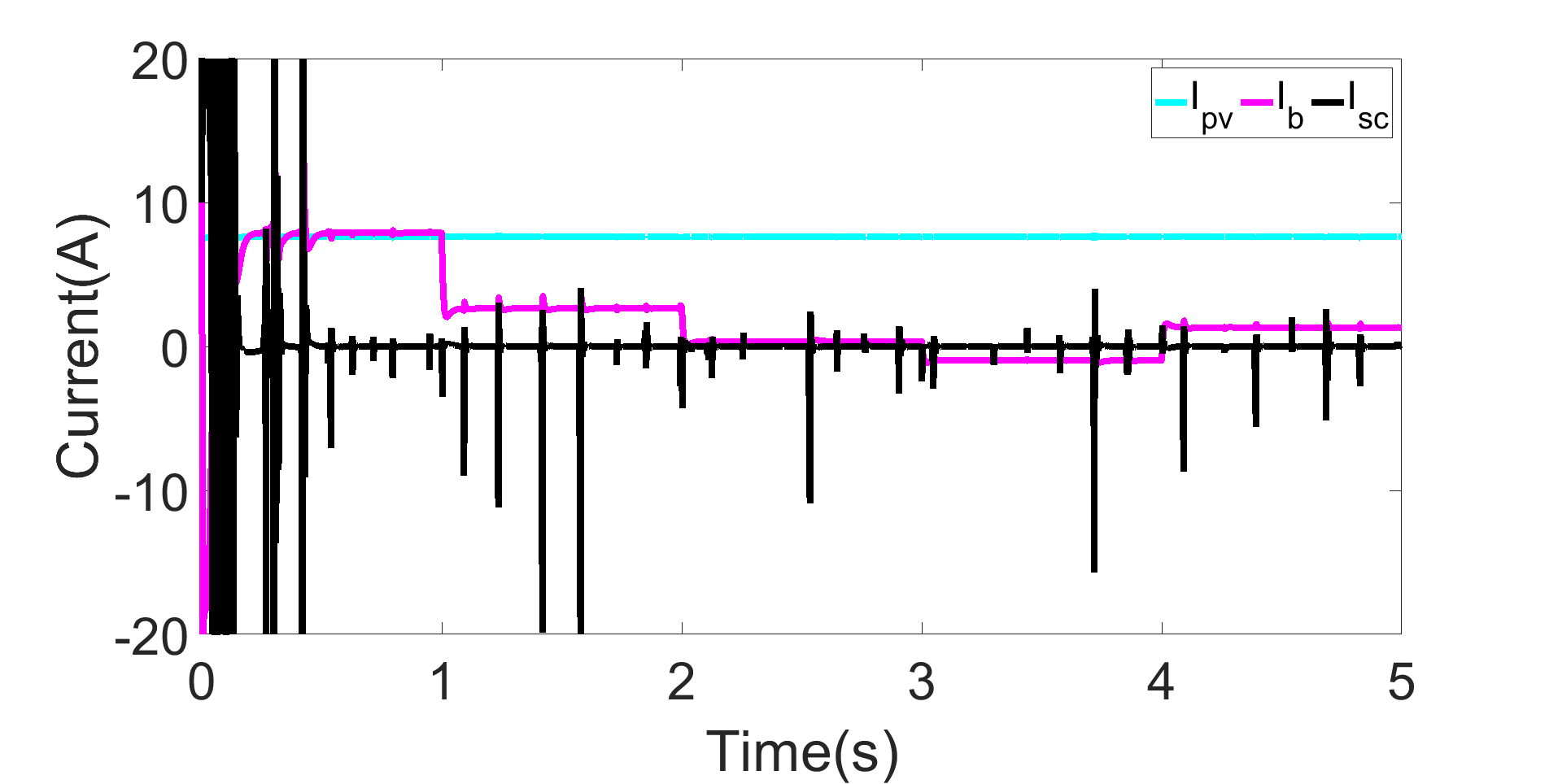}
\caption{\scriptsize{Proposed Control and Unknown Disturbances: Currents.}}
\label{fig:loadchange_currents_distest}
\end{subfigure}
\\
\begin{subfigure}{.32\textwidth}
  \centering
\includegraphics[width =  5cm,height=3cm]{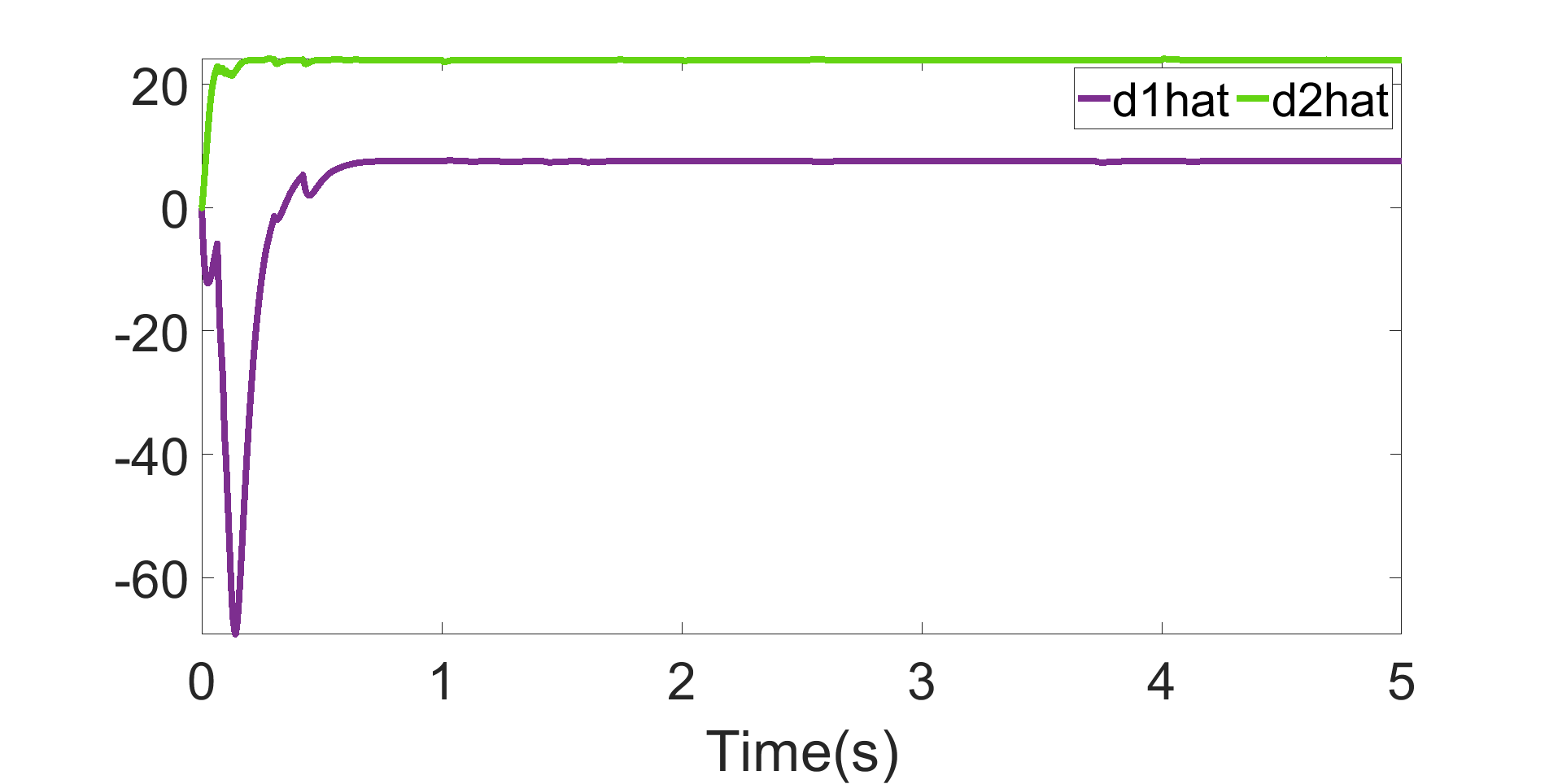}
\caption{\scriptsize{Proposed Control: $d_1~d_2$ Estimation.}}
\label{fig:loadchange_d1d2hat}
\end{subfigure}
\begin{subfigure}{.32\textwidth}
  \centering
\includegraphics[width =  5cm,height=3cm]{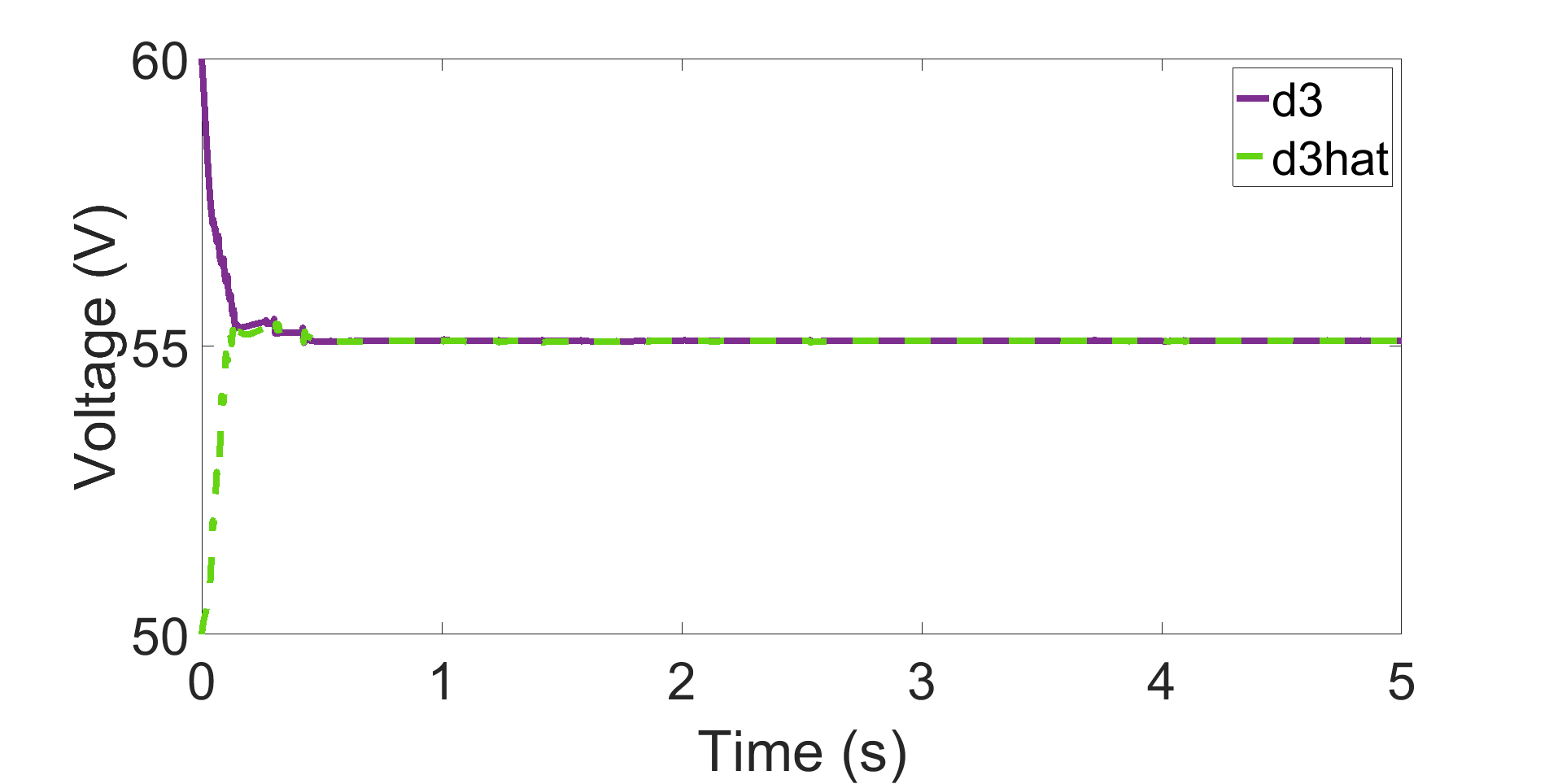}
\caption{\scriptsize{ Proposed Control: $d_3$ Estimation.}}
\label{fig:loadchange_d3}
\end{subfigure}
\begin{subfigure}{.32\textwidth}
  \centering
\includegraphics[width =  5cm,height=3cm]{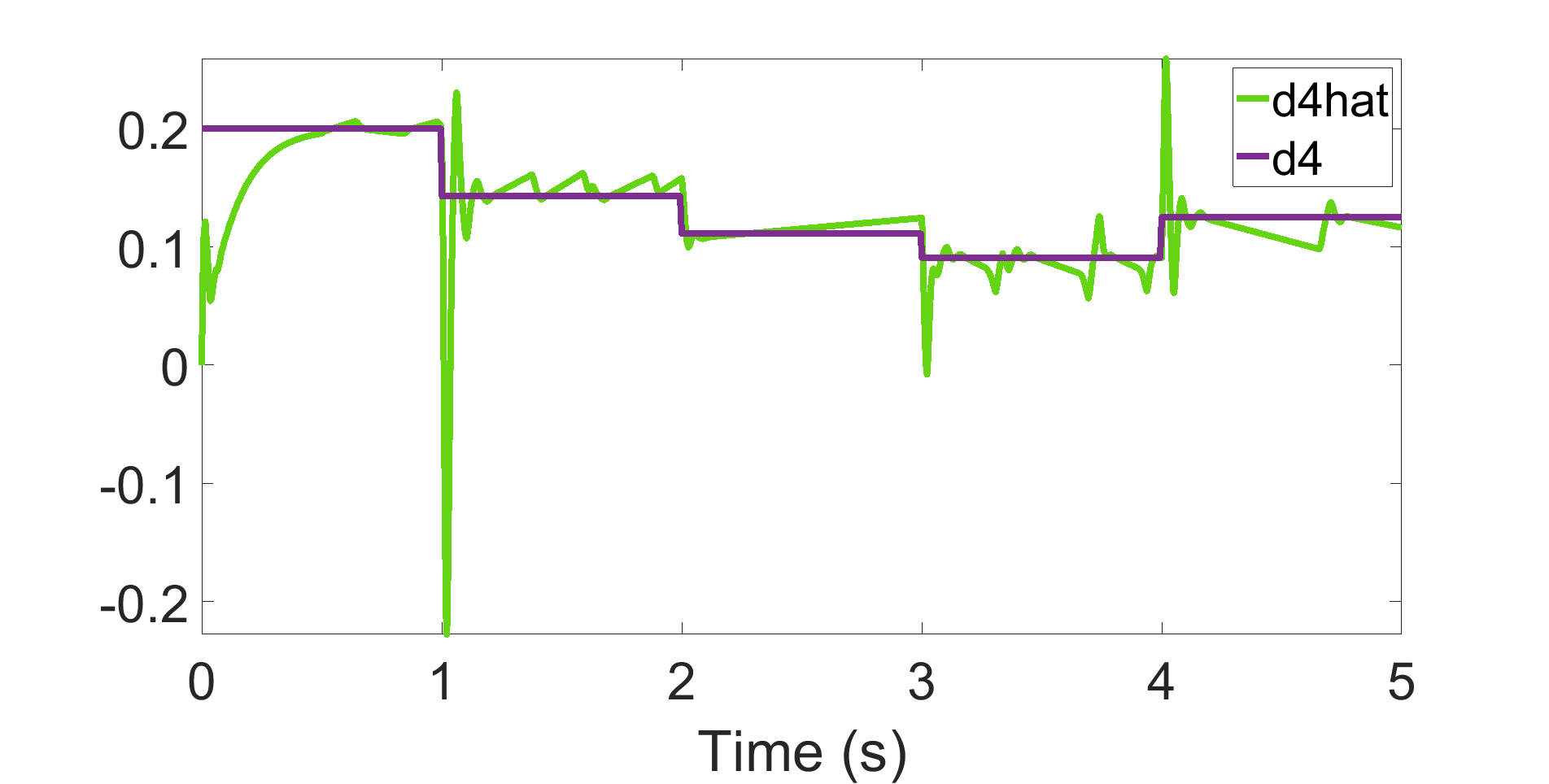}
\caption{\scriptsize{Proposed Control: $d_4$ Estimation.}}
\label{fig:loadchange_d4hat}
\end{subfigure}
\caption{Case-1: Validation of the proposed controller against state-of-the-art controller \cite{iovinetase17}(Backstepping) for load change. }
\label{fig:exp_results}
\end{figure}

This case is meant to test the developed controller when the atmospheric conditions remain constant and only the DCSSMG load changes. The irradiance and temperature are maintained steadily at 1000 $W/m^2$ and 25 $^{o}$C. The value of $x_{1ref}$ as per the  PV characteristic is deemed as $26.31~V$. The load resistance is varied from 5$\Omega$ to 7$\Omega$, 9$\Omega$, 11$\Omega$ and then to 8$\Omega$ at 1$s$, 2$s$, 3$s$ and 4$s$ respectively as in Fig. \ref{fig:loadchange_loadresistance} which results in the value of $x_{4ref}$ being set to 23.21 $V$, $23.75~V$, $23.96~V$, $24.09~V$ and $23.86~V$ respectively. When load is changed, it gets reflected in the corresponding disturbance $d_4$ of the system model. Figures \ref{fig:loadchange_outputs_bstep},\ref{fig:loadchange_inputs_bstep},\ref{fig:loadchange_currents_bstep} show the output voltages, input duty ratio and DCSSMG branch currents when the baseline backstepping algorithm is applied and all the disturbances are known. Figures \ref{fig:loadchange_outputs_bstepunknown}, \ref{fig:loadchange_inputs_bstepunkown},\ref{fig:loadchange_currents_bstepunkown} show the same when the baseline backstepping algorithm is applied and the disturbance values are not known. For a similar unknown situation, when the proposed controller is applied, the results obtained can be seen in figures \ref{fig:loadchange_outputs_distest},\ref{fig:loadchange_inputs_distest},\ref{fig:loadchange_currents_distest}. Finally, figures \ref{fig:loadchange_d1d2hat},\ref{fig:loadchange_d3} and \ref{fig:loadchange_d4hat} represent the evolution of observer value through time.
\vspace{-0.7cm}
\subsection{Case-2: Change in Temperature}
\begin{figure}[H]
\captionsetup[subfigure]{aboveskip=-1pt,belowskip=-1pt}
\centering
\begin{subfigure}{0.32\textwidth}
  \centering
\includegraphics[width =  5cm,height=3cm]{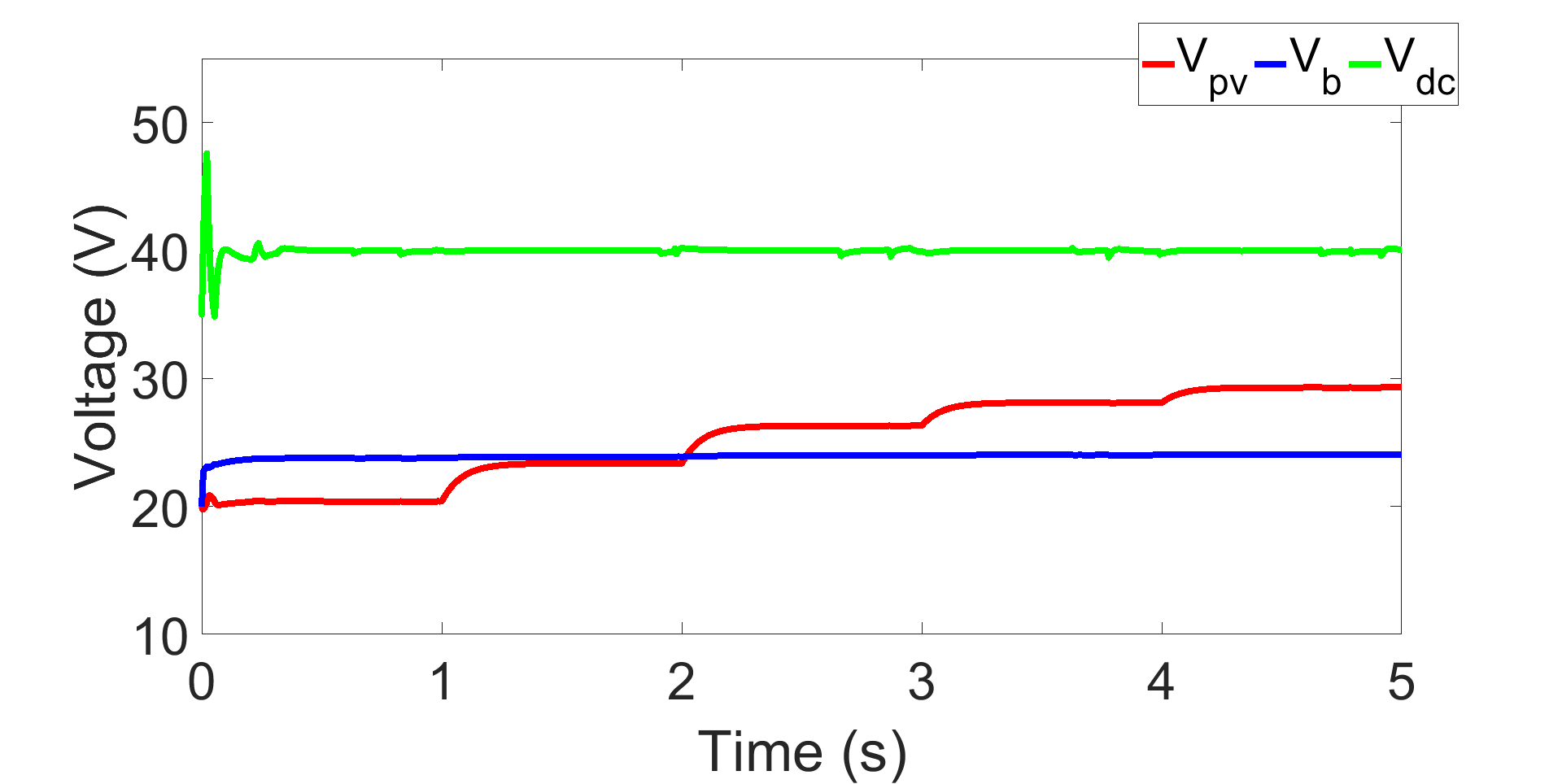}
\caption{\scriptsize{Backstepping and Known Disturbances: Outputs.}}
\label{fig:tempchange_outputs_bstep}
\end{subfigure}
\begin{subfigure}{.32\textwidth}
  \centering
\includegraphics[width =  5cm,height=3cm]{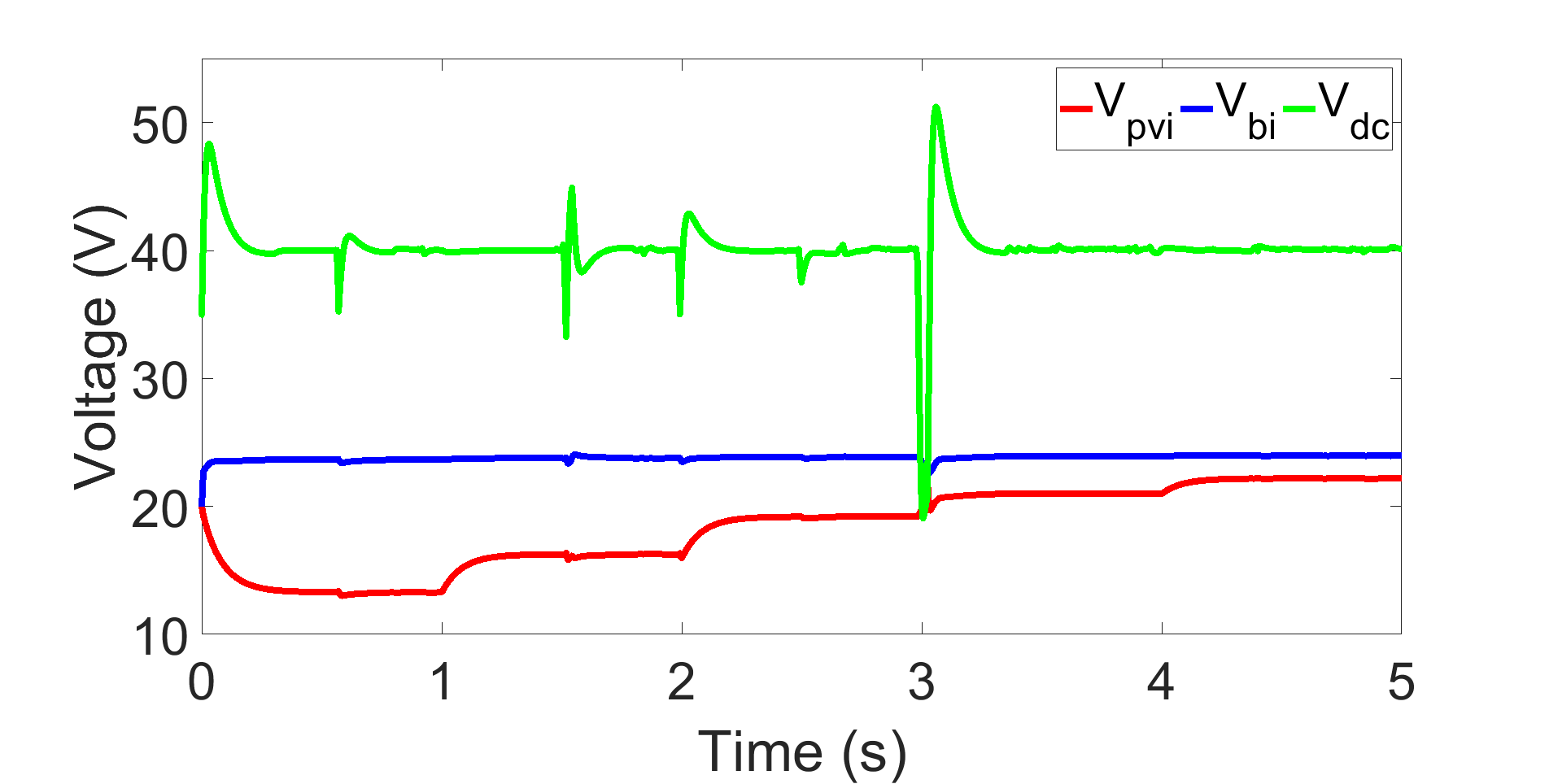}
\caption{\scriptsize{Backstepping and Unknown Disturbances: Outputs.}}
\label{fig:tempchange_outputs_bstepunknown}
\end{subfigure}
\begin{subfigure}{.32\textwidth}
  \centering
\includegraphics[width =  5cm,height=3cm]{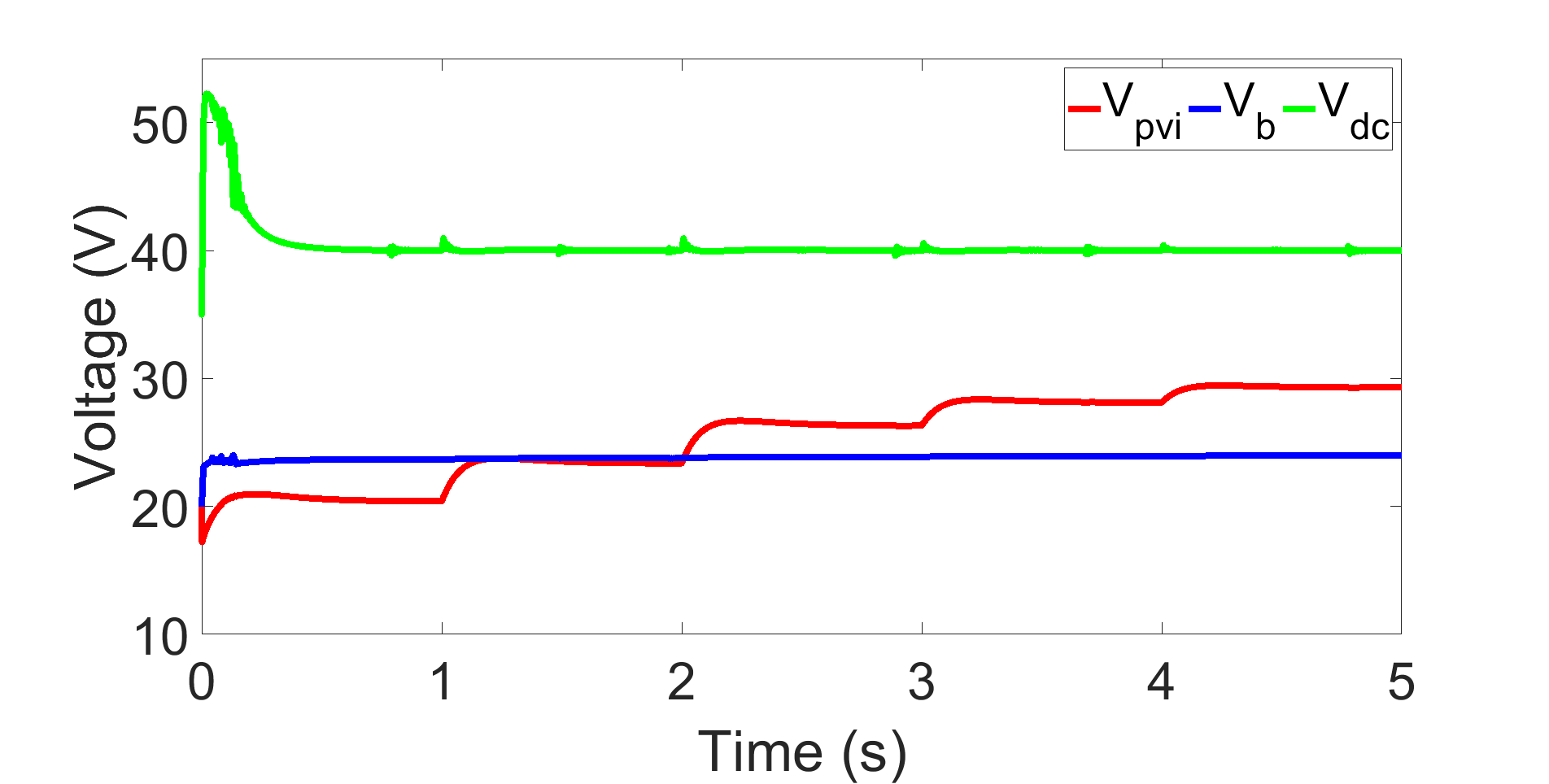}
\caption{\scriptsize{Backstepping and Unknown Disturbances: Outputs.}}
\label{fig:tempchange_outputs_distest}
\end{subfigure}\\
\begin{subfigure}{.32\textwidth}
  \centering
\includegraphics[width =  5cm,height=3cm]{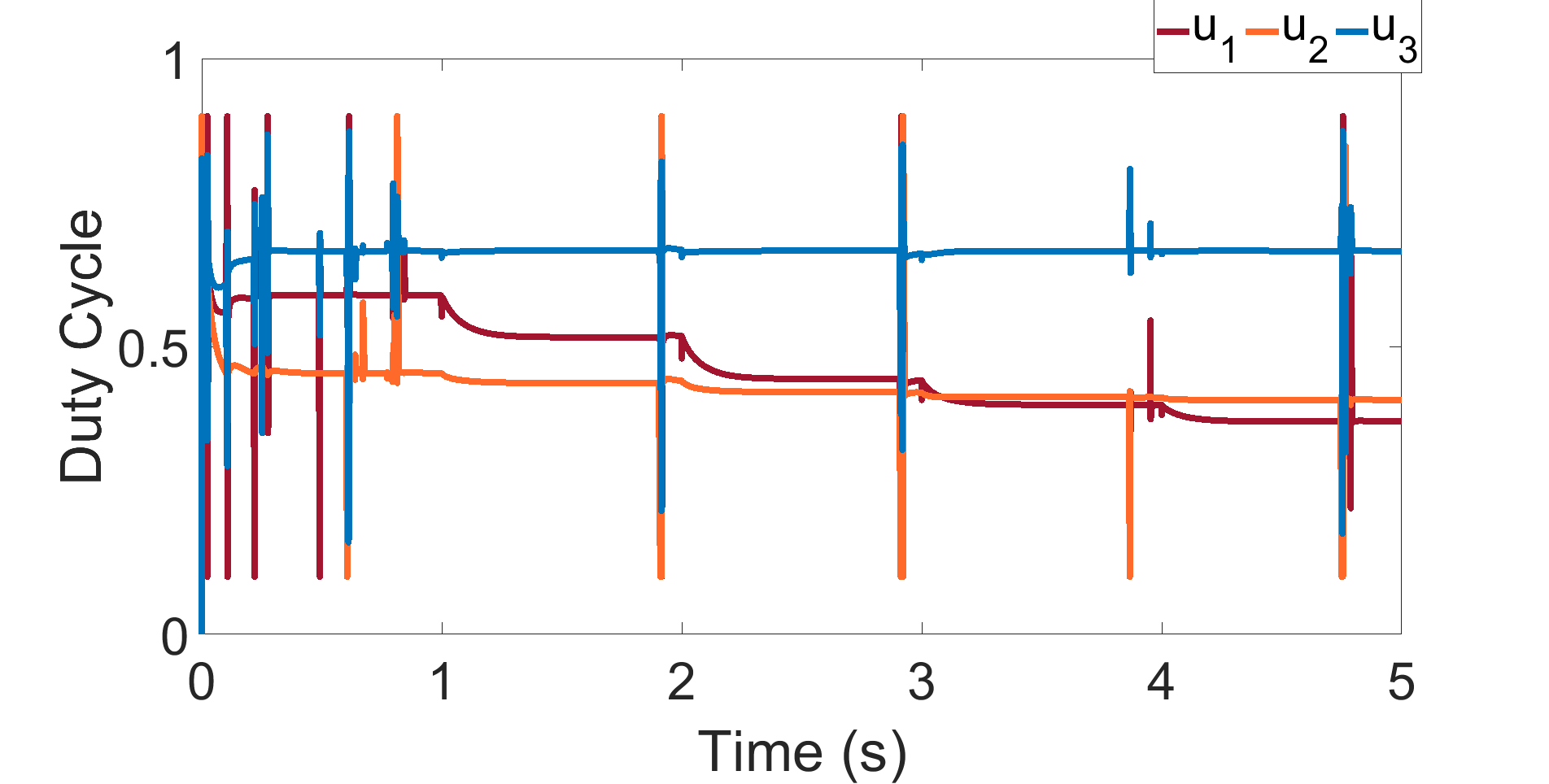}
\caption{\scriptsize{Backstepping and Known Disturbances: Inputs.}}
\label{fig:tempchange_inputs_bstep}
\end{subfigure}
\begin{subfigure}{.32\textwidth}
  \centering
\includegraphics[width =  5cm,height=3cm]{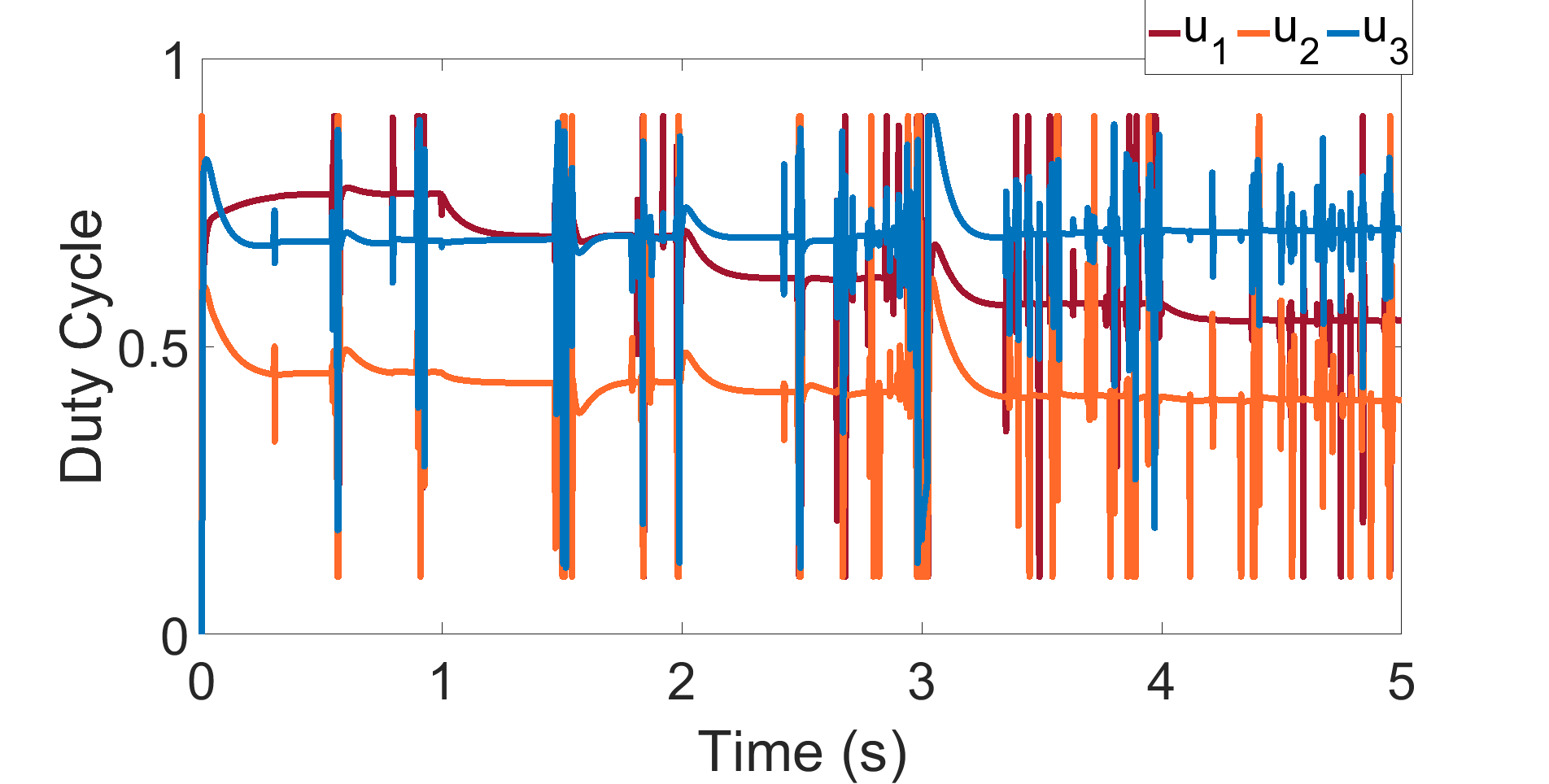}
\caption{\scriptsize{Backstepping and Unknown Disturbances: Inputs.}}
\label{fig:tempchange_inputs_bstepunknown}
\end{subfigure}
\begin{subfigure}{.32\textwidth}
  \centering
\includegraphics[width =  5cm,height=3cm]{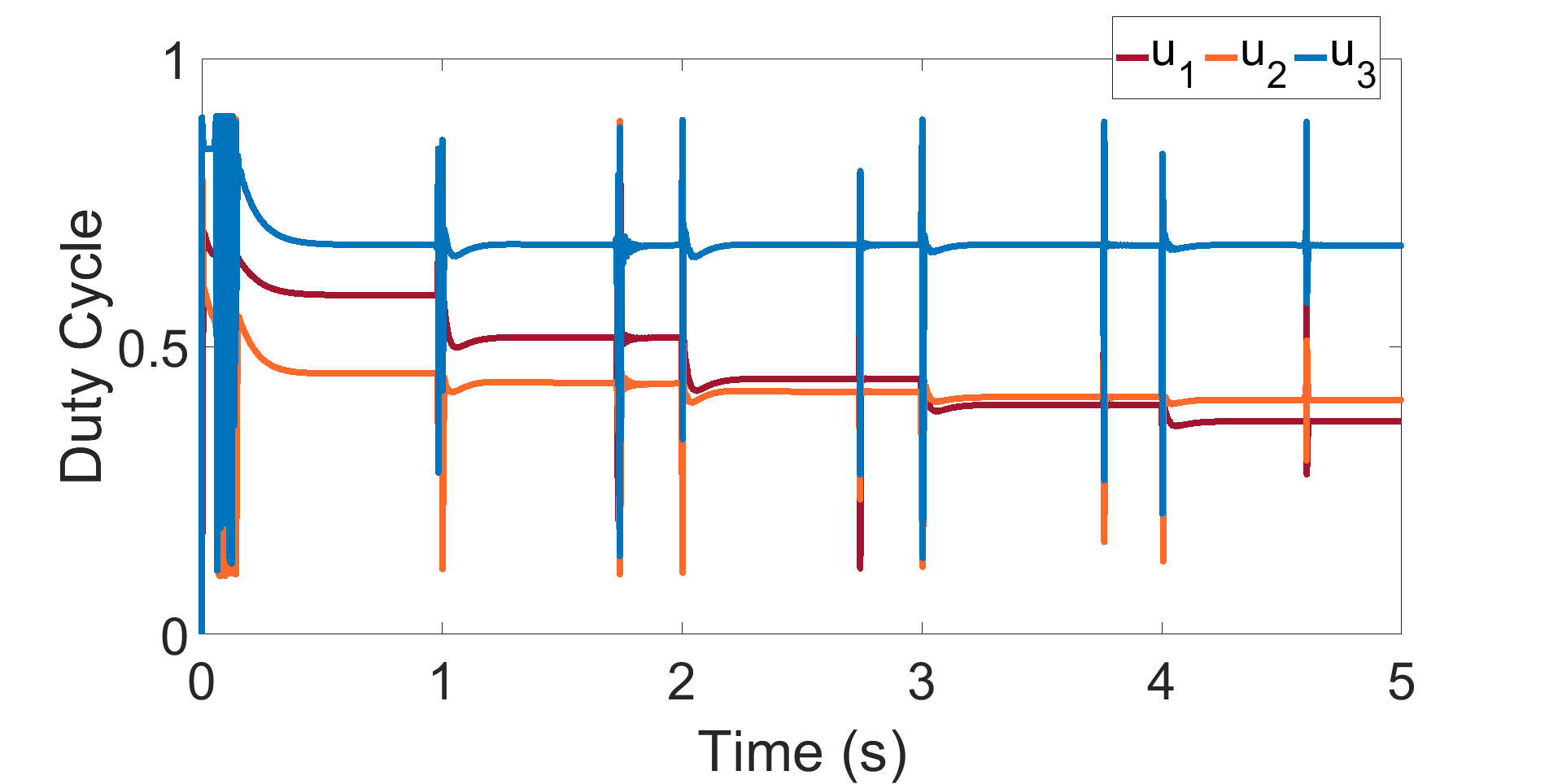}
\caption{\scriptsize{Proposed control and Unknown Disturbances: Inputs.}}
\label{fig:tempchange_inputs_distest}
\end{subfigure}\\
\begin{subfigure}{.32\textwidth}
  \centering
\includegraphics[width =  5cm,height=3cm]{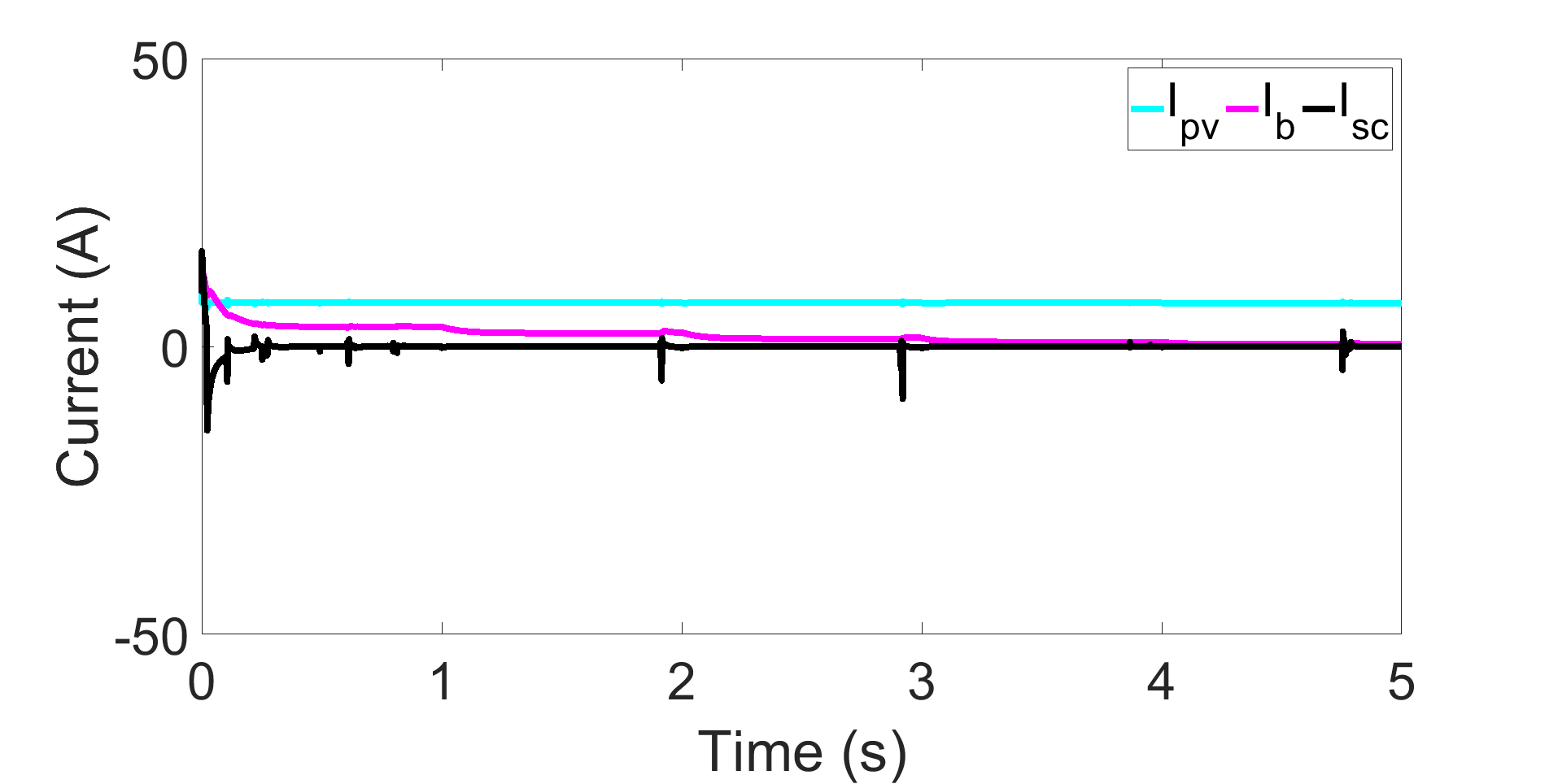}
\caption{\scriptsize{Backstepping and Known Disturbances: Currents.}}
\label{fig:tempchange_currents_bstep}
\end{subfigure}
\begin{subfigure}{.32\textwidth}
  \centering
\includegraphics[width =  5cm,height=3cm]{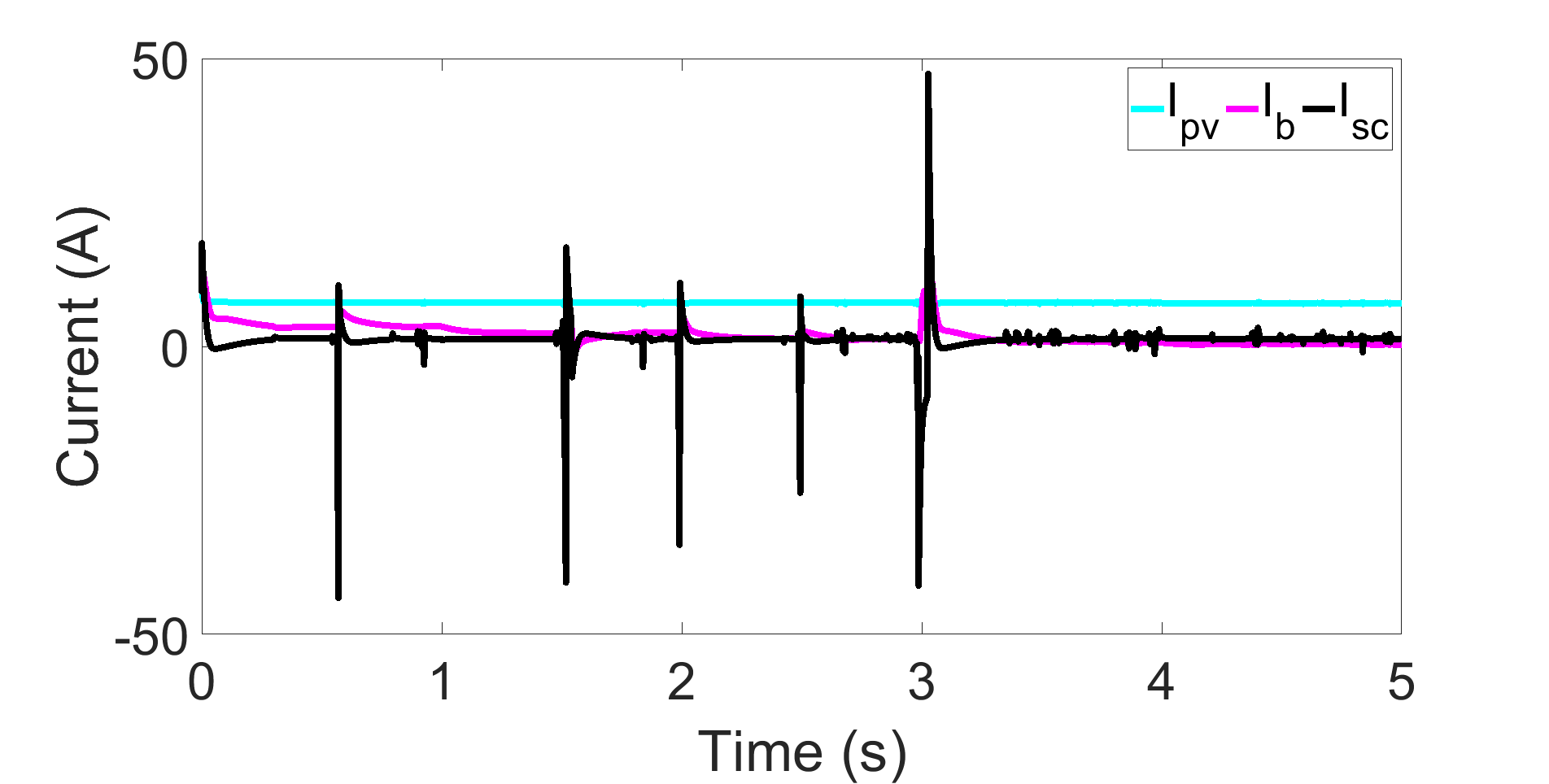}
\caption{\scriptsize{ Backstepping and Unknown Disturbances: Currents.}}
\label{fig:tempchange_currents_bstepunknown}
\end{subfigure}
\begin{subfigure}{.32\textwidth}
  \centering
\includegraphics[width =  5cm,height=3cm]{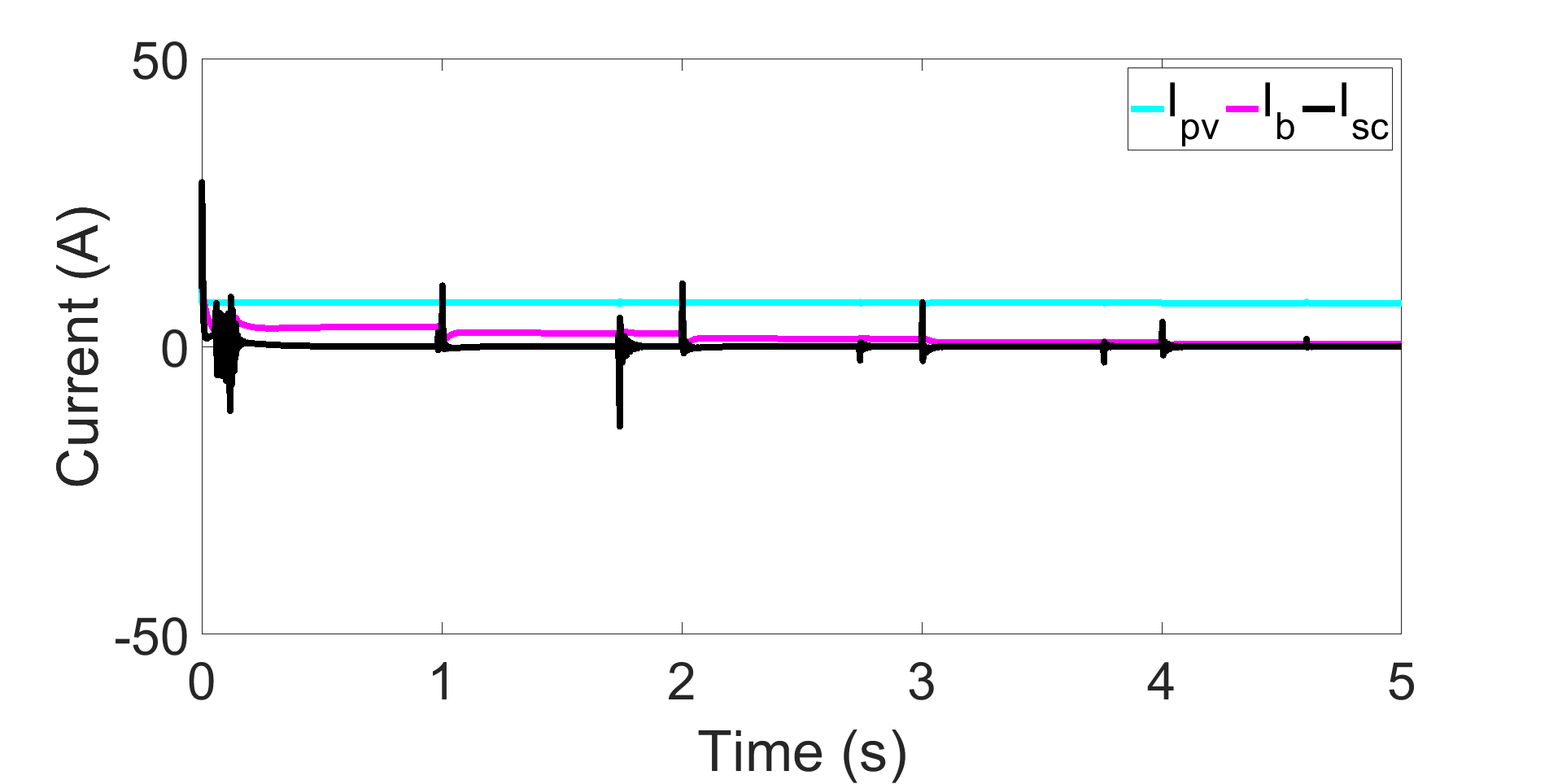}
\caption{\scriptsize{Proposed control and Unknown Disturbances: Currents.}}
\label{fig:tempchange_currents_distest}
\end{subfigure}\\
\begin{subfigure}{.32\textwidth}
  \centering
\includegraphics[width =  5cm,height=3cm]{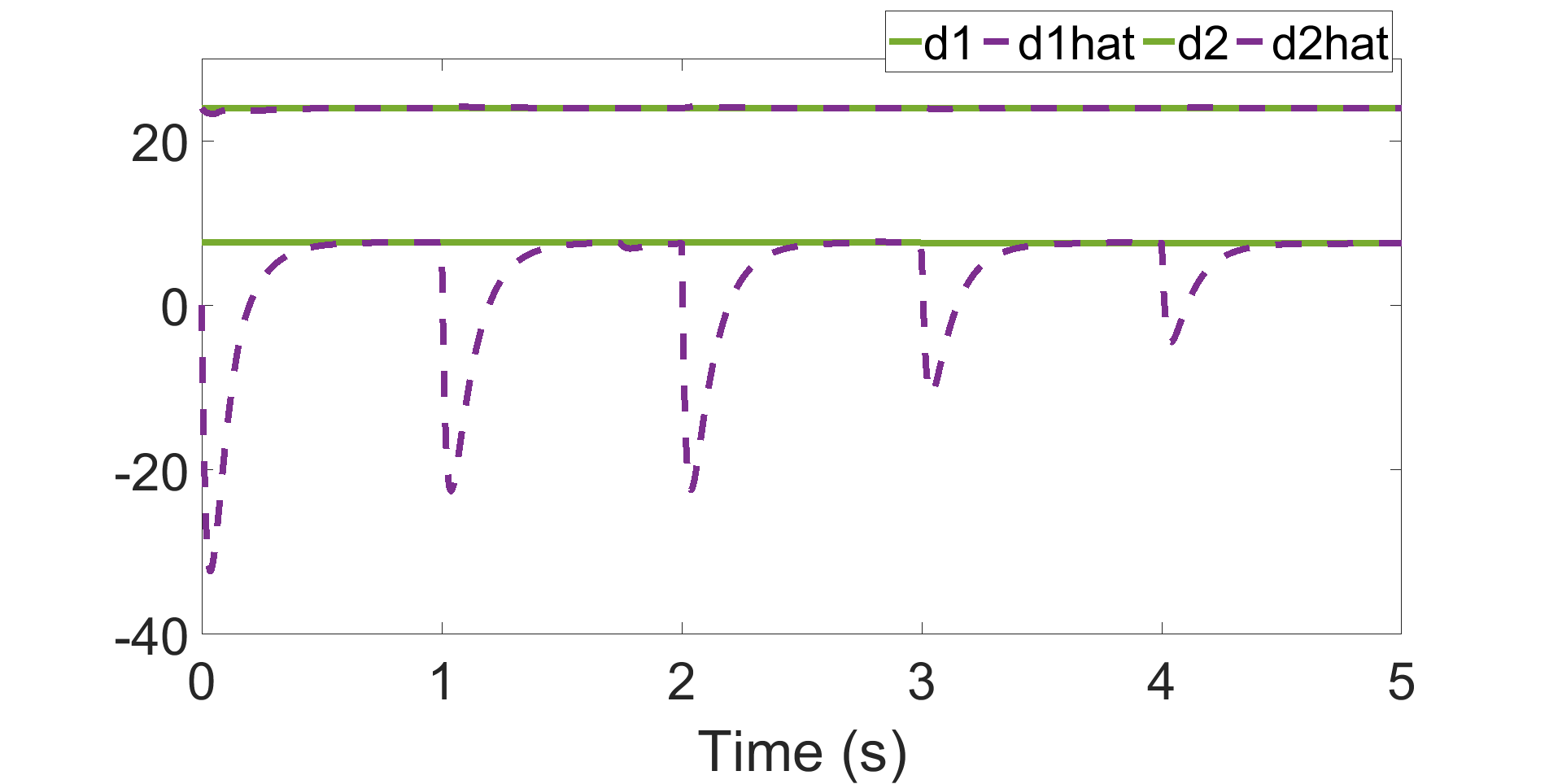}
\caption{\scriptsize{Proposed Control: $d_1,~d_2$ Estimation.}}
\label{fig:tempchange_d1d2hat_distest}
\end{subfigure}
\begin{subfigure}{.32\textwidth}
  \centering
\includegraphics[width =  5cm,height=3cm]{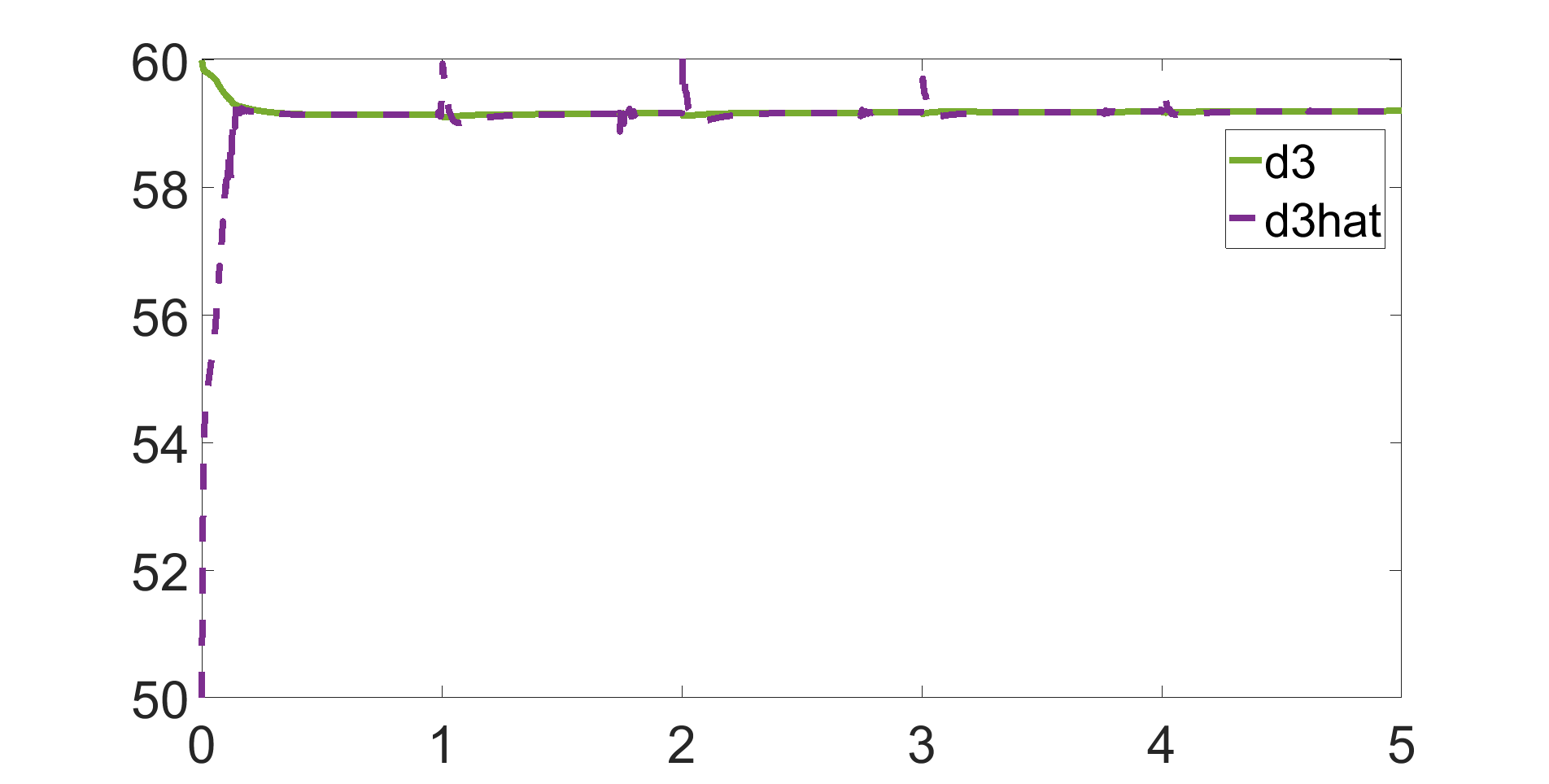}
\caption{\scriptsize{Proposed Control: $d_3$ Estimation.}}
\label{fig:tempchange_d3hat_distest}
\end{subfigure}
\begin{subfigure}{.32\textwidth}
  \centering
\includegraphics[width =  5cm,height=3cm]{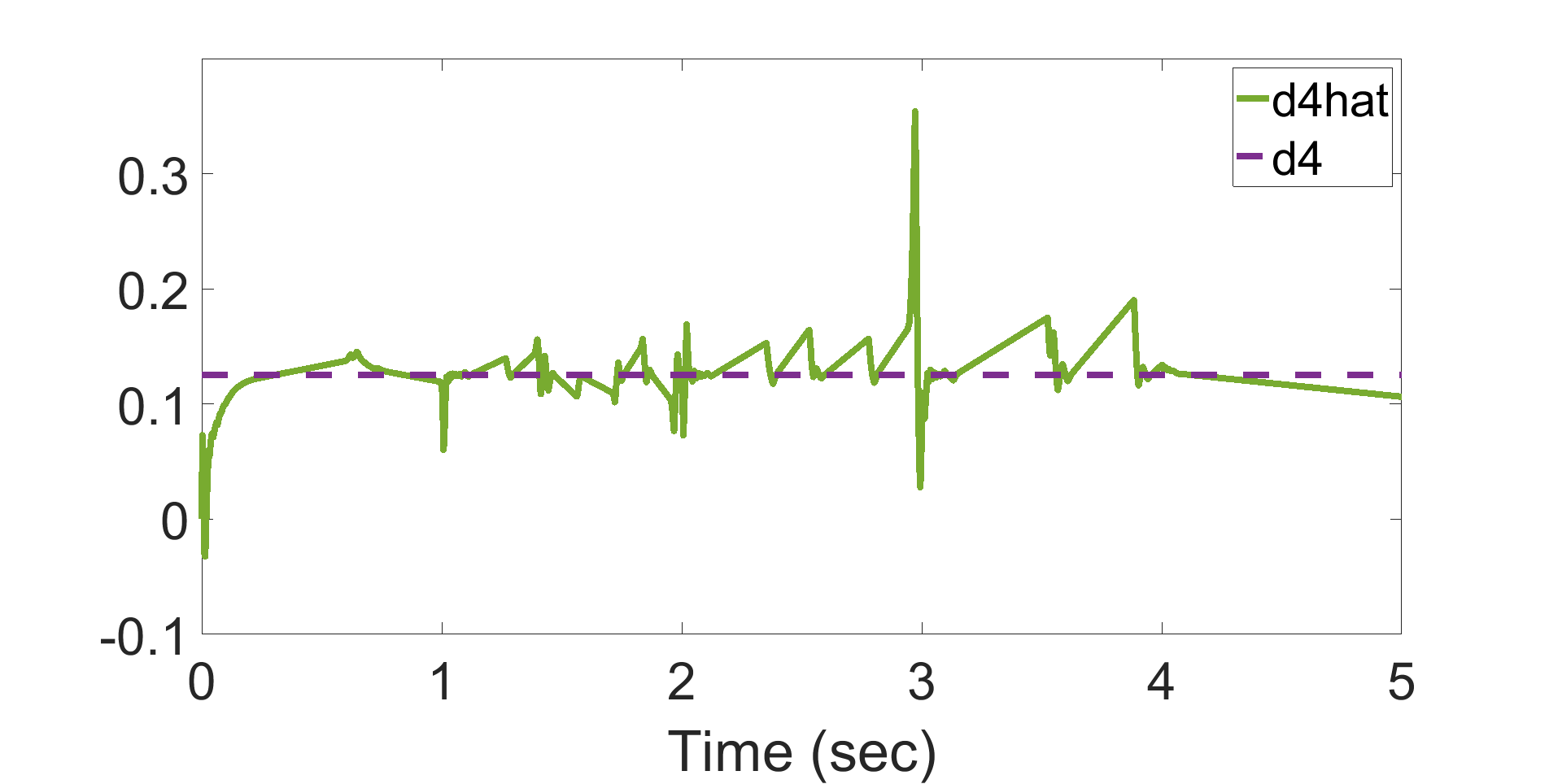}
\caption{\scriptsize{Proposed Control: $d_4$ Estimation.}}
\label{fig:tempchange_d4hat_distest}
\end{subfigure}
\caption{Case-2: Validation of the proposed controller against state-of-the-art controller \cite{iovinetase17}(Backstepping) for temperature change.}
\label{fig:exp_results}
\end{figure}

The power obtained from the photovoltaic cells becomes less than the maximum power possible when temperature rises. The values of the MPPT voltage and current also reduce with incerase in temperature. To demonstrate the effect of temperature the load power and irradiance are steadily maintained at 200$W$, 1000$W/m^2$ respectively as shown in Fig.\ref{fig:tempchange_tempvariation}. The temperature is varied from 75$^{o}$ C to 50$^{o}$ C and to 25$^{o}$ C, 10$^{o}$ C and 0$^{o}$ C respectively, at 1$s$, 2$s$, 3$s$ and 4$s$ respectively. According to the temperature, $x_{1ref}$ is calculated using PV characteristic and  set to $20.39~V$, $23.36~V$, $26.3~V$, $28.11~V$ and $29.29~V$ respectively.  Accordingly, the $x_{4ref}$ value is also set as $23.66~V$, $23.76~V$, 
$23.86~V$, $23.92~V$, $23.9598~V$ respectively. When temperature is changed, it gets reflected in the corresponding disturbance $d_1$ of the system model.  Figures \ref{fig:tempchange_outputs_bstep} ,\ref{fig:tempchange_inputs_bstep} and \ref{fig:tempchange_currents_bstep}s hows the output voltages, input duty ratio and DCMG branch currents when the baseline backstepping algorithm is applied and all the disturbance values are known. Similarly, figures \ref{fig:tempchange_outputs_bstepunknown}, \ref{fig:tempchange_inputs_bstepunknown} and \ref{fig:tempchange_currents_bstepunknown} shows the same when the baseline backstepping algorithm is applied and the disturbance values are not known. When the proposed controller is applied in case of unknown disturbance value, the results obtained can be seen in figures \ref{fig:tempchange_outputs_distest}, \ref{fig:tempchange_inputs_distest} and \ref{fig:tempchange_currents_distest}. Finally figures \ref{fig:tempchange_d1d2hat_distest},\ref{fig:tempchange_d3hat_distest} and \ref{fig:tempchange_d4hat_distest} represent the evolution of disturbance observer estimates through time.

\subsection{Case-3: Change in Irradiance}
\begin{figure*}[!ht]
\captionsetup[subfigure]{aboveskip=-1pt,belowskip=-1pt}
\centering
\begin{subfigure}{0.32\textwidth}
  \centering
\includegraphics[width =  5cm,height=3cm]{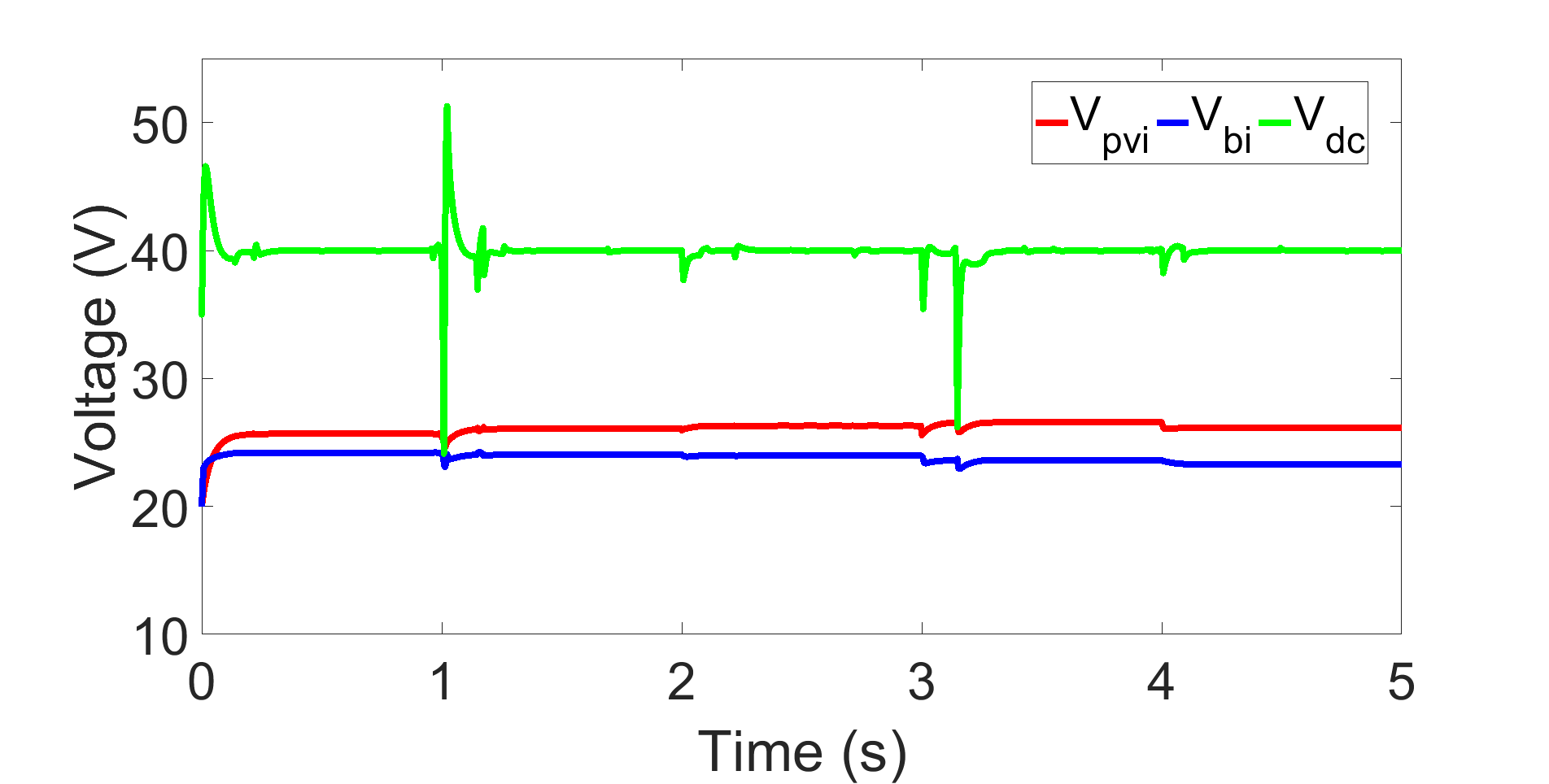}
\caption{\scriptsize{Backstepping and Known Disturbances: Outputs}}
\label{fig:irrchange_outputs_bstep}
\end{subfigure}
\begin{subfigure}{.32\textwidth}
  \centering
\includegraphics[width =  5cm,height=3cm]{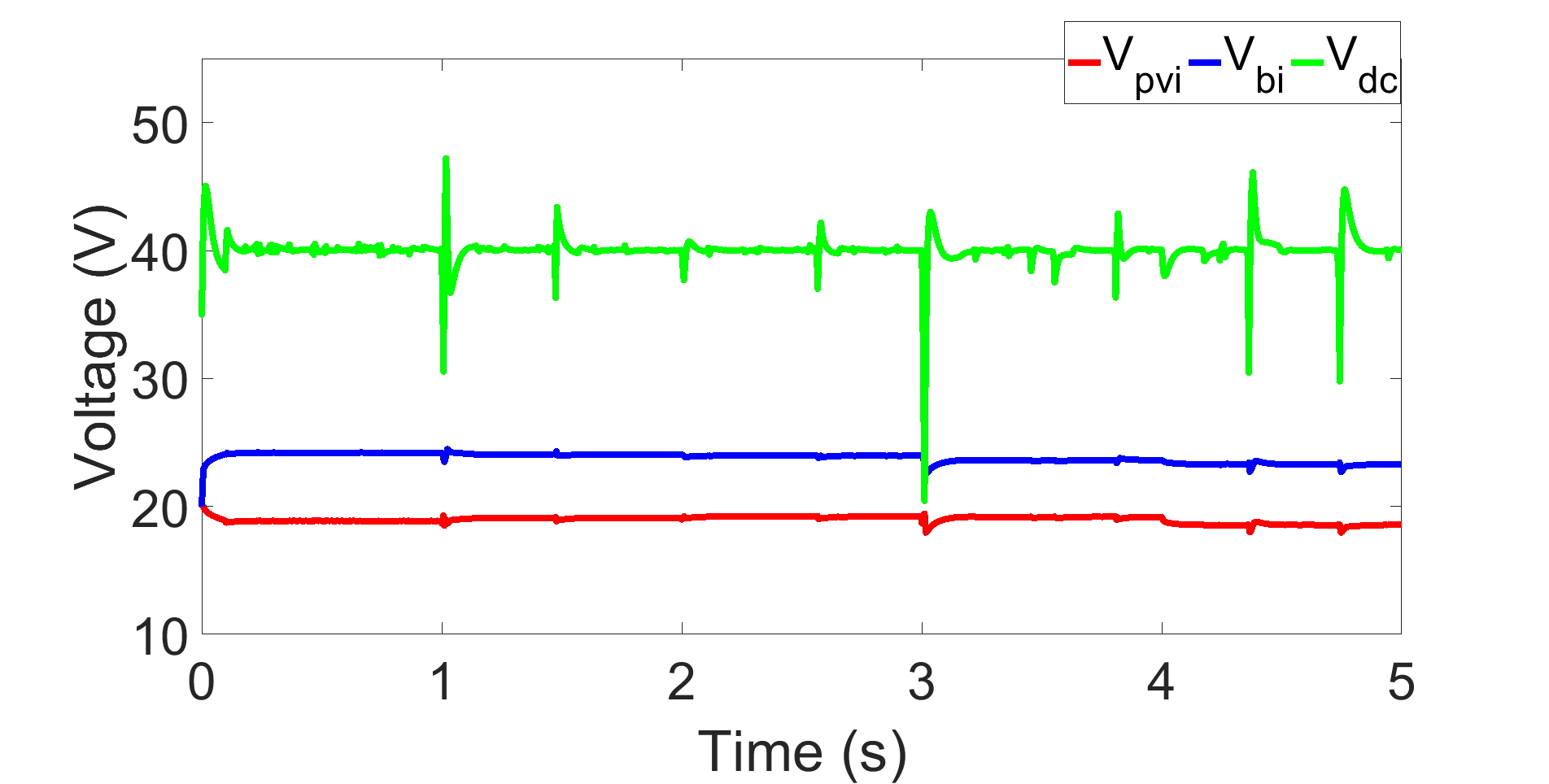}
\caption{\scriptsize{Backstepping and Unknown Disturbances: Outputs}};
\label{fig:irrchange_outputs_bstepunkown}
\end{subfigure}
\begin{subfigure}{.32\textwidth}
  \centering
\includegraphics[width =  5cm,height=3cm]{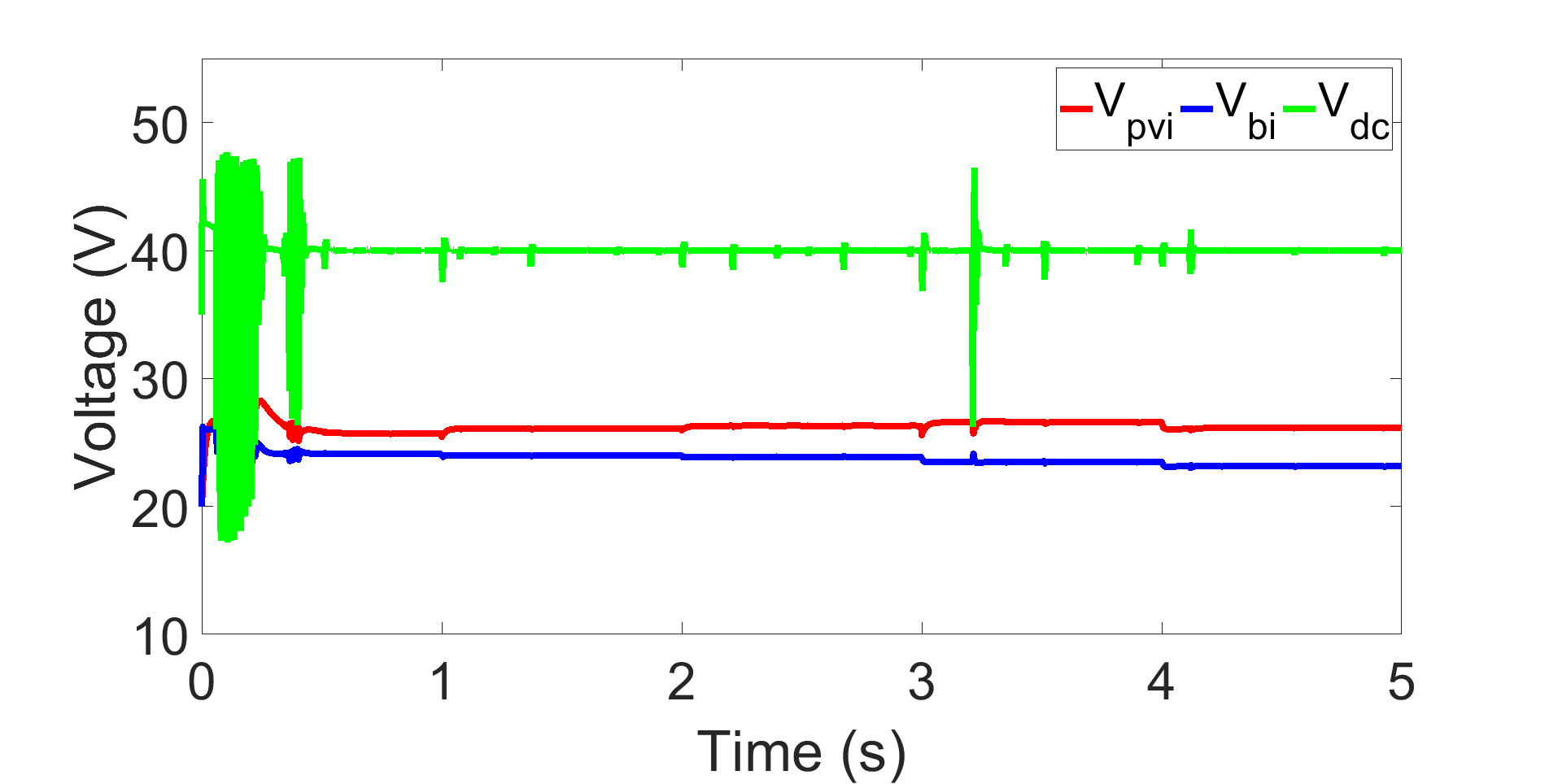}
\caption{\scriptsize{Backstepping and Unknown Disturbances: Outputs}};
\label{fig:irrchange_outputs_distest}
\end{subfigure}\\
\begin{subfigure}{.32\textwidth}
  \centering
\includegraphics[width =  5cm,height=3cm]{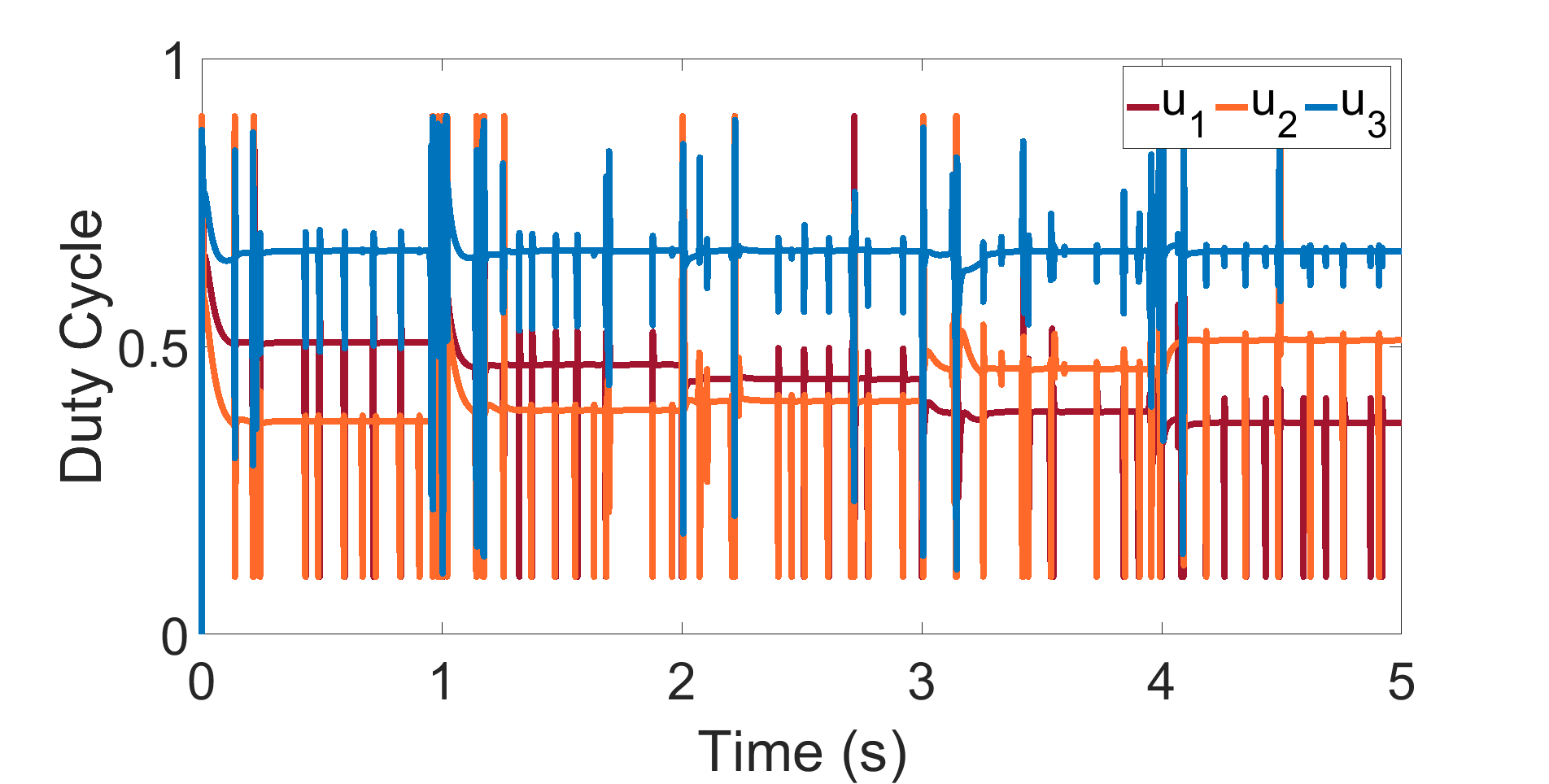}
\caption{\scriptsize{Backstepping and Known Disturbances: Inputs}}
\label{fig:irrchange_inputs_bstep}
\end{subfigure}
\begin{subfigure}{.32\textwidth}
  \centering
\includegraphics[width =  5cm,height=3cm]{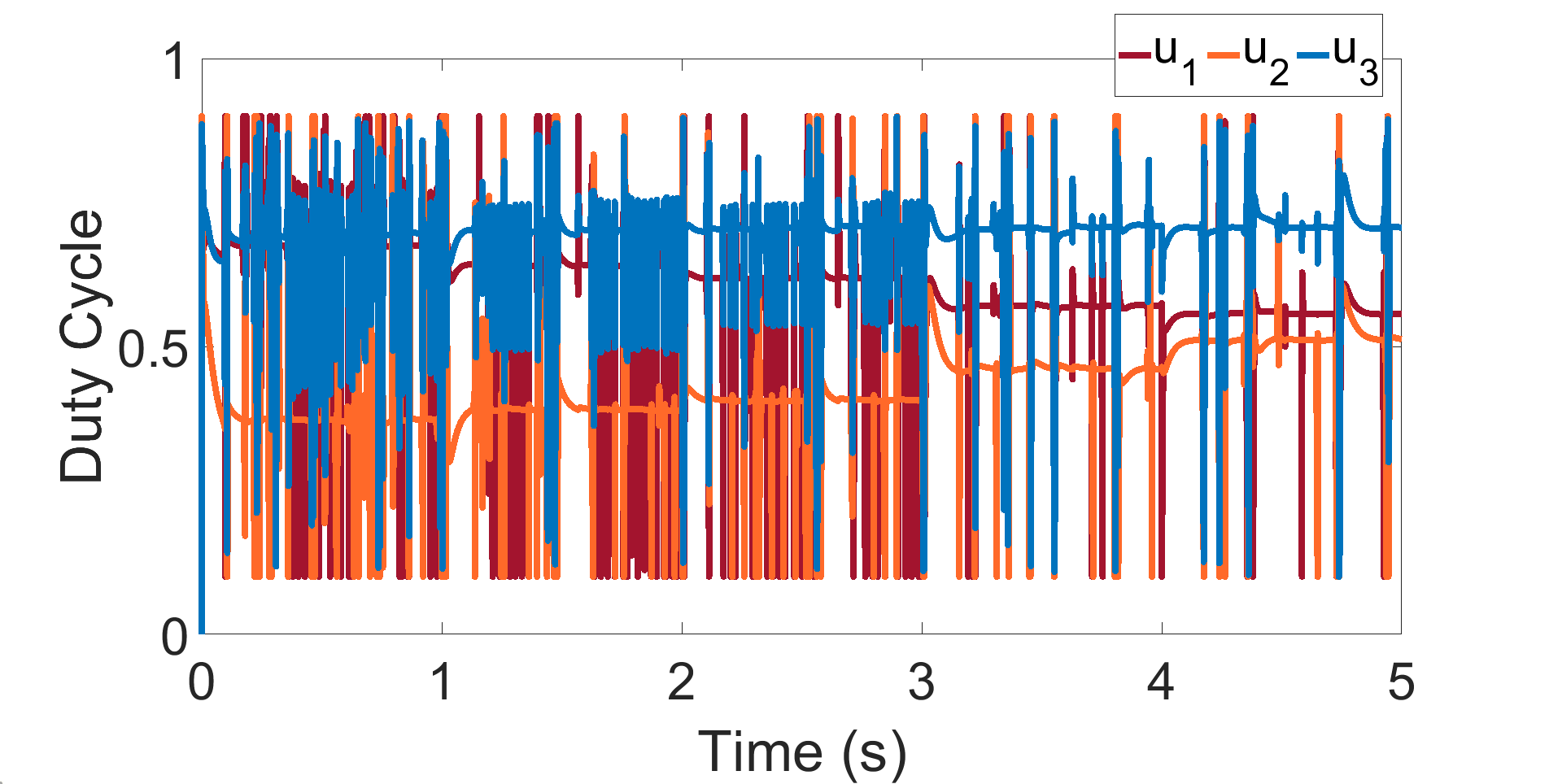}
\caption{\scriptsize{Backstepping and Unknown Disturbances: Inputs}}
\label{fig:irrchange_inputs_bstepunknown}
\end{subfigure}
\begin{subfigure}{.32\textwidth}
  \centering
\includegraphics[width =  5cm,height=3cm]{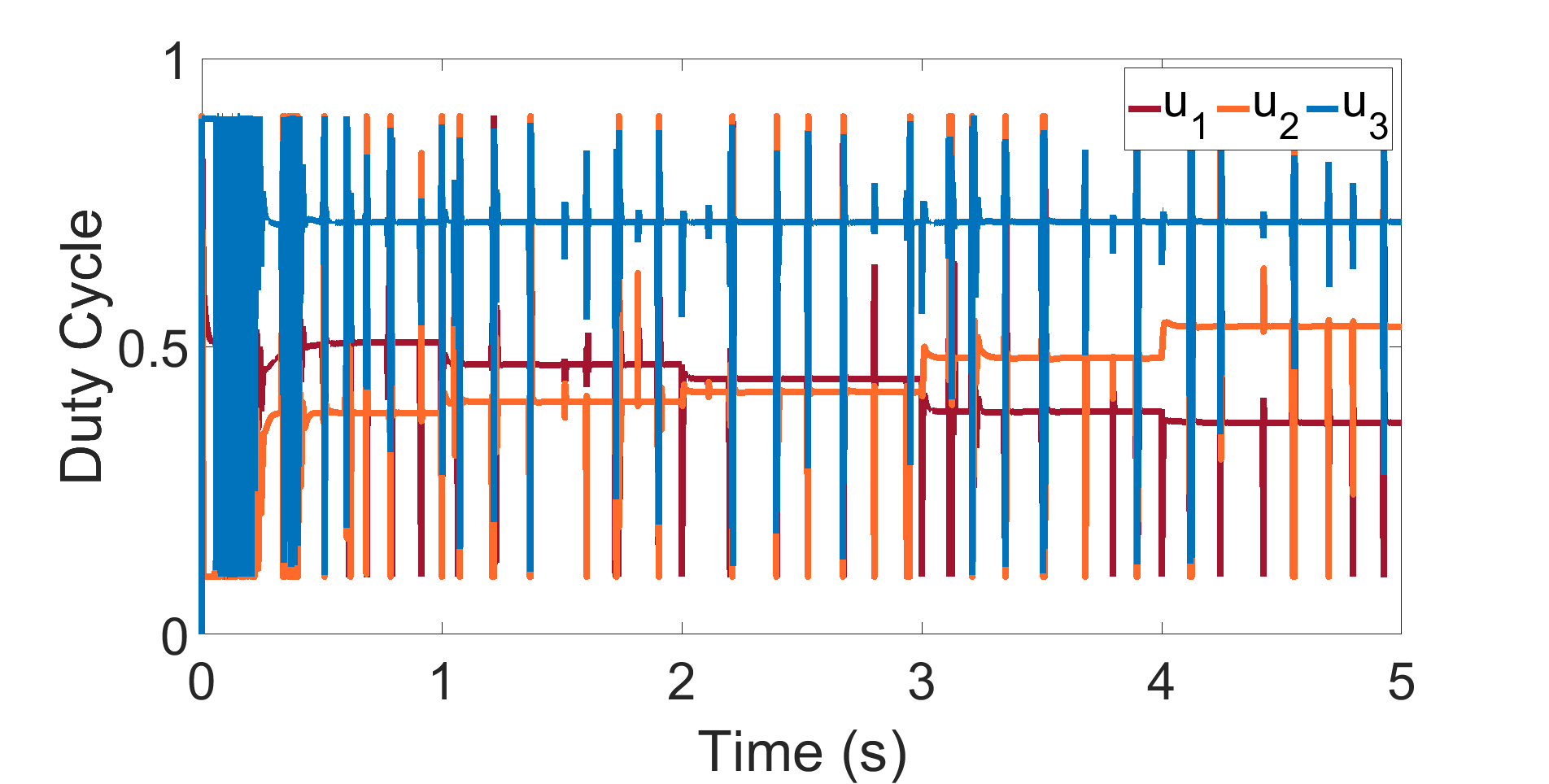}
\caption{\scriptsize{Proposed control and Unknown Disturbances: Inputs.}}
\label{fig:irrchange_inputs_distest}
\end{subfigure}\\
\begin{subfigure}{.32\textwidth}
  \centering
\includegraphics[width =  5cm,height=3cm]{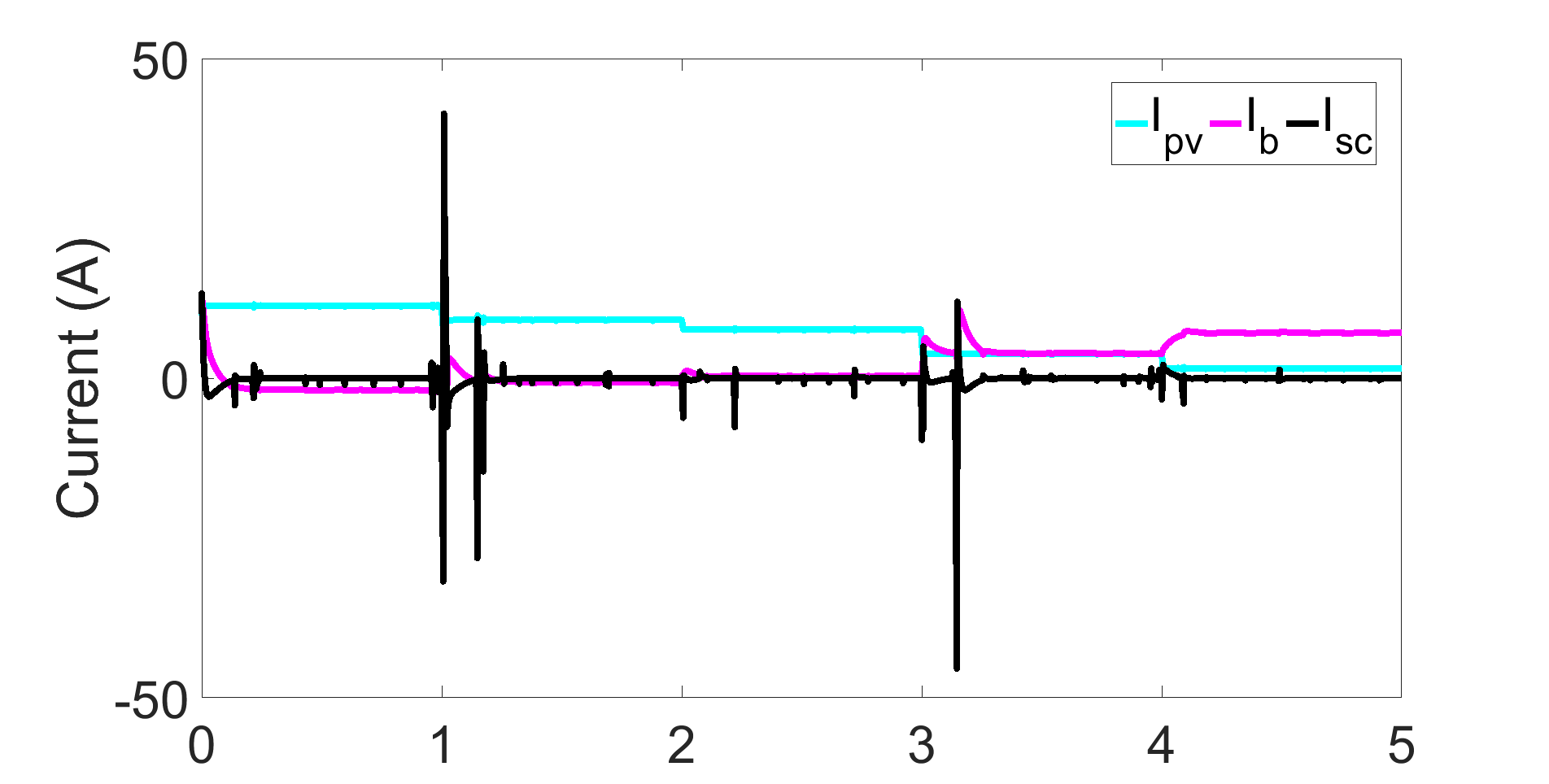}
\caption{\scriptsize{Backstepping and Known Disturbances: Currents.}}
\label{fig:irrchange_currents_bstep}
\end{subfigure}
\begin{subfigure}{.32\textwidth}
  \centering
\includegraphics[width =  5cm,height=3cm]{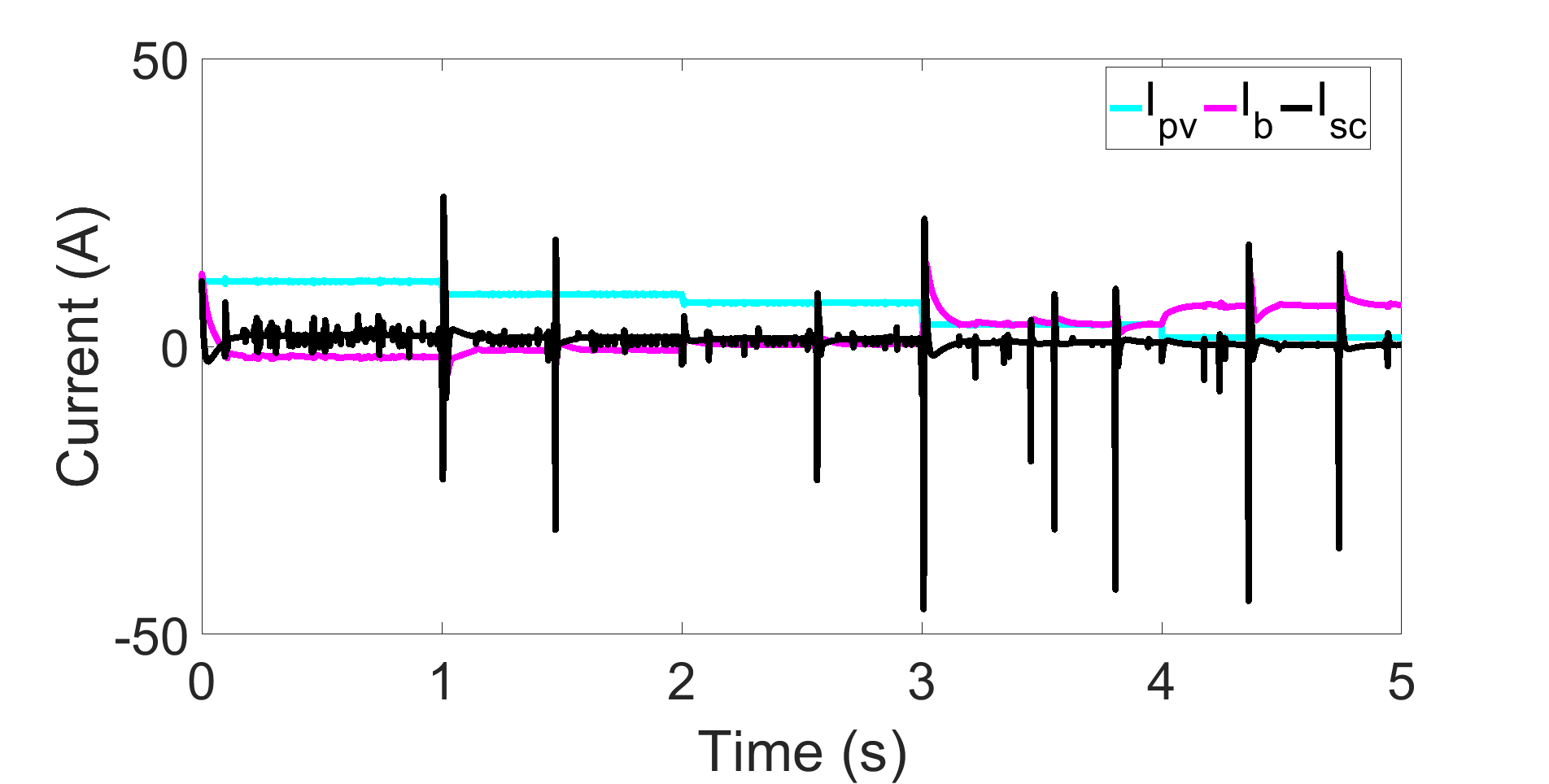}
\caption{\scriptsize{ Backstepping and Unknown Disturbances: Currents}}
\label{fig:irrchange_currents_bstepunknown}
\end{subfigure}
\begin{subfigure}{.32\textwidth}
  \centering
\includegraphics[width =  5cm,height=3cm]{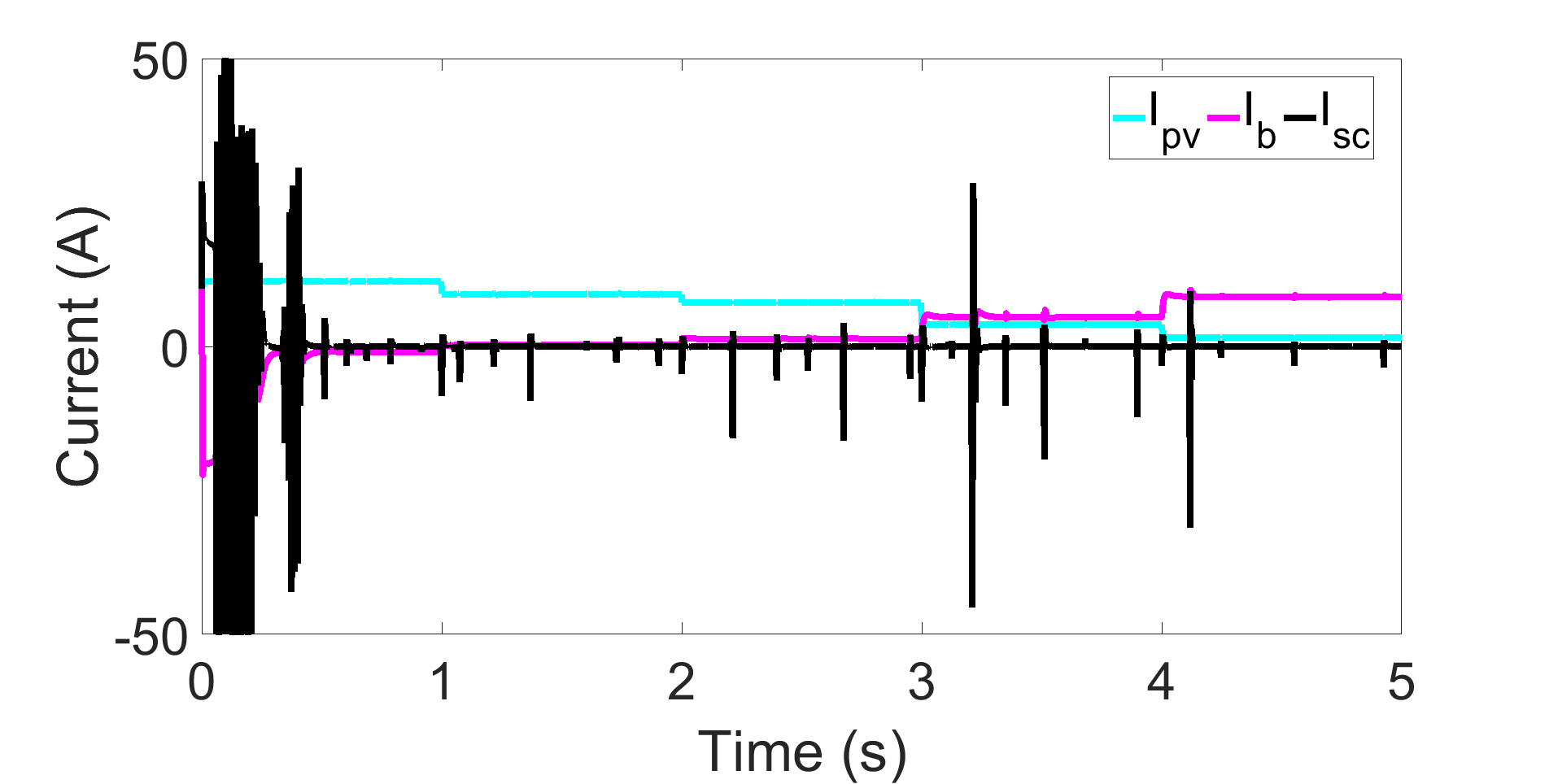}
\caption{\scriptsize{Proposed control and Unknown Disturbances: Currents.}}
\label{fig:irrchange_currents_distest}
\end{subfigure}\\
\begin{subfigure}{.32\textwidth}
  \centering
\includegraphics[width =  5cm,height=3cm]{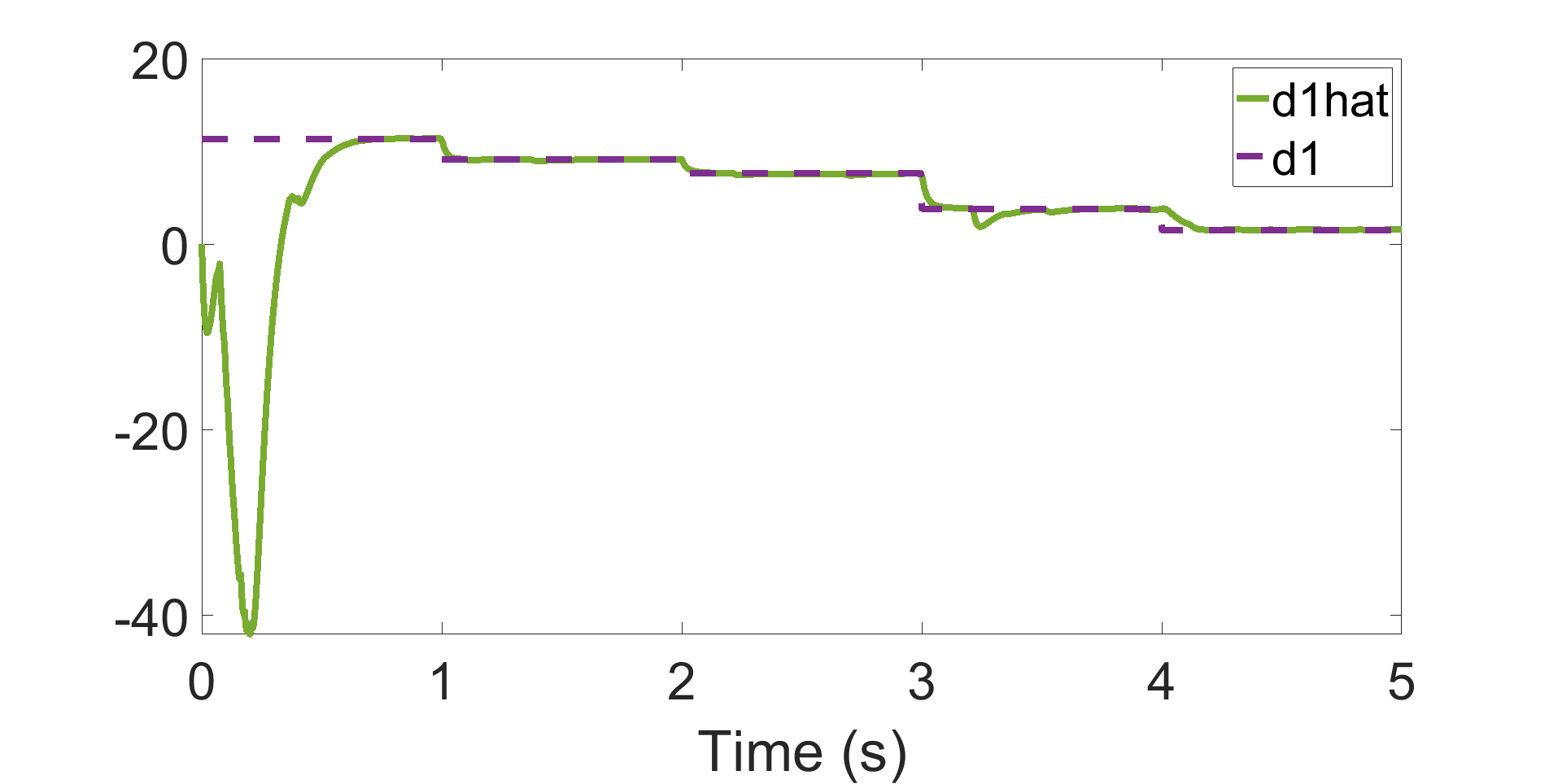}
\caption{\scriptsize{Proposed Control: $d_1,~d_2$ Estimation.}}
\label{fig:irrchange_d1_distest}
\end{subfigure}
\begin{subfigure}{.32\textwidth}
  \centering
\includegraphics[width =  5cm,height=3cm]{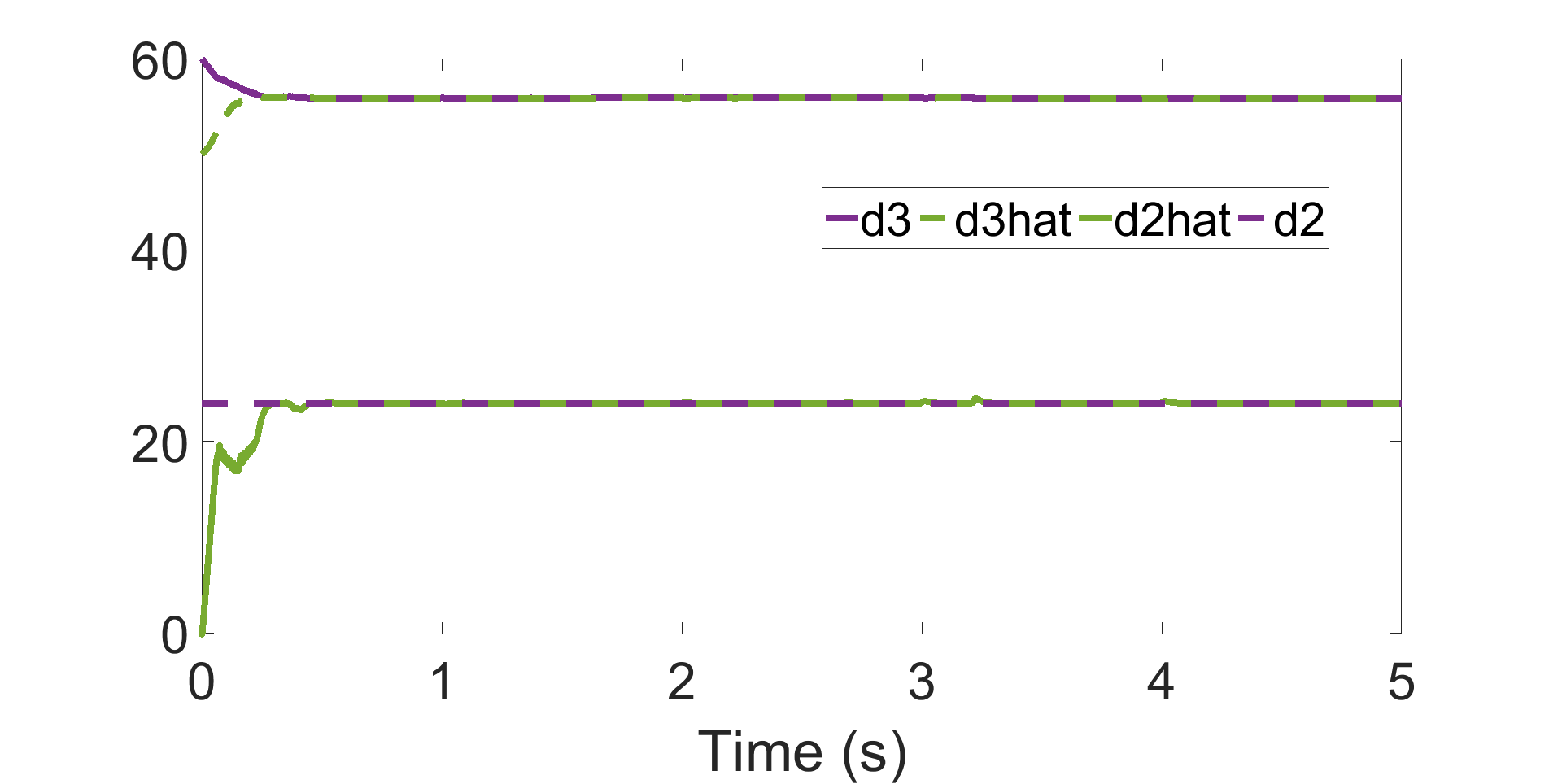}
\caption{\scriptsize{Proposed Control: $d_3$ Estimation.}}
\label{fig:irrchange_d2d3hat_distest}
\end{subfigure}
\begin{subfigure}{.32\textwidth}
  \centering
\includegraphics[width =  5cm,height=3cm]{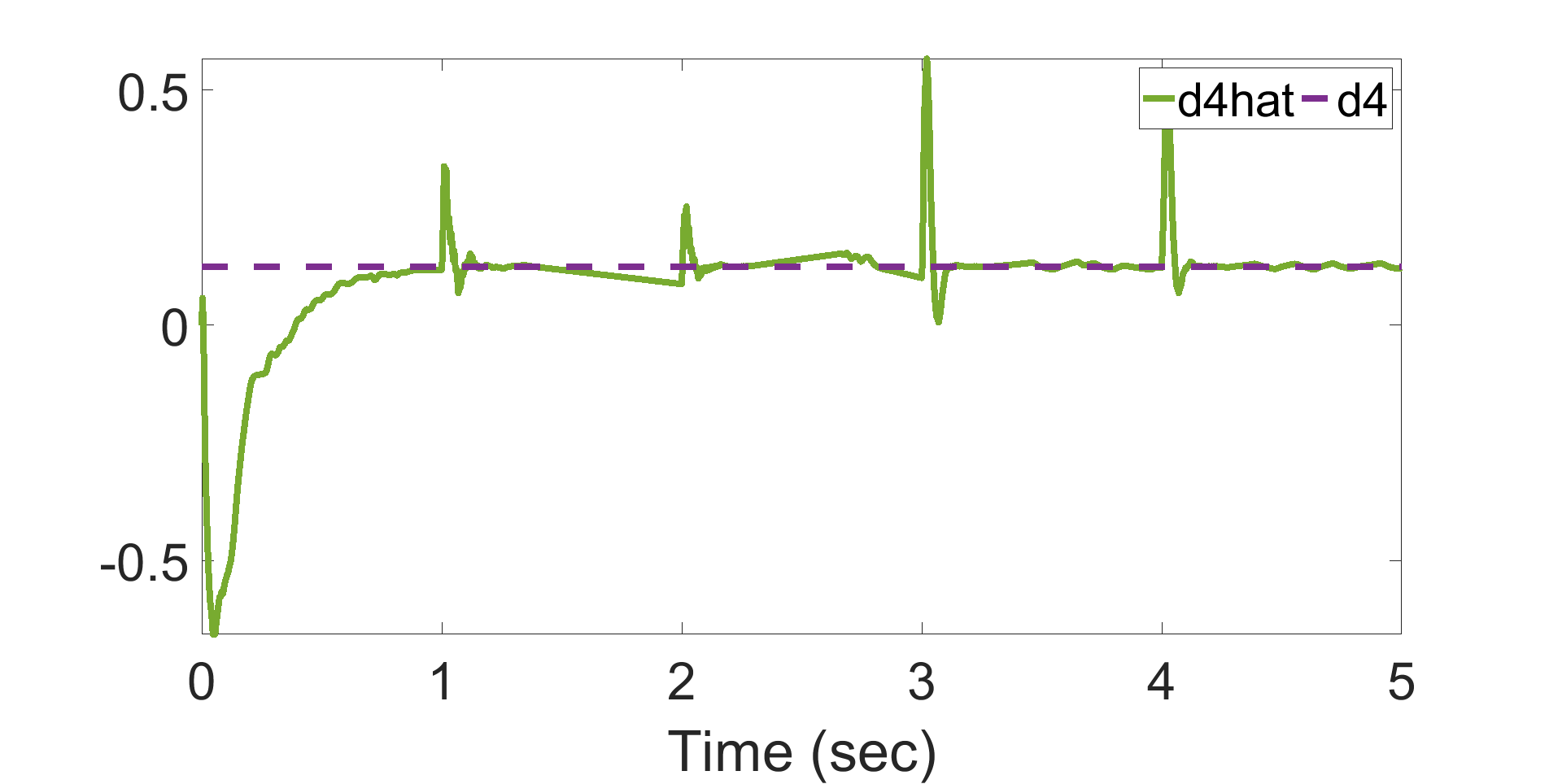}
\caption{\scriptsize{Proposed Control: $d_4$ Estimation.}}
\label{fig:irrchange_d4hat_distest}
\end{subfigure}
\caption{Case-3: Validation of the proposed controller against state-of-the-art controller \cite{iovinetase17}(Backstepping) for Irradiance change . }
\label{fig:exp_results3}
\end{figure*}

To demonstrate the effect of irradiance on the DCSSMG, the solar power incident on the PV panel is varied while temperature and load are maintained steadily at 25$^{o}$C and 200$W$. The irradiance changes for every 1 second from 1500$W/m^2$ to 1200$W/^2$ and then to 1000$W/m^2$, 500$W/m^2$ and 200$W/m^2$ at 1$s$, 2$s$, 3$s$ and 4$s$ respectively as shown in Fig.\ref{fig:irrchange_irrprofile}. As mentioned above, $x_{1ref}$ is set as $25.67~V$, $26.05~V$, $26.3~V$, $26.51~V$ and $26.55~V$ respectively and $x_{4ref}$ is set as $24.09~V$, $23.9738~V$, $23.8679~V$, $23.4889~V$ and $23.1336~V$ accordingly. When irradiance is changed, it is reflected in the corresponding disturbance $d_1$ of the system model. The output voltages, duty cycles and branch current profiles of the DCMG system when baseline backstepping algorithm is applied with known disturbance values are shown in figures \ref{fig:irrchange_outputs_bstep}, \ref{fig:irrchange_inputs_bstep} and \ref{fig:irrchange_currents_bstep}. The same variables are plotted for the backstepping algorithm in figures \ref{fig:irrchange_outputs_bstepunkown}, \ref{fig:irrchange_inputs_bstepunknown} and \ref{fig:irrchange_currents_bstepunknown} when the disturbances value is unknown. The proposed algorithm is applied for same change in irradiance when the disturbance values are unknown an the results are recorded in figures \ref{fig:irrchange_outputs_distest}, \ref{fig:irrchange_inputs_distest} and \ref{fig:irrchange_currents_distest}. The disturbance estimation carried out can be seen in figures \ref{fig:irrchange_d1_distest}, \ref{fig:irrchange_d2d3hat_distest} and \ref{fig:irrchange_d4hat_distest}. 

\subsection{Discussion}
     It is seen that the state of the art backstepping controller performs well in all the cases when the disturbance values are known. However, in absence of real-time disturbance data, the performance of the backstepping controller denigrates. In all the three cases it is seen that the control effort is maximized when disturbance value is unknown to the backstepping controller. Moreover, in cases-2 and 3, the supercapacitor current deviates to a non-zero value to ensure DC grid voltage regulation when disturbance values are unknown. A lot of transient is seen in DC grid voltages in all the three cases when the disturbance values become unknown. The supercapacitor current shows more of high frequency component whereas battery current is seen to reach steady state very smoothly. This shows how supercapacitor helps to reduce high frequency load on the battery in the presence of transients. 
    
    It is seen that the proposed algorithm takes a settling time between $70ms$ to $150ms$ as opposed to the backstepping algorithm whose settling time ranges between $60ms$ to $140ms$ when the disturbance values are known. Most of the disturbance observers in all the cases are seen to converge to the actual values below $100ms$. In case-2, the disturbance observer related to output current seem to take around $200ms$ to converge to the exact value but this does not affect MPPT or voltage control in any manner. In all cases, the disturbance observer of $d_4$ keeps deviating slightly from the actual value but DC voltage regulation and MPPT is very smooth. It is seen that disturbance observer $d_3$ accurately estimates even small changes in supercapacitor voltage for all the cases in real-time.
    The disturbance observers may be initialized at any random value. They eventually converge to the right estimate. In the inital $400ms$ transients are observed due to this phenomena. However, in other cases the observers converge before 100$ms$.

\section{Summary} \label{sec:aobssummary}
In this work, a DCSSMG with PV array, battery and supercapacitor is chosen and an adaptive observer based back-stepping control strategy is developed for the same. The overall stability of the DCMG with proposed controller is analyzed through the formulation of composite Lyapunov function consisting of many Lyapunov functions from different subsystems during the different stages of the design process and the stability of various states and estimators is proven through rigorous mathematical analysis.
The back-stepping controller designed in this work ensures fast and smooth MPP tracking, power balance and DC grid voltage control when the sensors for measuring system variables like PV array output current, battery voltage, supercapacitor voltage and load current like are absent. The simulation results show higher efficacy of the proposed control strategy compared to the state-of-the art model-based backstepping controller especially when the sensor input is missing.
  
\chapter{Adaptive Neural Controller for Unknown DCSSMGs with Unknown Disturbances}\label{chapter:DirectPerturbMPPT}\index{Direct Perturbation MPPT} 
\chaptermark{Adaptive Neural Controller}
In this chapter, a detailed depiction on synthesizing a back-stepping based controller for the DCSSMG system is given. This controller does not require the system model apriori for its implementation. Further, this technique ensures stability of the DCSSMG voltage against sensor-malfunction and atmospheric changes. This technique can also be used to lessen the sensor-count in the DCSSMG thereby rendering the DCSSMG more economic to deploy. 

\section{Introduction }\index{IntroMGStandalone}

\begin{figure}[!h]
\centering
\includegraphics[scale=0.55]{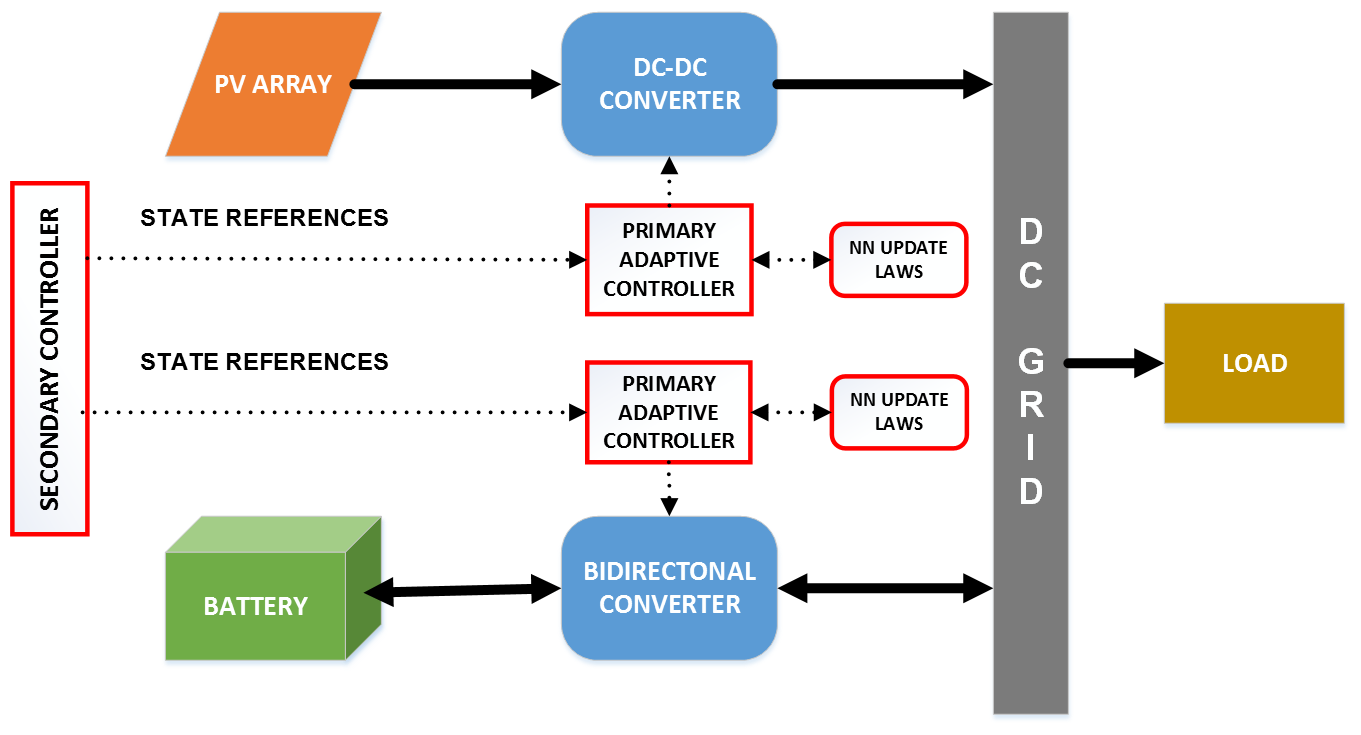}
\caption{Control Structure of DCSSMG System}
\label{fig:spvdg1_ch3}
\end{figure}

  In the existing literature, nonlinear controllers have been proposed for improving the dynamic performance in power electronics based systems. However, the stability of these controllers clearly relies on the knowledge of exact parameter values of the DCSSMG (i.e. resistance, inductance, capacitance) and environmental disturbances (irradiance, temperature and load). The addition of expensive sensors to measure these disturbances/parameters also adds up to the entire cost of the system \cite{isgtpaper}. \textcolor{black}{Conventional adaptive controllers use observers which are augmented with controllers to avoid placing too many sensors for control purposes. For instance, the authors in \cite{puretsteobs1} use current observers to estimate PV and load current and eliminate the corresponding sensors to reduce cost. These estimated currents are used to design a dead-beat controller. Similarly, \cite{puretsteadaptive1} develops an adaptive control strategy to estimate the best linear model representation of a wave energy converter. Although adaptive, this technique can be more effective if it can capture the underlying non-linearities in an online manner. \cite{selfdobsiecon20} uses a total of three disturbance observers for estimating the PV output current, battery voltage and load admittance in a non-linear isolated microgrid embedded with PV array and battery to develop an effective back-stepping controller. However, this algorithm cannot control output voltage when the controller is not provided with intrinsic system parameter information like converter resistance, inductance and capacitance.} In such cases, the operation of the entire system needs to be halted to re-calibrate the control parameters. The adaptive back-stepping algorithm developed in \cite{overviewtkr2} updates the value of both system parameters and disturbances in an online manner. However, all these controllers are model-based which can only estimate certain parameters of the full model and thus cannot be used if disturbances or system model is entirely unknown.  Furthermore, any errors in system modeling cannot be compensated using these controllers. 
 
 \textcolor{black}{The assumption in classical adaptive techniques is that the system nonlinearity is perfectly known.  In most practical situations, exact knowledge of the nonlinearities present in the system is not possible. Hence, the neural network based intelligent adaptive control is used in such situations so as to capture the detailed intricate system model in real-time.} For instance, \cite{nnref11} uses RBF neural networks to tune the controller gains of the bidirectional DC-DC converter for controlling the DCSSMG voltage. The PID gains are updated in an online fashion whenever sudden disturbances occur. In\cite{nnref12}, a feed-forward NN generates a supplementary control signal which is added to the conventional PI controller for managing power flow from/to the battery via a dual active bridge converter. Although these controllers make the system more adaptive, they do not identify the nonlinear system model.  Similarly, the authors of \cite{nnref14hybridmg} use a combination of RBFNNs and ADALINES to cater to multiple control goals in a hybrid AC-DC network. The authors of \cite{puretstennadaptive1} also propose the use of a feed-forward multilayered neural network for optimal control of wave energy converters.
 
However, in all these works, the controllers for various subsystems are developed for specific subsystems without considering their electrical interconnections. Moreover, the greatest drawback in these works is that the weight update laws are derived based on the back-propagation technique which does not have conclusive mathematical proof of convergence. This would mean that these techniques can be used mostly for off-line purposes and the associated NNs may diverge when used in an online manner for model estimation and control. 

The development of intelligent controllers for an unknown DCSSMG system under unknown disturbances is an unexplored and important problem. Hence, in this work, a model-free adaptive neural controller for an unknown DCSSMG system is developed to work in presence of unknown disturbances \cite{korukonda2021adaptive}. A comprehensive stability proof for the DCSSMG system is derived including various subsystems and their interconnections with the help of Lyapunov stability for achieving the uniformly ultimately boundedness (UUB) of all states. This property is a necessity especially for controllers to be applied for real-time system estimation and control purposes. Finally, a careful analysis of the proposed adaptive neural controller when compared with state of the art adaptive and nonlinear controllers is presented when real-time information regarding system model parameters and disturbances is unknown.

\textcolor{black}{Section \ref{sec:ancmodel} presents the overall system model of the DCSSMG system used and defines the exact control problem that is addressed in this chapter. Section \ref{sec:ancdesign} describes the proposed control technique along with proof of uniformly ultimate boundedness of the states of the DCSSMG system. Section \ref{sec:ancresults} delineates the various simulation studies performed on the system at hand in the presence of unknown disturbances and unknown system model. Section \ref{sec:ancsummary} summarizes the work done in this chapter. }

\section{Background and Problem Formulation} \label{sec:ancmodel}
\textcolor{black}{This section delineates the various subsystems present in the DCSSMG system along with the state-space model. It further describes the formulation of adaptive primary control design problem that is addressed in this chapter.}


\begin{figure}[!h]
\centering
\includegraphics[scale=0.55]{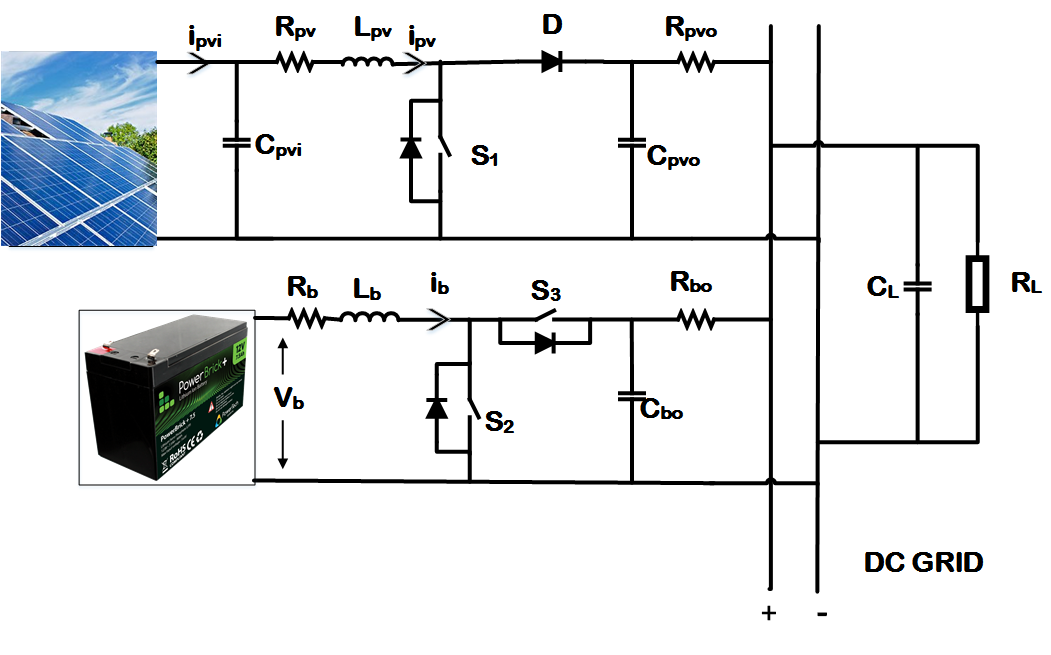}
\caption{Circuit Diagram of the DCSSMG System}
\label{fig:spvdg2}
\end{figure}

\subsection{System Description}
A low voltage DCSSMG is proposed for this work. It uses locally available renewable distributed energy sources like solar array to feed its loads. The PV array serves the load through a DC-DC boost converter. The duty cycle of this boost converter is decided by an MPPT algorithm. The DC load is parallelly connected to a battery through a bi-directional power converter. This converter allows the power to flow in and out of the battery according to the load conditions.

Fig.\ref{fig:spvdg2} depicts the overall circuit diagram of the DCSSMG system. This section models the DCSSMG as per the given circuit which is then used for designing the controller.  

The PV array generates an output current whose equation is described by \eqref{eq:ipv} as discussed in \cite{self1},  
\begin{align}\label{eq:ipv}
\scalebox{.9}{$
i_{pvi}=n_{p}I_{g}-n_{p}I_{s}\left(e^{\dfrac{q(v_{pv}+i_{pv}R_{s})}{n_{s}pKT}}-1\right)-\dfrac{v_{pv}+i_{pv}R_{s}}{R_{sh}}$}
\end{align}
The PV output current $i_{pv}$ and the maximum power point of the PV array are intricately dependent on temperature and irradiance. It can be observed in the  state-space model, a scaled version of the PV array output current is denoted by $D_1$. The voltage across the output capacitor $C_{pvi}$ of the PV array is described as $x_1$ while the input capacitor $C_{pvo}$ voltage is modeled as $x_2$. The PV inductor $L_{pv}$ current is further termed $x_3$. Change in temperature and irradiance affect $i_{pvi}$ and are thus modeled into disturbance $d_1$.
Change in temperature and irradiance affect $i_{pvi}$ and are thus modeled into disturbance $D_1$. 

The battery is assumed to be a continuous DC voltage source with voltage $V_b$ and is modeled into the state-space as disturbance $D_2$. The voltage across battery output capacitor $C_{bo}$ voltage is termed as $x_5$ and the battery inductor sports a current designated as $x_4$, The grid voltage  of the DCSSMG which is measured at capacitor $C_L$ is denoted as $x_6$ and the resistive load admittance $\frac{1}{R_L}$ is termed as $D_4$.

Following the state space averaging technique as described in Chapter 2, the following state space equations have been derived,
 \begin{eqnarray}
 \dot{x_1}&=& - g_1x_3 + D_1 \nonumber \\
 \dot{x_2}&=& \frac{x_3}{C_{pvo}} - \frac{x_2}{R_{pvo}C_{pvo}} + \frac{x_6}{R_{pvo}C_{pvo}}- g_2u_{1} \nonumber \\
 \dot{x_3}&=& \frac{x_1}{L_{pv}} - \frac{x_2}{L_{pv}} - \frac{x_{3}R_{pv}}{L_{pv}} + g_3u_1 \nonumber \\  
  \dot{x_4}&=& D_2 - \frac{x_{4}R_{b}}{L_{b}} - \frac{x_5}{L_{b}} + g_4u_2  \label{eq:ssmodel}\\  
 \dot{x_5}&=& \frac{x_4}{C_{bo}} - \frac{x_5}{R_{bo}C_{bo}} + \frac{x_6}{R_{bo}C_{bo}}- g_5u_{2} \nonumber \\
 \dot{x_6}&=& \frac{x_2}{C_{L}R_{pvo}} -\frac{x_6}{C_LR_{pvo}}- \frac{x_6}{C_LR_{bo}} +g_6x_5 + c_6D_3\nonumber 
 \end{eqnarray}
 where 
 \begin{align}
  x =& [V_{pvi}~~V_{pvo}~~i_{pv}~~i_b~~V_{bo}~~V_{dc} ] \nonumber \\
 D_1 =& \frac{i_{pvi}}{C_{pvi}} \hspace{1cm} D_2= \frac{V_b}{L_b}  \hspace{1cm} D_3= \frac{1}{C_LR_L} \nonumber \\
 g_1=&\frac{1}{C_{pvi}} ~~ g_2=\frac{x_3}{C_{pvo}}~~g_3=\frac{x_2}{L_{pv}} ~~g_4=\frac{x_5}{L_b}  \nonumber \\
 g_5=&\frac{x_4}{C_{bo}}~~g_6=\frac{1}{C_LR_{bo}}~~c_6=-x_6 \nonumber
 \end{align}
The output of the DCSSMG system is $y=[x_1\; x_6].$

\subsection{Problem Formulation}
\textcolor{black}{As seen in Fig.1, there are two control levels for the DCSSMG system- primary and secondary. The references of the primary controllers are assumed to be known and made available from the secondary level. Thus, the current work is concerned only with designing the adaptive neural controller at the primary level.}

\textcolor{black}{As per the system description above, given $y_d= [x_{1ref} ~~ x_{6ref}]$,
the control objective is to design the controllers $u_1$, $u_2$ such that the output $y$ converges to $y_d$ when the system parameters and disturbances are unknown. It is to be noted that the disturbances $D_1$, $D_2$, $D_3$ and system parameters although unknown are assumed to be bounded.}

\begin{figure*}[!t]
\centering
\includegraphics[scale=0.41]{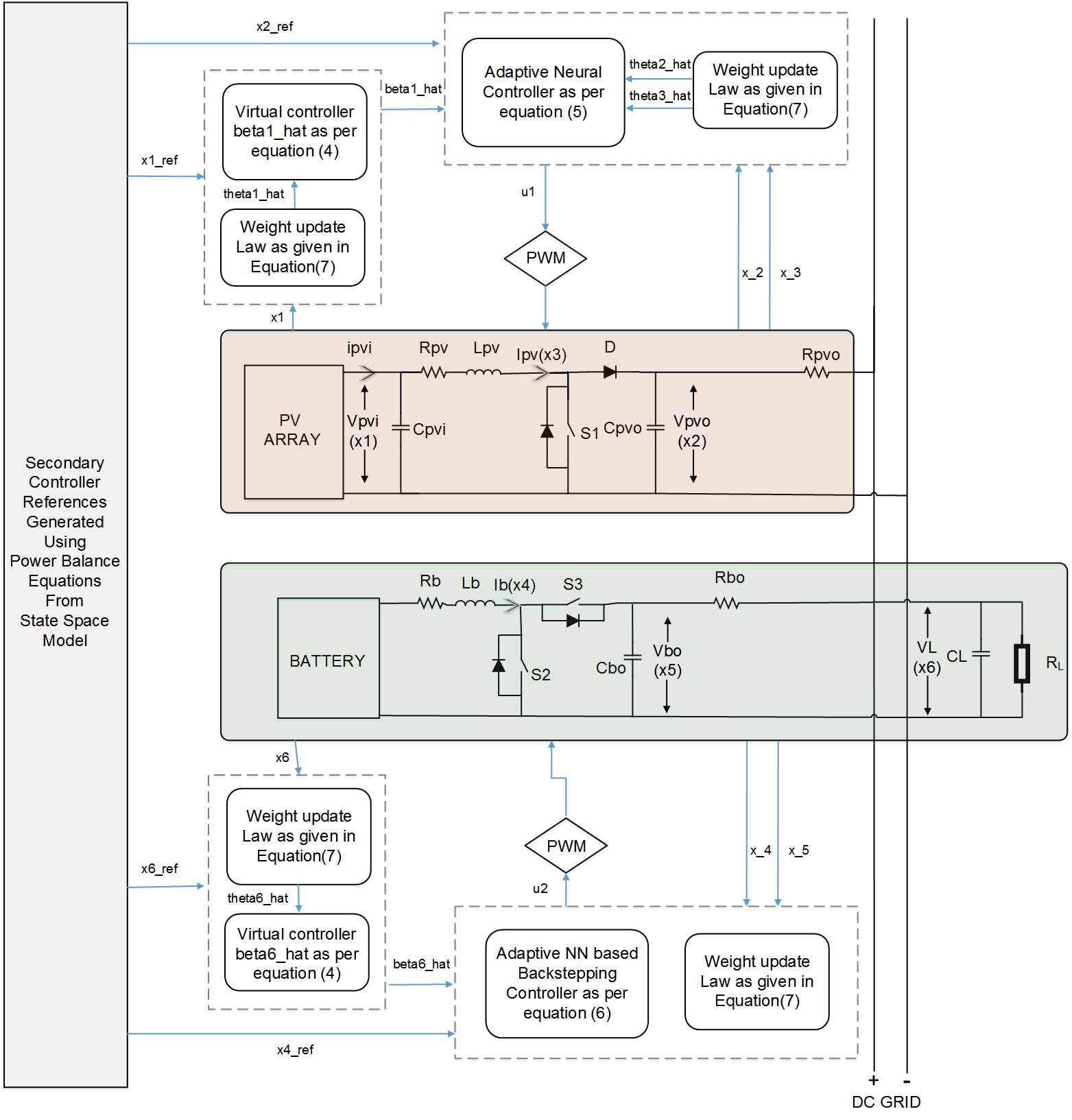}
\caption{Proposed Adaptive Neural Controller implementation diagram}
\label{fig:control}
\end{figure*}

\section{Proposed Controller Design} \label{sec:ancdesign}
This section details the design of primary adaptive controllers which regulate the DCSSMG system outputs to the reference values provided by the secondary controller. 
\textcolor{black}{Fig. 3 represents the detailed schematics for the developed ANC technique. The secondary controller sets output voltage reference $x_{1ref}$ and $x_{6ref}$ for the DCSSMG system which are generated using maximum power point tracking (MPPT) algorithm and power balance equations of the DC grid. 
A design methodology inspired from backstepping theory  is followed for breaking down the complete DCSSMG system into PV and battery subsystems indicated in orange and green in Fig.3 respectively. The individual primary ANC controller for each sub-system needs to follow the desired references from the secondary controller so as to ensure the desired operation of the DCSSMG system. }

\textcolor{black}{For instance, in the PV subsystem, the secondary reference $x_{1ref}$ is processed by the virtual controller using equations (12) and (37) to generate the virtual controller reference $\widehat{\beta}_1$ which serves as reference for state $x_3$. Since the reference for state $x_2$ is already known from the secondary controller, the errors in both these states are processed by the two neural networks as given in equation (31) to generate the duty cycle for the PV subsystem to maintain the output $x_1$. }

\textcolor{black}{Similarly, the battery subsystem is responsible for maintaining the DC votlage control $x_6$. For this purpose, it receives the secondary control references $x_{6ref}$ and $x_{4ref}$ based on power balance criteria. Depending on the state error $e_6$ , the virtual control input $\widehat{\beta}_6$ for state $x_5$ is generated with the help of an RBF neural network using equation (33). Later this virtual reference and the secondary reference received $x_{4ref}$ are used to generate state errors $e_4$ and $e_5$ .These errors are further utilized by two other neural networks using the controller expression presented in equation (34) to generate the duty cycle $u_2$ for the battery subsystem.  It is observed that the weight update law (37) is responsible continuous learning of the NN weights for all the 6 NNs to update their models in real-time thereby providing adaptivity to the proposed ANC controller against unknown changes in system parameters and disturbances. Algorithm one delineates the step-wise implementation procedure for the PV subsystem.} 
\subsection{Controller Design}
The controllers $u_1$ and $u_2$ are designed based on neural networks and back-stepping methodology. The dynamics of each state are captured using a NN which is modeled as 
\begin{align}
f_i= \mathbf{W}^{T}_i\mathbf{\phi}_i + \epsilon_i, i=1,,2,...,6  ~~|\epsilon_i|\leq\eta_i  \label{eq:nnmodel}
\end{align}. 
For the PV and battery subsystems , a virtual controller is first designed NNs 1 and 6 as shown below
\begin{align}
   \widehat{\beta}_i= \frac{\widehat{\theta}_i}{2\eta^2_i}\phi^T_i\phi_ie_i  , ~~ i=1,6 \label{eq:betahat}
\end{align}
where $e_1= x_1-x_1{ref}$ and $e_6= x_6-x_1{6ref}$. The final duty cycle is further designed for each subsystem using the following expressions,
\begin{align}
    u_1={(e_{2}-e_{3})}^{-1}\Bigg(\frac{\widehat{\theta}_{3}}{2\eta^2_{3}}\phi^T_{3}\phi_{3}e^2_{3} +\frac{\widehat{\theta}_{2}}{2\eta^2_{2}}\phi^T_{2}\phi_{2}e^2_{2}\Bigg) \label{eq:u1} 
\end{align}
\begin{align}
        u_2={(e_{5}-e_{4})}^{-1}\Bigg(\frac{\widehat{\theta}_{4}}{2\eta^2_{4}}\phi^T_{4}\phi_{4}e^2_{4} +\frac{\widehat{\theta}_{5}}{2\eta^2_{5}}\phi^T_{5}\phi_{5}e^2_{5}\Bigg) \label{eq:u2}
\end{align}
where $e_2=x_2-x_{2ref}$, $e_3=x_3-\beta_1$, $e_4=x_4-x_{4ref}$ and $~e_5=x_5-\beta_6$ and the weight update laws of all the NNs are designed as, 
\begin{align}
\dot{\widehat{\theta}}_{k} = \frac{\gamma_{k}\phi^T_{k}\phi_{k}e^2_k}{2\eta^2_{k}} - p_k\widehat{\theta}_k, ~~k=1,2,...,6 \label{eq:betahatdot}
\end{align}
\textcolor{black}{
 In our case, for all practical purposes, the control laws is free from singularity due to $e_2=e_3$ or $e_5=e_4$ . Since the control laws $u_1$ and $u_2$ are duty cycles with values in the range 0.1 to 0.9, which is ensured in implementation using limiting conditions on $u_1$ and $u_2$}  ,
 \begin{align}
 u_i &= 0.1 ~~if ~~ u_i < 0.1, \nonumber\\
 u_i &= u_i ~~if ~~0.1\leq u_i\leq0.9  ~~, i=1,2\\
 u_i &= 0.9 ~~if~~ u_i > 0.9 \nonumber 
 \end{align}

The following assumptions are made to derive the complete stability proof of the proposed controllers.\\
\textit{Assumption 1:}
For $i = 1,2,\ldots, n$, the magnitude of disturbance $D_i$ is bounded by an unknown smooth function $h_i$ as  $\left |{ {D_{i}\left ({{t} }\right)} }\right | \le {h_{i}} $.\\
\noindent  
\textit{Assumption 2:} 
 There are unknown constants $b_{mi}$ and $b_{Mi}$ that bound the value of $g_i$ as given 
  \begin{equation*} 0 < {b_{mi}} \le \left |{ {g_{i}} }\right | \le {b_{Mi}} < \infty, \quad \forall {\bar x_{i}} \in {R^{i}}. \end{equation*}
\begin{Theorem}
For the task of MPPT and DC voltage regulation control in the DCSSMG system given by \eqref{eq:ssmodel}, and the chosen controller inputs given by $u_1$ and $u_2$ in \eqref{eq:u1} and \eqref{eq:u2} where the virtual inputs and the weight update laws are given by \eqref{eq:betahat}
 and \eqref{eq:betahatdot} respectively, then the closed loop system error of the DCSSMG system is uniformly ultimately bounded (UUB) and hence the regulation error converges to satisfactory performance.
 The maximum theoretical bounds of the regulation error bounds are shown in Section \ref{sect:error_bound}.
\end{Theorem}
The extensive proof of the adaptive neural controller design has been shown in the next subsection.

\subsection{Stability Analysis}
\label{sect:error_bound}
\textit{Proof:} The proof uses Lyapunov stability and backstepping theory \cite{purebstep1} for designing adaptive controllers using neural networks. A positive definite composite Lyapunov function is chosen consisting of Lyapunov functions from different subsystems like the PV subsystem and the battery subsystem. The following two assumptions have been made to support the derivation,

In the process to follow, the controller for the PV boost converter is designed first followed by a controller for the bidirectional converter for the battery. 
Consider the dynamics of the state $x_1$ as $\dot{x}_1=-g_1x_3+D_1$. Define the error $e_1= x_1-x_{1ref}$. Since $x_{1ref}$ is constant, the derivative $\dot{e}_1=\dot{x}_1$. The following Lyapunov function is chosen as 
\begin{align}
V_1=  \frac{1}{2g_1}e^2_1 + \frac{1}{2\gamma_1} \widetilde{\theta}_1^2
\end{align}
where  $\widetilde{\theta}_1=\theta_1-\widehat{\theta}_1$. Upon differentiation, we get,
\begin{align}
\dot{V}_1&= e_1g^{-1}_1\dot{e}_1=e_1g^{-1}_1\dot{x}_1 \nonumber \\
&= -e_1x_3+ e_1g^{-1}_1D_1 - \frac{1}{\gamma_1}\widetilde{\theta}_1\dot{\widehat{\theta}}_1
\end{align}
\begin{align}
e_1g^{-1}_1D_1 \leq |e_1|\psi_1h_1 
\leq \frac{e^2_1\psi^2_1h^2_1}{2a^2_{11}} + \frac{a^2_{11}}{2}
\end{align}
where $h_1=|max(D_1)|$ and $\psi_1= max(g^{-1}_1)$.
Hence,
\begin{align}
\dot{V}_1 &\leq -e_1x_3+\frac{e^2_1\psi^2_1h^2_1}{2a^2_{11}} + \frac{a^2_{11}}{2} - \frac{1}{\gamma_1}\widetilde{\theta}_1\dot{\widehat{\theta}}_1 \nonumber \\
&\leq e_1\Bigg[ - x_3 +\frac{e_1\psi^2_1h^2_1}{2a^2_{11}}\Bigg]  +\frac{a^2_{11}}{2}- \frac{1}{\gamma_1}\widetilde{\theta}_1\dot{\widehat{\theta}}_1 \nonumber \\
&\leq e_1\Bigg[ (\beta_1 - x_3) -\beta_1 + \frac{e_1\psi^2_1h^2_1}{2a^2_{11}}\Bigg]+\frac{a^2_{11}}{2}- \frac{1}{\gamma_1}\widetilde{\theta}_1\dot{\widehat{\theta}}_1 \nonumber 
\end{align}
Now, in order to cancel out nonlinearities, $\beta_1$ is chosen as follows,
 \begin{align}
  \beta_1= k_1e_1 + \frac{e_1\psi^2_1h^2_1}{2a^2_{11}}
 \end{align}
This results in 
\begin{align} \label{eq:v1dot0}
\dot{V}_1 \leq -k_1e^2_1-(x_3-\beta_1)e_1 + \frac{a^2_{11}}{2}- \frac{1}{\gamma_1}\widetilde{\theta}_1\dot{\widehat{\theta}}_1
\end{align}
Since, the system parameters are not known apriori, the virtual control $\beta_1$ is estimated using a neural network as 
\begin{align}
 \beta_1= W^{T}_1\phi_1(x_1) + \epsilon_1    \label{eq:nn1}
\end{align}
 It can be inferred that 
\begin{align}
    W^T_1\phi_1 \leq  \frac{\theta_1}{2\eta^2_1}\phi^T_1(x_1)\phi_1(x_1) + \frac{\eta^2_1}{2}
\end{align}
where $ \theta_1= ||W^T_1W_1||$ and $|\epsilon_1| \leq \eta_1$. This value is now substituted in the expression of $(\beta_1-x_3)e_1$ and the equality is transformed to an inequality using sum of squares due to which the following expression is obtained,
\begin{align}\label{eq:b1x3e1}
(\beta_1-x_3)e_1 \leq& -x_3e_1 + \epsilon_1e_1 + \frac{\theta_1}{2\eta^2_1}\phi^T_1\phi_1e^2_1 + \frac{\eta^2_1}{2} \nonumber \\  
\leq& -x_3e_1 + \frac{e^2_1}{2} + \frac{\epsilon^2_1}{2}+ \frac{\eta^2_1}{2}  + \frac{\theta_1}{2\eta^2_1}\phi^T_1\phi_1e^2_1 
\end{align}
 On expanding $\theta= \widetilde{\theta}+\widehat{\theta}$, substituting in expression in \eqref{eq:b1x3e1} and further substituting it in the expression of $\dot{V}_1$ given in \eqref{eq:v1dot0},  
\begin{align}
\dot{V}_1 \leq & -\Bigg(k_1-\frac{1}{2}\Bigg)e^2_1 -x_3e_1 + \frac{\theta_1}{2\eta^2_1}\phi^T_1\phi_1e^2_1 \nonumber \\&  + \frac{\eta^2_1}{2} + \frac{\epsilon^2_1}{2}+\frac{a^2_{11}}{2} - \frac{1}{\gamma_1}\widetilde{\theta}_1\dot{\widehat{\theta}}_1\nonumber \\
 \leq& -\Bigg(k_1-\frac{1}{2}\Bigg)e^2_1 -x_3e_1 + \frac{\widetilde{\theta}_1}{2\eta^2_1}\phi^T_1\phi_1e^2_1 \nonumber \\   & + \frac{\widehat{\theta}_1}{2\eta^2_1}\phi^T_1\phi_1e^2_1 + \frac{\eta^2_1}{2} + \frac{\epsilon^2_1}{2} - \frac{1}{\gamma_1}\widetilde{\theta}_1\dot{\widehat{\theta}}_1+\frac{a^2_{11}}{2} \label{eq:v1dot1}
\end{align}
A new variable $\widehat{\beta}_1$, which will function as the virtual input for the $x_1$ dynamics is now chosen as 
\begin{align} 
    \widehat{\beta}_1= \frac{\widehat{\theta}_1}{2\eta^2_1}\phi^T_1\phi_1e_1 \label{eq:beta1hat}
\end{align}
Further noting that the expression $\frac{\widehat{\theta}_1}{2\eta^2_1}\phi^T_1\phi_1e^2_1$ equals $\widehat{\beta}_1e_1$, \eqref{eq:v1dot1} can be rewritten as,
\begin{align} \label{eq:v1dot2}
\dot{V}_1 \leq& -\Bigg(k_1-\frac{1}{2}\Bigg)e^2_1 +(\widehat{\beta}_1-x_3)e_1 + \frac{\widetilde{\theta}_1}{2\eta^2_1}\phi^T_1\phi_1e^2_1 \nonumber \\ &  + \frac{\eta^2_1}{2} + \frac{\epsilon^2_1}{2} - \frac{1}{\gamma_1}\widetilde{\theta}_1\dot{\widehat{\theta}}_1+\frac{a^2_{11}}{2}\nonumber \\
\leq& -\Bigg(k_1-\frac{1}{2}\Bigg)e^2_1 +(\widehat{\beta}_1-x_3)e_1 + \frac{\eta^2_1}{2} + \frac{\epsilon^2_1}{2}+\frac{a^2_{11}}{2}  \nonumber \\
& + \frac{\widetilde{\theta}_1}{\gamma_1}\Bigg[\frac{\gamma_1\phi^T_1\phi_1e^2_1}{2\eta^2_1} - \dot{\widehat{\theta}}_1\Bigg]
\end{align}
Due to interconnected dynamics between $x_1$, $x_2$ and $x_3$, this derivative expression $V_1$ further dictates controller design process in the upcoming steps.  

Next, the dynamics of $x_2$ and $x_3$ are $\dot{x}_2= f_2(x_2,x_3,x_6)-g_2(x_3)u_1$ and $\dot{x}_3= f_3(x_1,x_2,x_3)+g_3(x_2)u_1$ are considered for deciding on the next Lyapunov candidate. The Lyapunov function for these two states is chosen as follows,
\begin{align}
 V_{2,3}=& \frac{1}{2g_2(x_3)}e^2_2 +\frac{1}{2g_3(x_2)}e^2_3 + \frac{1}{2\gamma_2} \widetilde{\theta}^2_2 + \frac{1}{2\gamma_{3}}\widetilde{\theta}^2_{3} \\ 
 & \text{where}~~ e_2= x_2-x_{2ref} \hspace{1cm} e_3= x_3-\widehat{\beta}_1  \nonumber 
\end{align}

On differentiating $\dot{V}_{2,3}$, the following expression ensues,  
\begin{align} \label{eq:v23dot11}
\dot{V}_{2,3} =& g^{-1}_2(x_3)e_2\dot{x}_2-\frac{\dot{g}_2}{g^2_2}e^2_2+g^{-1}_3(x_2)e_3(\dot{x}_3-\dot{\beta}_1)  -\frac{\dot{g}_3}{g^2_3}e^2_3  \nonumber \\ &-  \frac{1}{\gamma_2}\widetilde{\theta}_2\dot{\widehat{\theta}}_2 - \frac{1}{\gamma_3}\widetilde{\theta}_3\dot{\widehat{\theta}}_3  \nonumber \\
=& e_2f_2g^{-1}_2 -e_2u_1  -\frac{\dot{g}_2}{g^2_2}e^2_2  + e_3f_3g^{-1}_3 + e_3u_1 -e_3g^{-1}_3\dot{\widehat{\beta}}_1 \nonumber \\ &-\frac{\dot{g}_3}{g^2_3}e^2_3 - \frac{1}{\gamma_2}\widetilde{\theta}_2\dot{\widehat{\theta}}_2 - \frac{1}{\gamma_3}\widetilde{\theta}_3\dot{\widehat{\theta}}_3  
\end{align}

It is known that  $\widehat{\beta}_1$ is a function of $x_1$ and $\widehat{\theta}_1$. Therefore, the following derivative of $\widehat{\beta}_1$ can be written,
\begin{align} \label{eq:beta1hatdot}
\dot{\widehat{\beta}}_1=& \frac{\partial{\widehat{\beta}_1}}{\partial{x_1}} \dot{x}_1  + \frac{\partial{\widehat{\beta}_1}}{\partial{\widehat{\theta}}_1} \dot{\widehat{\theta}}_1 \nonumber \\ 
= & \frac{\partial{\widehat{\beta}_1}}{\partial{x_1}}[-g_1x_3+D_1] +  \frac{\partial{\widehat{\beta}_1}}{\partial{\widehat{\theta}}_1} \dot{\widehat{\theta}}_1 \nonumber \\  
= & \omega_1 + \frac{\partial{\widehat{\beta}_1}}{\partial{x_1}}D_1 
\end{align}
where $\omega_1$= $-\frac{\partial{\widehat{\beta}_1}}{\partial{x_1}}g_1x_3 +  \frac{\partial{\widehat{\beta}_1}}{\partial{\widehat{\theta}}_1} \dot{\widehat{\theta}}_1$.
On further substituting this value in the expression \eqref{eq:v23dot11}, the following is obtained,
\begin{align}
\dot{V}_{2,3}= & e_2f_2g^{-1}_2 -e_2u_1  -\frac{\dot{g}_2}{g^2_2}e^2_2  + e_3f_3g^{-1}_3 + e_3u_1  -e_3g^{-1}_3\omega_1 \nonumber \\ &-e_3g^{-1}_3\frac{\partial{\widehat{\beta}_1}}{\partial{x_1}}D_1 -\frac{\dot{g}_3}{g^2_3}e^2_3 - \frac{1}{\gamma_2}\widetilde{\theta}_2\dot{\widehat{\theta}}_2 - \frac{1}{\gamma_3}\widetilde{\theta}_3\dot{\widehat{\theta}}_3   
\end{align}
Using sum of squares it is possible to write the following terms as,
\begin{align}
-\frac{\dot{g}_2}{g^2_2}e^2_2 \leq &\frac{\dot{g}^2_2\psi^4_2e^4_2}{2a^2_{21}} + \frac{a^2_{21}}{2}  \label{eq:ssq231}\\ 
-\frac{\dot{g}_3}{g^2_3}e^2_3 \leq &\frac{\dot{g}^2_3\psi^4_3e^4_3}{2a^2_{31}} + \frac{a^2_{31}}{2} \label{eq:ssq232}\\
-e_3g^{-1}_3\frac{\partial{\widehat{\beta}_1}}{\partial{x_1}}D_1 \leq &  \frac{h^2_1 \psi^2_3m^2_{32}e^2_3}{2a^2_{32}} + \frac{a^2_{32}}{2} \label{eq:ssq233}\
\end{align}
where $m_{32}= \frac{\partial{\widehat{\beta}_1}}{\partial{x_1}}$.
Substituting the terms \eqref{eq:ssq231}, \eqref{eq:ssq232} and \eqref{eq:ssq233} in $\dot{V}_{2,3}$,  
\begin{align} \label{eq:v23dot14}
\dot{V}_{2,3} \leq&  e_2f_2g^{-1}_2 + e_3f_3g^{-1}_3 - e_3g^{-1}_3\omega_1 + (e_3-e_2)u_1 \nonumber \\
& - \frac{1}{\gamma_2}\widetilde{\theta}_2\dot{\widehat{\theta}}_2 - \frac{1}{\gamma_3}\widetilde{\theta}_3\dot{\widehat{\theta}}_3  + \frac{\dot{g}^2_2\psi^4_2e^4_2}{2a^2_{21}} + \frac{a^2_{21}}{2} +\frac{\dot{g}^2_3\psi^4_3e^4_3}{2a^2_{31}} \nonumber \\
& + \frac{a^2_{31}}{2} + \frac{h^2_1 \psi^2_3m^2_{32}e^2_3}{2a^2_{32}} + \frac{a^2_{32}}{2}
\end{align}
In the term  $(e_3-e_2)u_1$, two variables $\beta_2$ and $\beta_3$ are introduced and the whole expression is rewritten as, 
\begin{align} \label{eq:e3e2u1}
(e_3-e_2)u_1 = e_3(u_1-\beta_3) + e_3\beta_3 -e_2(u_1-\beta_2) -e_2\beta_2
\end{align}
On substituting the value of $(e_3-e_2)u_1$ from \eqref{eq:e3e2u1} into \eqref{eq:v23dot14} and rearranging, the following expression is obtained, 
\begin{align} \label{eq:v23dot15}
\dot{V}_{2,3} \leq &  e_2\Bigg[f_2g^{-1}_2 + \frac{\dot{g}^2_2\psi^4_2e^3_2}{2a^2_{21}}-\beta_2 \Bigg] +e_3(u_1-\beta_3) +\frac{a^2_{21}}{2} \nonumber \\
 &-e_2(u_1-\beta_2) - \frac{1}{\gamma_2}\widetilde{\theta}_2\dot{\widehat{\theta}}_2 - \frac{1}{\gamma_3}\widetilde{\theta}_3+\dot{\widehat{\theta}}_3 +\frac{a^2_{31}}{2}+\frac{a^2_{32}}{2} \nonumber \\ & + e_3\Bigg[f_3g^{-1}_3-g^{-1}_3\omega_1+\frac{\dot{g}^2_3\psi^4_3e^3_3}{2a^2_{31}}+\frac{h^2_1 \psi^2_3m^2_{32}e_3}{2a^2_{32}}+ \beta_3 \Bigg]  
\end{align}

The previously introduced variables $\beta_2$ and $\beta_3$ are chosen to cancel out the nonlinearities in \eqref{eq:v23dot15} as given below, 
\begin{align}
\beta_2=& k_2e_2+ f_2g^{-1}_2 + \frac{\dot{g}^2_2\psi^4_2e^3_2}{2a^2_{21}} \\
\beta_3=& -k_3e_3+ e_3 -f_3g^{-1}_3+g^{-1}_3\omega_1 -\frac{\dot{g}^2_3\psi^4_3e^3_3}{2a^2_{31}}-\frac{h^2_1 \psi^2_3m^2_{32}e_3}{2a^2_{32}} 
\end{align}
The expression of $\dot{V}_{2,3}$ gets transformed into a much simpler form as,
\begin{align} \label{eq:v23dot16}
\dot{V}_{2,3} \leq & e_3(u_1-\beta_{3})-e_2(u_1-\beta_{2})+e_1e_3-k_2e^2_2-k_3e^3_2 \nonumber \\& - \frac{1}{\gamma_2}\widetilde{\theta}_2\dot{\widehat{\theta}}_2- \frac{1}{\gamma_3}\widetilde{\theta}_3\dot{\widehat{\theta}}_3 + \frac{a^2_{31}}{2}  + \frac{a^2_{32}}{2} + \frac{a^2_{21}}{2} 
\end{align}

There are many unknown parameters in the expressions of $\beta_2$ and $\beta_3$. Hence, two neural networks, 
\begin{align}
\beta_2 &= W^{T}_2\phi_2(x_1,x_2,x_3,x_6, \widehat{\theta}_6) + \epsilon_2 , \nonumber \\
\beta_3 &= W^{T}_3\phi_3(x_1,x_3,x_3,\widehat{\theta}_1) + \epsilon_3 \label{eq:beta2beta3}
\end{align}
are employed to estimate these values. 

Noting these expressions, and using sum of squares, the following bounds are defined for $(u_1- \beta_{3})e_3$ and $(u_1- \beta_{2})e_2$, 
\begin{align}
 (u_1- \beta_{3})e_3 \leq & e_3u_1 + \frac{\widehat{\theta}_{3}}{2\eta^2_{3}}\phi^T_{3}\phi_{3}e^2_3  + \frac{e^2_3}{2}+ \frac{\epsilon^2_{3}}{2}+ \frac{\eta^2_{3}}{2} \nonumber\\ &+ \frac{\widetilde{\theta}_{3}}{2\eta^2_{3}}\phi^T_{3}\phi_{3}e^2_3\\ 
-(u_1- \beta_{2})e_2 \leq & -e_2u_1 + \frac{\widehat{\theta}_{2}}{2\eta^2_{2}}\phi^T_{2}\phi_{2}e^2_2 \nonumber  + \frac{e^2_2}{2}+ \frac{\epsilon^2_{2}}{2}+ \frac{\eta^2_{2}}{2} \\ &+ \frac{\widetilde{\theta}_{2}}{2\eta^2_{2}}\phi^T_{2}\phi_{2}e^2_2 
\end{align}

On substituting these bounds in \eqref{eq:v23dot16}, the following inequality is reached at,
\begin{align}
\dot{V}_{2,3} \leq& e_1e_3+ (e_3-e_2)u_1+\frac{\widehat{\theta}_{3}}{2\eta^2_{3}}\phi^T_{3}\phi_{3}e^2_3+\frac{\widehat{\theta}_{2}}{2\eta^2_{2}}\phi^T_{2}\phi_{2}e^2_2  \nonumber \\  &+ \frac{\widetilde{\theta}_{3}}{2\eta^2_{3}}\phi^T_{3}\phi_{3}e^2_3+ \frac{\widetilde{\theta}_{2}}{2\eta^2_{2}}\phi^T_{2}\phi_{2}e^2_2 \nonumber -k_2e^2_2-k_3e^2_3  + \frac{e^2_2}{2}\\ & + \frac{e^2_3}{2} + \frac{\epsilon^2_{3}}{2}+ \frac{\epsilon^2_{2}}{2} + \frac{\eta^2_{3}}{2}+ \frac{\eta^2_{2}}{2}  + \frac{a^2_{31}}{2}  + \frac{a^2_{32}}{2}  + \frac{a^2_{21}}{2} \nonumber \\  & -\frac{1}{\gamma_{2}}\widetilde{\theta}_{2}\dot{\widehat{\theta}}_{2} -\frac{1}{\gamma_{3}}\widetilde{\theta}_{3}\dot{\widehat{\theta}}_{3} 
\end{align}
To obtain the Lyapunov derivative of the entire PV subsystem, the $\dot{V}_1$ is combined with that of $\dot{V}_{2,3}$ to give rise to $\dot{V}_{1,2,3}$ whose expression is given below, 
\begin{align} \label{eq:v123dot1}
\dot{V}_{1,2,3} \leq& -\Sigma_{i=1}^3\Bigg(k_i-\frac{1}{2}\Bigg)e^2_i + \Sigma_{i=1}^3 \frac{\eta^2_i}{2}+\frac{\epsilon^2_i}{2} + \frac{a^2_{11}}{2} \nonumber \\
 & +\frac{a^2_{21}}{2}+ \Sigma_{i=1}^2 \frac{a^2_{3i}}{2}+ \Sigma^3_{i=1}\frac{\widetilde{\theta}_i}{\gamma_i}\Bigg(\frac{\gamma_i\phi^T_i\phi_ie^2_i}{2\eta^2_i}-\dot{\widehat{\theta}}_i\Bigg)  \nonumber \\
 & + (e_3-e_2)u_1 +\frac{\widehat{\theta}_{3}}{2\eta^2_{3}}\phi^T_{3}\phi_{3}e^2_3 +\frac{\widehat{\theta}_{2}}{2\eta^2_{2}}\phi^T_{2}\phi_{2}e^2_2 \nonumber
\end{align}
The controller $u_1$ for the boost converter of the PV array is chosen in terms of the neural network parameters which capture the dynamics related to both the states $x_2$ and $x_3$ as given below:
\begin{align}
 u_1=-(e_3-e_2)^{-1}\Bigg(\frac{\widehat{\theta}_{3}}{2\eta^2_{3}}\phi^T_{3}\phi_{3}e^2_3 +\frac{\widehat{\theta}_{2}}{2\eta^2_{2}}\phi^T_{2}\phi_{2}e^2_2\Bigg)
\end{align}
On substituting the value of $u_1$, the composite Lyapunov derivative of the solar PV subsystem is obtained as follows,
\begin{align}
\dot{V}_{1,2,3} \leq & -\Sigma_{i=1}^3\Bigg(k_i-\frac{1}{2}\Bigg)e^2_i + \Sigma_{i=1}^3 \frac{\eta^2_i}{2}+\frac{\epsilon^2_i}{2} + \frac{a^2_{11}}{2} \nonumber \\
 & +\frac{a^2_{21}}{2}+ \Sigma_{i=1}^2 \frac{a^2_{3i}}{2}+ \Sigma^3_{i=1}\frac{\widetilde{\theta}_i}{\gamma_i}\Bigg(\frac{\gamma_i\phi^T_i\phi_ie^2_i}{2\eta^2_i}-\dot{\widehat{\theta}}_i\Bigg) 
\end{align}
This composite Lyapunov derivative will later be added to the Lyapunov derivative of the battery subsystem to analyse the stability of the entire system.

A similar backstepping based procedure is performed on the states $x_6$, $x_4$ and $x_5$, of the battery subsystem and the following control laws are designed, 
\begin{align}
\widehat{\beta}_6 =& -\frac{\widehat{\theta}_6}{2\eta^2_6}\phi^T_6\phi_6e_6 \\
u_2 =&-(e_4-e_5)^{-1}\Bigg(\frac{\widehat{\theta}_{4}}{2\eta^2_{4}}\phi^T_{4}\phi_{4}e^2_4 +\frac{\widehat{\theta}_{5}}{2\eta^2_{5}}\phi^T_{5}\phi_5e^2_5\Bigg)
\end{align}

A composite Lyapunov function for the battery subsystem is chosen as given below, $V_{4,5,6}=\Sigma_{i=4}^6\frac{1}{2g_i}e^2_i+ \frac{1}{2\gamma_i}\widetilde{\theta}^2_i$ whose derivative is shown in the following expression, 
\begin{eqnarray}
\dot{V}_{4,5,6} &\leq& -\Sigma_{i=4}^6\Bigg[\Bigg(k_i-\frac{1}{2}\Bigg)e^2_i + \frac{\eta^{2}_{i}}{2} + \frac{\epsilon^2_{i}}{2}\Bigg] + \frac{a^2_{61}}{2}\nonumber \\
& & + \Sigma^{2}_{i=1} \frac{a^2_{4i}}{2}+ \Sigma^{2}_{i=1} \frac{a^2_{5i}}{2} +  \Sigma^{6}_{i=4}\frac{\widetilde{\theta}_i}{\gamma_i}\Bigg(\frac{\gamma_i\phi^T_i\phi_ie^2_i}{2\eta^2_i}-\dot{\widehat{\theta}}_i\Bigg) 
\end{eqnarray}

Adding the Lyapunov derivative expressions of the two subsystems, $\dot{V}_{1,2,3}$ and $\dot{V}_{4,5,6}$, the derivative of the Lyapunov of the entire system $\dot{V}$ becomes,
\begin{eqnarray}
\dot{V} &\leq& -\Sigma_{i=1}^6\Bigg[\Bigg(k_i-\frac{1}{2}\Bigg)e^2_i + \frac{\eta^{2}_{i}}{2} + \frac{\epsilon^2_{i}}{2}\Bigg] \nonumber \\
& &+  \Sigma^{6}_{i=1}\frac{\widetilde{\theta}_i}{\gamma_i}\Bigg(\frac{\gamma_i\phi^T_i\phi_ie^2_i}{2\eta^2_i}-\dot{\widehat{\theta}}_i\Bigg)  + \frac{a^2_{11}}{2}+ \frac{a^2_{21}}{2}+ \frac{a^2_{61}}{2} \nonumber \\
& &+ \Sigma^{2}_{i=1} \Bigg(\frac{a^2_{3i}}{2}+\frac{a^2_{4i}}{2}+ \frac{a^2_{5i}}{2}\Bigg) \label{eq:vdot1} 
\end{eqnarray}

The following update law is chosen for NN weights $\widehat{\theta}_i$,
\begin{align}
\dot{\widehat{\theta}}_{i} = \frac{\gamma_{i}\phi^T_{i}\phi_{i}e^2_i}{2\eta^2_{i}} - p_i\widehat{\theta}_i \label{eq:thetahatdot}
\end{align}
which leads to, 
\begin{align} \label{eq:vdot}
\dot{V} \leq  &-\Sigma_{i=4}^6\Bigg[\Bigg(k_i-\frac{1}{2}\Bigg)e^2_i + \frac{\eta^{2}_{i}}{2} + \frac{\epsilon^2_{i}}{2}+\frac{p_i}{\gamma_i}\widehat{\theta}_i\widetilde{\theta}_i \Bigg]\nonumber \\ & + \frac{a^2_{11}}{2}+ \frac{a^2_{21}}{2}+ \frac{a^2_{61}}{2} + \Sigma^{2}_{i=1} \Bigg(\frac{a^2_{3i}}{2}+\frac{a^2_{4i}}{2}+ \frac{a^2_{5i}}{2}\Bigg) 
\end{align}

It must be noted that $ \widehat{\theta}_i\widetilde{\theta}_i\leq  \frac{\theta^2_i}{2}-\frac{\widetilde{\theta}^2_i}{2} $. This value is substituted in the expression of $\dot{V}$ given in \eqref{eq:vdot} and the following inequality is reached,
\begin{align}
  \dot{V} 
  \leq& -a_0\Sigma^6_{i=1}\Bigg(\frac{e^2_i}{b_{mi}}+\frac{\widetilde{\theta}^2_i}{2\gamma_i}\Bigg)+ b_0\nonumber \\
    a_0 =& min\Bigg[\Bigg(k_i-\frac{1}{2}\Bigg)b_{mi},p_i\Bigg]\nonumber  ~~\text{where}~~ i=1,2,...,6\\
  b_0 =& \Sigma_{i=1}^6 \Bigg(\frac{\eta^2_i}{2} + \frac{\epsilon^2_i}{2} +\frac{p_i}{2\gamma_i}{\theta}^2_i \Bigg) \nonumber \\  & + \frac{a^2_{11}}{2}+ \frac{a^2_{21}}{2}+ \frac{a^2_{61}}{2} + \Sigma^{2}_{i=1} \Bigg(\frac{a^2_{3i}}{2}+\frac{a^2_{4i}}{2}+ \frac{a^2_{5i}}{2}\Bigg) \nonumber
\end{align}

For this system, applying Assumption 2, it can be written that,
$\frac{1}{g_i}\leq \frac{1}{b_{mi}} \text{~~and~~}  -\frac{1}{b_{mi}} \leq -\frac{1}{g_i} $ which is equivalent to 
\begin{align}
  -\Bigg(\frac{1}{b_{mi}}\Bigg)e^2_i \leq -\Bigg(\frac{1}{g_{i}}\Bigg)e^2_i 
\end{align}
Hence, it can easily be deduced that,
\begin{align}
\dot{V}\leq -a_0\Sigma^6_{i=1}\Bigg(\frac{1}{g_{i}}\Bigg)e^2_i +b_0     
\end{align}
From these expressions it can be easily deduced that, 
\begin{align}
\dot{V} \leq -a_0V + b_0    
\end{align}

This results in 
\begin{equation}\label{eq:uub} 
V \le \left ({{V\left ({{t_{0}} }\right) - \frac {b_{0}}{a_{0}}} }\right){e^{ - {a_{0}}(t - {t_{0}})}} + \frac {b_{0}}{a_{0}} \le V\left ({{t_{0}} }\right) + \frac {b_{0}}{a_{0}} 
\end{equation}
This final expression shows that all the system states attain ultimately uniform  boundedednes (UUB) and thus remain stable.

\subsection{Notes on bounds and gain selection}
\textcolor{black}{Considering the definition of overall Lyapunov function $V$ of the DCSSMG system as,
\begin{align}
    V= \Sigma^6_{i=1} \frac{1}{2g_i}e^2_i + \frac{1}{2\gamma_i}\widetilde{\theta}^2_i
\end{align}
and the bounds on $|g_i|$ as given in Assumption 2, the following can be written,
\begin{align}
    \frac{1}{2b_M}\Sigma^6_{i=1}e^2_i \leq \Sigma^6_{i=1} \frac{1}{2g_i}e^2_i \leq V \leq B \label{eq:bounds1}
\end{align}
 where $B=V(t_{0}) + \frac {b_{0}}{a_{0}}$ and $b_M=max(b_{Mi})$
 The following expression can be obtained as already discussed,
 \begin{align}
     \frac{1}{2\gamma_M}\Sigma^6_{i=1}\widetilde{\theta}^2_i \leq \Sigma^6_{i=1} \frac{1}{2\gamma_i}\widetilde{\theta}^2_i \leq V \leq B  \label{eq:bounds2}
 \end{align}
 The inequalities \eqref{eq:bounds1} and \eqref{eq:bounds2} result in the following bounds in errors, 
 \begin{align}
     \Sigma^6_{i=1}e^2_i \leq 2b_M B, \quad \Sigma^6_{i=1}\widetilde{\theta}^2_i \leq 2\gamma_M B 
 \end{align}
These expressions show that the sum of errors in states and sum of errors in the NN weights are bounded within the values specified by \eqref{eq:bounds1} and \eqref{eq:bounds2}.  Also, the following expression can be written combining Assumption 2 and \eqref{eq:uub},
 \begin{align}
     \Bigg(\frac{1}{b_{Mi}}\Bigg)e^2_i \leq \frac{1}{2g_i}e^2_i \leq V \leq V(t_0)+ \frac{b_0}{a_0}
 \end{align}
 Applying limit on this expression, the following bound can be derived for individual errors as follows,
 \begin{align}
     \lim_{t\to\infty} e^2_i \leq b_{Mi} \Bigg(\frac{b_0}{a_0}\Bigg)
 \end{align}
 It can be verified that $\epsilon_i,~\theta_i,b_{Mi}$ are constants whereas, $k_i,p_i,\eta_i,\gamma_i,a_i$ are tunable gains which are adjusted so that for any $\Delta>0$, the limit of error $\lim_{t\to\infty} |e_i|. \leq \Delta$ for all $t\geq t_0 +T$.
  \begin{align}
      b_0 =&  3\Bigg(\eta^2_i + \epsilon^2_i +\frac{p_i}{\gamma_i}{\theta}^2_i \Bigg) + 6a^2 \nonumber \\
       a_0 =& min\Bigg[\Bigg(k_i-\frac{1}{2}\Big)b_{mi,min},p_i\Bigg]\nonumber \\
     \lim_{t\to\infty} e^2_i \leq & b_{Mi} \Bigg(\frac{ 3\Big(\eta^2_i + \epsilon^2_i +\frac{p_i}{\gamma_i}{\theta}^2_i \Big) + 6a^2}{min\Big[\Big(k_i-\frac{1}{2}\Big)b_{mi,min},p_i\Big]}\Bigg)
 \end{align}}
\section{Simulation Studies} \label{sec:ancresults}
The current section studies the performance of the developed ANC controller through thorough simulations performed using MATLAB 2018b software. A total of 6 radial basis function neural networks (NNs) are considered where each NN with 20 nodes captures the dynamic model of respective states. \textcolor{black}{The optimal number of nodes in the NN were chosen through sensitivity analysis of the output steady state error with respect to the number of nodes. Increasing the number of nodes to 20 resulted in lesser neural network reconstruction error thereby improving the steady state performance. Thus, this number was fixed for the simulation studies.}
\vspace{-0.5cm}
\begin{table}
\begin{tabular}{c}
\includegraphics[width=1.0\textwidth]{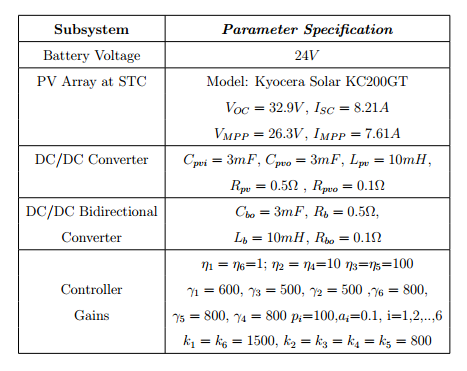}
\end{tabular}
\caption{DCSSMG Specifications}
\label{table:syspar}
\end{table}

The reference output of the DCSSMG system is $y_d=[x_{1ref} x_{6ref}]$ where $x^{6ref}$ is the DCSSMG voltage and $x_{1ref}$ is the reference for the PV array output voltage. It is to be noted that $x_{6ref}$ is always set to 40$V$ and $x_{1ref}$ is the MPPT voltage set by the secondary controller depending on the PV characteristic of the chosen PV array.  The details of different subsystems of the DCSSSMG system are given in Table \ref{table:syspar}. The efficacy of the proposed ANC controller is compared with that of a state of the art back-stepping based controller \cite{iovinetase17} (BS) and a classical adaptive back-stepping based controller (ABS).

\subsection{Case-1: Change in Load}
\begin{figure}[H]
\captionsetup[subfigure]{aboveskip=-1pt,belowskip=-1pt}
\centering
\begin{subfigure}{0.45\textwidth}
  \centering
\includegraphics[width =  7.5cm,height=5cm]{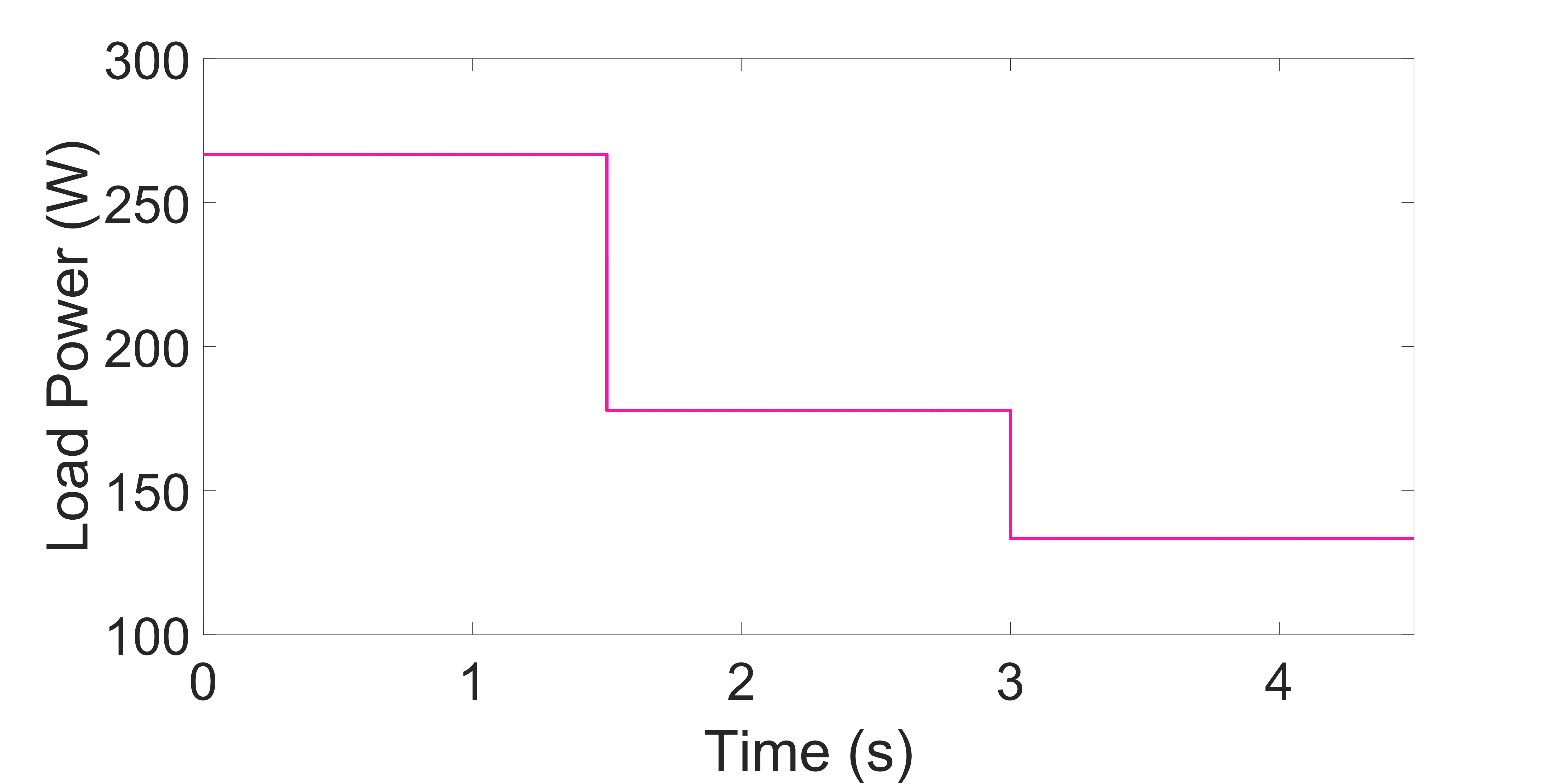}
\caption{\scriptsize{Case 1: Change in Load Power}}
\label{fig:loadpower_plot_lc2}
\end{subfigure}
\begin{subfigure}{.45\textwidth}
  \centering
\includegraphics[width =  7.5cm,height=5cm]{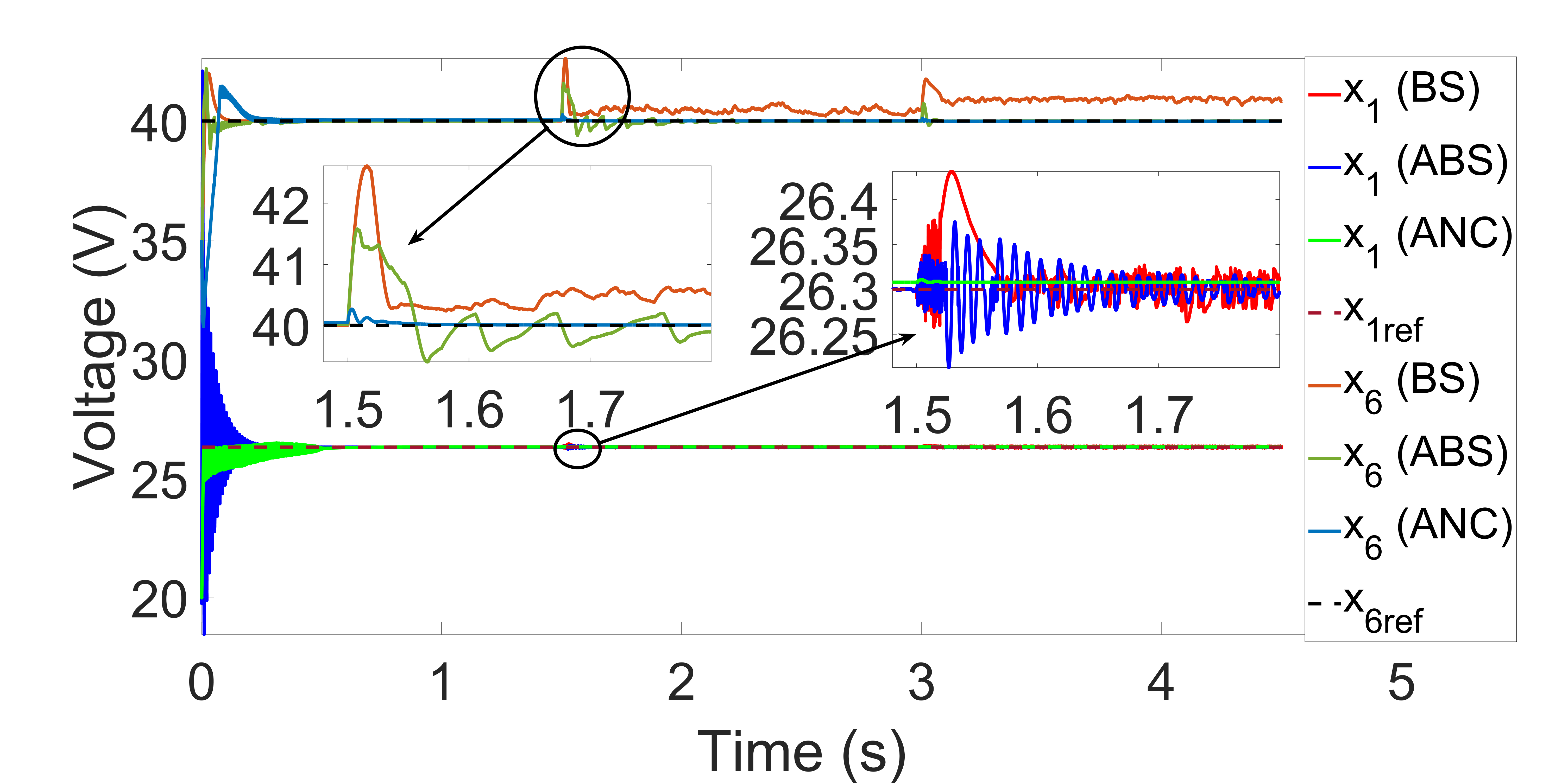}
\caption{\scriptsize{Case 1: Output voltage comparison}}
\label{fig:outputs_lc2}
\end{subfigure}\\
\begin{subfigure}{.45\textwidth}
  \centering
\includegraphics[width =  7.5cm,height=5cm]{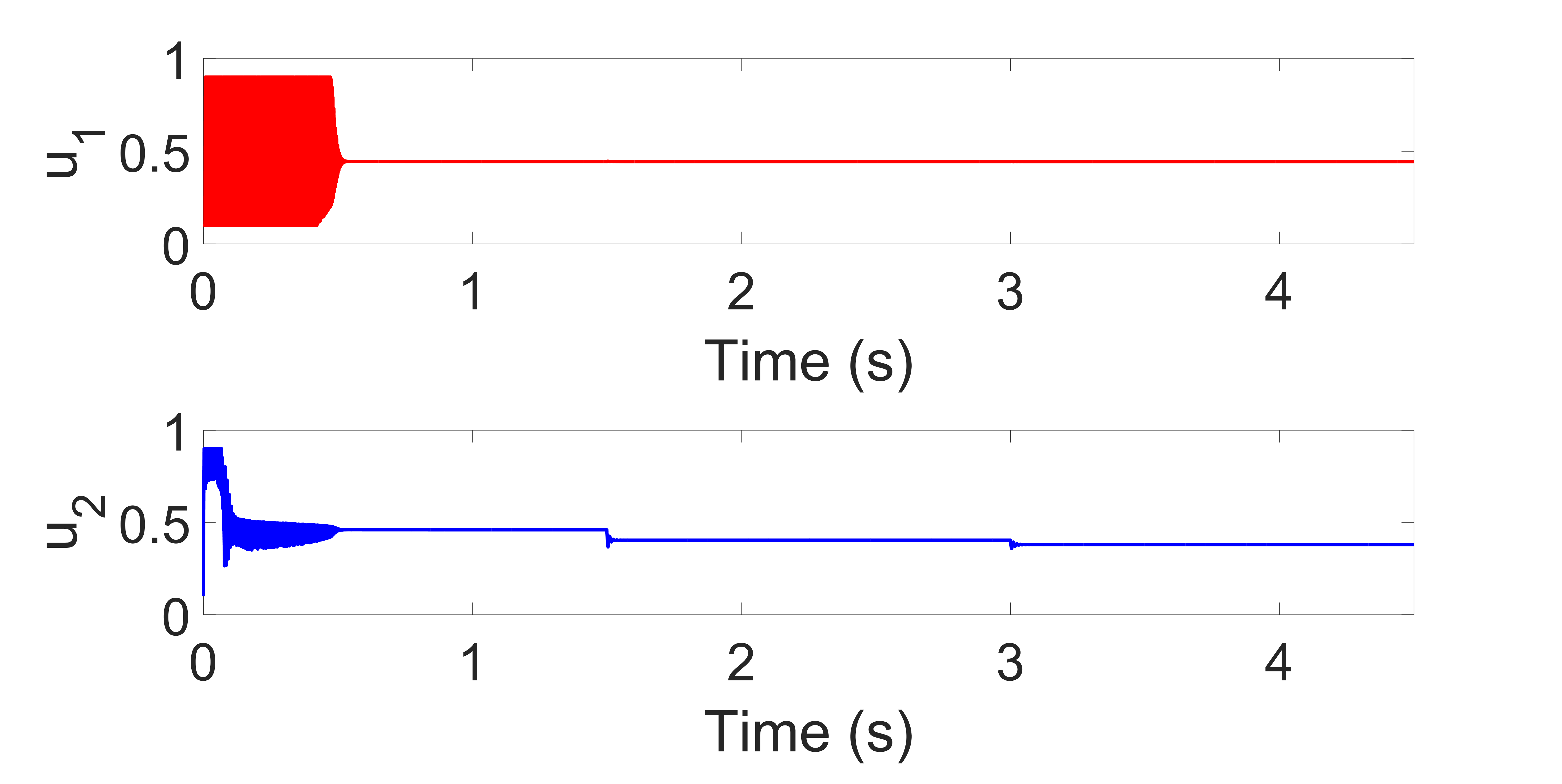}
\caption{\scriptsize{Case 1: Duty cycle - Proposed ANC Controller}}
\label{fig:u_adaptive_lc2}
\end{subfigure}
\begin{subfigure}{.45\textwidth}
  \centering
\includegraphics[width =  7.5cm,height=5cm]{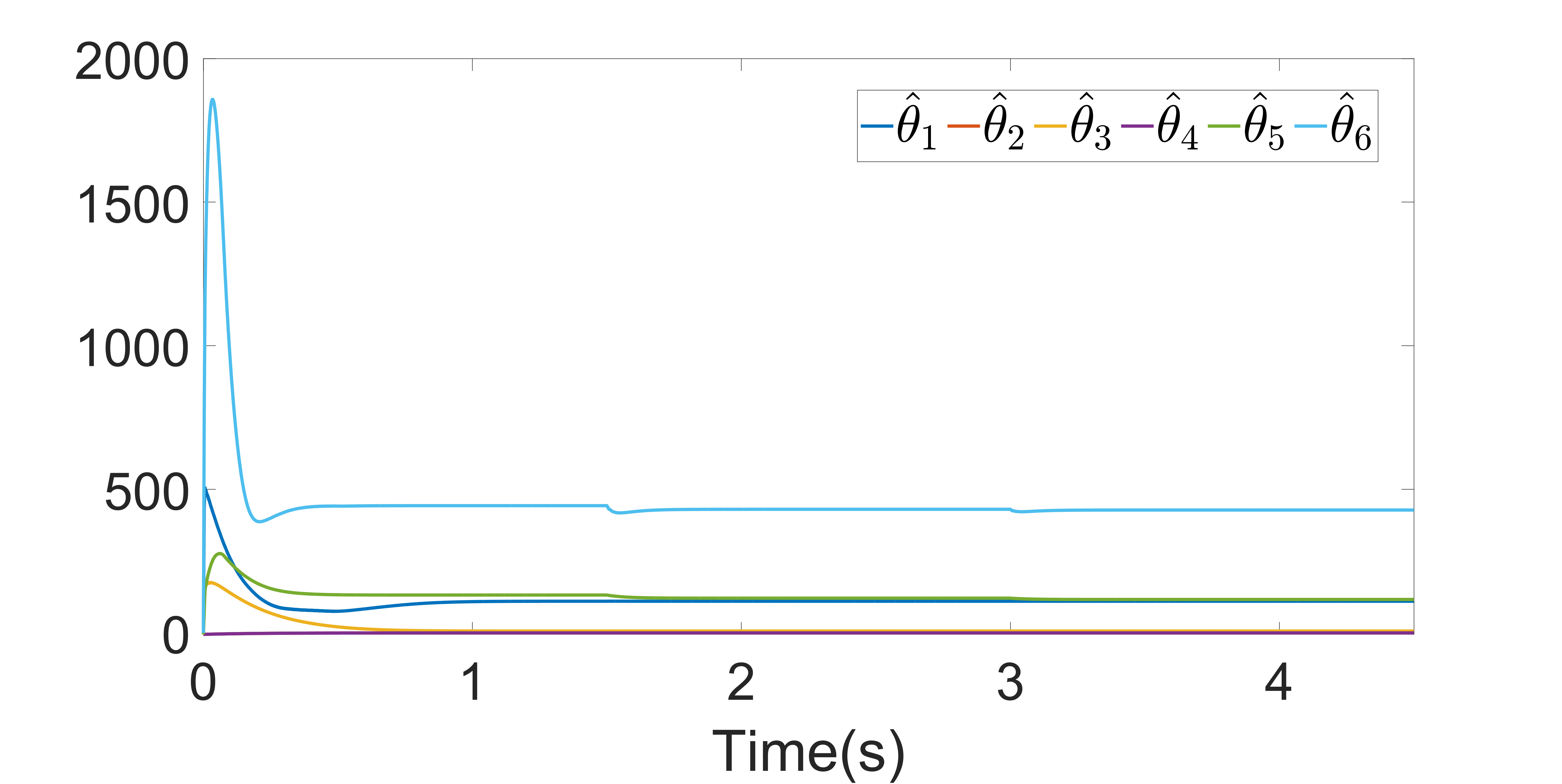}
\caption{\scriptsize{Case 1: NN weights - ANC Controller}}
\label{fig:thetahat_adaptive_lc2}
\end{subfigure}\\
\begin{subfigure}{.45\textwidth}
  \centering
\includegraphics[width =  7.5cm,height=5cm]{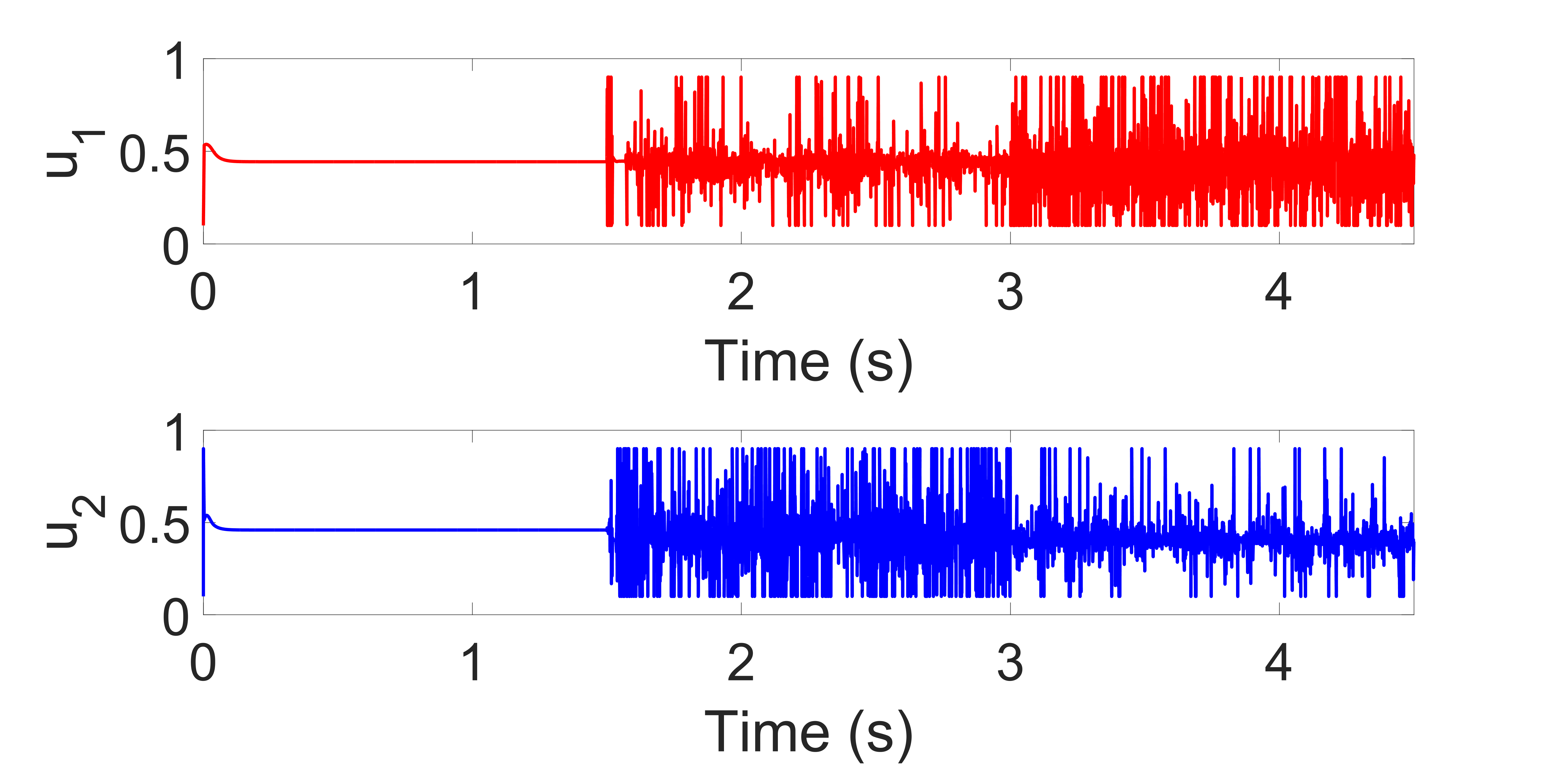}
\caption{\scriptsize{Case 1: Duty cycle - BS Controller.}}
\label{fig:u_alessio_ns_lc2}
\end{subfigure}
\begin{subfigure}{.45\textwidth}
  \centering
\includegraphics[width =  7.5cm,height=5cm]{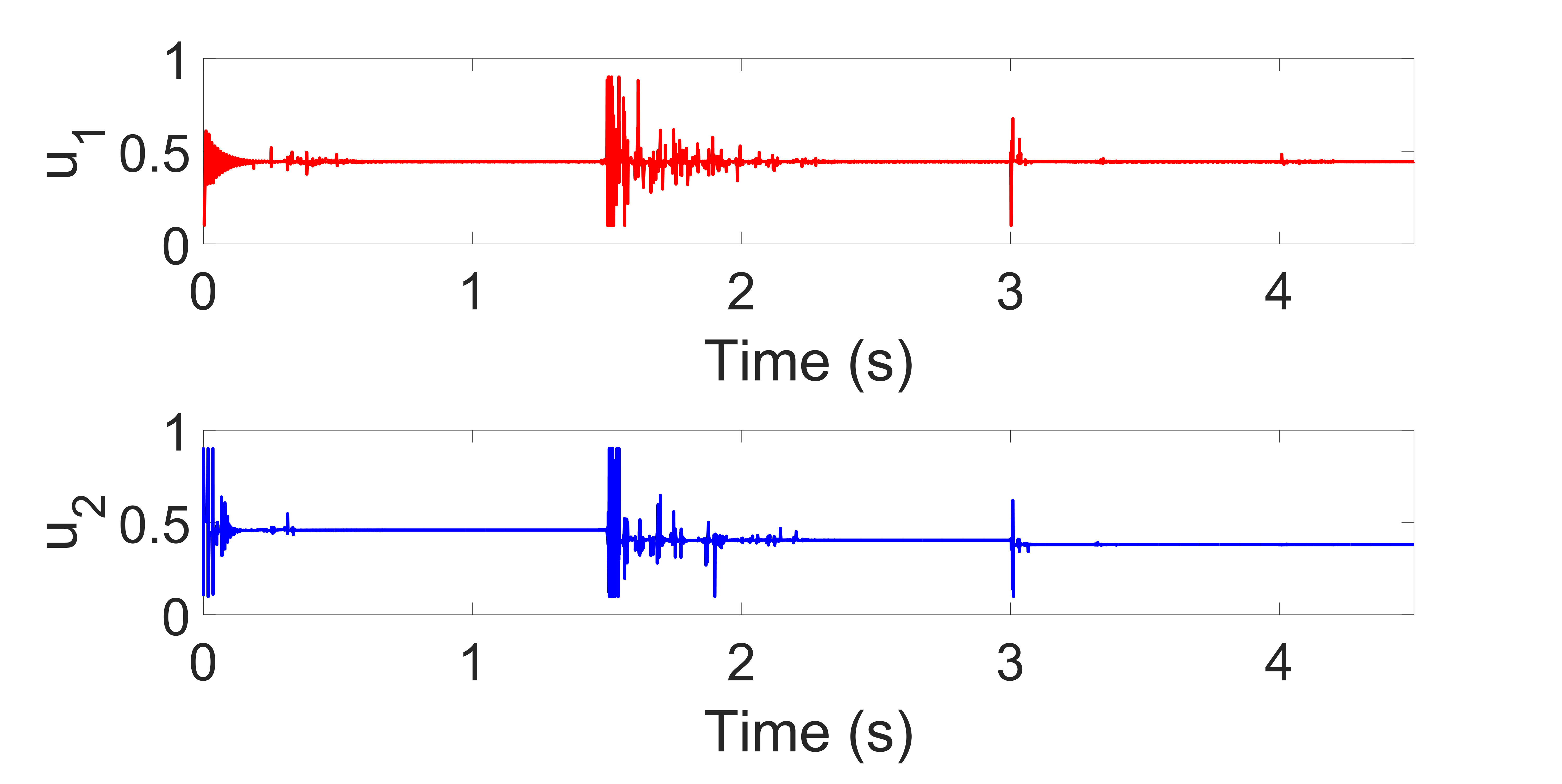}
\caption{\scriptsize{Case 1: Duty cycle - ABS Controller.}}
\label{fig:u_mahmud_lc2}
\end{subfigure}
\caption{Case-1: Validation of the proposed Adaptive Neural Controller (ANC) against state-of-the-art controllers- Adaptive Back-stepping (ABS) controller\cite{overviewtkr2} and Backstepping controller(BS)\cite{iovinetase17} for change in load. }
\end{figure}

This case is meant to test the developed controller when the atmospheric conditions remain constant and only the DCSSMG load changes. The irradiance and temperature are maintained steadily at 1000 $W/m^2$ and 25 $^{o}$C. The value of $x_{1ref}$ as per the  PV characteristic is deemed as $26.31~V$. However, the load resistance is set from 6$\Omega$ to 9$\Omega$ and 12$\Omega$, at 0$s$, 1.5$s$ and 3$s$ respectively whose corresponding load power is shown in Fig.\ref{fig:loadpower_plot_lc2}. 
When load is changed, it gets reflected in the corresponding disturbance $D_3$ of the system model. Fig.\ref{fig:outputs_lc2} shows the comparative variation in MPPT voltage and the DC grid voltage of the DCSSMG system when the three different controllers including proposed ANC controller is used. It is assumed that the exact value of disturbance $D_3$ is known to all the three controllers from 0-1.5 $s$ and it becomes unknown after 1.5$s$.  Fig.\ref{fig:u_adaptive_lc2}, Fig.\ref{fig:u_alessio_ns_lc2}, and Fig.\ref{fig:u_mahmud_lc2} show how the control inputs are generated using ANC, BS and ABS controllers. Fig.\ref{fig:thetahat_adaptive_lc2} shows how the NN weights get updated when load gets changed in the presence of ANC controller. 

\textcolor{black}{It is seen in Fig\ref{fig:outputs_lc2} that the output voltages are well regulated by the BS controller from 0$s$ to 1.5$s$ when the exact value of the load is known. When the load changes at 1.5$s$, and the BS controller does not receive the updated value of load, the DC grid voltage ($x_6$) does not settle at the given reference value. Fig.\ref{fig:u_alessio_ns_lc2} shows how the input duty cycle gets distorted due to unknown load value. Opposed to this, the ABS controller estimates the load value thereby achieving improved DC grid voltage control ($x_6$). However, a lot of oscillation is seen in controlling both the output states $x_1$ and $x_6$ using ABS controller. This oscillation stays as long as the load value does not converge to its exact value. This is also reflected in the duty cycle generated by the ABS controller as shown in Fig.\ref{fig:u_mahmud_lc2}. }

\textcolor{black}{Opposed to the BS and ABS controllers, the proposed ANC controller facilitates the output states $x_1$ and $x_6$ to settle way faster and with reduced oscillations. When the value of load changes or becomes unknown, the NNs present in the ANC controller update their weights according to changes in the system and provide accurate system model to the controller for better performance. The updation of weights especially $\widehat{\theta}_6$ which captures the changes in $D_3$ can be observed in Fig.\ref{fig:thetahat_adaptive_lc2}. Even the peak overshoot was reduced to almost one-eight value as compared to the BS and ABS algorithms as can be seen in Table \ref{tab:comp_perf}. }
\vspace{-0.5cm}
 \subsection{Case-2: Change in Temperature}
\begin{figure}[H]
\captionsetup[subfigure]{aboveskip=-1pt,belowskip=-1pt}
\centering
\begin{subfigure}{0.45\textwidth}
  \centering
\includegraphics[width =  7.5cm,height=5cm]{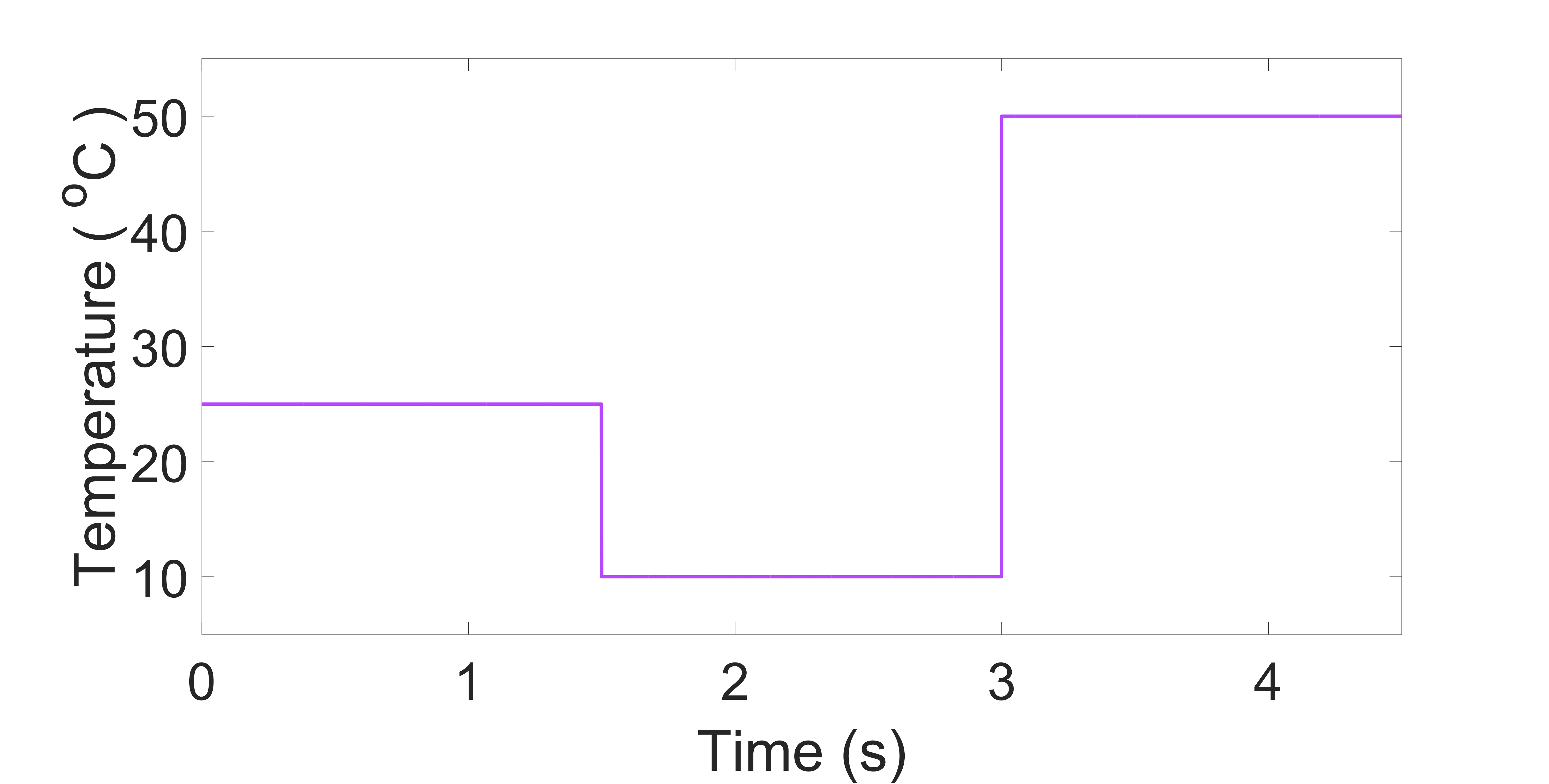}
\caption{\scriptsize{ Case 2: Change in Temperature}}
\label{fig:temp_plot_tc2}
\end{subfigure}
\begin{subfigure}{.45\textwidth}
  \centering
\includegraphics[width =  7.5cm,height=5cm]{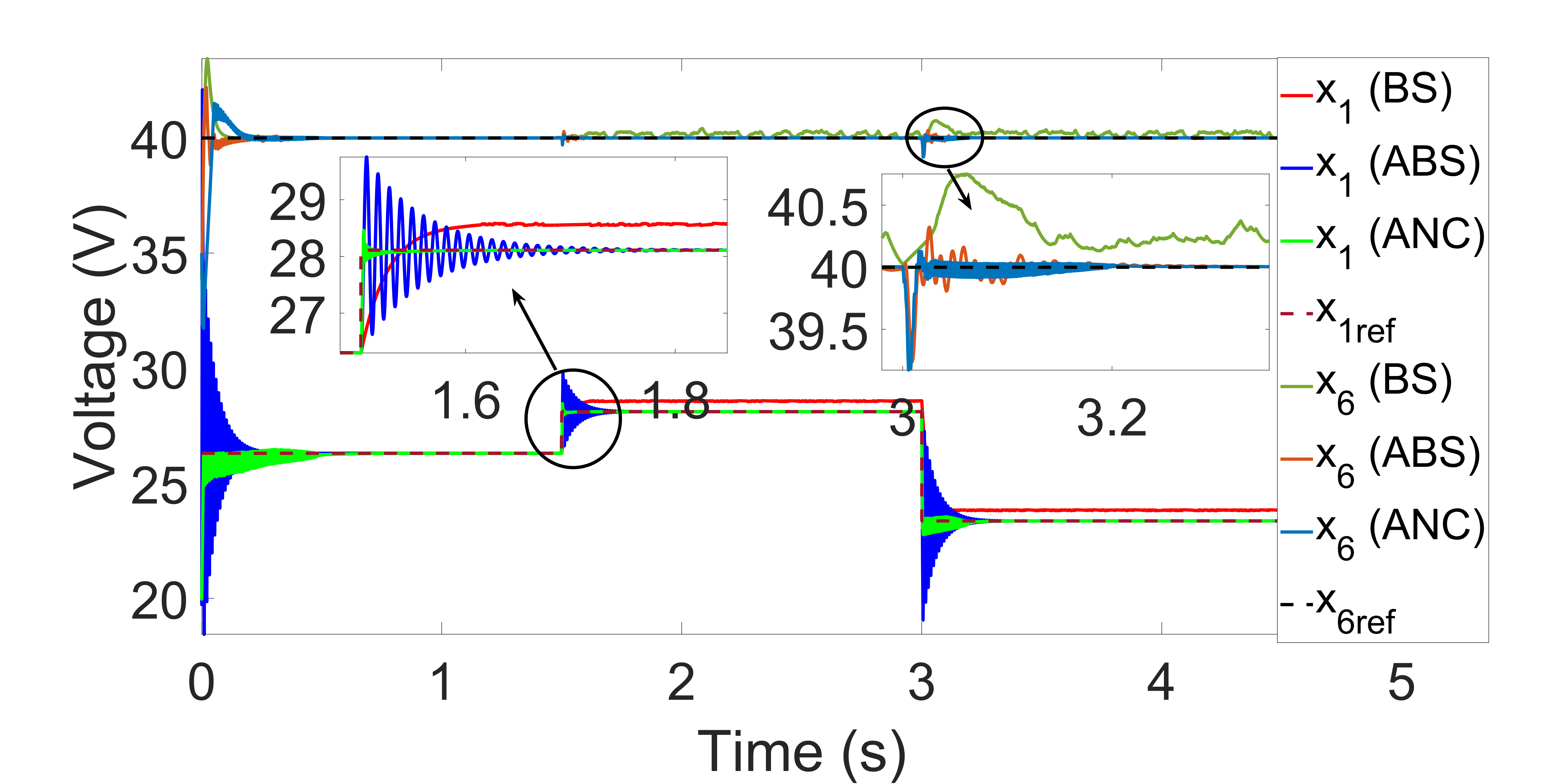}
\caption{\scriptsize{Case 2: Output voltage comparison}}
\label{fig:outputs_tc2}
\end{subfigure}\\
\begin{subfigure}{.45\textwidth}
  \centering
\includegraphics[width =  7.5cm,height=5cm]{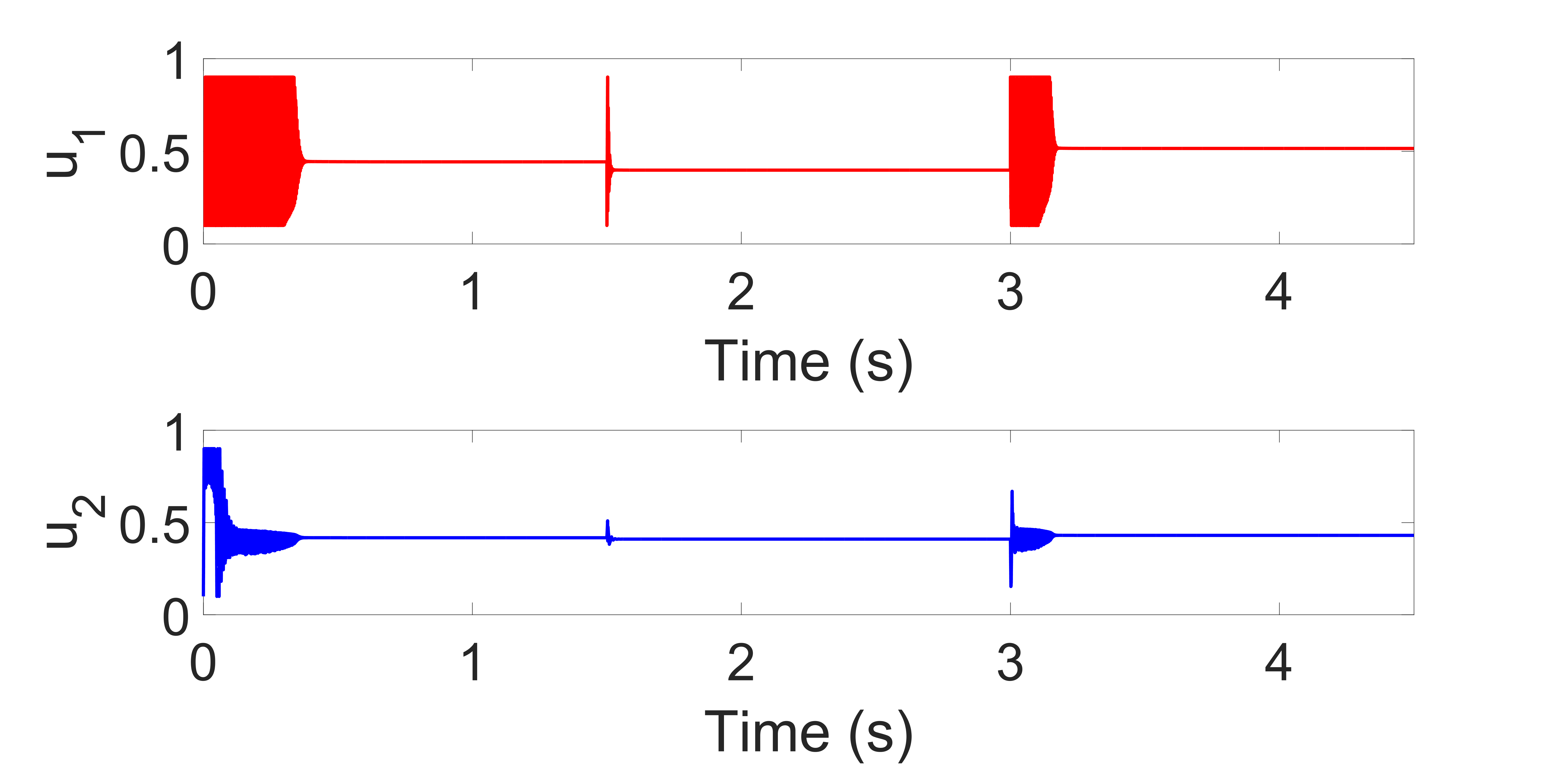}
\caption{\scriptsize{Case 2: Duty cycle - Proposed ANC Controller Controller}}
\label{fig:u_adaptive_tc2}
\end{subfigure}
\begin{subfigure}{.45\textwidth}
  \centering
\includegraphics[width =  7.5cm,height=5cm]{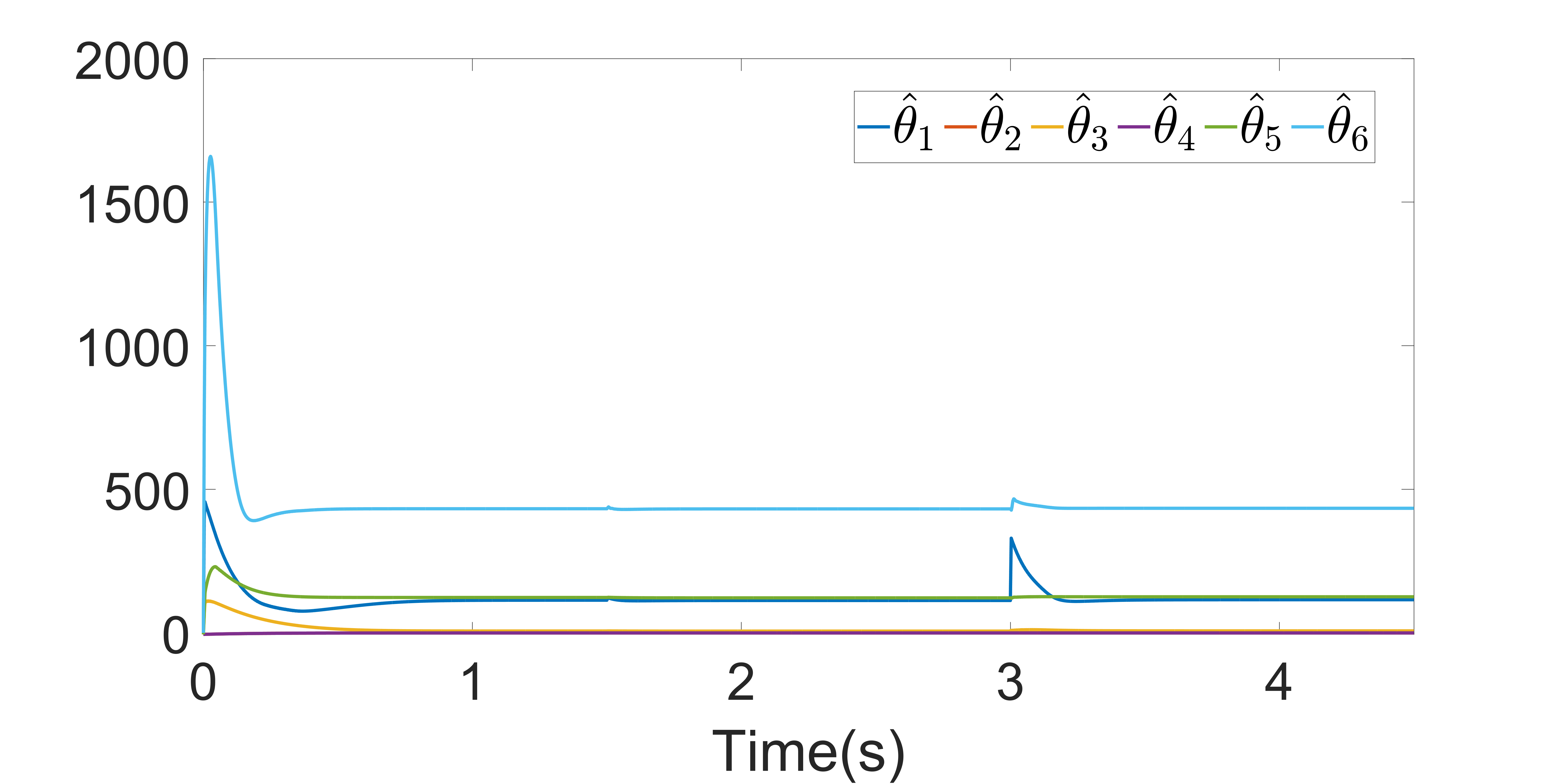}
\caption{\scriptsize{Case 2: NN weights - ANC Controller}}
\label{fig:thetahat_adaptive_tc2}
\end{subfigure}\\
\begin{subfigure}{.45\textwidth}
  \centering
\includegraphics[width =  7.5cm,height=5cm]{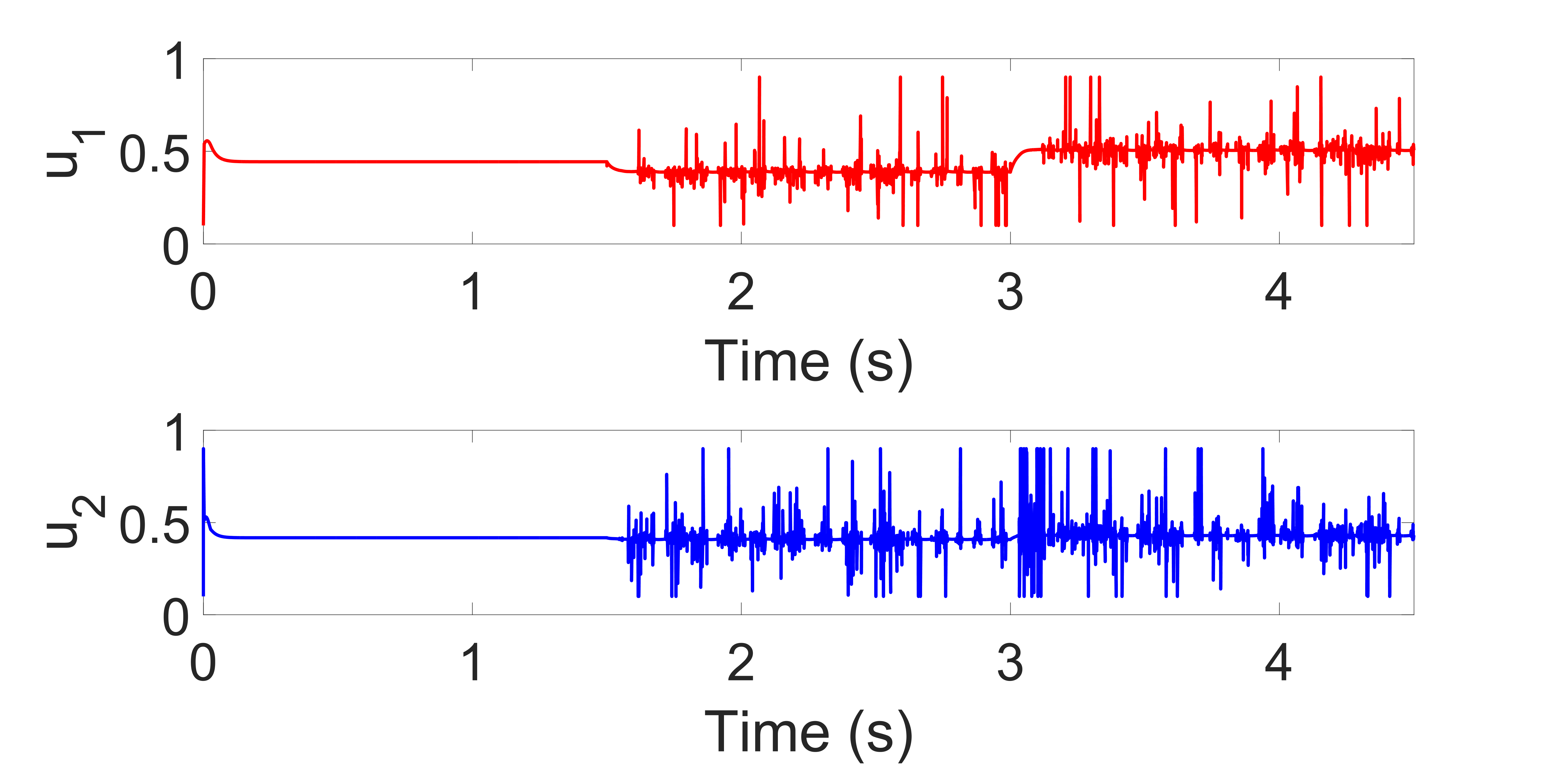}
\caption{\scriptsize{Case 2: Duty cycle - BS Controller.}}
\label{fig:u_alessio_ns_tc2}
\end{subfigure}
\begin{subfigure}{.45\textwidth}
  \centering
\includegraphics[width =  7.5cm,height=5cm]{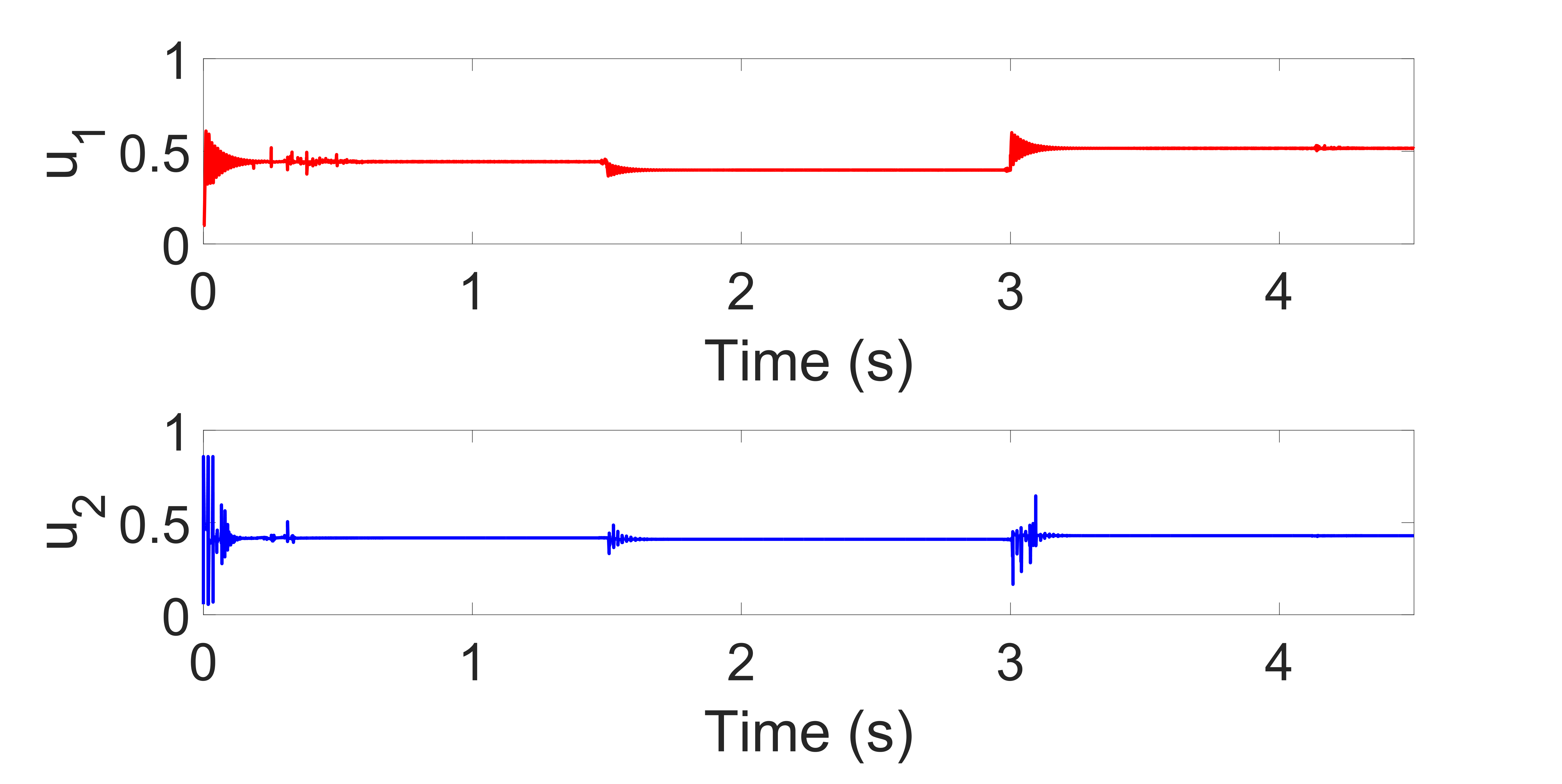}
\caption{\scriptsize{Case 2: Duty cycle - ABS Controller.}}
\label{fig:u_mahmud_tc2}
\end{subfigure}
\caption{Case-2: Validation of the proposed Adaptive Neural Controller (ANC) against state-of-the-art controllers- Adaptive Back-stepping (ABS) controller \cite{overviewtkr2}and Backstepping controller(BS)\cite{iovinetase17} for change in temperature. }
\end{figure}

\textcolor{black}{The power obtained from the photovoltaic cells becomes less than the maximum power possible when temperature rises. The values of the MPPT voltage and current also reduce with incerase in temperature. To demonstrate the effect of temperature the load power and irradiance are steadily maintained at 200$W$, 1000$W/m^2$ respectively as shown in Fig.\ref{fig:temp_plot_tc2}. The temperature is set to 25$^{o}$ C to 15$^{o}$ C and to 50$^{o}$ C respectively, at 0 $s$, 1.5$s$ and 3$s$ respectively. According to the temperature, $x_{1ref}$ is calculated using PV characteristic and set to $26.3~V$, $28.11~V$ and $23.36~V$ respectively. When temperature is changed, it gets reflected in the corresponding disturbance $D_1$ of the system model. The temperature is initially assumed to be known to all the controllers but it becomes unknown after 1.5$s$. The MPP voltage and DC grid voltage profiles are shown in Fig.\ref{fig:outputs_tc2} when each of the three controllers - BS, ABS and ANC are used in the DCSSMG system.  Fig.\ref{fig:u_adaptive_tc2} , Fig.\ref{fig:u_alessio_ns_tc2} and Fig.\ref{fig:u_mahmud_tc2} show how the control inputs are generated when ANC, BS and ABS controllers are applied in the DCSSMG system. Fig.\ref{fig:thetahat_adaptive_tc2} shows how the NN weights get evolve when temperature is changed in the presence of ANC controller. }
\textcolor{black}{From Fig.\ref{fig:outputs_tc2} it can be seen that the BS controller generates a steady state error in both the output voltage states $x_1$ and $x_6$ when the exact value of $D_1$ becomes unknown due to change in temperature at 1.5$s$ which also leads to oscillating control input as shown in Fig.\ref{fig:u_alessio_ns_tc2}. Similar to the previous case, the ABS controller shows oscillation till the value of $D_1$ gets properly updated by its observer. The ANC controller also shows similar oscillations till the NN weights especially $\widehat{\theta}_1$ properly converges to capture the changes in $D_1$ as shown in Fig.\ref{fig:thetahat_adaptive_tc2}. However, it is seen that the oscillations in DC grid output voltage $x_6$ and MPP voltage $x_1$ are lesser for the ANC controller compared to ABS and BS controllers.}
 \subsection{Case-3: Change in Irradiance}
To demonstrate the effect of irradiance on the DCSSMG, the solar power incident on the PV panel is varied while temperature and load are maintained steadily at 25$^{o}$C and 200$W$.
\begin{figure}[H]
\captionsetup[subfigure]{aboveskip=-1pt,belowskip=-1pt}
\centering
\begin{subfigure}{0.45\textwidth}
  \centering
\includegraphics[width =  7.5cm,height=5cm]{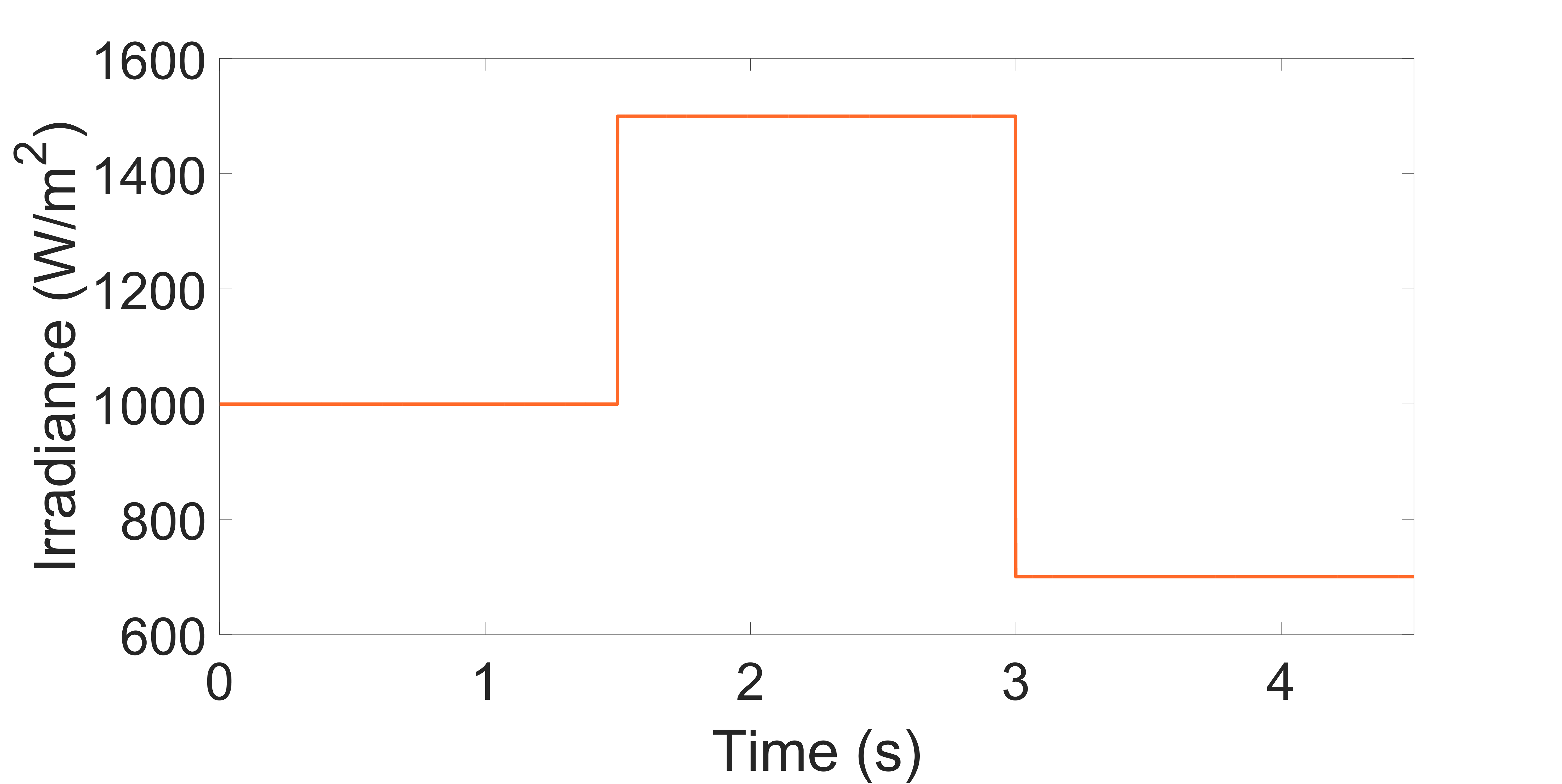}
\caption{\scriptsize{ Case 3: Change in Irradiancee}}
\label{fig:irradiance_ic2}
\end{subfigure}
\begin{subfigure}{.45\textwidth}
  \centering
\includegraphics[width =  7.5cm,height=5cm]{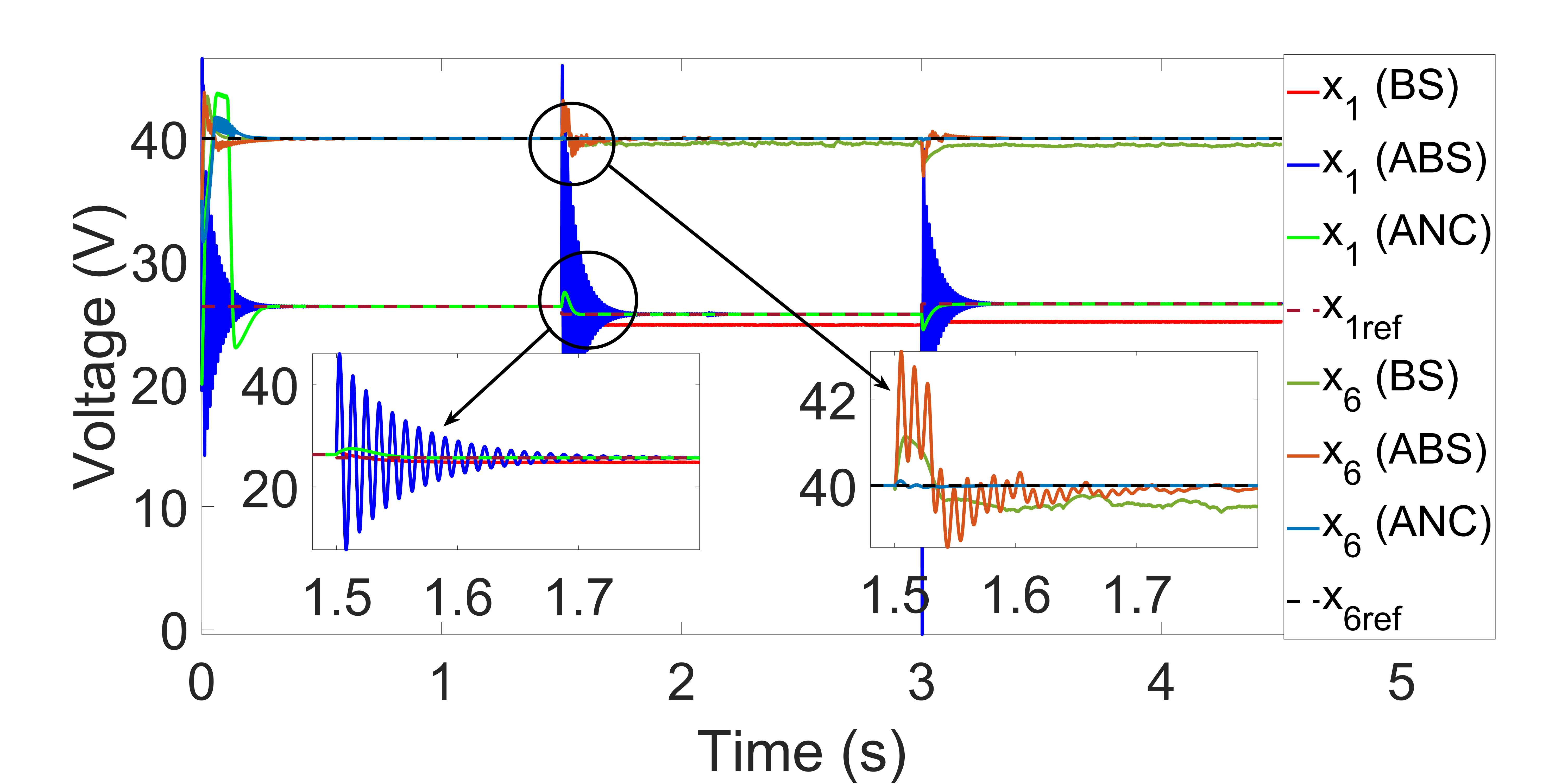}
\caption{\scriptsize{Case 3: Output voltage comparison}}
\label{fig:outputs_ic2}
\end{subfigure}\\
\begin{subfigure}{.45\textwidth}
  \centering
\includegraphics[width =  7.5cm,height=5cm]{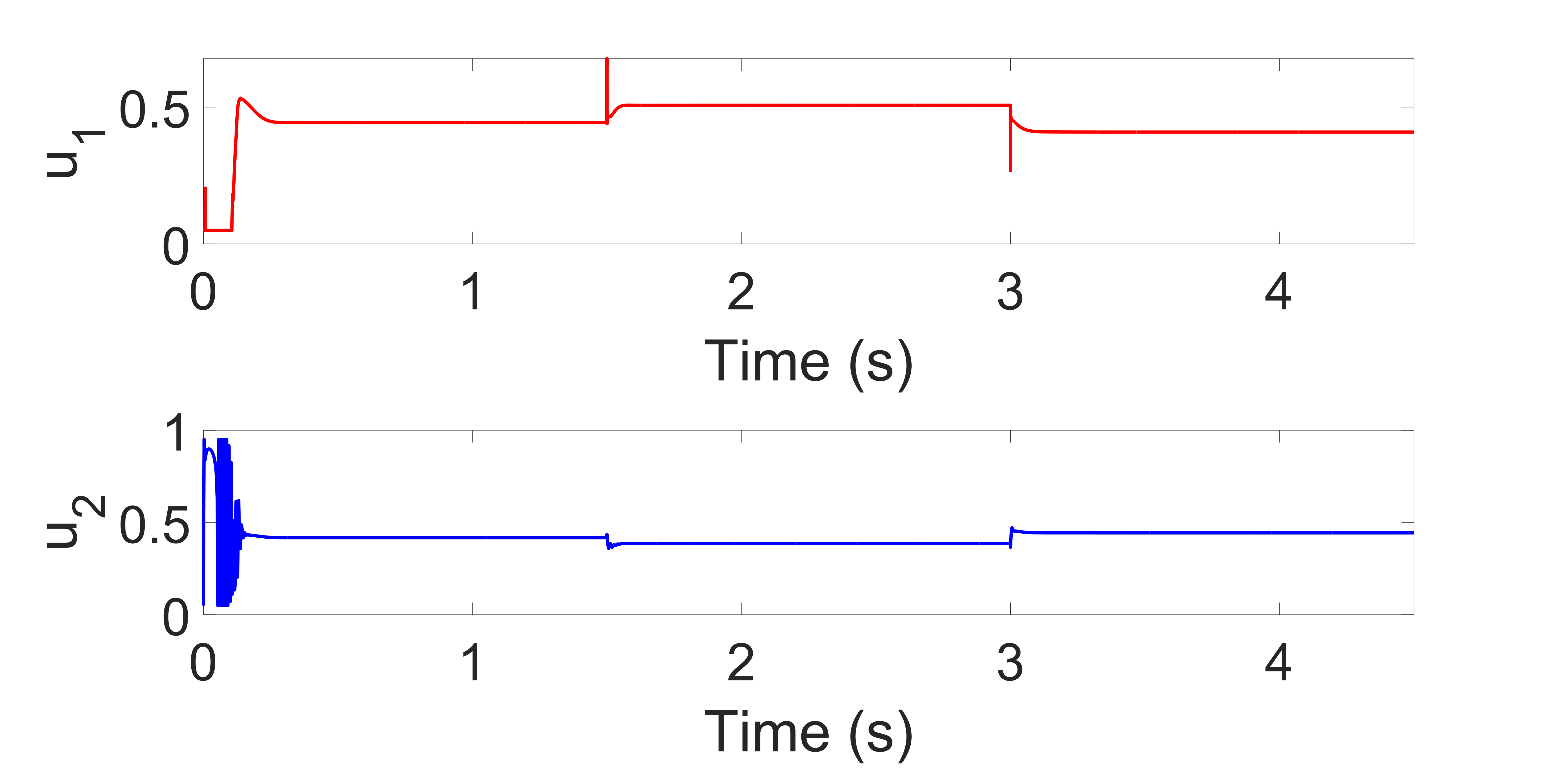}
\caption{\scriptsize{Case 3: Duty cycle - Proposed ANC Controller Controller}}
\label{fig:u_adaptive_ic2}
\end{subfigure}
\begin{subfigure}{.45\textwidth}
  \centering
\includegraphics[width =  7.5cm,height=5cm]{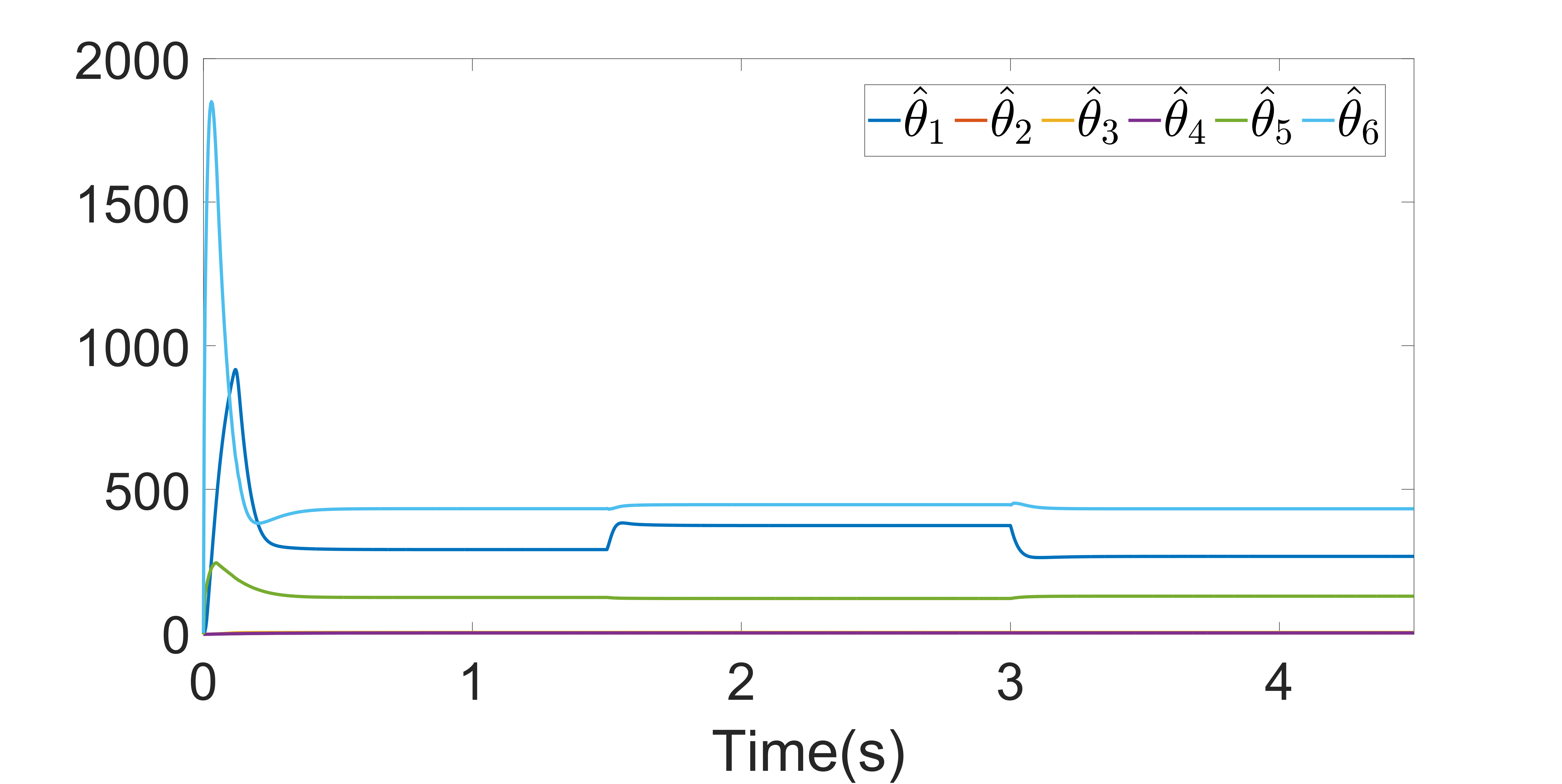}
\caption{\scriptsize{Case 3: NN weights - ANC Controller}}
\label{fig:thetahat_adaptive_ic2}
\end{subfigure}\\
\begin{subfigure}{.45\textwidth}
  \centering
\includegraphics[width =  7.5cm,height=5cm]{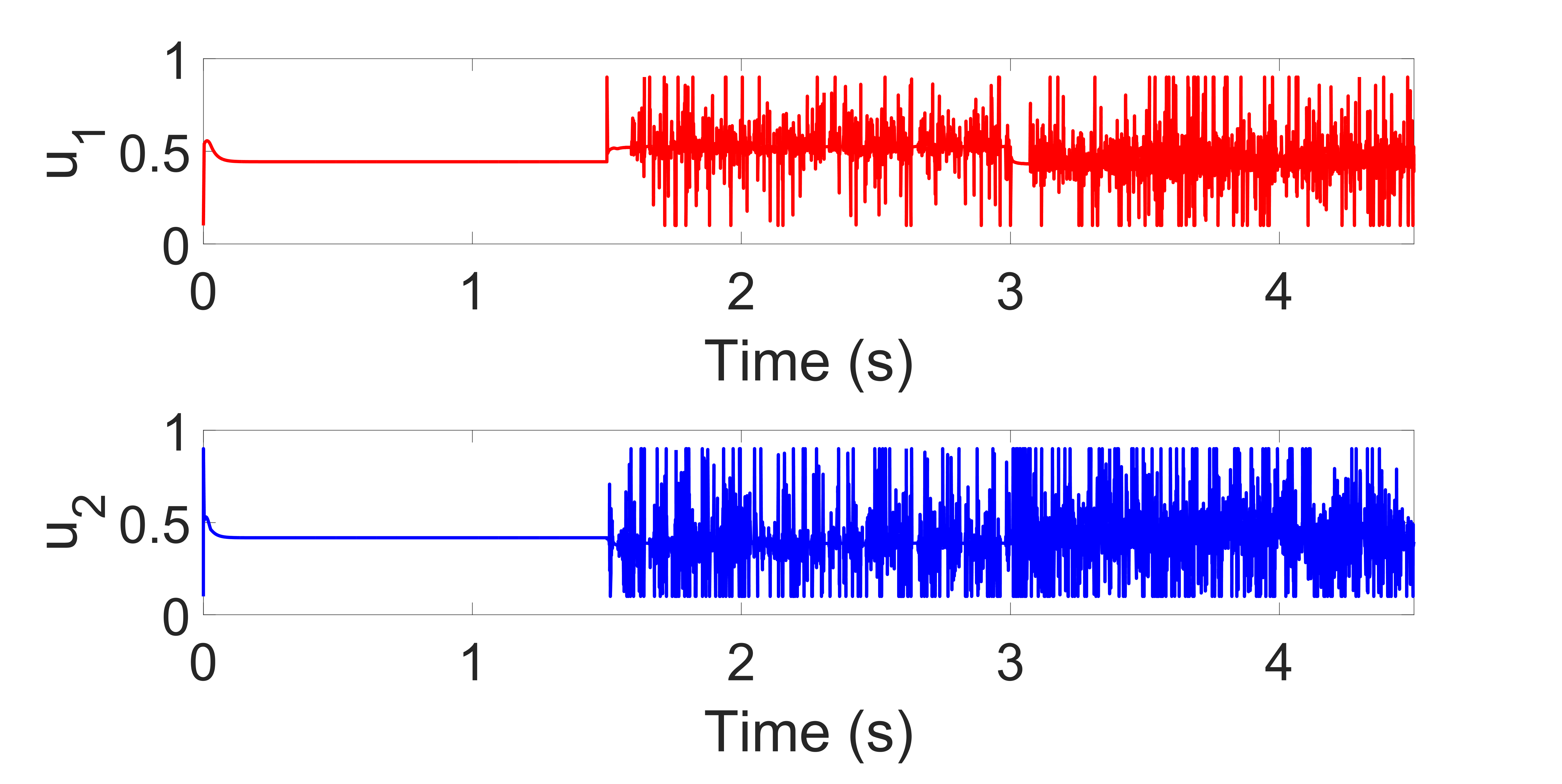}
\caption{\scriptsize{Case 3: Duty cycle - BS Controller.}}
\label{fig:u_alessio_ns_ic2}
\end{subfigure}
\begin{subfigure}{.45\textwidth}
  \centering
\includegraphics[width =  7.5cm,height=5cm]{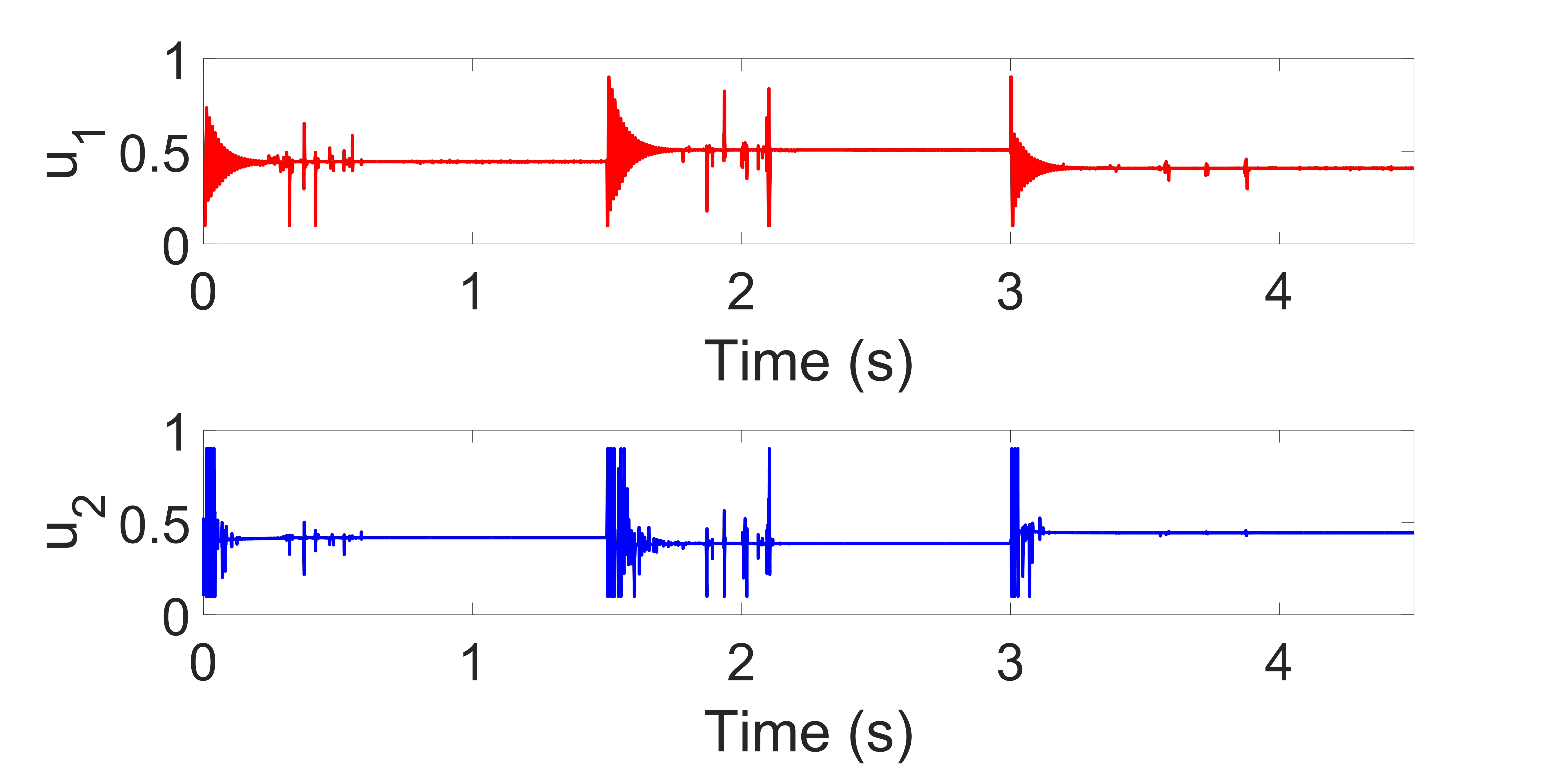}
\caption{\scriptsize{Case 3: Duty cycle - ABS Controller.}}
\label{fig:u_mahmud_ic2}
\end{subfigure}
\caption{Case-3: Validation of the proposed Adaptive Neural Controller (ANC) against state-of-the-art controllers- Adaptive Back-stepping (ABS) controller\cite{overviewtkr2} and Backstepping controller(BS)\cite{iovinetase17} for change in irradiance. }
\end{figure}
The solar power incident on the PV panel is varied from 1000$W/m^2$ to 1500$W/^2$ and then to 700$W/m^2$, 700$W/m^2$ and 500$W/m^2$ at 0$s$, 1.5$s$ and 3$s$  respectively as shown in Fig.\ref{fig:irradiance_ic2}. Correspondingly, $x_{1ref}$ is set as $26.3~V$, $25.67~V$ and $26.52~V$ respectively. As radiation falling on the solar panel changes, disturbance $D_1$ gets affected. When ANC controller is applied, the effect of change in solar radiation is captured by the NN weights as can be seen in Fig.\ref{fig:thetahat_adaptive_ic2}. With change in $D_1$, the MPP voltage and DC grid voltage get perturbed. Figures Fig.\ref{fig:u_adaptive_ic2} , Fig.\ref{fig:u_alessio_ns_ic2} and Fig.\ref{fig:u_mahmud_ic2} show how three different controllers - the proposed ANC controller, BS controller and ABS controller generate duty cycle values for this case and their corresponding outputs can be seen in Fig.\ref{fig:outputs_ic2}). 
\textcolor{black}{Similar to previous cases, the BS controller does not perform well when the value of $D_1$ becomes unknown due to change in irradiance as can be seen in figures\ref{fig:outputs_ic2} and \ref{fig:u_alessio_ns_ic2}. The ABS controller does not show  any steady state error due to accurate estimation of $D_1$ but settling time is quite high as seen in Table \ref{tab:comp_perf}. However, the proposed ANC controller reduces both the SSE and settling time while also reducing the overshoot. The change in $D_1$ is accurately captured by $\widehat{\theta}_1$ as shown in Fig \ref{fig:thetahat_adaptive_ic2}.  }
\subsection{Case-4: Change in System Parameter}
\begin{figure}[H]
\captionsetup[subfigure]{aboveskip=-1pt,belowskip=-1pt}
\centering
\begin{subfigure}{0.45\textwidth}
  \centering
\includegraphics[width =  7.5cm,height=5cm]{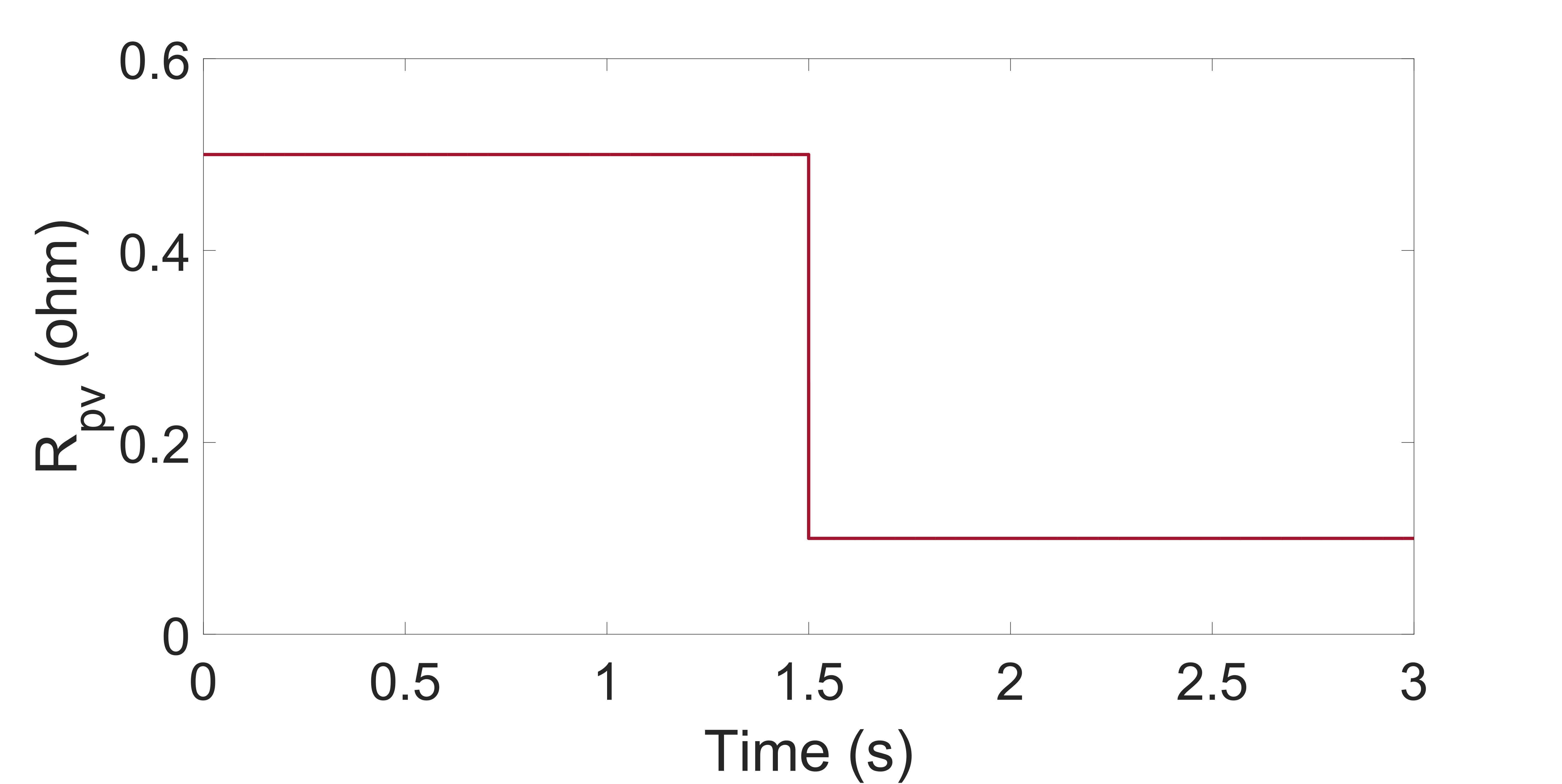}
\caption{\scriptsize{Case 4: Change in PV Converter Resistance}}
\label{fig:rpv_plot_pc2}
\end{subfigure}
\begin{subfigure}{.45\textwidth}
  \centering
\includegraphics[width =  7.5cm,height=5cm]{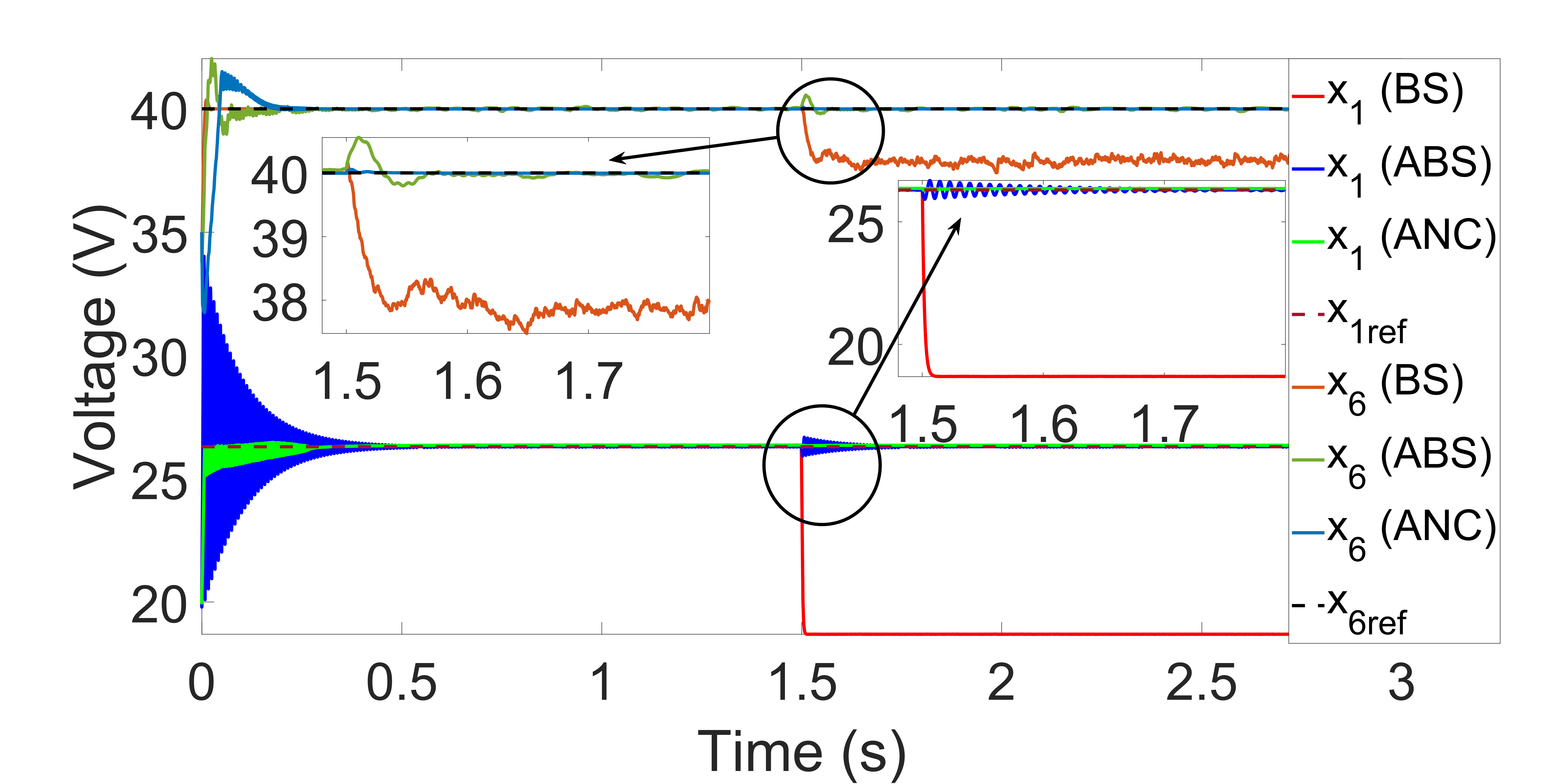}
\caption{\scriptsize{Case 4: Output voltage comparison}}
\label{fig:outputs_pc2}
\end{subfigure}\\
\begin{subfigure}{.45\textwidth}
  \centering
\includegraphics[width =  7.5cm,height=5cm]{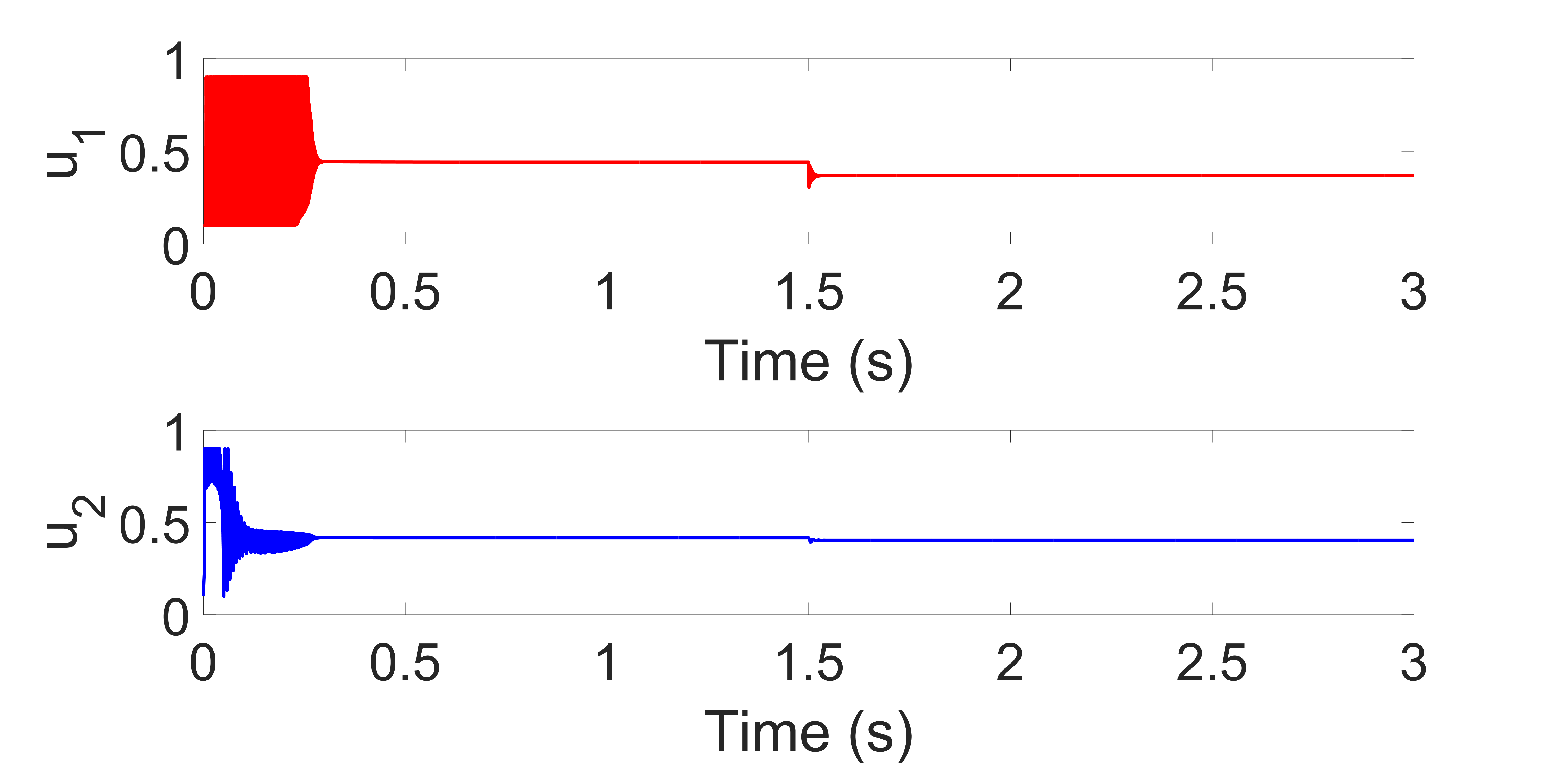}
\caption{\scriptsize{ Case 4: Duty cycle - Proposed ANC Controller }}
\label{fig:u_adaptive_pc2}
\end{subfigure}
\begin{subfigure}{.45\textwidth}
  \centering
\includegraphics[width =  7.5cm,height=5cm]{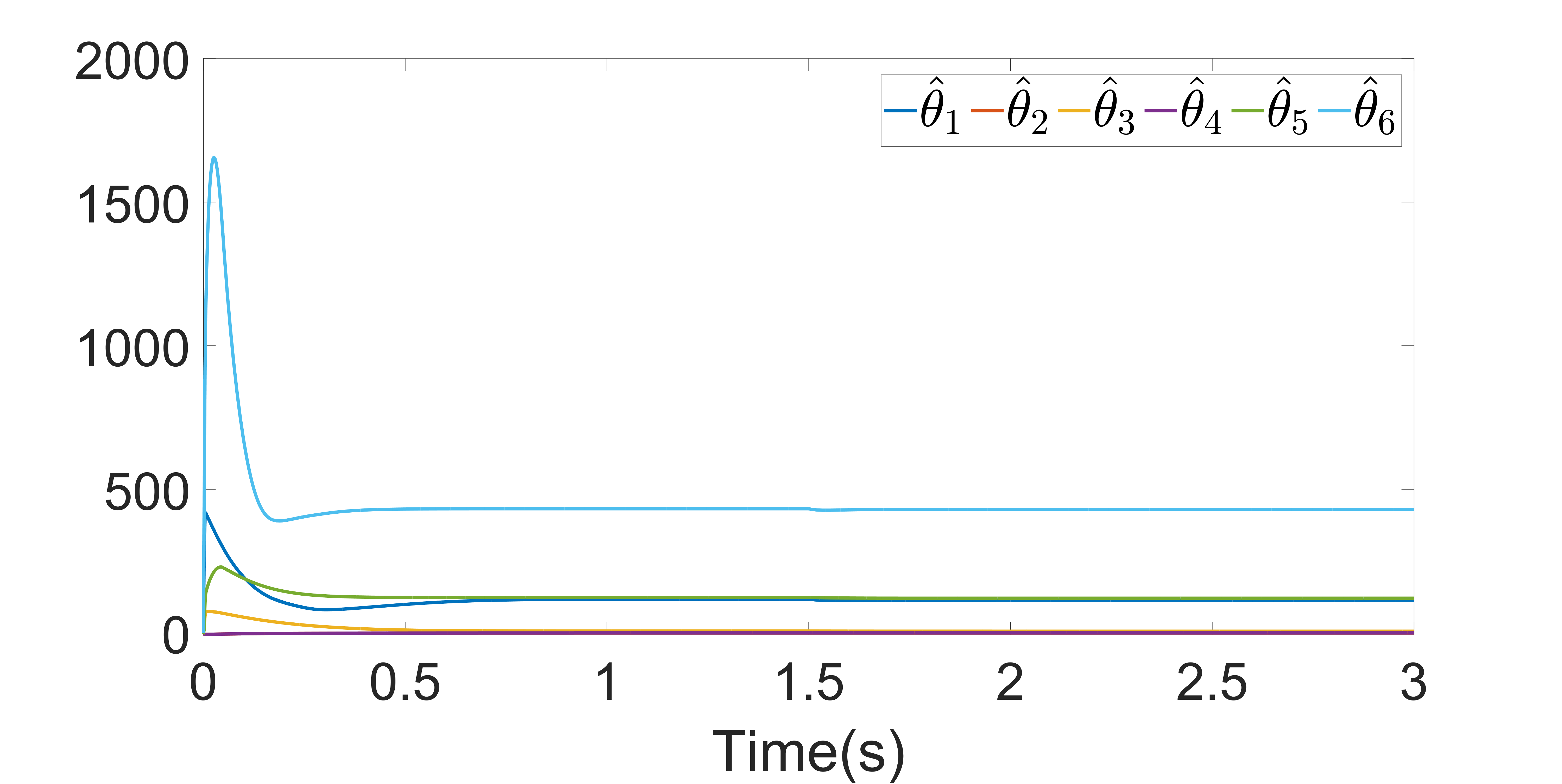}
\caption{\scriptsize{Case 4: NN weights - ANC Controller}}
\label{fig:thetahat_pc2}
\end{subfigure}\\
\begin{subfigure}{.45\textwidth}
  \centering
\includegraphics[width =  7.5cm,height=5cm]{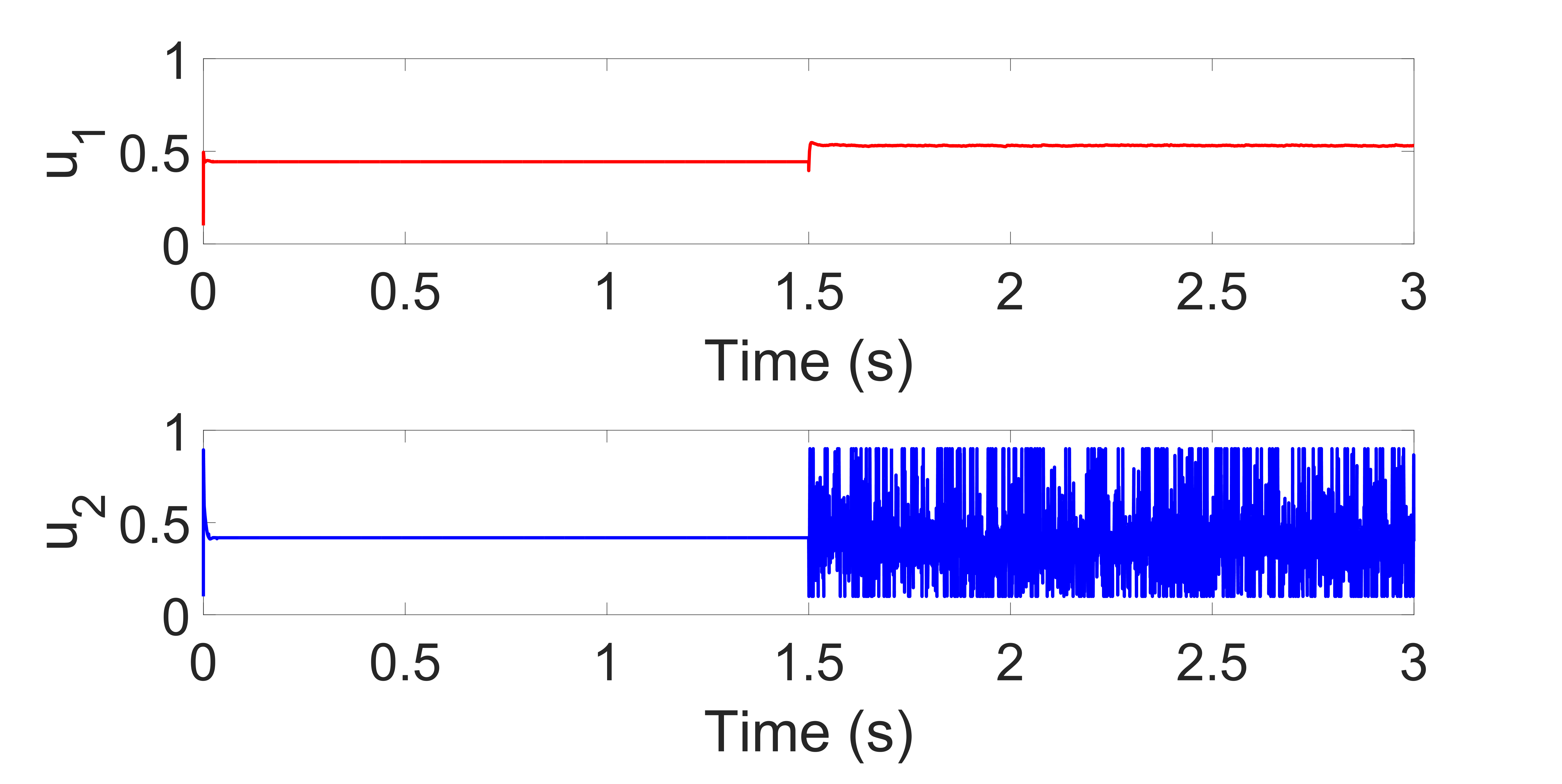}
\caption{\scriptsize{Case 4: Duty cycle - BS Controller.}}
\label{fig:u_alessio_ns_pc2}
\end{subfigure}
\begin{subfigure}{.45\textwidth}
  \centering
\includegraphics[width =  7.5cm,height=5cm]{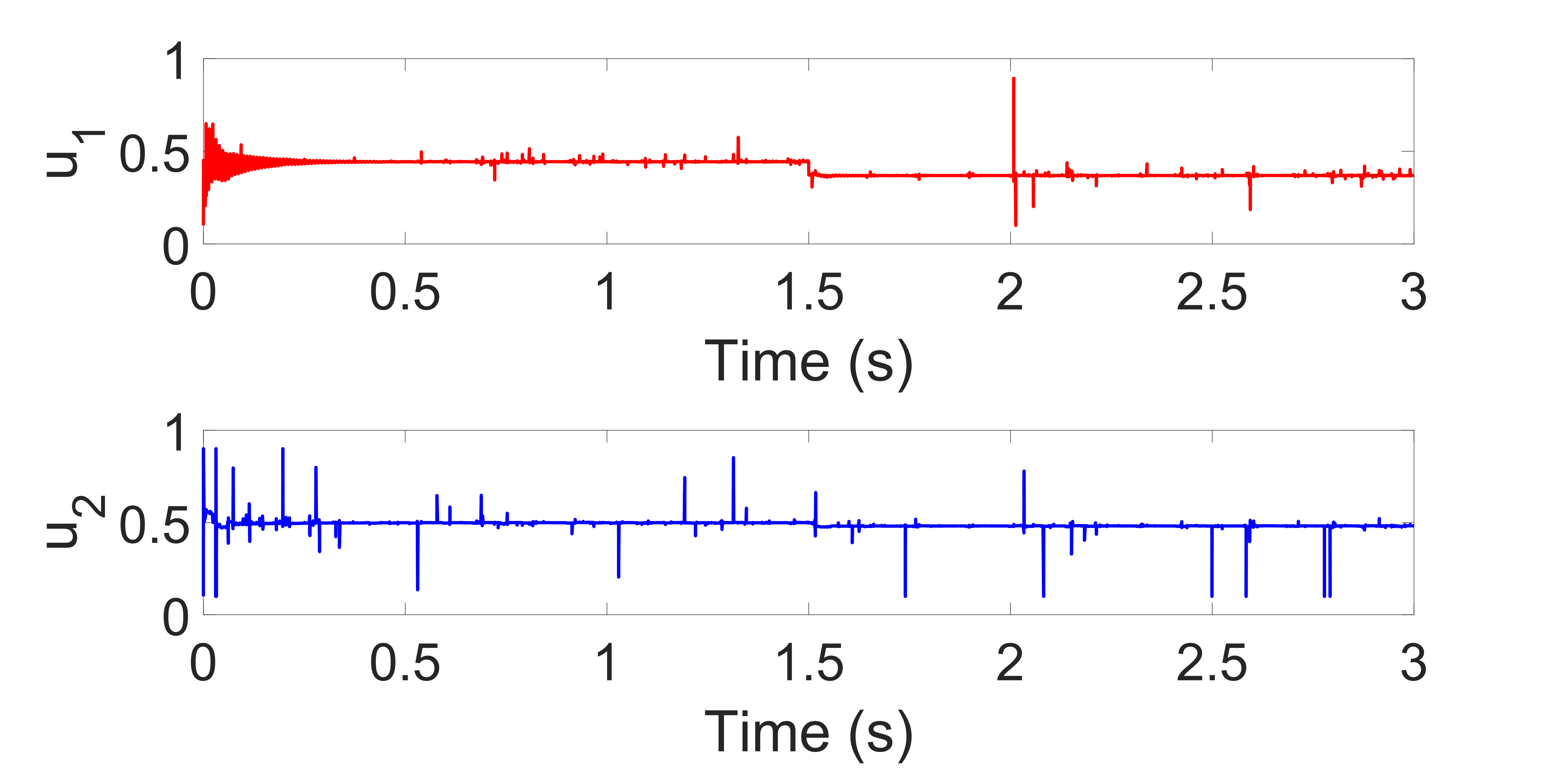}
\caption{\scriptsize{Case 4: Duty cycle - ABS Controller}}
\label{fig:u_mahmud_pc2}
\end{subfigure}
\caption{Case-4: Validation of the proposed Adaptive Neural Controller (ANC) against state-of-the-art controllers- Adaptive Back-stepping (ABS) controller \cite{overviewtkr2} and Backstepping controller(BS)\cite{iovinetase17} for change in PV converter resistance ($R_{pv}$). }
\end{figure}

\textcolor{black}{Many parameters that are present in the system model like converter resistance, converter inductance and capacitance sometimes vary in the course of operation and if their real-time values are not known to the controller, it can cause the system to operate in an undesirable way. In this case, it is assumed that the value of PV converter resistance $R_{pv}$ changes and is not known to the primary controller. For this case, the solar irradiance and temperature are considered to be 1000 $W/m^2$ and $25^oC$  which results in $x_{1ref}$ being set to a constant value $26.31~V$ as per the PV characteristic. The load is considered to be 200 $W$. The value of $R_{pv}$ is initially assumed to be 0.5 $\Omega$ and known to the controller in operation. Later at 1.5$s$, the value of $R_{pv}$ drops to 0.1$\Omega$ and becomes unknown to the controller as shown in Fig.\ref{fig:rpv_plot_pc2}. Similar to previous cases in Fig.\ref{fig:outputs_pc2}, we have shown the comparison in output voltages when different controllers are implemented with the DCSSMG system when system parameter $R_{pv}$ changes. Also, the control inputs with different controllers are shown in figures\ref{fig:u_adaptive_pc2}, \ref{fig:u_alessio_ns_pc2} and \ref{fig:u_mahmud_pc2}. Moreover, the convergence of the NN weights can be seen in Fig.\ref{fig:thetahat_pc2}. }

\textcolor{black}{Figure.\ref{fig:outputs_pc2} shows how the voltage $x_1$ drops significantly by 6.8$V$ when the parameter $R_{pv}$ changes and becomes unknown to the BS controller at 1.5$s$. The ANC controller provides superior performance compared to that of BS and ABS controllers. The ABS control inputs are seen to have many spikes in their results. This is due to improper parameter estimation of the ABS algorithm when interconnections between the PV and battery subsystem are considered. The effect of $R_{pv}$ is captured by the ANC controller using $\widehat{\theta}_1$ as shown in Fig.\ref{fig:thetahat_pc2}}

\subsection{Case-5: Simultaneous Variations in Disturbances/Parameters}
\begin{figure}[H]
\captionsetup[subfigure]{aboveskip=-1pt,belowskip=-1pt}
\centering
\begin{subfigure}{0.45\textwidth}
  \centering
\includegraphics[width =  7.5cm,height=5cm]{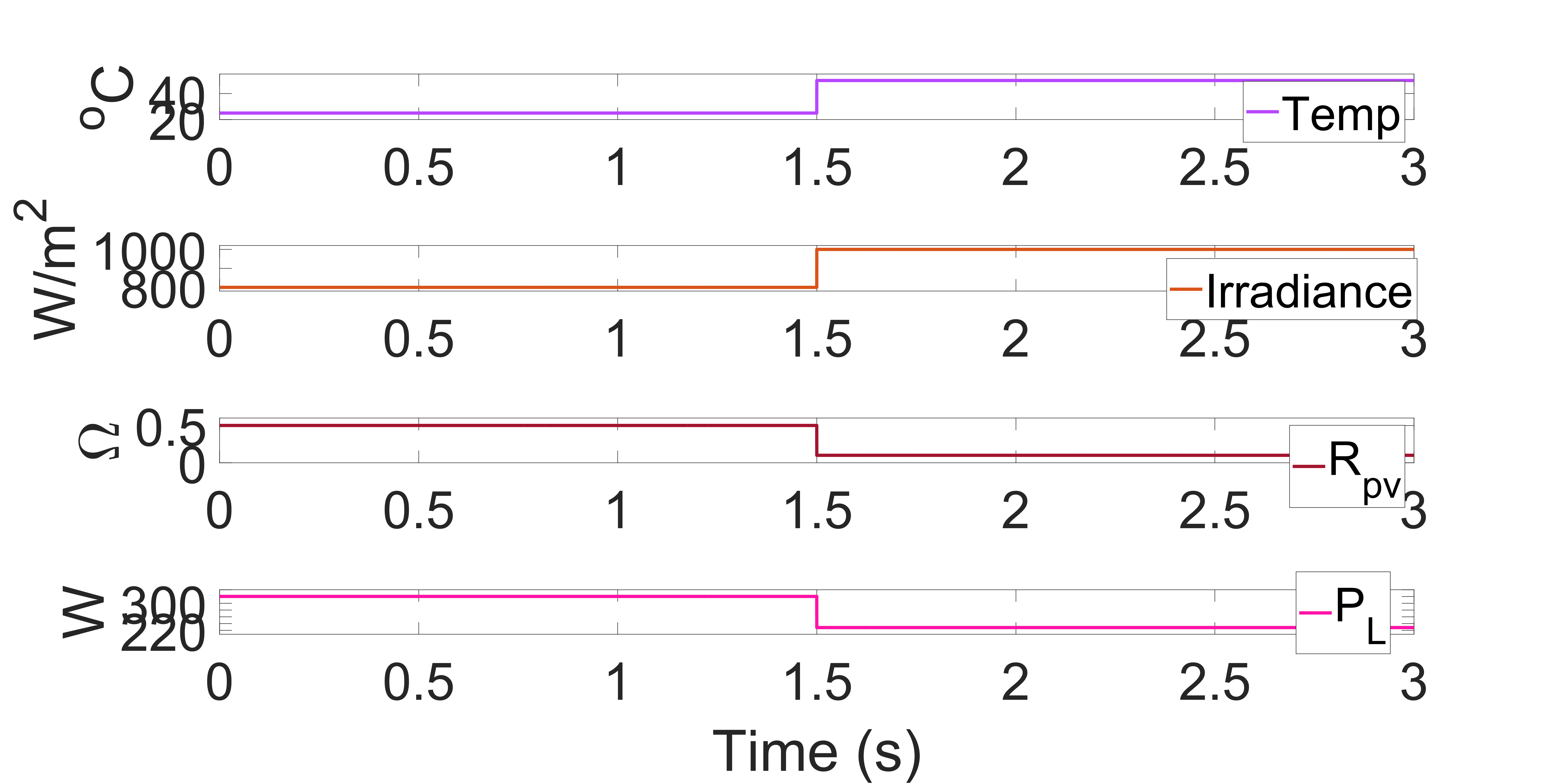}
\caption{\scriptsize{Case 5: Variation in Multiple Parameters}}
\label{fig:parameterplot_ac2}
\end{subfigure}
\begin{subfigure}{.45\textwidth}
  \centering
\includegraphics[width =  7.5cm,height=5cm]{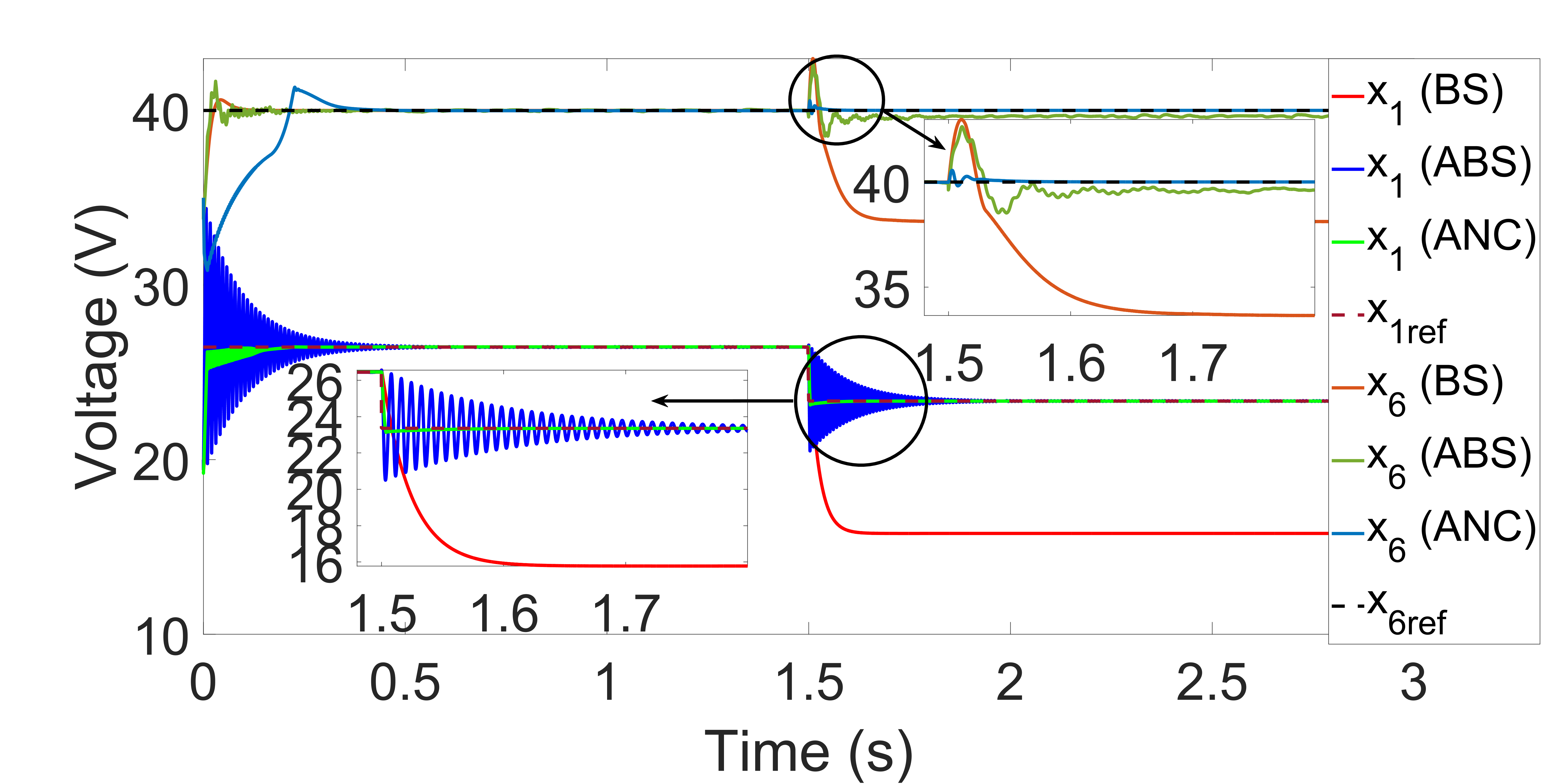}
\caption{\scriptsize{Case 5: Output voltage comparison}}
\label{fig:outputs_ac21}
\end{subfigure}\\
\begin{subfigure}{.45\textwidth}
  \centering
\includegraphics[width =  7.5cm,height=5cm]{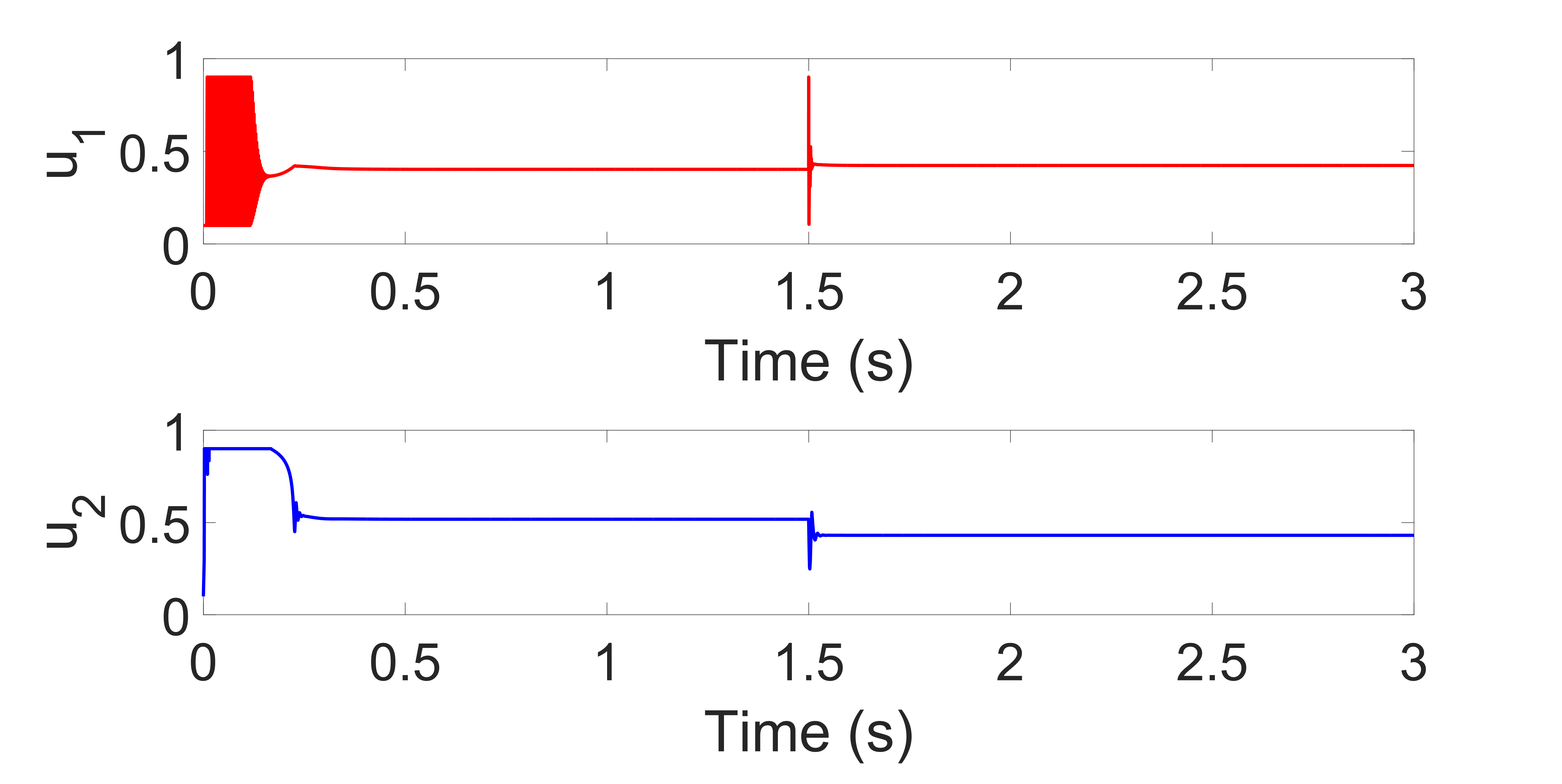}
\caption{\scriptsize{ Case 5: Duty cycle - Proposed ANC Controller Controller}}
\label{fig:u_adaptive_ac2}
\end{subfigure}
\begin{subfigure}{.45\textwidth}
  \centering
\includegraphics[width =  7.5cm,height=5cm]{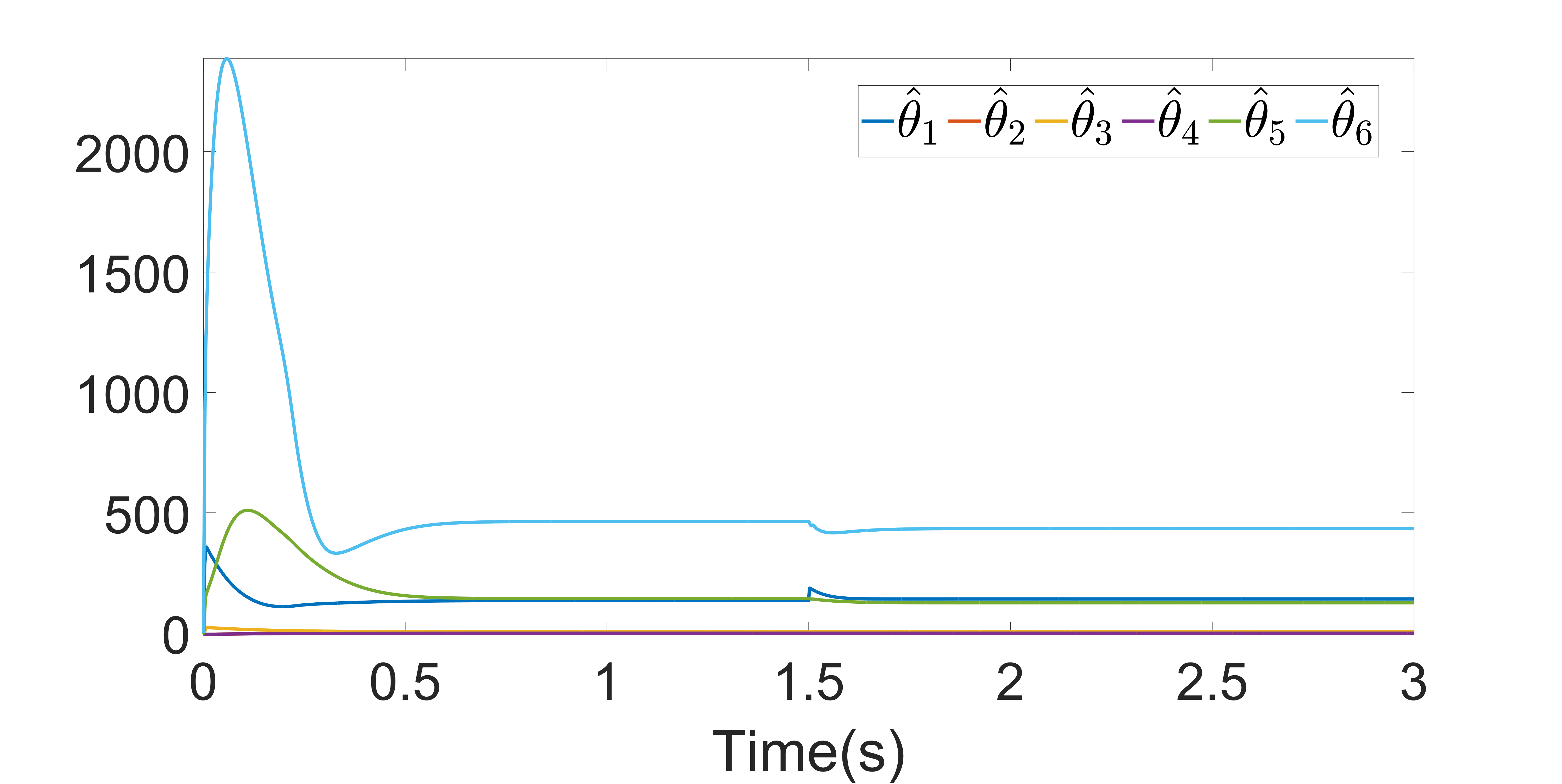}
\caption{\scriptsize{Case 5: NN weights - ANC Controller}}
\label{fig:thetahat_ac2}
\end{subfigure}\\
\begin{subfigure}{.45\textwidth}
  \centering
\includegraphics[width =  7.5cm,height=5cm]{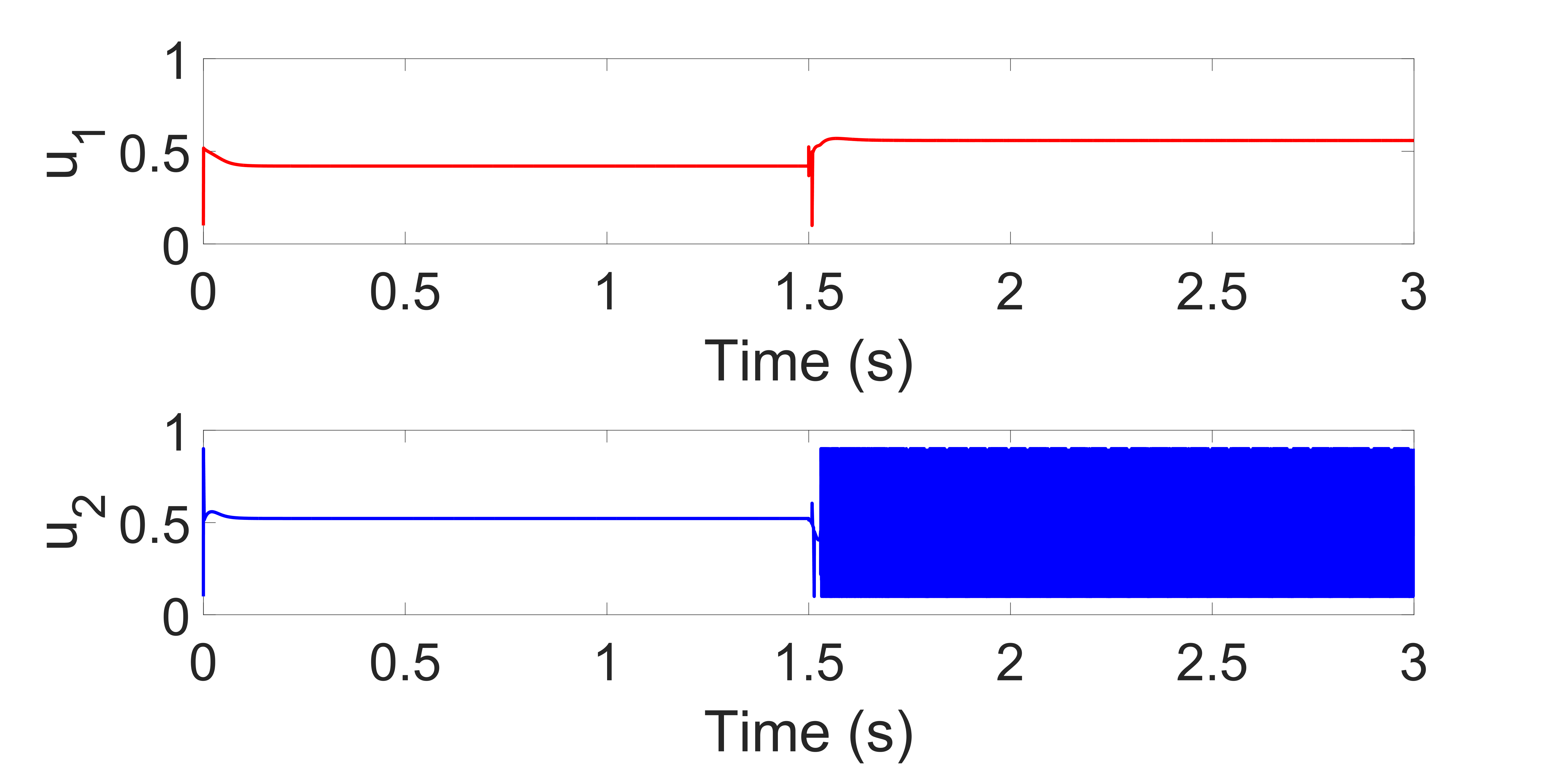}
\caption{\scriptsize{Case 5: Duty cycle - BS Controller.}}
\label{fig:u_alessio_ns_ac2}
\end{subfigure}
\begin{subfigure}{.45\textwidth}
  \centering
\includegraphics[width =  7.5cm,height=5cm]{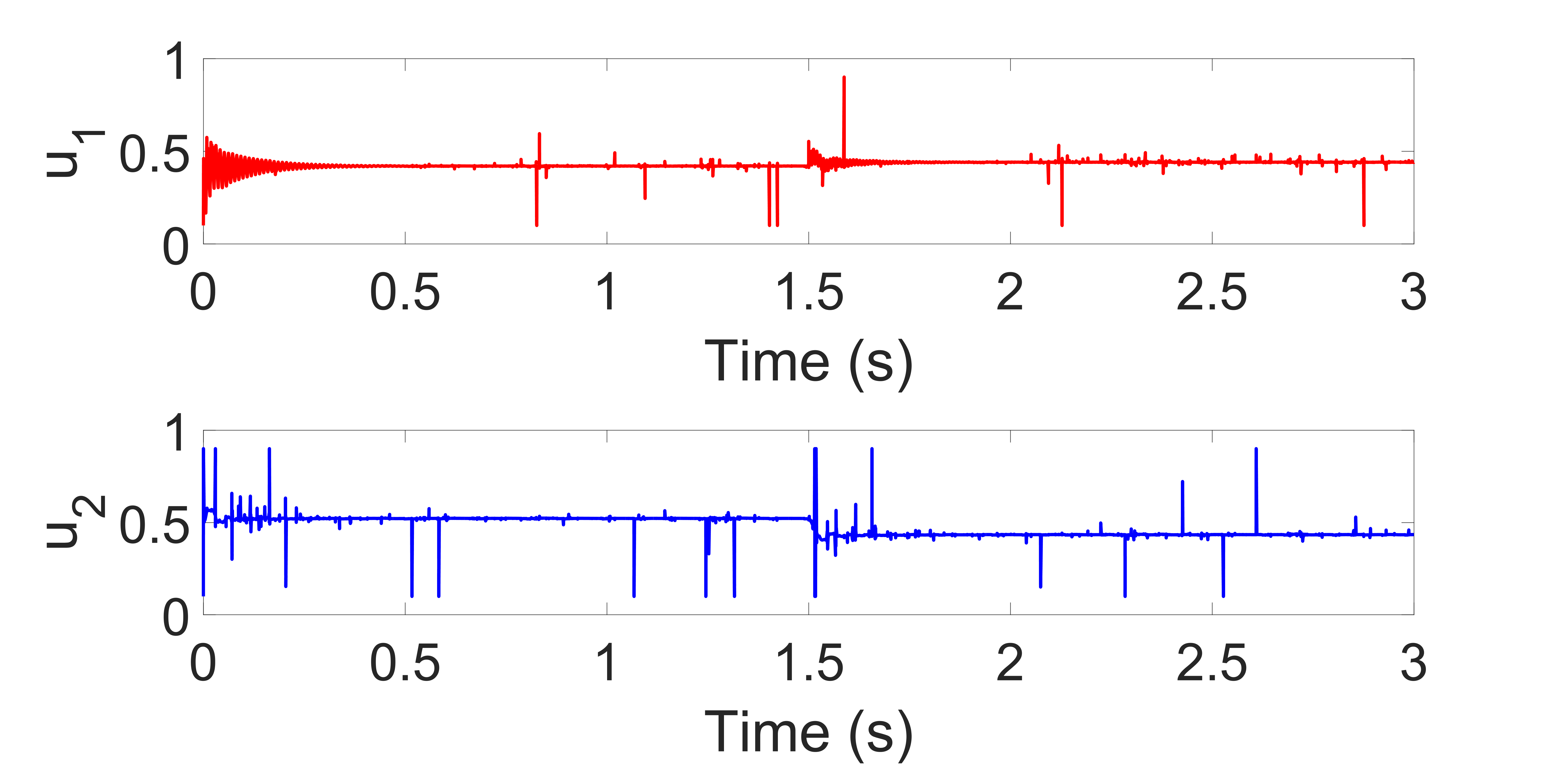}
\caption{\scriptsize{Case 5: Duty cycle - ABS Controller}}
\label{fig:u_mahmud_ac2}
\end{subfigure}
\caption{Case-5: Validation of the proposed Adaptive Neural Controller (ANC) against state-of-the-art controllers- Adaptive Back-stepping controller (ABS)\cite{overviewtkr2} and Backstepping controller(BS)\cite{iovinetase17} for change in multiple parameters. }
\end{figure}

\textcolor{black}{In this case, we simultaneously change multiple parameters and parameters and compare the working of three controllers including our proposed ANC controller. The temperature and irradiance is considered to be 25$^oC$ and 800$W/m^2$, load power is considered to be 320$W$ and the parameter $R_{pv}$ is considered to be 0.5$\Omega$ at time t=0$s$. From t=0$s$ to t=1.5$s$, these values remain constant and known to the primary controller. At t=1.5$s$, the values of temperature, irradiance, load power and $R_{pv}$ simultaneously change to 50$^oC$, 100$W/m^2$ 228.57$W$ and 0.1$\Omega$ as shown in Fig.\ref{fig:parameterplot_ac2}. The comparative performance of the three controllers at hand and their control inputs including NN weight profile when using the ANC controller are shown in figures Fig.\ref{fig:outputs_ac21}-\ref{fig:u_mahmud_ac2}. }

\textcolor{black}{It can be seen from Fig.\ref{fig:outputs_ac21} that when BS controller is used, both the output states $x_1$ and $x_6$ drop by 7.6$V$ and 6.3$V$ respectively showing that the BS controller totally fails in handling voltage regulation in the presence of multiple disturbance and parameter variations. Additionally, for this case, the ABS controller also gets deviated from its reference point by 0.32 $V$. It shows that the ABS controller is not able to handle changes in multiple unknown parameters in a satisfactory way. However, the overall changes in system dynamics at 1.5$s$ has been captured by all the NNs of the ANC controller.  }

\subsection{Discussion}
\textcolor{black}{The above figures and Table \ref{tab:comp_perf} show the comparative performance of the proposed ANC controller with that of the BS and ABS controllers. The BS controller has been designed using conventional nonlinear back-stepping based principles as given in  \cite{iovinetase17} and the ABS controller has been designed with the help of conventional lyapunov based adaptive estimators and back-stepping principles as in  \cite{overviewtkr2}. In all the cases, the information regarding system parameters and disturbances is made available to the primary controller from 0-1.5$s$ and after first transition at 1.5$s$, the controllers do not receive any updated information regarding the same. }

\textcolor{black}{It is seen that in all the cases, the ANC controller results in lower overshoot, lower rise and settling times with no steady state error. The ABS controller also achieves desired output control in cases where unknown disturbances and parameters change individually. However, this controller leads to error in steady state DC grid voltage when there is simultaneous change in multiple disturbances/parameters as shown in Case-5. Moreover, it suffers from high overshoot and large rise/settling times as compared to the ANC controller. The BS controller fails to achieve zero steady state error in all cases since it is not designed for handling unknown values of disturbances/parameters. It only works properly in all cases from 0$s$ to 1.5$s$ when the system model and disturbance values are known to the controller.  The settling time, rise time and peak overshoot of the BS controller is higher compared to that of ANC in most cases. Any discrepancy in the performance of the BS controller can be attributed to the fact that it does not receive updated values of varying disturbances and parameters and hence, does not deliver as designed.}

\begin{table}[H] 
\label{table:results}
\begin{center}
\caption{Comparative Performance Analysis of various controllers}
\begin{tabular}{ |c|c|c|c|c|c|c| } 
\hline 
\multirow{2}{5em}{Simulation Case} &\multirow{2}{4em}{Controller} & Output &Rise  & Settling  & Steady State &  Overshoot  \\
           &       & State &  Time (ms)     &   Time(ms)    &      Error (V)           &  ($\%$)         \\
\hline \hline
 \multirow{6}{5em}{Case-1 (Load Change) } & \multirow{2}{4em}{BS}  & $x_1$ & 29 & 75& 0 & 0.5\\  
 \cline {3-7} &   & $x_6$ & 14 & 40& 0.45& 5 \\ 
\cline {2-7}& \multirow{2}{4em}{ABS} & $x_1$ & 32 & 140& 0 & 0.26\\  
\cline {3-7}& & $x_6$ & 8 & 80& 0& 4 \\ 
\cline {2-7}& \multirow{2}{4em}{Proposed (ANC)} & $x_1$ & 4 & 20& 0 & 0.03\\  
\cline {3-7}& & $x_6$ & 3 & 33& 0& 0.7 \\ 
\hline \hline
 \multirow{6}{5em}{Case-2 (Temperature Change) } & \multirow{2}{4em}{BS}  & $x_1$ & -- & 100& 0.44 & 0\\  
 \cline {3-7} &   & $x_6$ & 60 & 140& 0.2& 1.87 \\ 
\cline {2-7}& \multirow{2}{4em}{ABS} & $x_1$ & 5 & 136& 0 & 5.3\\  
\cline {3-7}& & $x_6$ & 26 & 120& 0& 0.9 \\ 
\cline {2-7}& \multirow{2}{4em}{Proposed (ANC)} & $x_1$ & 3 & 30& 0 & 1.5\\  
\cline {3-7}& & $x_6$ & 17 & 100& 0& 0.05 \\ 
\hline \hline
 \multirow{6}{5em}{Case-3 (Irradiance Change) } & \multirow{2}{4em}{BS}  & $x_1$ & 1 & 60& 0.87 & 2.3\\  
 \cline {3-7} &   & $x_6$ & 10 & 92& 0.55& 2.95 \\ 
\cline {2-7}& \multirow{2}{4em}{ABS} & $x_1$ & 3 & 160& 0 & 19.4\\  
\cline {3-7}& & $x_6$ & 6 & 126& 0& 7 \\ 
\cline {2-7}& \multirow{2}{4em}{Proposed (ANC)} & $x_1$ & 11 & 45& 0 & 4.3\\  
\cline {3-7}& & $x_6$ & 6 & 30& 0& 0.25 \\ 
\hline \hline
 \multirow{6}{5em}{Case-4 ($ R_{pv}$ Change) } & \multirow{2}{4em}{BS}  & $x_1$ & 144 & 158& 6.8 & 0.93\\  
 \cline {3-7} &   & $x_6$ & 37 & 98& 2.06& 0.7 \\ 
\cline {2-7}& \multirow{2}{4em}{ABS} & $x_1$ & 7 & 81& 0 & 1.4\\  
\cline {3-7}& & $x_6$ & 12 & 71 & 0& 1.4 \\ 
\cline {2-7}& \multirow{2}{4em}{Proposed (ANC)} & $x_1$ & 1 & 20& 0 & 0.1\\  
\cline {3-7}& & $x_6$ & 6 & 20& 0& 0.25 \\ 
\hline \hline
 \multirow{6}{5em}{Case-5 (Multi- Parameter Change) } & \multirow{2}{4em}{BS}  & $x_1$ & 56 & 141& 7.6 & 0.1\\  
 \cline {3-7} &   & $x_6$ & 12 & 195& 6.3& 7.37 \\ 
\cline {2-7}& \multirow{2}{4em}{ABS} & $x_1$ & 3 & 195& 0 & 12.75\\  
\cline {3-7}& & $x_6$ & 12 & 105 & 0.32& 7.37 \\ 
\cline {2-7}& \multirow{2}{4em}{Proposed (ANC)} & $x_1$ & 3 & 64& 0 & 0.9\\  
\cline {3-7}& & $x_6$ & 3 & 60& 0& 1.35 \\ 
\hline
\end{tabular}
\label{tab:comp_perf}
\end{center}
\end{table}

\section{Summary} \label{sec:ancsummary}
In this chapter, an adaptive neural backstepping controller has been proposed for an unknown DCSSMG system with unknown disturbances. The overall stability of the DCSSMG system is analyzed through the formulation of composite Lyapunov function consisting of many Lyapunov functions from different subsystems of the DCSSMG during the different stages of the design process and the uniform ultimate boundedness of various states and weights of the neural networks is proven through rigorous mathematical analysis.This technique ensures efficient MPPT and voltage control of the DCSSMG system when the knowledge of the system parameters like inductance/capacitance and disturbances like irradiance, temperature and load are not available. The controller has been tested for four different scenarios when subjected to intermittencies and system model is not fed into the controller. In light of this, the proposed controller is highly desirable for the evolving practical requirements in the market especially when considered from user's frame of reference. \textcolor{black}{The provided results suggest that the proposed ANC controller handles simultaneous unknown changes in multiple disturbances/parameters very well to achieve record level reduction in rise-time, settling-time, steady state error and peak overshoot.}


\chapter{Hybrid Adaptive Cyber-Physical Framework for Densely Connected ACSSMG } \label{chapter:nonlinear}\index{Strict feedback}\index{Lyapunov Stability} \index{Nonlinear Controller}
\chaptermark{Hybrid Adaptive Framework}
\section{Introduction}
Cyber-physical systems (CPS) \cite{lee2008cyber} are complex systems that arise from a natural intermingling of physical systems with communication infrastructure and computation technologies. Today's embedded systems will be infused with multiple new capabilities like adaptability, scalability and reliability when CPS technologies are added to them. 
 
The Smart Standalone Microgrid systems are a result of development of many technologies like sensor networks, power electronics and communication technology. Due to the confluence of many technologies, the SSMGs prove to be an ideal candidate for employing CPS technologies \cite{khaitan2015design}. The fast acting renewable energy based distributed generators (DGs) provide both active power and local reactive power support making them resourceful in handling voltage sags \cite{7073775}. Moreover, the distribution system is increasingly being flooded with renewable energy technologies for which the existing infrastructure and control mechanisms are not prepared. Microgrid concepts formulate new approaches to rethink the integration and control of new technologies into the existing infrastructure \cite{lasseter2011smart}. These systems are pervaded by a plethora of smart sensors, controllers and communication infrastructure.

\textcolor{black}{The SSMG considered in this thesis is assumed to be isolated and equipped with nonlinear primary controllers, distributed secondary controllers, centralized tertiary controller and distributed demand response mechanisms \cite{selfself2}. For executing these functions, a communication architecture containing two levels of communication networks is needed. The two types of networks are decided based on data rate and range of coverage, and are given as follows:}
\begin{itemize}
\item \textcolor{black}{Home Area Network (HAN): This network is built around a customer premises like a home, a building or an industrial setup where data at the appliance or generator level is communicated to the primary controller. The communication necessities for these applications are data security, low cost , low power expense and can be provided by technologies that can provide upto 100kbps over distances ranging upto 100m. Technologies like ethernet, WiFi, Bluetooth, Zwave and Zigbee are generally suitable for this purpose.}

\item \textcolor{black}{Field Area Network (FAN): For applications related to optimal operation of the microgrid, generator coordinator, health monitoring and demand response pertaining to secondary level of control, communication technologies with a data rate of 100kbps to 10Mbps are necessary over a range of 10 km. For this, technologies like Zigbee Mesh networks, Wifi mesh networks, power line communications (PLC), cellular networks and coaxial cables are necessary}. 
\end{itemize}

\textcolor{black}{Moreover, for our hybrid SSMG structure, we will need computing devices at two levels of control- on the DCSSMG level and the ACSSMG level. Local controllers implemented on Digital Signal Processors and FPGAs may be used which cover both the primary and secondary distributed control functions. At the tertiary  level, the planning operations and optimization functions can be carried out by a central work stations with high computational power. 
}

\begin{figure}[!ht]
\centering
\includegraphics[width=0.7\textwidth,  keepaspectratio]{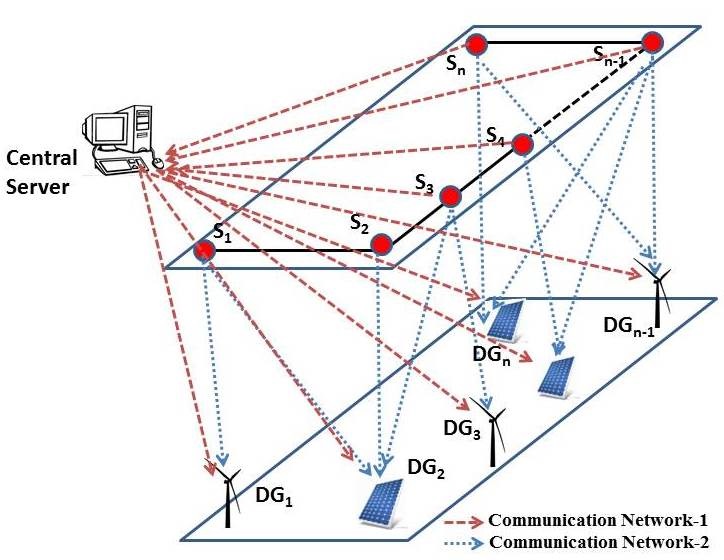}
\caption{Illustration of a Smart Grid with Communications}
\label{Fig:fig1}
\end{figure}

Consider a voltage control problem in smart grid scenario has been presented in Fig. \ref{Fig:fig1}. The diagram depicts an n bus grid where each bus is connected to a DG like solar, wind, etc. Sensors are placed at every point of common coupling (PCC) which is attached to a DCSSMGs. Traditionally, the voltage references of the IED in the respective DCSSMGs is appropriately adjusted to control the voltage at a particular PCC. \textcolor{black}{The controllers present in such a system would be pre-designed for a specific operating point. Due to employment of communication networks and ubiquitous sensing, a distributed mode of control can be easily adopted to provide more reliability to the AC SSMG. Contrary to the traditional control, each DCSSMG can now receive the voltage information from many other PCCs in order to regulate its own voltage. The tertiary controller, on the other hand, would keep track of different parameters in the system that are prone to variations and make short/long term forecasts on the same. This information can be used in an online manner to adaptively update the distributed controllers to tackle different scenarios during the course of operation.}

\textcolor{black}{In practice, to implement such CPS technologies in the ACSSMG, it is necessary to tackle issues pertaining individually to CCCP along with the ones that emerge from the combination of these domains. To promote flexibility in the ACSSMG and to accommodate different situations, the control framework needs to be more situation aware and adaptive so that system doesn't become unstable due to unanticipated phenomena. The adaptive controllers shall depend not only on the structure of the physical system but also on the properties of communication and computation worlds. This calls for hybrid CPS modeling frameworks which contain inputs and outputs belonging to multiple fields including CCCP.}

Communication has been used for enhancing the performance of power systems in terms of voltage and frequency control \cite{olivares2014trends}. The centralized type of control  \cite{rezaei2015robust,mehrizi2012constrained} which has been the conventional norm in microgrid control is slowly fading out in practise and distributed philosophy \cite{zhou2016consensus,sun2015multiagent,bidram2013secondary} is gaining adoption in advanced power grids to improve scalability and reliability. The authors in \cite{xin2011cooperative} use a cooperative control strategy for multiple solar plants using minimal communications. A distributed secondary control technique is developed in \cite{shafiee2013distributed} and implemented using a densely connected communication framework to augment the working of traditional droop. Improving on these works, a hybrid strategy has been adopted in \cite{liang2013stability} which uses different strategies in the presence of centralized communication network and otherwise. Certain works such as \cite{majumder2012power} focus on innovating communication protocols and routing techniques to improve power sharing in different configurations of the microgrid. Most of the works found in the literature are either specific to power domain or communication domain or computation domain even though they portray application to the smart grid. However, they do not model the inter-dependencies between parameters of various domains.

The work in \cite{li2012multicast} proposed a simple optimization technique to design controller gains based on communication topology for voltage control in smart grids. A greedy algorithm was used to route the connections between sensors and DGs based on controller stability. The authors in \cite{mishra2015generalized} provide a more generic algorithm to rope in large number of unexplored communication topologies and increase the stability of an ACSSMG. However, the different control parameters and communication constraints are assumed to be fixed. In practice, various physical parameters like load and communication parameters like delay keep changing simultaneously over the course of the day. \cite{7564533} tried to partially address this by modeling the smart grid system as hybrid system but it could only capture the variation in delays. \textcolor{black}{To our knowledge, no one has addressed the adaptive nature of a generic CPES framework having different types of dynamics.} 

To capture the interdependencies between communication and control and to design controllers adaptively for simultaneous variations in cyber physical parameters, this work develops a generic, hybrid and customizable framework to capture the dynamics of communication and control using the theory of hybrid switching systems. Optimization formulations using Common Lyapunov Function (CLF) to design controllers acknowledging variations in both physical and communication parameters in delay-free and delay existent systems. Furthermore, a BCD based technique for finding a solution to the developed formulations which are non-convex in nature.


 Section \ref{Sec:MIMO} of this chapter gives an idea of mathematical modeling and communication structure of the ACSSMG system. Section \ref{Sec: CLF} describes the proposed Common Lyapunov Function based optimization frameworks capable of designing controllers for multi-parametric variations of such systems while Section \ref{Sec:BCD} explains the BCD technique for solving the proposed optimization problems. Section \ref{Sec:descon} describes application of the developed techniques to the control of PCC voltages in ACSSMG in an adaptive manner and the results obtained upon application of these procedures on certain scenarios like load fluctuations, delay fluctuations and communication failure \textcolor{black}{in a 4-bus distribution feeder} have been presented in Section \ref{Sec:results}. Appropriate conclusions have been summarized in Section \ref{sec:hybridsummary}.
 
\section{Model of the ACSSMG}\label{Sec:MIMO}
The ACSSMG is replete with many comoponents like sensors, DCSSMGs which pertain to the physical functions of the system like voltages and currents. It is also equipped with a communication network which is influenced by parameters such as communication structure, link loss and bandwidth as shown in Fig. \ref{Fig:fig1}. This calls for novel modeling techniques which can embed heterogeneous parameters from both physical and communication spheres. This section describes various assumptions and models pertaining to physical and communication domains of the ACSSMG along with the distributed control strategy and the hybrid switching system representation used to model different operating conditions of the ACSSMG. 
\textcolor{black}{The detailed application of this modeling framework is described in Section \ref{Sec:descon} for a multi-input multi-ouput (MIMO) voltage control problem of an $n$ bus distribution feeder.}
 
\subsection{Physical Model}
{
Let the ACSSMG contains $n_s$ number of nodes at which sensors are placed. Further, let there be $n_c$ number of controllers which are controlling the voltages at required points.} Every DCSSMG is equipped with a secondary controller which senses the voltage at PCC
comes up with a strategy to vary the voltage output of DCSSMG achieve proper voltage regulation at the PCCs. The MIMO system model of the system can be considered in the form as below: 

\begin{equation}
\left.
\begin{aligned}
&\mathbf{\dot{x}}(t)=\mathbf{Ax}(t)+\mathbf{Bu}(t)~~\\
&\mathbf{y}(t)=\mathbf{Cx}(t)\label{Eq:sys}
\end{aligned}
\right\}
\end{equation}
where
$\mathbf{x}(t)$ is the state vector of length $n$, $\mathbf{u}(t)$ is the control vector of length $n_c$ and $\mathbf{y}(t)$ represents the observation vector from sensors of length $n_s$. Thus,

\begin{eqnarray*}
\mathbf{A} \in \mathbb{R}^{n \times n},~\mathbf{B}\in \mathbb{R}^{n \times n_c}~\mathbf{C} \in \mathbb{R}^{n_s \times n} \nonumber
\end{eqnarray*}
\textcolor{black}{Most of the dynamic control problems in ACSSMG can be modeled into state space format.} Even, if the system may be non-linear, it can be linearized around its equilibrium point. The observations in the system are considered to be noiseless.

\subsection{Communication Network}
A wireless communication network is assumed to pervade through the ACSSMG. The sensors present at each PCC contain transmitter nodes for relaying information to the controllers and the controllers have receivers or receiver nodes to receive the voltage information from the transmitter nodes. Both these types of nodes are spread along two communication networks one across the tertiary control level and the other across the secondary control level.

Network-1 facilitates the sensor nodes at various PCCs to provide their information to the central tertiary controller containing receiver nodes. The central server/tertiary controller tracks the data of various parameters of the ACSSMG and forecasts their future values to design situation aware distributed controllers adaptively.

 Network-2 consists of interconnections between the sensors and controllers which will be operational on a regular basis. Since there are $n_s$ number of PCCs placed with sensors, this networks contains $n_s$ transmitter nodes and since $n_c$ DCSSMGs are present, it contains $n_c$ number of receiver nodes.

The communication network is assumed to be densely connected which means that every controller receives information from all the sensors. Data is assumed to be transferred continuously between the nodes which means that the sampling rate is good and there is no packet loss. 

\subsection{Distributed Controller}
A distributed state-feedback controller is designed for the secondary control of ACSSMG which is given by,  
\begin{equation}\label{Eq:controller} 
\mathbf{u}(t)=\mathbf{K{x}}(t) 
\end{equation}
 with $\mathbf{K}\in \mathbb{R}^{n_c \times n_s}$ showing the structure of the communication network between different transmitter and receiver nodes at the secondary control level in Network-2. \textcolor{black}{This control structure allows inputs from multiple sensors to be available to individual distributed controllers}. In case, there exists a communication link between receiver placed at secondary controller $i$ and transmitter placed at PCC $j$, then $\mathbf{K}_{ij}$ would not be zero. In the absence of this connection $\mathbf{K}_{ij}$ would assume a zero value.  \textcolor{black}{For full communication structure subscribed in this work, all the elements of $\mathbf{K}$ matrix would be non-zero.} Substituting \eqref{Eq:controller} in \eqref{Eq:sys}, results in
\begin{equation} \label{Eq:sys3}
\mathbf{\dot{x}}(t)= \mathbf{\bar A x}(t)   
\end{equation}
where $~\mathbf{\bar{A}}= \mathbf{A+BK}$  represents closed loop system matrix. The open loop stability of the ACSSMG without the secondary controllers is represented by the eigenvalues of the matrix $\mathbf{A}$. The stability of the closed loop system  can be understood by studying the eigenvalues  $\mathbf{\bar A}$. The presence of non-negative eigen values for the closed loop system $\mathbf{\bar A} $ signify system stability failing which the system is deemed to be unstable.


\subsection{Delay Representation in the MIMO model}
Delay existing in between various sensors and controllers modeled in the following manner \cite{li2012multicast}, \\
\begin{equation} \label{Eq:ddelay}
\mathbf{\dot{x}}(t)= \mathbf{Ax}(t)+ \underset{(j,i)\epsilon \mathbb{R}}{\Sigma} \mathbf{b}(:,i) \mathbf{k}(j,:) \mathbf{Cx}(t-d_{ji})
\end{equation}
where $(j,i)~\epsilon~ \mathbb{R} $ there exists a communication link between sensor $j$ and controller $i$ is established, $\mathbf{b}(:,i)$ is the $i^{th}$ column of the matrix $\mathbf{B}$, $\mathbf{k}(j,:)$ is the $j^{th}$ row of matrix $\mathbf{K}$, and $d_{ji}$ is the delay between sensor $j$ and controller $i$. The wireless networks can send information in order of milliseconds whereas the dynamics of DCSSMGs is generally in the order of tens to hundreds of milliseconds. \\
Here, 
\begin{align}
&\mathbf{x}(t-d_{ji}) \nonumber \\
&= \mathbf{x}(t) - d_{ji}\mathbf{\dot{x}}(t) + o(d_{ji}^2) \nonumber \\
&\approx \mathbf{x}(t) - d_{ji} (\mathbf{Ax}(t)+ \underset{(j,i)\epsilon \mathbb{R}}{\Sigma} \mathbf{b}(:,i)\mathbf{k}(j,:) \mathbf{x}(t)) \nonumber  \\
&= (\mathbf{I- d}_{ji}(\mathbf{Ax}(t)+ \underset{(j,i)\epsilon \mathbb{R}}{\Sigma} \mathbf{b}(:,j) \mathbf{k}(i,:))))\mathbf{x}(t) \label{Eq:delayy}
\end{align}\\
By neglecting the higher order terms and replacing \eqref{Eq:ddelay} in \eqref{Eq:delayy}, the dynamics of the system with small delay can be written as: 
\begin{equation}\label{Eq: dmodel}
\mathbf{\dot{x}}(t)\approx \mathbf{\bar{A} x}(t) 
\end{equation}
where $\mathbf{\bar{A}} = \mathbf{(I-BDK)(A+BK)}$. The delay matrix $\mathbf{D}$ is given by, 
\begin{equation}
\mathbf{D}_{ji} =\begin{cases}
    &d_{ji} \text{,  if the link between $j$ and $i$ exists}\\
    &0 \text{, otherwise}
  \end{cases}
\end{equation}
\subsection{Switching System Representation}
The model shown in \eqref{Eq:sys} represents the working of the ACSSMG only for a particular value/small range of parameters. However, the variation of the parameters during the course of operation is quite large \cite{parameterref}.\textcolor{black}{For instance, the variation in residential loads does not follow a specific pattern like the industrial loads which are generally goal oriented. The SSMG when setup in a residential area can experience great fluctuations in the load since residential load in an isolated microgrid  is highly dependent on the eccentricities and personal decisions of the individuals living in the area. For example, if a family decides to take a vacation, the load may suddenly drop or if the members of the family fall sick, there can be a sudden increase in the load. Thus load variations can be large especially in case of isolated microgrids as opposed to grid connected microgrids or networked microgrids where many variations can be predicted due to law of large numbers.}

In such cases, the overall system can alternatively be modeled as a hybrid system consisting of $n$ switching subsystems: 
\begin{equation}\label{Eq:swmodel}
\left.
\begin{aligned}
\mathbf{\dot{x}}&= \mathbf{A}_1\mathbf{x} +\mathbf{B}_1\mathbf{u}; \hspace{0.25cm}~t_0\leq t<t_1 \\
\mathbf{\dot{x}}&=\mathbf{A}_2\mathbf{x} +\mathbf{B}_2\mathbf{u}; \hspace{0.25cm}~t_1\leq t<t_2 \\
       &\vdots                    \\
\mathbf{\dot{x}}&=\mathbf{A}_n\mathbf{x} +\mathbf{B}_n\mathbf{u}; \hspace{0.25cm}~t_{n-1}\leq t<t_n \\
\end{aligned}
\right\}
\end{equation}
where each subsystem captures the set of physical and communication parameters operational for a particular duration of the day. For example, $\mathbf{A}_1$ can represent the variation of load from 5 a.m-7 a.m while $\mathbf{A}_2$ can represent the variation of load between 7 a.m-9 a.m and so on. 

\section{CLF Based Controller Design} \label{Sec: CLF}
Due to an existent extensive communication network ranging over long distances, the voltage profile at the PCCs is decided not only by the physical parameters but also by many communication parameters. For this purpose, the switching representation of the system as presented in \eqref{Eq:swmodel} has been adopted which can consider multiple scenarios as multiple subsystems. 
A CLF \cite{lunze2009handbook,4782010,mahmoud2010switched} can exist in such a way that it stabilizes all the subsystems considered in the switching representation simultaneously. Thus, the controller devised keeping the Common Lyapunov Function in perspective can stabilize all the scenarios that the CLF can. The conditions for the availability of such a function can be shaped into a LMI formulation where different electrical loading conditions or different delay conditions are represented as various switching subsystems. 

It is to be noted that for our purpose, we shall consider only two subsystems at a time since increase in the number of subsystems reduce the likeliness of existence of the CLF. The forecast tool will be of much use in deciding the two subsystems to be considered for a particular time of the day. The optimization frameworks which have been derived for no-delay systems and delay systems basing on this concept have been presented in this section.   
\subsection{Systems without Delay}
For designing controllers on the basis of CLF, we assume two closed loop switching subsystems $\mathbf{\dot{x}=A}_{K_1}\mathbf{x}$ and $\mathbf{\dot{x}=A}_{K_2}\mathbf{x}$ where $\mathbf{A}_{K_1}= \mathbf{A}_{1}+\mathbf{BK}$ and $\mathbf{A}_{K_2}=\mathbf{A}_{2}+\mathbf{BK}$ satisfy the common Lyapunov stability criterion.

Considering a Lyapunov function $\mathbf{V}_1=\mathbf{x}^T\mathbf{P}_1\mathbf{x}$ for $\mathbf{A}_{K_1}$ and applying $\mathbf{\dot V}_1 \prec 0$ which means
\begin{equation}\label{Eq:trio}
\mathbf{{A}}^T_{K_1}\mathbf{P}_1+\mathbf{P}_1\mathbf{A}_{K_1}= \mathbf{-P}_0,\hspace{0.2cm} \mathbf{P}_0 \succ0,\hspace{0.2cm} \mathbf{P}_1\succ 0
\end{equation}
Similarly, considering the Lyapunov function $\mathbf{V}_2=\mathbf{x}^T\mathbf{P}_2\mathbf{x}$  for the system $\mathbf{\dot{x}=A}_{K_2}\mathbf{x}$ and applying $\mathbf{\dot{V}}_2\prec0$ for stability, 
\begin{equation}\label{Eq:trioo}
\mathbf{{A}}^T_{K_2}\mathbf{P}_2+\mathbf{P}_2\mathbf{A}_{K_2}= -\mathbf{P}_1,\hspace{0.2cm} \mathbf{P}_1 \succ0,\hspace{0.2cm} \mathbf{P}_2\succ0
\end{equation}
Here, the matrix $\mathbf{P}_1$ holds a major position in forming the connection between the two subsystems.
Now, consider the time derivative of the second Lyapunov function $\mathbf{V}_2$ :
\begin{flalign}
\mathbf{\dot{V}}_2 &= \mathbf{x}^T\mathbf{A}_{K_2}^T\mathbf{P}_2\mathbf{x+x}^T\mathbf{P}_2\mathbf{A}_{K_2}\mathbf{x}
=\mathbf{-x}^T\mathbf{P}_1\mathbf{x}  
\end{flalign}
Substituting \textcolor{black}{\eqref{Eq:trioo} in \eqref{Eq:trio}} results in: 
\begin{align}
{\mathbf{A}^T_{K_1}}(\mathbf{A}^T_{K_2}\mathbf{P}_2+\mathbf{P}_2\mathbf{A}_{K_2}) + {(\mathbf{A}^T_{K_2}}\mathbf{P}_2+\mathbf{P}_2\mathbf{A}_{K_2})\mathbf{A}_{K_1} = \mathbf{P}_0 \hspace{0.2cm} \label{Eq:eq47}
\end{align}
Upon applying the assumption
\begin{equation} \label{Eq:trioo2}
\mathbf{A}_{K_1}\mathbf{A}_{K_2}=\mathbf{A}_{K_2}\mathbf{A}_{K_1}
\end{equation}
in \eqref{Eq:eq47} and realigning, it can bee seen that, 
\begin{equation}
\mathbf{A}^T_{K_2}(\mathbf{A}^T_{K_1}\mathbf{P}_2+\mathbf{P}_2 \mathbf{A}_{K_1}) + (\mathbf{A}^T_{K_1}\mathbf{P}_2+\mathbf{P}_2 \mathbf{A}_{K_1})\mathbf{A}_{K_2} = \mathbf{P}_0 ,
\end{equation}
which means
\begin{align}
\mathbf{A}^T_{K_2}\mathbf{\bar{P}}+\mathbf{\bar{P}}{\mathbf{A}_{K_2}}=\mathbf{P}_0 \label{Eq:trio1}
\end{align}
where $\mathbf{\bar{P}}=\mathbf{A}^T_{K_1}\mathbf{P}_2+\mathbf{P}_2\mathbf{A}_{K_1}$.
From \eqref{Eq:trio1}, if $\mathbf{P}_0\succ 0$ and $\mathbf{A}_{K_2}$ is stabilizing then, $\mathbf{\bar{P}} \prec 0$ which means that $\mathbf{P}_2$ is also stabilizing for $\mathbf{A}_{K_1}$. Adding a $\gamma$ parameter, this condition can be rewritten as:
\begin{equation}\label{Eq:trioo1}
\mathbf{A}^T_{K_1}\mathbf{P}_2+\mathbf{P}_2\mathbf{A}_{K_1}+ \gamma \mathbf{I}\prec0
\end{equation} Thus, under the conditions mentioned above, $\mathbf{V}_2$ becomes the common Lyapunov function for both $\mathbf{A}_{K_1}$ and $\mathbf{A}_{K_2}$. To limit the control effort, the following condition is also taken in to account.
\begin{equation} \label{Eq:Kbound}
\mathbf{\| K \|}_2\leq \rho
\end{equation}

Thus, if a controller $\mathbf{K}$ and matrix $\mathbf{P_2}$ can be obtained while satisfying assumptions  \eqref{Eq:trioo}, \eqref{Eq:eq47}, \eqref{Eq:trioo2}, \eqref{Eq:trioo1} and \eqref{Eq:Kbound}, it ensures that the controller can stabilize the system when switched from $\mathbf{A_{K_1}}$  to  $\mathbf{A_{K_2}}$  or vice versa. Formulating these assumptions as constraints, the optimization problem can be written in the following way:
\begin{align}
& \underset {{K,P_2}}{\max}~~~\gamma  \nonumber  \\
s.t.~~ 
&\mathbf{A}^T_{K_1} \mathbf{P}_2 + \mathbf{P}_2 \mathbf{A}_{K_1} + \gamma \mathbf{I} \prec0,  \nonumber \\
& \text{condition } \eqref{Eq:eq47} \text{ holds}, \nonumber \\
& \mathbf{A}_{K_1}\mathbf{A}_{K_2} = \mathbf{A}_{K_2}\mathbf{A}_{K_1}, \nonumber \\
& \mathbf{A}_{K_2}^T\mathbf{P}_2+\mathbf{P}_2\mathbf{A}_{K_2}\prec0, \nonumber \\
&\mathbf{\| K \|}_2\leq \rho. \label{Eq:opt11}
\end{align} 
\subsection{Systems with Delay}
For this case, let the two switching system dynamics under consideration be: 
\begin{eqnarray}
\mathbf{\dot{x}}= \mathbf{A}_1\mathbf{x+Bu + f}_1\mathbf{(x)} ,\hspace{0.2cm} {\|\mathbf{f}_1(\mathbf{x})\|^2_2} \preceq {\alpha^2_1}{\|\mathbf{x}\|^2_2} \label{Eq:delay1} \\
\mathbf{\dot{x}}= \mathbf{A}_2\mathbf{x+Bu + f}_2\mathbf{(x)} ,\hspace{0.2cm} \|\mathbf{f}_2(\mathbf{x})\|^2_2 \preceq {\alpha^2_2}\|\mathbf{x}\|^2_2 \label{Eq:delay2}
\end{eqnarray}
Considering a common controller $\mathbf{u=Kx}$ for both the system dynamics in \eqref{Eq:delay1} and \eqref{Eq:delay2} results in the following switched system representation: 
\begin{equation}\label{Eq:swmodel1}
\left.
\begin{aligned}
\mathbf{\dot{x}= A}_{K_1}\mathbf{x + f}_1(\mathbf{x}) ~~~~\\
\mathbf{\dot{x}= A}_{K_2}\mathbf{x+ f}_2\mathbf{(x)}  ~~~~
\end{aligned}
\right\}
\end{equation}

Here, $ \mathbf{A}_{K_1}= \mathbf{A}_1 +\mathbf{BK}$ and $\mathbf{A}_{K_2}= \mathbf{A}_2 + \mathbf{BK}$  show the system design for two different physical parameters whereas 
$\mathbf{f}_1(\mathbf{x})= \mathbf{-BD}_1\mathbf{KA}_{K_1}\mathbf{x}$ and $\mathbf{f}_2(\mathbf{x})= \mathbf{-BD}_2\mathbf{KA}_{K_2}\mathbf{x}$ 
signify different delays present in the two subsystems. 

It is assumed that the dynamics of the two subsystems mentioned in \eqref{Eq:delay1} and \eqref{Eq:delay2} are stabilized by Lyapunov functions $\mathbf{V}_1(\mathbf{x})=\mathbf{x}^T\mathbf{P}_1\mathbf{x}$ and $\mathbf{V}_2(\mathbf{x})=\mathbf{x}^T\mathbf{P}_2\mathbf{x}$ respectively where $\mathbf{P}_1,\mathbf{~P}_2\succ0$. The time derivative of $\mathbf{V}_2$ can be written as,

\begin{align}
\mathbf{\dot{{V}}}_2(\mathbf{x})&= \mathbf{x}^T\mathbf{P}_2\mathbf{\dot{x}} + \mathbf{\dot{x}}^T\mathbf{P}_2\mathbf{x} \nonumber \\
            &= \mathbf{x}^T\mathbf{P}_2(\mathbf{A}_{K_2}\mathbf{x} + \mathbf{f}_2(\mathbf{x})) +(\mathbf{x}^T\mathbf{A}^T_{K_2}+\mathbf{f}^T_2(\mathbf{x}))\mathbf{P}_2\mathbf{x}  
\end{align}
This equation can be rewritten as 
\begin{eqnarray}
\mathbf{\dot{V}}_2(\mathbf{x})= \mathbf{y}^T\mathbf{F}_2\mathbf{y}  \label{Eq:delay3} 
\end{eqnarray}
where 
\begin{equation}
\centering
\mathbf{F}_2=
\begin{bmatrix}
{\mathbf{A}_{K_2}}^T \mathbf{P}_2 + \mathbf{P}_2{\mathbf{A}_{K_2}} &  & \mathbf{P}_2 \\
&         & \\
\mathbf{P}_2 &   & 0
\end{bmatrix} 
\text{and }
\mathbf{y}=
\begin{bmatrix}
\mathbf{x} & \mathbf{f}_2(\mathbf{x})
\end{bmatrix}\nonumber 
\end{equation}
It is also known that  ${\|\mathbf{f}_2(\mathbf{x})\|^2_2} \leq {\alpha^2_2}{\|\mathbf{x}\|^2_2}$. This can be easily rewritten as 
\begin{eqnarray}
 \mathbf{y}^T\mathbf{G}_2\mathbf{y} \leq 0 
\end{eqnarray}
where 
\begin{equation}
\centering
\mathbf{G}_2=
\begin{bmatrix}
-\alpha_2^2\mathbf{I}  & 0 \\
0 & \mathbf{I}
\end{bmatrix}
\end{equation}
Now, for stability, $\dot{\mathbf{V}_2}=\mathbf{y}^T\mathbf{F}_2\mathbf{y}<0$ must exist and it will happen if there exists a $\mathbf{y}^T\mathbf{G}_2\mathbf{y}\leq0$ such that 
$\mathbf{F}_2-\gamma_2 \mathbf{G}_2<0$ which results in 

\begin{eqnarray}
\centering
\begin{bmatrix}
\mathbf{A}^T_{K_2}\mathbf{P}_2+\mathbf{P}_2\mathbf{A}_{K_2}+\gamma_2\alpha^2_2\mathbf{I} & & \mathbf{P}_2 \\
& & \\
\mathbf{P}_2 & & -\gamma_2 \mathbf{I}
\end{bmatrix}
\prec 0 ~~ \label{Eq:red44}
\end{eqnarray}
Following a similar procedure for $\mathbf{V}_1$ gives rise to
\begin{eqnarray}
\centering
\begin{bmatrix}
\mathbf{A}^T_{K_1}\mathbf{P}_1+\mathbf{P}_1\mathbf{A}_{K_1}+\gamma_1\alpha^1_2\mathbf{I} & & \mathbf{P}_1 \\
& & \\
\mathbf{P}_1 & & -\gamma_1 \mathbf{I}
\end{bmatrix}
\prec 0 ~~ \label{Eq:redd45}
\end{eqnarray}

The following assumptions have been made in similar terms to that of \eqref{Eq:trio} and \eqref{Eq:trioo} after adding terms to represent uncertainty 
\begin{align}
&\mathbf{A}^T_{K_1}\mathbf{P}_1+\mathbf{P}_1\mathbf{A}_{K_1}+\gamma_1{\alpha^2_1}\mathbf{I=-P}_0,~\mathbf{P}_0\succ0, ~\mathbf{P}_1\succ0 \label{Eq:delay5} \\ 
&\mathbf{A}^T_{K_2}\mathbf{P}_2+\mathbf{P}_2\mathbf{A}_{K_2}+\gamma_2{\alpha^2_2}\mathbf{I=-P}_1, ~\mathbf{P}_1\succ0,~\mathbf{P}_2\succ0  \label{Eq:delay6}
\end{align}
It is seen that \eqref{Eq:delay5} and \eqref{Eq:delay6} when applied to \eqref{Eq:red44} and \eqref{Eq:redd45} give rise to the following:
\begin{eqnarray}
\centering
\begin{bmatrix}
-\mathbf{P}_{i-1} & & \mathbf{P}_i \\
& & \\
\mathbf{P}_i & & -\gamma_i \mathbf{I}
\end{bmatrix}
\prec 0 ~~,~ i=1, 2. \label{Eq:red48}
\end{eqnarray}
Also, substituting \eqref{Eq:delay6} into \eqref{Eq:delay5} :
\begin{align}
\hspace{0.75cm} \gamma_1{\alpha^2_1\mathbf{I+A}^T_{K_1}(\mathbf{A}^T_{K_2}\mathbf{P}_2+\mathbf{P}_2\mathbf{A}_{K_2}}+ \gamma_2{\alpha^2_2}\mathbf{I})+~~~~\nonumber \hspace{2.2cm} \\
\hspace{0.75cm}(\mathbf{A}^T_{K_2}\mathbf{P}_2+\mathbf{P}_2\mathbf{A}_{K_2}+\gamma_2\alpha^2_2\mathbf{I})\mathbf{A}_{K_1}=\mathbf{P}_0   \hspace{3.0cm}  \label{Eq:delay7} 
\end{align}
Substituting another assumption,  
\begin{equation}
\mathbf{A}_{K_1}\mathbf{A}_{K_2}= \mathbf{A}_{K_2}\mathbf{A}_{K_1} \label{Eq:delay8} 
\end{equation} into \eqref{Eq:delay7} and rearranging we obtain,
\begin{align}
 \mathbf{A}^T_{K_2}\mathbf{P}+\mathbf{PA}_{K_2}-\mathbf{\bar{P}}=0 \label{Eq: 58} \hspace{2.5cm}
\end{align}
where $\mathbf{\bar{P}=P}_0 + \gamma_1\alpha^2_1(\mathbf{A}^T_{K_2}+\mathbf{A}_{K_2})-\gamma_2\alpha^2_2(\mathbf{A}^T_{K_1}+\mathbf{A}_{K_1}) $
and $ \mathbf{P= A}^T_{K_1}\mathbf{P}_2+\mathbf{P}_2\mathbf{A}_{K_1}+ \gamma_1 \alpha_1^2\mathbf{I}$. Now, if, 
\begin{equation}\label{Eq:delay10}
\centering
\mathbf{\bar{P}}\succ0 
\end{equation}
then from \eqref{Eq: 58} we find that 
\begin{equation}
\centering
\mathbf{A}^T_{K_2}\mathbf{P+PA}_{K_2}\succ0
\end{equation}
Since $\mathbf{A}_{K_2}$ is stabilizing, $\mathbf{P}\prec0$ which means
\begin{align}
\mathbf{A}^T_{K_1}\mathbf{P}_2+\mathbf{P}_2\mathbf{A}_{K_1}+ \gamma_1 \alpha_1^2\mathbf{I}&\prec0
\end{align}
that is, the Lyapunov function $\mathbf{V}_2$ stabilizes the system $\mathbf{A}_{K_1}$ also becoming the CLF for the two systems mentioned in \eqref{Eq:delay1} and \eqref{Eq:delay2}.
After converting the assumptions \eqref{Eq:Kbound},  \eqref{Eq:delay5}, \eqref{Eq:delay6}, \eqref{Eq:red48}, \eqref{Eq:delay8}, and \eqref{Eq:delay10} into constraints, the optimization problem for finding the value of $\mathbf{K}$ is formulated as below:
\begin{align}
& \underset{K,P_1,P_2}{\max} \hspace{0.3cm} \gamma_1 + \gamma_2 \hspace{4.2cm} \nonumber \\
 s.t.~~
 & \text{conditions in } \eqref{Eq:red48} \text{ hold} \nonumber\\
&\mathbf{A}^T_{K_2}\mathbf{P}_2 + \mathbf{P}_2\mathbf{A}_{K_2}+ \gamma_2 {\alpha^2_2} \mathbf{I = -P}_1 \hspace{1.2cm}\nonumber \\
&\mathbf{A}^T_{K_1}\mathbf{P}_1 + \mathbf{P}_1{\mathbf{A}_{K_1}}+ \gamma_1 {\alpha^2_1} \mathbf{I = -P}_0 \nonumber \hspace{1.2cm} \\
&\mathbf{A}_{K_1}\mathbf{A}_{K_2}= \mathbf{A}_{K_2}\mathbf{A}_{K_1} \nonumber \\
&\mathbf{\bar{P}} \succ 0 ,~ \mathbf{\| K \|}_2 \leq \rho,\nonumber\\  &\mathbf{P}_1 \succ 0, ~\mathbf{P}_2 \succ 0, \nonumber \\
& \gamma_1> 0 ,~\gamma_2> 0  \label{Eq:opt21}
\end{align}
Solving the optimization in \eqref{Eq:opt21} would result in a controller $\mathbf{K}$, that can stabilize the systems with two different parametric conditions as mentioned in \eqref{Eq:delay1} and \eqref{Eq:delay2}.

\section{Block Coordinate Descent Technique} \label{Sec:BCD}
The optimization problems in \eqref{Eq:opt11} and \eqref{Eq:opt21} are non-convex and difficult to solve within a reasonable amount of time. An approximate solution may however be obtained by utilizing the BCD \cite{anjos2012handbook} philosophy. The idea here is to carry out the optimization with respect to smaller blocks of  variables, in an iterative fashion. Interestingly, for the problems at hand, it will be shown that these smaller subproblems are convex and easy to solve. The following algorithms delineate the procedure adopted for the same. 
\alglanguage{pseudocode}
\begin{algorithmic}[1]
\State Initialize $\mathbf{A}^{(0)}_{K_n} := \mathbf{A}_n-\beta \mathbf{I} $, $n=1,~2$, $\mathbf{P}_0:=\mathbf{I}$, $\rho=5$.
\State Find $\mathbf{\gamma}^{(0)}$ and $\mathbf{P}^{(0)}_2$ by solving \eqref{Eq:opt11} using constraints mentioned in \eqref{Eq:trioo}, \eqref{Eq:eq47}, and \eqref{Eq:trioo1}.
\State \textbf{do}
\State Initialize $\mathbf{P}^{i}_2=\mathbf{P}^{i-1}_2$, $\gamma^{(i)}=\gamma^{(i-1)}$
\State Replace $\mathbf{A}^{i}_{K_2}$ with $\mathbf{A}^{i-1}_{K_2}$ in \eqref{Eq:eq47}.
\State Solve for $\mathbf{K}^{(i)}$ using \eqref{Eq:Kbound}, \eqref{Eq:trioo}, \eqref{Eq:eq47}, \eqref{Eq:trioo2}, and  \eqref{Eq:trioo1}.
\State Find $\mathbf{\gamma}^{(i)}$ and $\mathbf{P}^{(i)}_2$ by solving \eqref{Eq:opt11} using constraints mentioned in \eqref{Eq:trioo}, \eqref{Eq:eq47}, and \eqref{Eq:trioo1}.
\State \textbf{while }{$\gamma^{(i-1)}-\gamma^{(i)} > 10^{-6}$}
\State \textbf{end while}
\end{algorithmic}
For instance, the optimization problem in \eqref{Eq:opt11} is divided into two parts. First, stable $\mathbf{A}_{K_n}$ matrices are assumed in the initialization process and the optimization process is carried out only with constraints pertaining to $\gamma \text{ and } \mathbf{P}_2$. With these values fixed, the controller $\mathbf{K}$ is obtained by solving \eqref{Eq:opt11} only with respect to constraints pertaining to $\mathbf{K}$ as shown in step-6. Further, $\gamma$ and $\mathbf{P_2}$ is obtained which is further utilized to find $\mathbf{K}$. Thus, the iterations continue, till the value of $\gamma$ converges.
\alglanguage{pseudocode}
\begin{algorithmic}[1]
\State Initialize $\mathbf{A}^{(0)}_{K_n} := \mathbf{A}_n-\beta \mathbf{I} $, $n=1,~2$, $\mathbf{P}_0:=\mathbf{I}$, $\rho=5$.
\State Find $\mathbf{\gamma_1}^{(0)}$, $\mathbf{\gamma_2}^{(0)}$ $\mathbf{P}^{(0)}_1$ and $\mathbf{P}^{(0)}_2$ by solving \eqref{Eq:opt21} using constraints mentioned in \eqref{Eq:delay5}, \eqref{Eq:delay6}, and \eqref{Eq:red48}.
\State \textbf{do}
\State Initialize $\mathbf{P}^{i}_1=\mathbf{P}^{i-1}_1$, $\mathbf{P}^{i}_2=\mathbf{P}^{i-1}_2$, $\gamma_1^{(i)}=\gamma_1^{(i-1)}$, $\gamma_2^{(i)}=\gamma_2^{(i-1)}$
\State Replace $\mathbf{A}^{i}_{K_2}$ with $\mathbf{A}^{i-1}_{K_2}$ in \eqref{Eq:eq47}.
\State Solve for $\mathbf{K}^{(i)}$ using \eqref{Eq:Kbound}, \eqref{Eq:delay5}, \eqref{Eq:delay6}, \eqref{Eq:red48}, \eqref{Eq:delay8}, and \eqref{Eq:delay10}.
\State Using the value of $\mathbf{K}^{(i)}$ obtained in the previous step, Find ${\gamma_1}^{(i)}$,${\gamma_2}^{(i)}$ $\mathbf{P}^{(i)}_1$ and $\mathbf{P}^{(i)}_2$ by solving \eqref{Eq:opt21} using constraints mentioned in \eqref{Eq:delay5}, \eqref{Eq:delay6}, and \eqref{Eq:red48}..
\State \textbf{while }{$\gamma^{(i-1)}-\gamma^{(i)} > 10^{-6}$} 
\State \textbf{end while}
\end{algorithmic}
The first two steps of the algorithms are meant to improve the feasibility of solution for the next optimization in the process. The controller achieved will be able to stabilize the bus voltages for two zones simultaneously due to CLF based formulation. But the two zones should be chosen with proper discretion depending on the zones in which the system will operate in a particular period of time. 
\section{Application to Voltage Control in ACSSMG}\label{Sec:descon}
This section describes the application of the general system model and control design previously developed for the purpose of controlling the bus voltages in the ACSSMG scenario. 

\subsection{Description of the ACSSMG}
\begin{figure}[!ht]
\centering
\includegraphics[width=0.7\textwidth,  keepaspectratio]{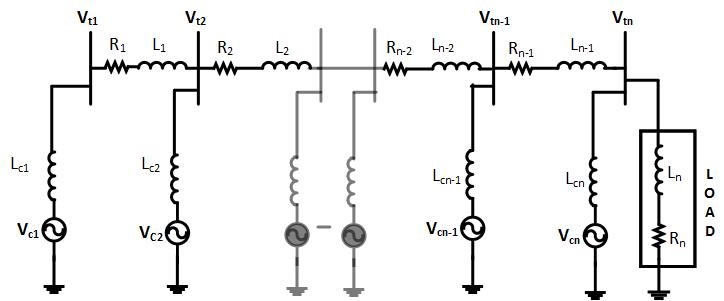}
\caption{Single line diagram of the n-bus system connected to multiple DCSSMGs.}
\label{Fig:physical grid1}
\end{figure}
\textcolor{black}{Fig. \ref{Fig:physical grid1} shows an $n$-bus distribution feeder} where bus voltages $\mathbf{V}_t= [V_{t_1},...,V_{t_n}]'$ are controlled with the help of DCSSMGs which are modeled as voltage sources $\mathbf{V}_c= [V_{c_1},...,V_{c_n}]'$.  \textcolor{black}{The buses present in the grid may be connected to a substation whose effect is ignored in this work to effectively portray the contribution of DCSSMGs.} The control response of DCSSMGs is much faster than conventional sources due to low inertia and fast acting power electronic controllers. 

Upon applying Kirchoff's current law at each bus in Laplace domain and converting them to time domain it is possible to arrive at the state space representation of the system represented in \eqref{Eq:sys}.
Applying nodal analysis at bus-1:
\begin{align}
\frac{V_{t_1}-V_{c_1}}{sL_{c_1}}+\frac{V_{t_1}-V_{t_2}}{R_1+sL_1}&=0 \label{Eq:SysDer10}\\
\Big(\frac{L_1}{L_{c_1}}+1\Big) sV_{t_1}- sV_{t_2}+& \frac{R_1}{L_{c_1}}V_{t_1} -\frac{R_1}{L_{c_1}}V_{c_1} - \frac{L_1}{L_{c_1}} sV_{c_1} =0  
\end{align}
Upon converting this to time domain:
\begin{align}
\Big(\frac{L_1}{L_{c_1}}+1\Big) \dot V_{t_1}-\dot V_{t_2}+ \frac{R_1}{L_{c_1}}V_{t_1}  -\frac{R_1}{L_{c_1}}V_{c_1} - \frac{L_1}{L_{c_1}} \dot V_{c_1} =0 \label{Eq:SysDer1} 
\end{align}
In a similar way, the nodal analysis is carried out at all the buses and the final time domain equations can be written as follows:
\textcolor{black}{
\begin{equation}\label{Eq:sys00}
\mathbf{T} \dot{\mathbf{V}}_t{(t)}= \mathbf{T}_1\mathbf{V}_t(t) + \mathbf{T}_2\mathbf{V}_c(t)+\mathbf{T}_3\dot{\mathbf{V}}_c{(t)}
\end{equation}
where 
\begin{align}
\mathbf{T}&=
\begin{bmatrix} 
    \frac{L_{1}}{L_{c1}}+1 & -1 & 0 & \dots & 0  \\
    \frac{L_{2}}{L_{c1}}  & \frac{L_{2}}{L_{c2}}+1 & -1 & \dots & 0 \\
    \vdots  &   &   & \dots  & \\
    \frac{L_{n}}{L_{c1}}  & \frac{L_{n}}{L_{c2}} & \frac{L_{n}}{L_{c3}} & \dots & \frac{L_{n}}{L_{cn}}+1
\end{bmatrix} \label{Eq:T}
\end{align}
\begin{align}
\\\mathbf{T}_1&=
\begin{bmatrix}
    -\frac{R_{1}}{L_{c1}} & 0 & 0 &  \dots & 0\\
    -\frac{R_{2}}{L_{c1}} & -\frac{R_{2}}{L_{c2}} & 0 & \dots & 0 \\
     \vdots  &   &   & \dots  & \\
    -\frac{R_{n}}{L_{c1}} & -\frac{R_{n}}{L_{c2}} & -\frac{R_{n}}{L_{c3}}& \dots &-\frac{R_{n}}{L_{cn}} 
\end{bmatrix} \label{Eq:T1}
\end{align}
\begin{align}
\mathbf{T}_2&=-\mathbf{T}1 \label{Eq:T2}
\end{align}
\begin{align}
\\\mathbf{T}_3&=
\begin{bmatrix}
 \frac{L_{1}}{L_{c1}} & 0 & 0 &  \dots & 0\\
 \frac{L_{2}}{L_{c1}} & \frac{L_{2}}{L_{c2}} & 0 & \dots & 0 \\
  \vdots  &   &   & \dots  & \\
 \frac{L_{n}}{L_{c1}} & \frac{L_{n}}{L_{c2}} & \frac{L_{n}}{L_{c3}}& \dots & \frac{L_{n}}{L_{cn}}\end{bmatrix}\label{Eq:T4}
\end{align}\\
Since the system is linear, Equation \eqref{Eq:sys00} assumes the following form:
\begin{equation}
\mathbf{T} \Delta \dot{\mathbf{V}}_t(t)= \mathbf{T}_1\Delta \mathbf{V}_t(t) + \mathbf{T}_2 \Delta \mathbf{V}_c(t)+ \mathbf{T}_3 \Delta\dot{\mathbf{V}}_c(t) \label{Eq:sys1}
\end{equation}
Equation \eqref{Eq:sys1} can further be rewritten as :
\begin{equation}\label{Eq:sys2}
\Delta \dot{\mathbf{V}}_t{(t)}= \mathbf{A'} \Delta \mathbf{V}_t(t) + \mathbf{B'} \Delta \mathbf{V}_c(t)+ \mathbf{M'} \Delta \dot{\mathbf{V}_c}(t)
\end{equation}
where 
\begin{equation}
 \mathbf{A'}=\mathbf{T}^{-1}\mathbf{T}_{1},\hspace{0.1cm} \mathbf{B'}=\mathbf{T}^{-1}\mathbf{T}_2 , \mathbf{M'}=\mathbf{T}^{-1}\mathbf{T}_{3}\\ 
\end{equation}
$\Delta \mathbf{V}_t(t)=\mathbf{V}_t-\mathbf{V}_{ref}$ and $\Delta \mathbf{V}_c$ represents the change in DCSSMG voltages required for bus voltages to approach $\mathbf{V}_{ref}$. For the purpose of voltage control, $\Delta \mathbf{V}_c$ is the control input $u$ and the following state is chosen, 
\begin{equation} 
\mathbf{x}(t)=\Delta{\mathbf{V}_t(t)}-\mathbf{M'} \Delta{\mathbf{V}_c(t)}   
\end{equation}  
and applied onto \eqref{Eq:sys2} to arrive at the state space representation of the system represented in \eqref{Eq:sys} where $\mathbf{A=A'} \text{and} \mathbf{~B=A'M'+B'}$.}

\textcolor{black}{Sensors placed at various buses collect the information on respective bus voltages ${V_{t_i}} i=1,...n$. As full communication architecture has been assumed, each sensor relays its information to all the DCSSMGs present in the ACSSMG. Using this information, each IED needs to generate a proper voltage ${V_{c_i}} i=1,...n$ such that the bus voltages reach their reference values $\mathbf{V_{ref}}$. The CLF based hybrid distributed controller design methodology designs a $\mathbf{K}$ matrix that assure quick convergence of bus voltages when different parameters like load and communication delay change or communication link is lost.}

\subsection{Functioning of the Central/Tertiary Controller} 
The optimization formulations and the developed BCD technique are implemented in the tertiary controller of the ACSSMG and the designed controller gains are sent to the secondary levels to update the existing controllers based on various cyber and physical parameter predictions like climate, load and network predictions. The update mechanism may be time triggered or event triggered. In time-triggered updation, new controllers are computed for every time slot of 2-3 hours based on latest predictions and sent to the lower controllers. In event-triggered updation, the controllers will be updated whenever a particular parameter gets changed beyond acceptable levels prescribed by the microgrid operator. The following procedure is adopted for adaptively designing the CLF based controllers:
\begin{figure}[!ht]
\centering
\includegraphics[width=0.65\textwidth]{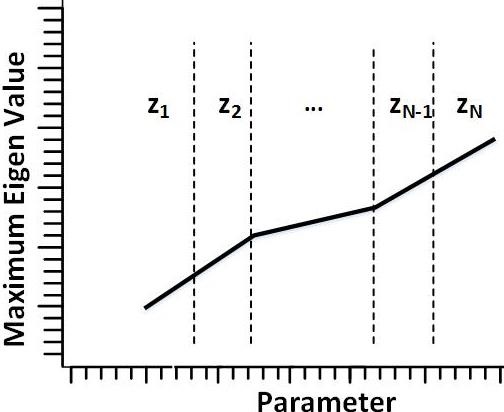}
\caption{Segregated zones}
\label{Fig:maxeigvsparameter}
\end{figure}
\alglanguage{pseudocode}
\begin{algorithmic}[1]
\State Make an eigen value chart with maximum eigen value on the y axis and the cyber/physical parameter on the x axis.
\State Dismember the chart into uniform smaller zones $z_1,~z_2,...,~z_N$ on the basis of maximum eigen value as shown in Fig. \ref{Fig:maxeigvsparameter}.
\State Find the system matrices with worst eigen values for all $N$ individual zones $\mathbf{A}_1,\mathbf{A}_2,\cdots,\mathbf{A}_N$. These are obtained by plugging in the worst case parameter values of individual zones.
\State Select the particular zone in which the grid is operating during the current time slot and a neighbor zone where the grid will operate in the next time slot as per the forecasted data, say $z_1$ and $z_2$ .
\State Solve the appropriate CLF formulation as presented in Section \ref{Sec:BCD} and obtain the corresponding $\mathbf{K}$ matrix.
\end{algorithmic}
\section{Simulation Configuration} \label{Sec:results}

A sample 4-bus system is used for demonstrating voltage control whose parameters have been given in Table-\ref{tab:table1}.
The value of $R_4$ and $L_4$ vary as the ACSSMG load change over period of time. For all the cases $L_4$ has been taken to be $0.0148$. Also, the value of $L_{c_n}=0.001$ where $n=1,2,3,4$.


\begin{table}[ht!]
  \centering
  \caption{Grid Parameters}
  \label{tab:table1}
  \begin{tabular}[t]{|cc|cc|}
    \hline 
    Parameter & Value & Parameter & Value \\
    \hline \\ [-1.0em]
    $R_1$ &  0.175  & $L_1$ & 0.0005  \\
    $R_2$ &  0.1667 & $L_2$ & 0.0004  \\
    $R_3$ &  0.2187 & $L_3$ & 0.0006  \\  
    \hline
  \end{tabular}
\end{table}
The different system matrices used for different loading conditions $R_4= 0.001$ and $0.5$ are given as follows: 
\begin{align}\label{Eq:A1} 
\mathbf{A}_1 =
\begin{bmatrix}
 -150.438 & -63.001 & -21.476  & -0.000 \\
  -50.657 & -94.501 & -32.214 &  -0.000 \\
   35.604 &   9.197 & -53.691  & -0.001 \\
  155.010 & 138.913 & 100.579  & -0.003 \\
\end{bmatrix}
\end{align}
\begin{align}\label{Eq:A2}
\mathbf{A}_2=
\begin{bmatrix}
   -152.3020 & -64.8651 & -23.3403 &  -1.8645 \\
  -53.4530 & -97.2977 & -35.0105  & -2.7968 \\
   30.9449 &   4.5372 & -58.3508  & -4.6613 \\
  144.7589 & 128.6619 &  90.3282  &-10.2548 \\
\end{bmatrix}
\end{align}

\section{Numerical Results}

In this section the different CLF controllers designed for voltage control are applied in a 4-bus ACSSMG. Cases like variation in load, variation in delay and loss of communication link have been explored for demonstrating the efficacy of the proposed method. CVX optimization tool \cite{cvx1}\cite{cvx2} in MATLAB is used for solving the different optimization formulations.
\begin{figure}[!ht]
\begin{subfigure}{.95\textwidth}
\begin{center}
\includegraphics[width=0.95\linewidth]{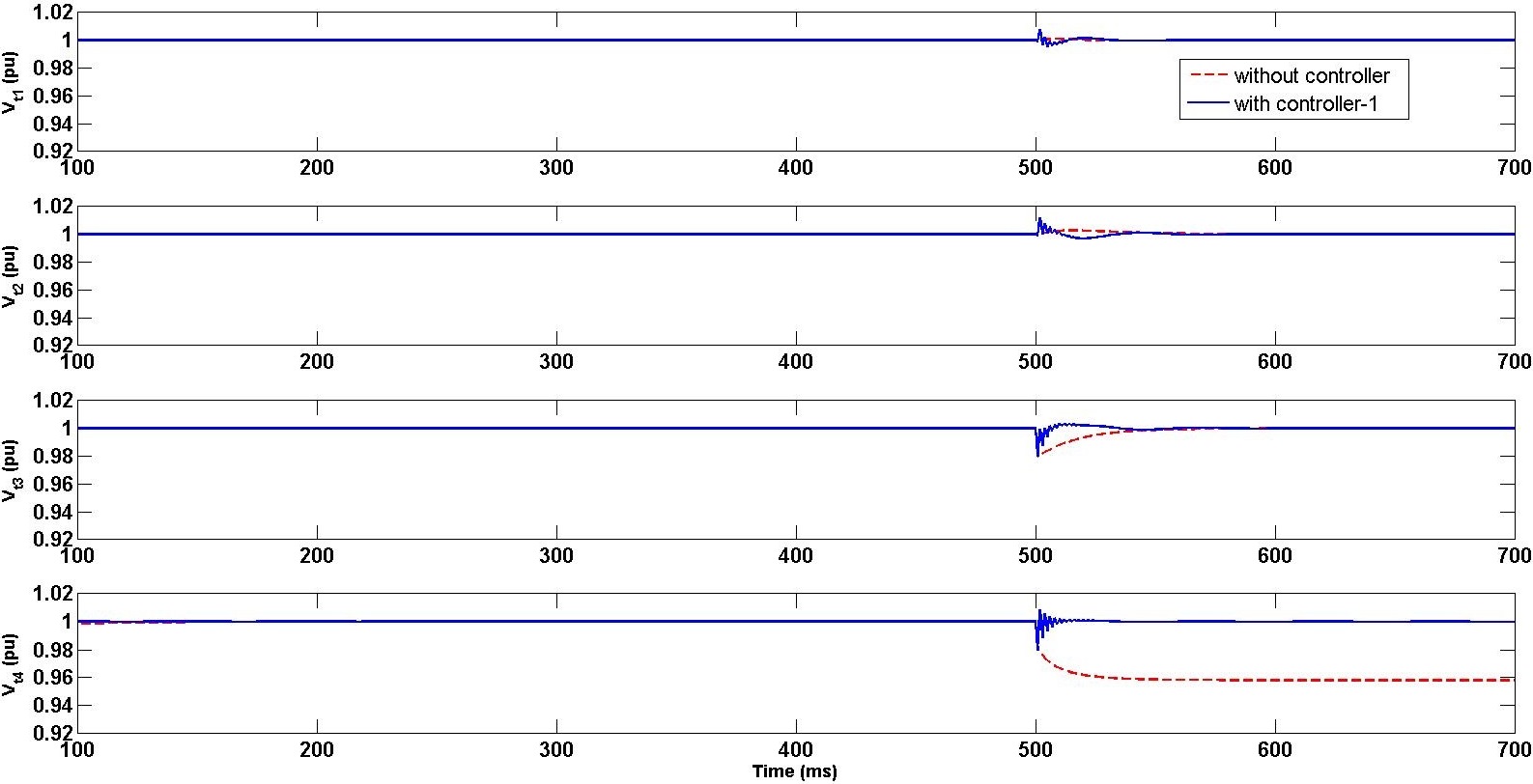}
\label{Fig:result1}
\caption{}
\end{center}
\end{subfigure}
\begin{subfigure}{.95\textwidth}
\begin{center}
\includegraphics[width=0.95\linewidth]{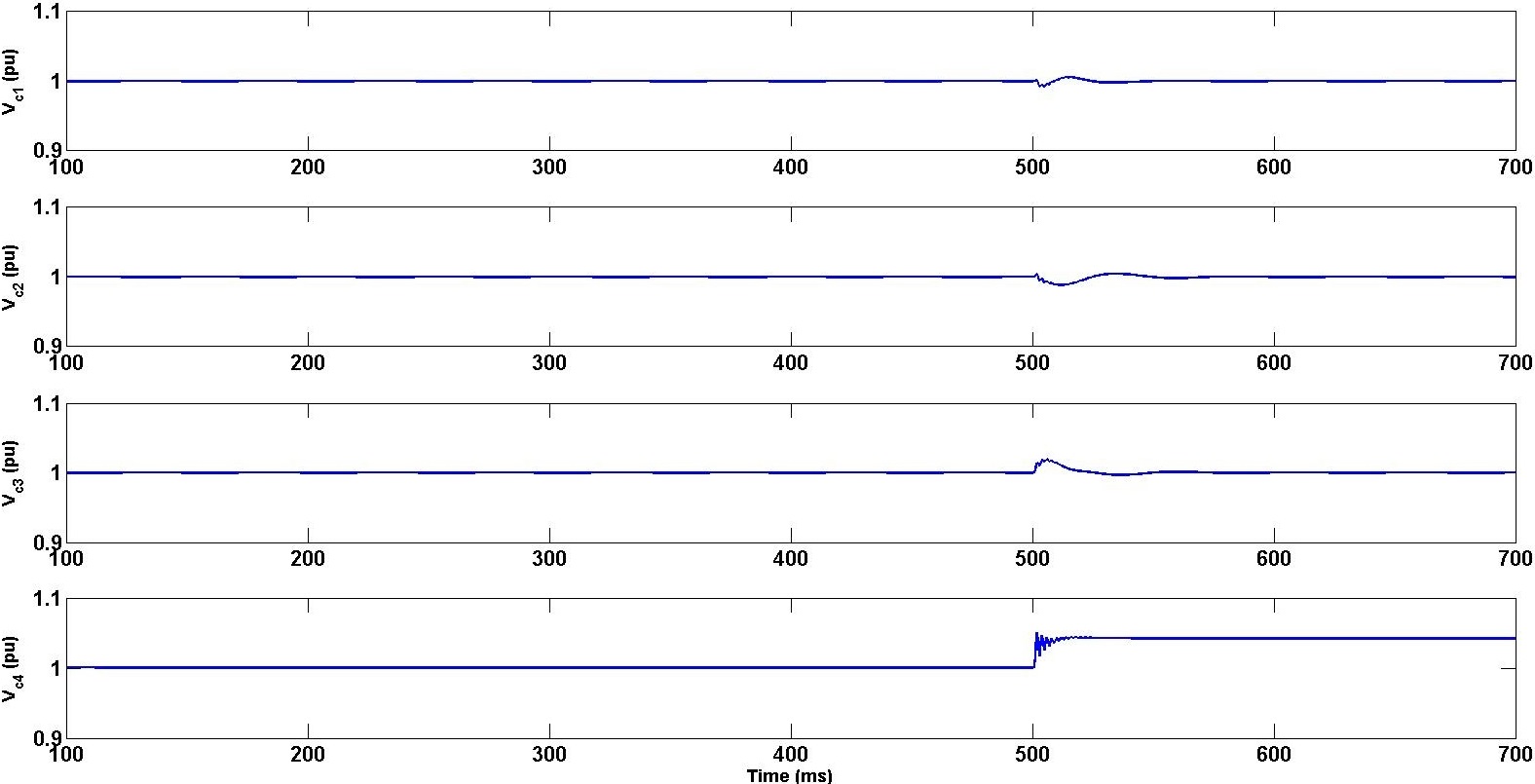}
\caption{}
\label{fig:result11}
\end{center}
\end{subfigure}
\caption{Results for change in load at t=500ms
(a) Comparison of bus voltages between grid with controller-1 and uncompensated grid.
(b) The magnitude of IED controller voltages in presence of controller-1}
\label{Fig:simpleresult1}
\end{figure}
\subsection{Change in Load Resistance} 
The load on the power system is modeled as a resistance and reactance. As described previously, two zones of operation $\mathbf{A}_1$ and $\mathbf{A}_2$ with different load resistances have been selected and controller has been designed. The $\mathbf{K}$ matrix designated as controller-1 can comfortably take care of voltage control within these two neighboring zones as well as maintain system stability while switching between the two zones of operation. When, the system is predicted to go beyond the bound of the selected two zones, the controller gets updated by the central server. The result for this case is shown in \eqref{Eq:resultaA111} and \eqref{Eq:resultaA112}.
\begin{align}\label{Eq:resultaA111}
\mathbf{K}= 
\begin{bmatrix}
1.5063  &  0.4404  &  0.5500  & -1.5126  \\
   0.5038 &   0.7212 &   0.5915 &  -1.4257 \\
  -0.0774  &  0.2346 &   1.0336  & -1.3883 \\
  -1.6643 &  -1.7800 &  -0.7057  & -1.7950 \\
\end{bmatrix}  
\end{align}    

\begin{equation}\label{Eq:resultaA112}
\left.
\begin{aligned}
\text{Maximum eigenvalue of}~  \mathbf{A}_1 &=   -0.0039  \\
\text{Maximum eigenvalue of}~  \mathbf{A}_{K_1} &=  -2.3859  \\
\text{Maximum eigenvalue of}~  \mathbf{A}_2 &=  -18.8524  \\
\text{Maximum eigenvalue of}~  \mathbf{A}_{K_2} &=  -21.5665  \\
 \| \mathbf{K} \|_2 &=  3.1688\\  
\end{aligned}
\right\}
\end{equation}
Fig.\ref{fig:result1} shows the bus voltages when load is switched from R=0.5 to R=0.01 at time t=500 ms.Fig.   \ref{fig:result11} shows the IED voltages in the presence of controller-1 when load is switched from R=0.5 to R=0.01 at time t=500 ms.
The results indicate that voltage at bus-4 reduces due to switching in the absence of controller which is taken care by addition of controller into the system.


\subsection{Change in communication delay and load}
If, it is possible to know the region of maximum delay bound, the controllers can be designed for the maximum delay predicted during a particular period of the day along with the maximum delay predicted for the upcoming period of the day. Along with this, the load variation is also an inevitable change. The controller-2 designed using optimization formulation for delay case has demonstrated that it can stabilize the bus voltages for simultaneous change in load and delay profile.
The following are the results obtained with controller-2 when two subsystems have been considered- one with maximum delay of $0.5~ms$ operating in region $A_1$ and the other with maximum delay of $1~ ms$ operating in region $A_2$. 
\begin{equation}
\mathbf{D_1} =
\begin{bmatrix}
    0.2500 &   0.5000 &   0.4375 &   0.3125 \\
    0.3750 &   0.3125 &   0.2500 &   0.4375 \\
    0.5000 &   0.4375 &   0.3125 &   0.2500 \\
    0.3750 &   0.2500 &   0.4375 &   0.3125 \\
\end{bmatrix} \times ~10^{-3}
\end{equation}
\begin{figure}[H]
\centering
\begin{subfigure}{0.95\textwidth}
\includegraphics[width=0.95\linewidth]{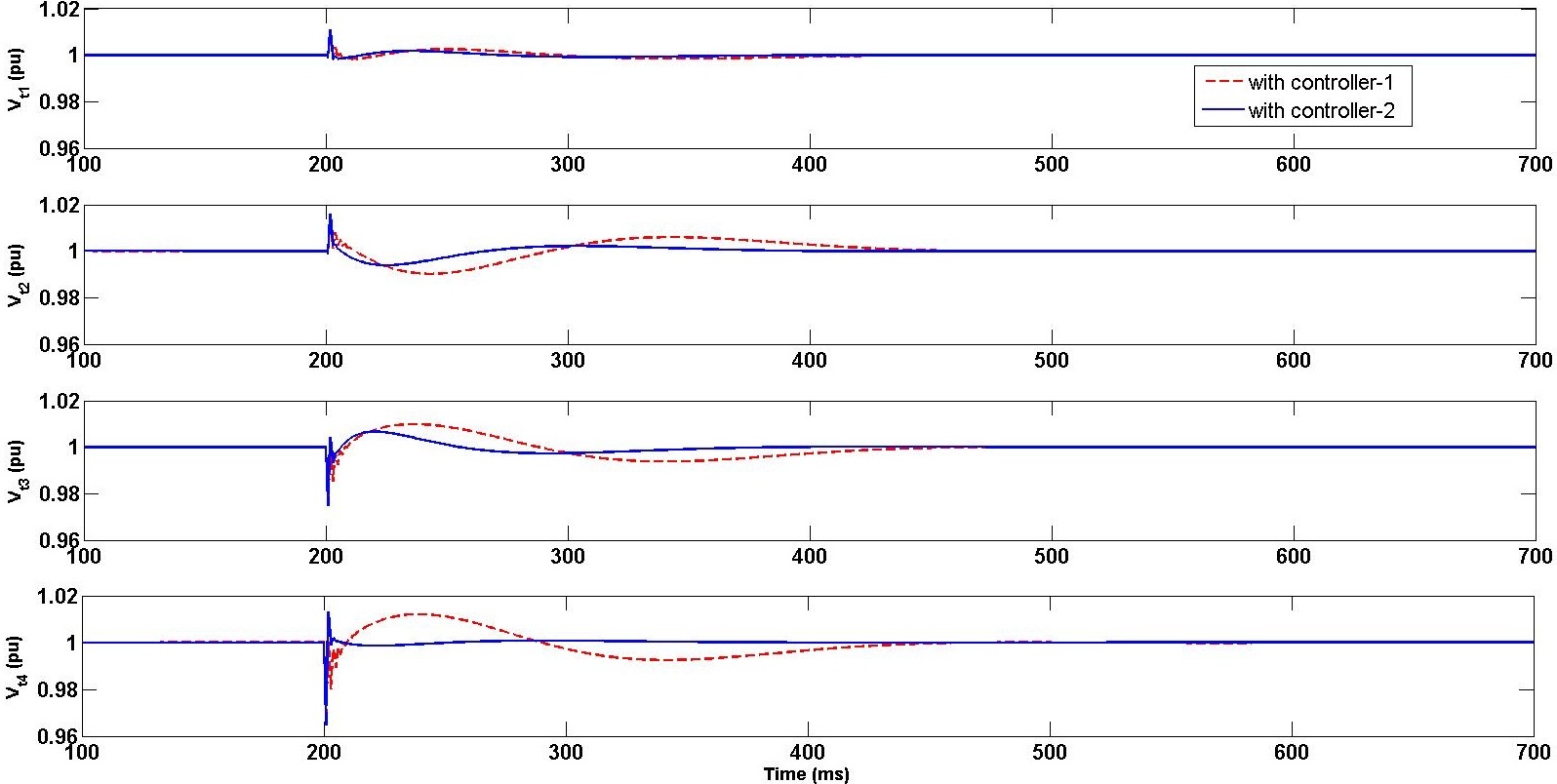}
\caption{}
\label{fig:result2}
\end{subfigure}
\begin{subfigure}{0.95\textwidth}
\includegraphics[width=0.95\linewidth]{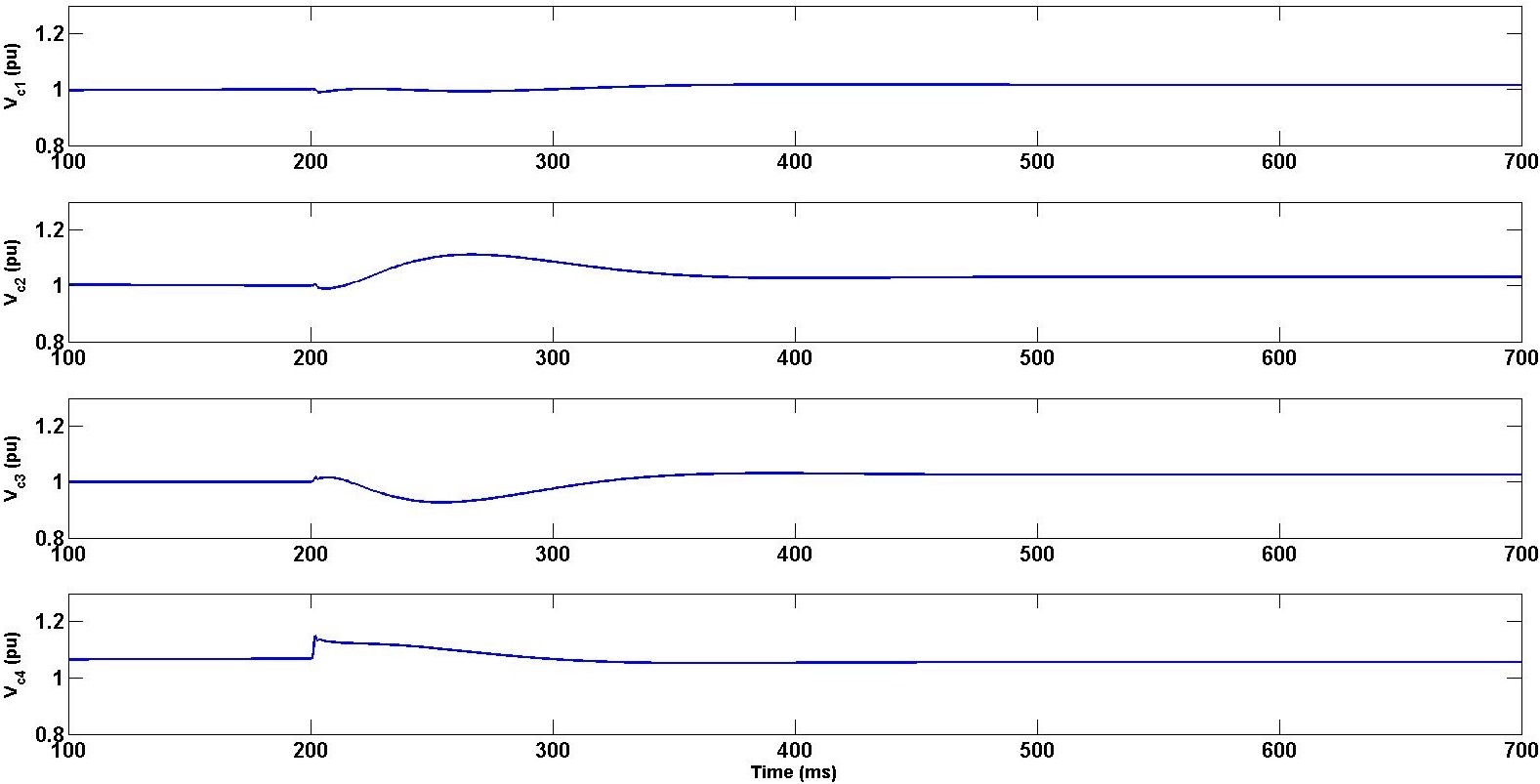}
\caption{}
\label{fig:result21}
\end{subfigure}
\caption{Results for change in load and delay at t=200ms: (a) Comparison of bus voltages between the situation with controller-1 designed only for load variation and the situation with controller-2 designed for both load and delay variation. (b) The magnitude of IED controller voltages for system with controller-2 designed for both load and delay variations.}
\label{Fig:delayresult1}
\end{figure}
\begin{equation}
\mathbf{D_2} =
\begin{bmatrix}
    0.5000 &   1.0000 &   0.8750 &   0.6250 \\
    0.7500 &   0.6250 &   0.5000 &   0.8750 \\
    1.0000 &   0.8750 &   0.6250 &   0.5000 \\
    0.7500 &   0.5000 &   0.8750 &   0.6250 \\
\end{bmatrix} \times ~10^{-3}
\end{equation}

\begin{equation}
\mathbf{K} =
\begin{bmatrix}
0.9661  & -0.0401 &  -0.6552  & -1.7199 \\
   -0.1347 &   0.4414 &  -0.4786 &  -1.7313 \\
   -0.6294 &  -0.4689 &  -0.0069 &  -1.6075 \\
   -2.2338 &  -2.1026  & -2.1243 &  -2.0097 \\
\end{bmatrix}
\end{equation}
\begin{equation}
\left.
\begin{aligned}
\text{Maximum eigenvalue of}~  \mathbf {A}_1 &=   -0.0039  \\
\text{Maximum eigenvalue of}~  \mathbf {A}_{K_1} &=  -2.3938  \\
\text{Maximum eigenvalue of}~  \mathbf {A}_2 &=  -18.8524  \\
\text{Maximum eigenvalue of}~  \mathbf {A}_{K_2} &=  -21.9100  \\
 \|\mathbf{ K} \|_2 &=  4.6872 \\ 
\end{aligned}
\right\}
\end{equation}
Fig.\ref{fig:result2} shows the bus voltages when both load is switched from R=0.5 to R= 0.01 and delay of 1ms is added at time t=200 ms. Fig.\ref{fig:result21} shows the IED controller voltages in the presence of controller designed for both load and delay variations for the same.
The results indicate that the controller-2 designed with delay considerations settles quickly after the disturbance compared to the controller-1 designed without delay considerations.


\subsection{Failure of a communication link}
The following are the details of the controller-3 designed for a case where there has been communication failure between sensor-3 and IED-3. Moreover, the system parameters have also been changed due to change in load resistance from $\mathbf{A_1}$ to $\mathbf{A_1}$. 
\begin{equation}
\mathbf{K} =
\begin{bmatrix}\label{Eq:ak111}
 1.3076   & 0.4948 &  -0.7865  & -1.3994 \\
     0.1251 &   1.1195  & -0.3809 &  -1.5391 \\
    -0.2523 &  -0.2388 &  -0.0000  & -1.2732 \\
    -1.8282 &  -1.2753 &  -2.3808  & -1.6920 \\ 
\end{bmatrix}
\end{equation} 
\begin{figure}[H]
\centering
\begin{subfigure}{0.95\textwidth}
\includegraphics[width=0.95\linewidth]{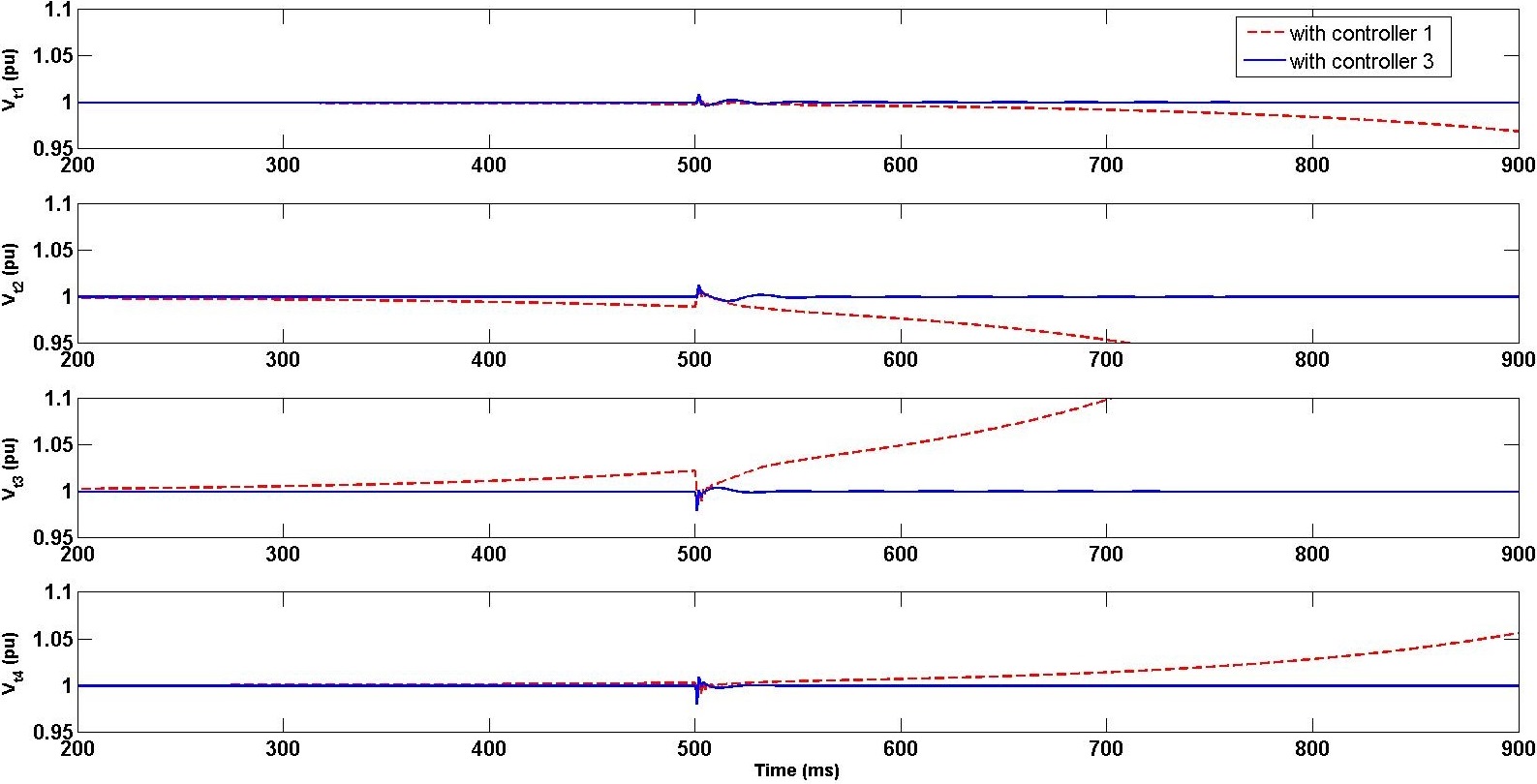}
\caption{}
\label{fig:result3}
\end{subfigure}
\begin{subfigure}{0.95\textwidth}
\includegraphics[width=0.95\linewidth]{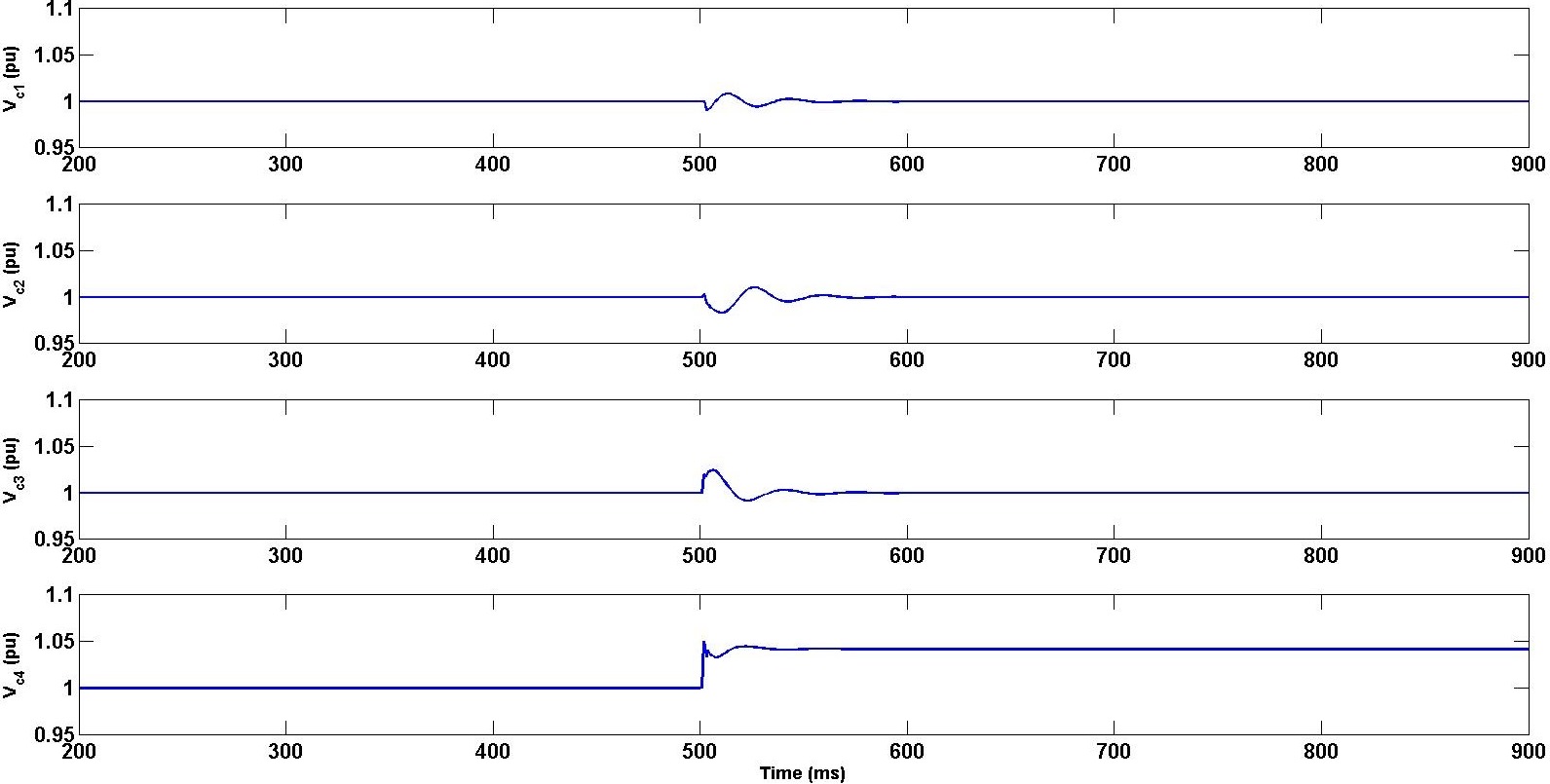}
\caption{}
\label{fig:result31}
\end{subfigure}
\caption{Results for change in load and loss of communication link at t=500 ms (a) Comparison of bus voltages between the grid with controller-1 designed only for load variation and grid with controller-3 designed for both load variation and communication link loss (b) The magnitude of DG controller voltages in the ACSSMG with controller-3 designed for both load and communication link loss.}
\label{Fig:linklossresult1}
\end{figure}
\begin{equation}\label{Eq:ak112}
\left.
\begin{aligned}
 \text{Maximum eigenvalue of}~  \mathbf {A}_1 &=  -0.0039 \\
\text{Maximum eigenvalue of}~  \mathbf {A}_{K_1} &= -3.0426 \\
\text{Maximum eigenvalue of}~  \mathbf {A}_2 &=  -18.8524 \\
\text{Maximum eigenvalue of}~  \mathbf {A}_{K_2} &= -22.1455 \\ 
\| \mathbf{K} \|_2&= 3.8840 \\
\end{aligned}
\right\}
\end{equation}
Fig.\ref{fig:result3} shows the bus voltages when load is switched from R= 0.5 to R= 0.01 at time t=500 ms for two controllers after experiencing a loss in communication link between sensor 3 and controller 3. The controller-1 designed without considerations of link loss makes the bus voltages unstable whereas the controller-3 designed with link loss consideration stabilizes the system after the disturbance. Fig.\ref{fig:result31} shows the DCSSMG controller voltages in the systems with controller-3 designed with communication link loss considerations when load is changed at t=500 ms.

\section{Summary} \label{sec:hybridsummary}
 This work presents a generic hybrid and adaptive framework for enhancing the distributed control in ACSSMG using communication under simultaneous variation of various electrical and communication parameters. The strength of this work has been the novel control frameworks from the Cyber-Physical perspectives. It has been shown that the variation of parameters for a sample \textcolor{black}{coordinated voltage control problem in a distribution feeder} can be modeled as hybrid system and the associated controllers can be designed using convex optimization. Delay and link loss can play a crucial role in Cyber-Physical energy systems and these issues have also been addressed. The overall control procedure not only performs well in situations with variation in multiple parameters but also increases the range of operation for these controllers. 
 
\chapter{Optimal Communication Design based Control Framework for Sparsely Connected ACSSMG} \label{chapter:distobs}\index{Disturbance Observer!Back-stepping Design}
\chaptermark{Optimal Communication Design}
\section{Introduction}

Consider the ACSSMG in Fig. \ref{Fig:fig1}.
The grid in the picture consists of n numbe of PCCs where each PCC is connected to a DCSSMG with DGs like  solar, wind, etc. A sensor is placed at every PCC and its voltage is sent to DCSSMGs for determining the operating voltages of their interlinking converters (IEDs). An established procedure for this problem would be to let each DG control the voltage of the respective bus with the help of data obtained from sensors placed at that particular bus. A set of distributed controllers using a central tertiary controller is proposed as an alternative to centralized control. These controllers can provide enhanced control using the information acquired from sensors present at multiple buses. Along with the perks obtained from incorporation of CPS into the grid, it is imperative that issues would arise pertaining to both, control theory and communications. The ACSSMG load may be subjected to change and so is the generation. Similarly, the communication network may experience issues with limited bandwidth, communication delay \cite{7458197} or a loss of communication. Hence, a new paradigm of designing cyber physical controllers is being explored very rapidly which can keep the ACSSMG stable under all these variations.

\begin{figure}[!ht]
\centering
\includegraphics[width=0.65\textwidth, keepaspectratio]{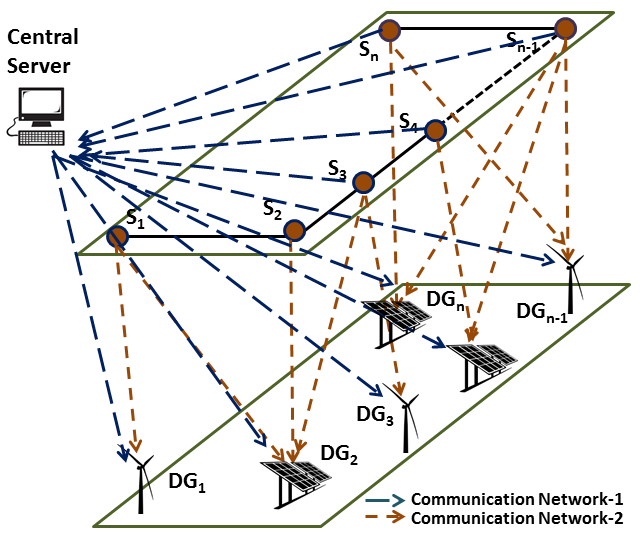}
\caption{Illustration of the ACSSMG}
\label{Fig:fig1}
\end{figure}


It has been noticed that design of communication network has considerable effect on the various physical parameters of the system \cite{selfself4}. Works like \cite{majumder2012power} \cite{nguyen2010distributed} develop communication routing algorithms and protocols to connect DGs and enhance power sharing in various microgrid configurations. The effect of different communication structures on distribution system protection performance has been studied in \cite{6840974}. These however, do not come up with any systematic strategy for control of a physical system variable based on communication topology design. The work in \cite{li2012multicast} proposed a simple optimization technique to design the communication topology for voltage control in the ACSSMG. A greedy algorithm was used to route the connections between sensors and DGs based on controller stability. But, this approach fails to consider many topologies while evaluating the most stable controller for a particular situation. In addition, both the constraints and parameters in the domains of communication and control remain fixed. In practice, various physical parameters like load and communication parameters like delay keep changing simultaneously over the course of the day. 

In order to overcome these issues, this work proposes a novel CBSCD framework to design controllers for ACSSMG.
This framework provides opportunity to rope in connection constraints that can be specified to define the boundaries of the communication network. Resource constraints like communication bandwidth to describe the connecting capabilities of various sensors and controllers can be added into the communication design process. Moreover, utility based constraints like cost, can be specified, so as to accommodate required operational demands of power utilities. Finally, the regular physical variable constraints like load requirement on the buses can be specified.

The idea behind developing this framework is to handle changes in multiple parameters of the ACSSMG more effectively. Appropriate constraints can be tuned to respond to any change in system. This way, it provides more options to respond to a change in system and this framework will be very useful if operator can select an appropriate constraint depending on his/her experience about the possible source of the problem. This way it is more effective than other conventional approaches.
The organization of this chapter is as follows. 

Section \ref{Sec:MIMO} gives a general idea of mathematical modeling and communication structure of the ACSSMG. Section \ref{Sec:LF} describes the Lyapunov function based optimization frameworks adopted for designing stable controllers in the presence and absence of delays. This follows Section \ref{Sec: GSCD} which explains the CBSCD methodology to design the controllers with most appropriate communication design. Section \ref{Sec:descon} describes application of the developed techniques to the context of voltage control in the ACSSMG and the results obtained upon application of these procedures on certain scenarios like load fluctuations, delay fluctuations and communication failure have been presented in Section \ref{Sec:results}. Appropriate conclusions with possible future work have been indicated in Section \ref{Sec:cfw}.

\section{Cyber Constraints in Communication Design}\label{Sec:MIMO}
To describe the communication structure shown in Fig. \ref{Fig:fig1} some standard parameters need be defined. Hence, this section delineates the communication structure and the various constraints that are used to describe it. 

A wireless communication network is assumed to pervade through the ACSSMG. The sensors present at each PCC contain transmitter nodes for relaying information to the controllers and the controllers have receivers or receiver nodes to receive the voltage information from the transmitter nodes. Both these types of nodes are spread along two communication networks one across the tertiary control level and the other across the secondary control level.

Network-1 facilitates the sensor nodes at various PCCs to provide their information to the central tertiary controller containing receiver nodes. The central server/tertiary controller tracks the data of various parameters of the ACSSMG and forecasts their future values to design situation aware distributed controllers adaptively.

 Network-2 consists of interconnections between the sensors and controllers which will be operational on a regular basis. Since there are $n_s$ number of PCCs placed with sensors, this networks contains $n_s$ transmitter nodes and since $n_c$ DCSSMGs are present, it contains $n_c$ number of receiver nodes.

The communication network is assumed to be densely connected which means that every controller receives information from all the sensors. Data is assumed to be transferred continuously between the nodes which means that the sampling rate is good and there is no packet loss. The routing between the nodes would be on a multicast \cite{7128386} \cite{kompella1993multicast} basis.

It is to be noted that the DCSSMGs in the power network are classified into two categories- peripheral DCSSMGs and central DCSSMGs. The peripheral DCSSMGs are the DCSSMGs located nearer to the boundary of the power network and which are capable of handling only lighter loads. The central DCSSMGs, on the other hand, are located in the central region of the power network and possess capacity to handle higher loads. Thus, in a distributed control scenario, the central DCSSMGs in general, make use of higher communication resources as they need to be connected to data from more number of sensors compared to peripheral ones. 
 
\begin{figure}[H]
    \centering
    \begin{subfigure}{0.45\textwidth}
    \includegraphics[width=\linewidth, height=5cm]{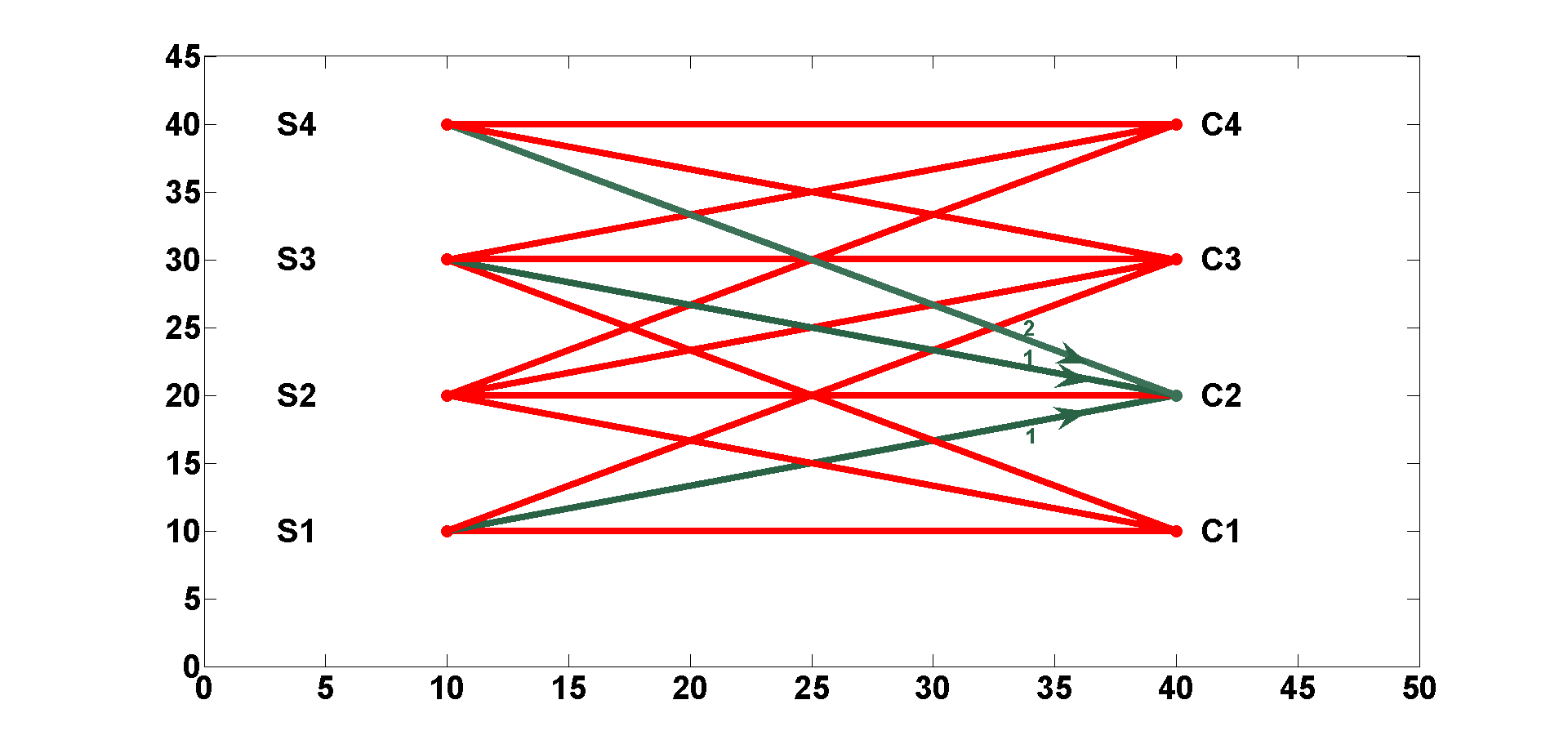}
    \caption{$cc=2$ connections}
    \label{fig:cc22}
    \end{subfigure}
    \begin{subfigure}{0.45\textwidth}
    \includegraphics[width=\linewidth, height=5cm]{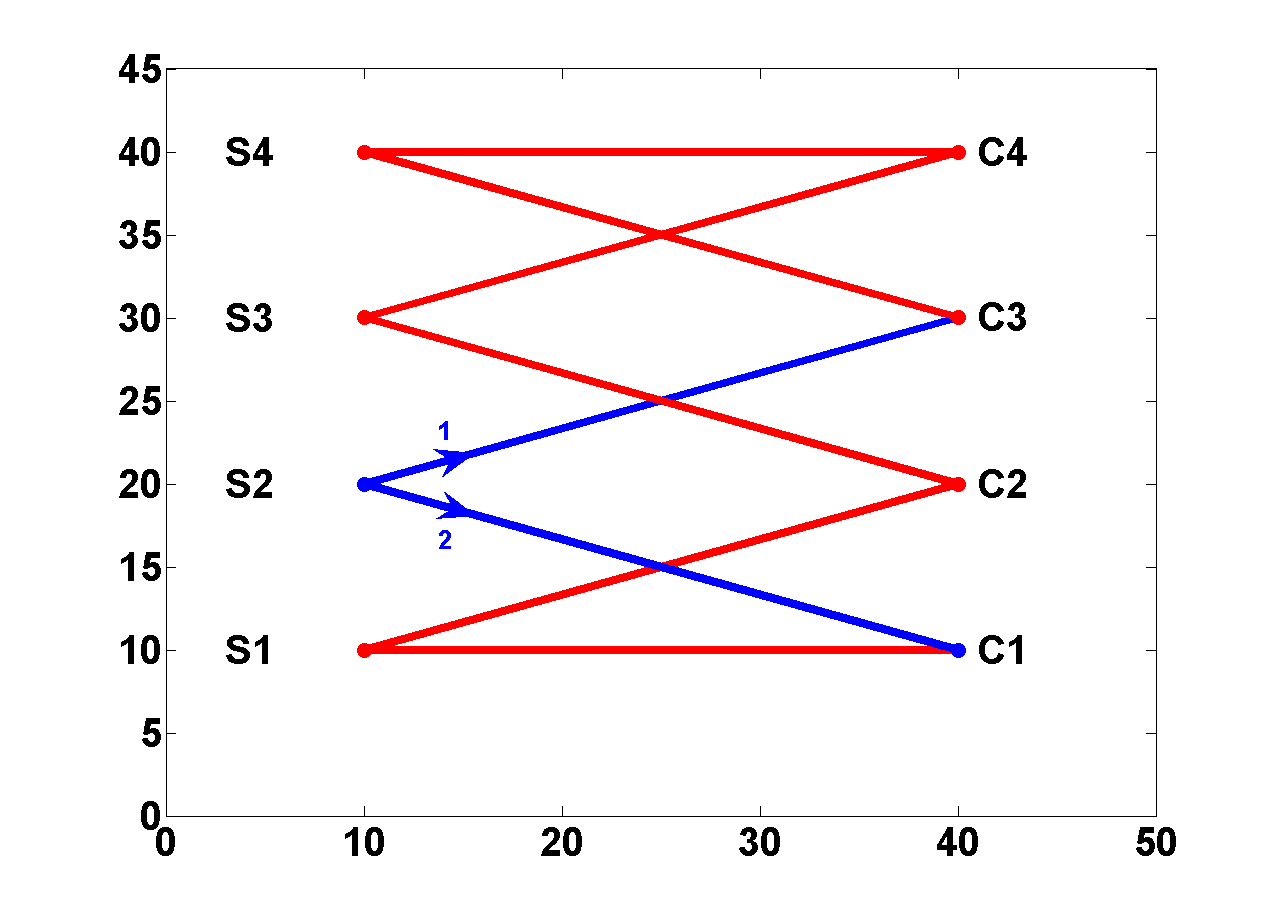}
    \caption{$bwc=2$ connections}
    \label{fig:bw22}
    \end{subfigure}
    \caption{Illustration of Communication Constraints}
\end{figure}

The topology of the communication network is defined by the following communication constraints:
\begin{enumerate}
\item Bandwidth Constraint: This constraint specifies the number of connections that must start from any node. For example, Fig. \ref{fig:bw22} shows a set of connections that can exist when $bwc=2$ with $bwc$ representing this constraint.
\item Connection Constraint: This constraint deals with the general topology of the communication network. It specifies the availability of a particular node to be connected to other nodes. The value of connection constraint $cc$ signifies the maximum number of p-neighbor nodes located on single side of the current bus from  which a particular node can receive information. The term p-neighbor node refers to the node in the previous layer which is situated in between the sensor to controller nodes connected to a bus other than the current bus. For instance, if $cc=2$, then the possible set of connections will be as shown in Fig. \ref{fig:cc22}. 
\item Peripheral Cost Constraint: This constraint $prc$ represents the maximum number of sensors that can be connected to a DCSSMG located in the periphery of the power network. This has been defined as per the assumption that DCSSMGs with less capacity will be installed on the boundary of the power network which can handle lesser loads compared to the central ones.
\item Central Cost Constraint: This constraint $cnc$ represents the minimum number of sensors that must be connected to central DCSSMG so that it can be operational. This means that this DCSSMG needs to be connected to atleast $cnc$ number of sensors to be switched on failing which it is shut down. 
This has been defined as per the assumption that DCSSMGs with more capacity will be installed in the central region of the power network and that they can handle higher loads compared to the peripheral ones.
\end{enumerate}

\section{Lyapunov based Optimization Formulations} \label{Sec:LF}
This section deals with the description of optimization formulations obtained using basic Lyapunov stability analysis for finding controllers for a given set of system parameters and communication design. It is to be noted that the physical system models used in this chapter have been adopted from the previous chapter since the ACSSMG setup used for both is the same. Conditions assuring stability have been formulated into an optimization framework using Linear Matrix Inequalities (LMIs). Two cases will be looked into, namely, the delay-free case and the small delay case.

\subsection{System without delay}
This formulation has been developed on the assumption that all the communication links are delay free. Considering the linear system model as in \eqref{Eq:sys} , and given a Lyapunov function $\mathbf{V(x)}=\mathbf{x}^T\mathbf{Px}~~(\mathbf{P}\succ 0)$, it is well known that equilibrium point goes to zero, if the following two inequalities hold simultaneously for all $\mathbf{x} \neq 0$
\begin{eqnarray}
\mathbf{V(x)}\succ 0 \label{Eq:v},~~ 
\mathbf{\dot{V}(x)}\prec 0 \label{Eq:v.}
\end{eqnarray}
The rate derivative of the Lyapunov function $\mathbf{V}$ for the system model \eqref{Eq:sys} is obtained as:
\begin{eqnarray}
\mathbf{\dot{V}(x)}=\mathbf{\dot{x}}^T\mathbf{Px}+\mathbf{x}^T\mathbf{P\dot{x}}=\mathbf{x}^T\mathbf{(\bar{A}}^T\mathbf{P} +\mathbf{P \bar{A})x} \label{Eq:Lyap}
\end{eqnarray}

Given that the matrix $\mathbf{P} \succ 0 $, from \eqref{Eq:Lyap} it can be inferred that $\mathbf{\dot{V} \prec 0}$, if the following condition holds true:
\begin{equation}
\mathbf{\bar{A}}^T\mathbf{P}+\mathbf{P \bar{A}} ~\prec~0 \label{Eq:Lyap2}
\end{equation}
For the computed $\mathbf{\bar{A}}$ to be more stable than $\mathbf{A}$, the matrix $\mathbf{P}$ is computed by selecting an arbitrary positive value of $\beta$ such that the following condition holds true:

\begin{equation}
\mathbf{(A-\beta I)}^T\mathbf{P}+ \mathbf{P(A-\beta I)}=-\mathbf{I} \label{Eq:ABPI}
\end{equation}

It should be noted that for different value of $\beta$, the controller solution will be different. By introducing a design parameter $\gamma$ into \eqref{Eq:sys3}, the optimization problem in the form of LMIs is obtained as:
\begin{align}
\underset {{K}}{\max}~~~&\gamma_1  \nonumber  \\
{s.t.\hspace{0.4cm}} 
&\mathbf{\bar{A}}^T\mathbf{P} +\mathbf{P\bar{A}}+\gamma \mathbf{I} \prec 0 \nonumber \\
&\mathbf{K}_{ij}=0, \text{if link between sensor $j$ and DG $i$ is absent} \nonumber \\
&\mathbf{\| K \|}_2\leq \rho 
\label{Eq:unknown}
\end{align}
Note that the stability margin increases with $\gamma$. The constraint,
\begin{equation}\label{Eq:Kbound}
\|\mathbf{K}\|_2\leq \rho
\end{equation} is incorporated so that the controller gains are bounded, where $\rho$ is a user defined scalar value.

\subsection{Systems with delay} \label{SSec:SND}
The communication networks can deliver the data in order of milliseconds whereas the dynamics of DGs is generally in the order of tens to hundreds of milliseconds. Since, the time scales are comparable, the effect of delay should be included into the system model. The dynamics with small delay as given in \eqref{Eq: dmodel} can be rewritten as. 
\begin{equation}
\centering
\mathbf{\dot{x}}= \mathbf{(A+BKC)x} + \mathbf{f(x)}
\end{equation}
where $\mathbf{f(x)}= \mathbf{-BDK(A+BKC)x}$ is considered as the uncertainty due to delay and  
\begin{equation}
\centering
\mathbf{f}^T\mathbf{(x)f(x)} \preceq \alpha^2 \mathbf{x}^T\mathbf{x} \label{Eq:normalpha}
\end{equation}
The norm bound on $\alpha$ can be found as, 
\begin{equation}
\centering
\alpha = \sqrt{mn}~\rho ~ \mathbf{\| B\|}_2 (\mathbf{\|A\|}_2 + \mathbf{\| B\|}_2 \rho)~\underset{j,i}{\max}\{d_{ji}\}
\end{equation}

Now, assuming that the dynamics of this system be stabilized by a Lyapunov function $\mathbf{V(x)}=\mathbf{x}^T\mathbf{Px}$  ($\mathbf{P}\succ0$)
\begin{align}
\mathbf{\dot{V}(x)} &= \mathbf{x}^T\mathbf{P}(\mathbf{A}_K\mathbf{x} + \mathbf{f(x)}) + \mathbf{(x)}^T\mathbf{A}^T_K+\mathbf{f}^T(\mathbf{x}))\mathbf{Px}  
\end{align}
where $\mathbf{A}_K=\mathbf{A}+\mathbf{BKC}$. 

This equation can be rewritten as $\mathbf{\dot{V}(x)}= \mathbf{y}^T\mathbf{Fy}  \label{Eq:delay3}$ 
where 
\begin{equation}
\centering
\mathbf{F}=
\begin{bmatrix}
\mathbf{{A}}^T_K \mathbf{P}+ \mathbf{P{A}}_K &  & \mathbf{P} \\
&         & \\
\mathbf{P} &   & 0
\end{bmatrix} 
\text{and }
\mathbf{y}=
\begin{bmatrix}
\mathbf{x} & \mathbf{f(x)}
\end{bmatrix}\nonumber 
\end{equation}
Also, \eqref{Eq:normalpha} can be rewritten as: 
$ \mathbf{y}^T\mathbf{Gy} \preceq 0$
where 
\begin{equation*}
\centering
\mathbf{G}=
\begin{bmatrix}
-\alpha^2\mathbf{I}  & 0 \\
0 & \mathbf{I}
\end{bmatrix}
\end{equation*}
Now, $\mathbf{\dot{V}=y}^T\mathbf{Fy}\prec0$ holds only if there exists a $\mathbf{y}^T\mathbf{Gy}\preceq0$ such that 
$\mathbf{(F-\gamma G)}\prec0$ and a positive $\gamma$, which results in 
\begin{align}\label{Eq:delayopt}
\begin{bmatrix}
\mathbf{A}^T_{K}\mathbf{P+PA}_{K}+\gamma\alpha^2 \mathbf{I} &  & \mathbf{P} \\
&   &  \\
\mathbf{P} &  & -\gamma\mathbf{I}
\end{bmatrix}&\prec0 \hspace{0.4cm}
\end{align} 

Thus, using the conditions as constraints \eqref{Eq:delayopt} along with \eqref{Eq:Kbound} to limit the controller gains, the optimization formulation for this case can be written as follows:
\begin{align}
&\underset {K}{\max} ~~ \gamma \nonumber \\
~~{s.t}~~ & \text{condition \eqref{Eq:delayopt} holds} \nonumber \\
& \mathbf{\| K \|}_2 \leq~\rho    
\end{align}	
\section{CBSCD Based Controller Design Methodology} \label{Sec: GSCD}
The distributed control structure in \ref{Eq:controller} shows a clear dependence of control on communication topology. Using the optimization frameworks described in Section \ref{Sec:LF}, it is possible to find the controller gains with maximum stability for a particular communication topology. A particular operating condition of the ACSSMG is dictated by a set of physical and communication parameters like load, bandwidth constraint and connection constraint. For any operating condition, there can exist many communication topologies. This section delineates the proposed CBSCD methodology which finds the most stable controllers with least set of operating communication resources for different operating conditions of the ACSSMG during the course of the day. The controllers are designed based on forecast data, which will be scheduled as per the variation in operating conditions. 

The overall methodology consists of two different parts which are explained in the following two subsections- the connection finding algorithm which finds the possible connections for a particular set of communication and cost constraints and the controller design procedure containing the resource minimization logic.  
 
\subsection{Connection Finding Algorithm}
A communication network design has to be carried out between the transmitter nodes from the sensors and the receiver  nodes of the DCSSMG controllers after considering differnt constraints like connection, bandwidth and utilities. Given a set of these constraints, the following connection finding algorithm is used to find the possible connection topologies that satisfy them : 

\alglanguage{pseudocode}
\begin{algorithmic}[1]
\State Initialize connection set $R$ as empty set.
\State Initialize the constraints $bwc$, $cc$, $prc$ and $cnc$ in the communication network.
\State Initialize set $Cost$ as all possible DCSSMG cost configurations for the given constraints.    
\State \textbf{for} All configurations in $Cost$ \textbf{do}
\State \quad Initialize connection set $R_1$ as empty set.
\State \quad List out the non-zero costs of individual DCSSMGs in increasing order and store in $cost$-$~~~~~ sortup$.
\State \quad Store the size of $cost-sortup$ in $count$.
\State \quad \textbf{while} $count ~>~ 0$
\State \quad \quad Create set $v$ for the first DCSSMG in $cost-sortup$ with all possible existing configurations $~~~~~~~~$ generated for the given cost.
\State \quad \quad \textbf{if} $count~ > ~bwc$
\State \quad \quad \quad Add feasible connections in $v$ to each connection in $R_1$ and update $R_1$.
\State \quad \quad\textbf{else}
\State \quad \quad \quad Add only the elements in $v$ which when added to $R_1$ result in a minimum of $bwc$ - $count + 1$ connections at all sensors and update $R_1$.
\State \quad \quad\textbf{end if}
\State \quad \quad Remove the current DCSSMG and update $cost-sortup$.
\State \quad \quad $count$:=$count$-$1$.
\State \quad \textbf{end while}
\State \quad Add the connection in $R_1$ to $R$.
\State \quad Remove the configuration from $Cost$.
\State \textbf{end for}
\end{algorithmic}

\begin{figure}[H]
\centering
\begin{subfigure}{0.48\textwidth}
\includegraphics[width=\linewidth, height=5cm]{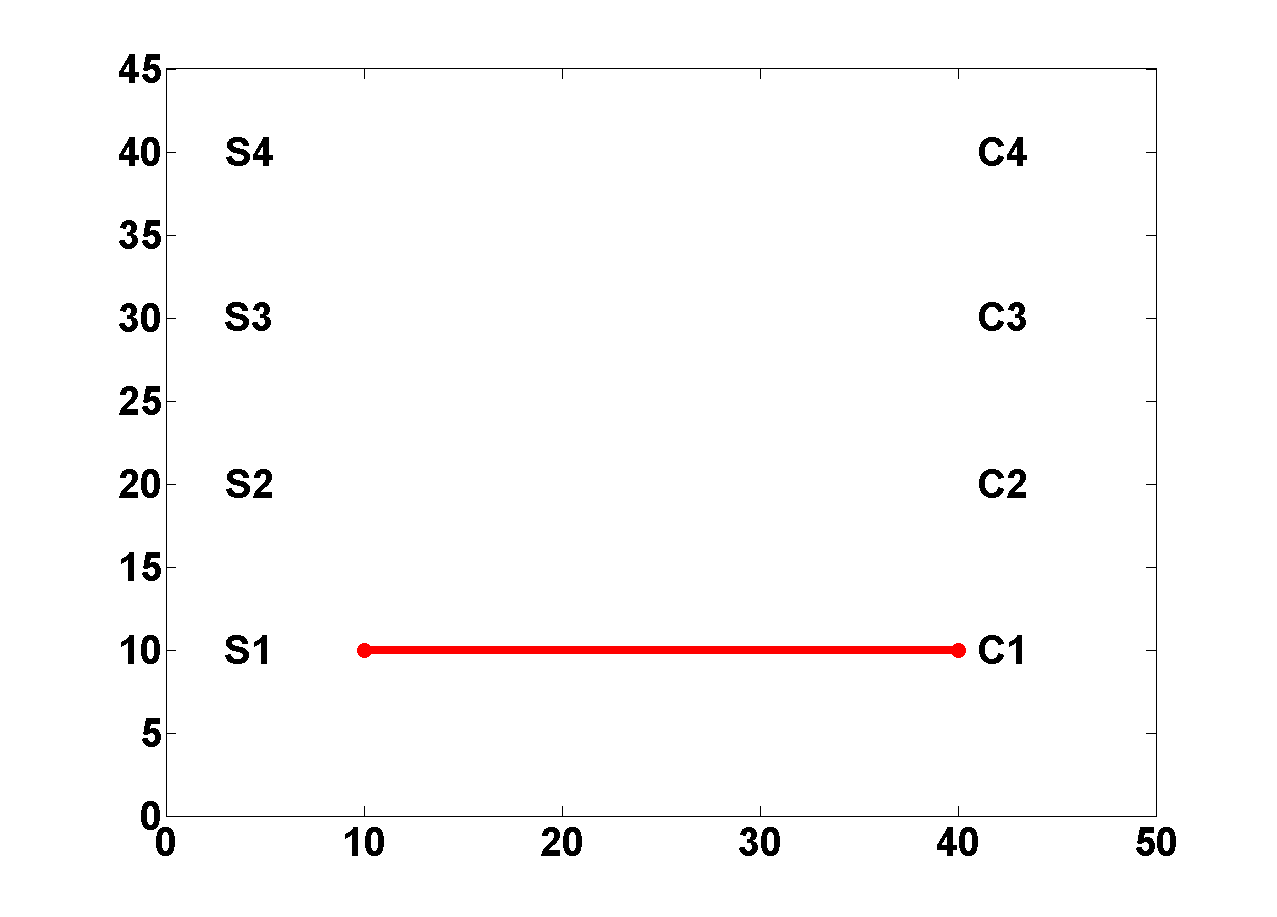}
\caption{DG-1 connection-1}
\label{fig:1_1}
\end{subfigure}
 \begin{subfigure}{0.48\textwidth}
\includegraphics[width=\linewidth, height=5cm]{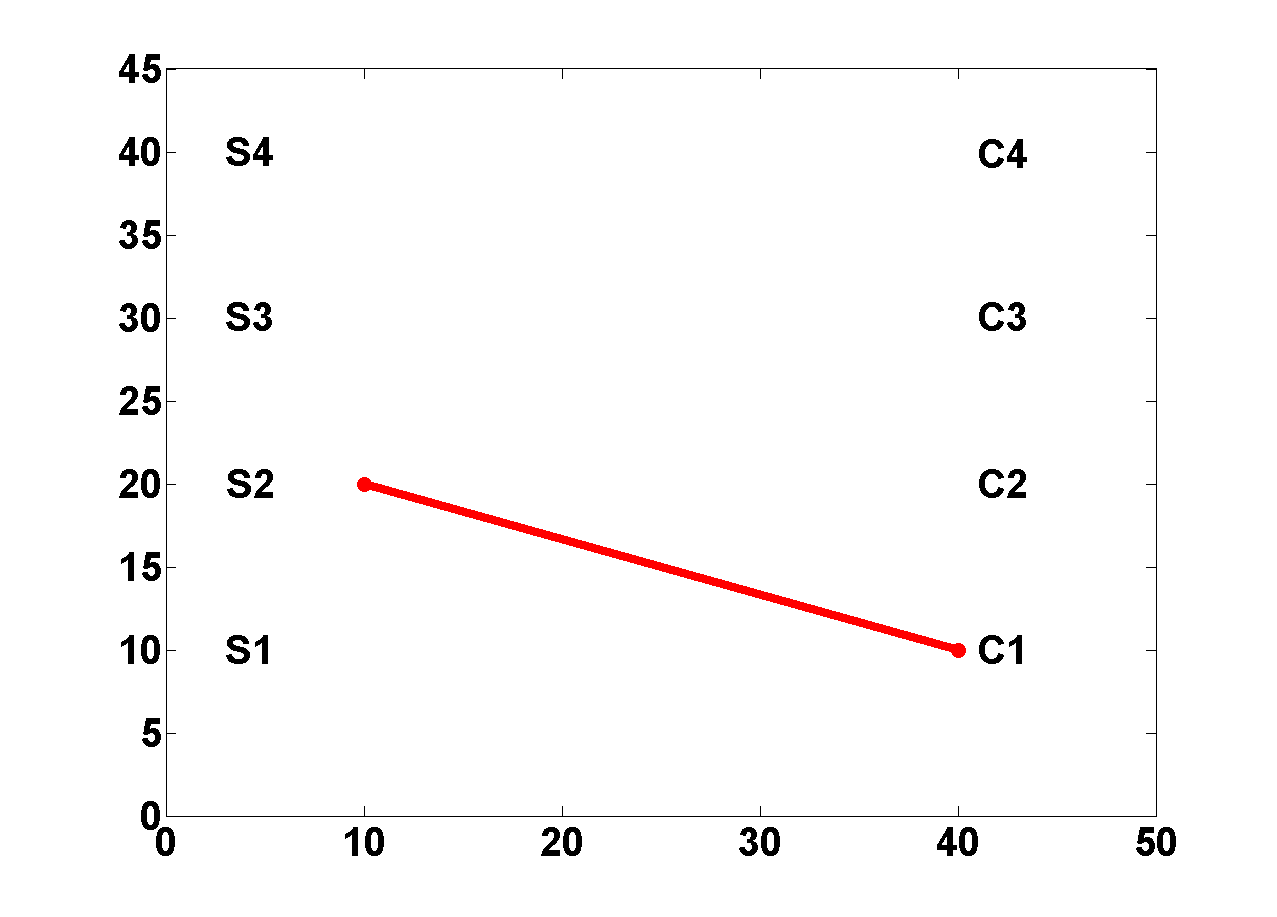}
\caption{DG-1 connection-2}
\label{fig:1_2}
\end{subfigure}\\
\begin{subfigure}{0.48\textwidth}
\includegraphics[width=\linewidth, height=5cm]{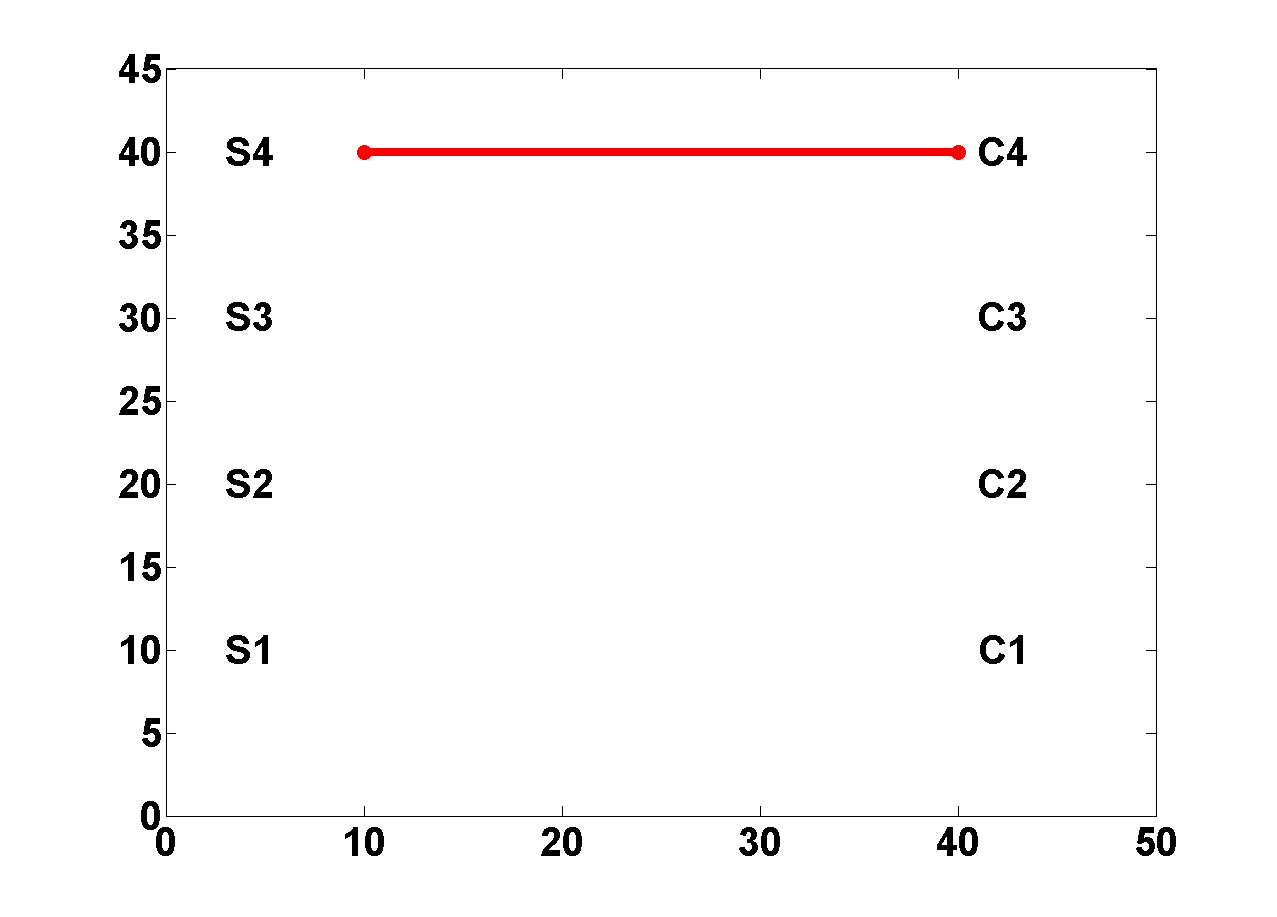}
\caption{DG-4 connection-1}
\label{fig:2_1}
\end{subfigure}
 \begin{subfigure}{0.48\textwidth}
\includegraphics[width=\linewidth, height=5cm]{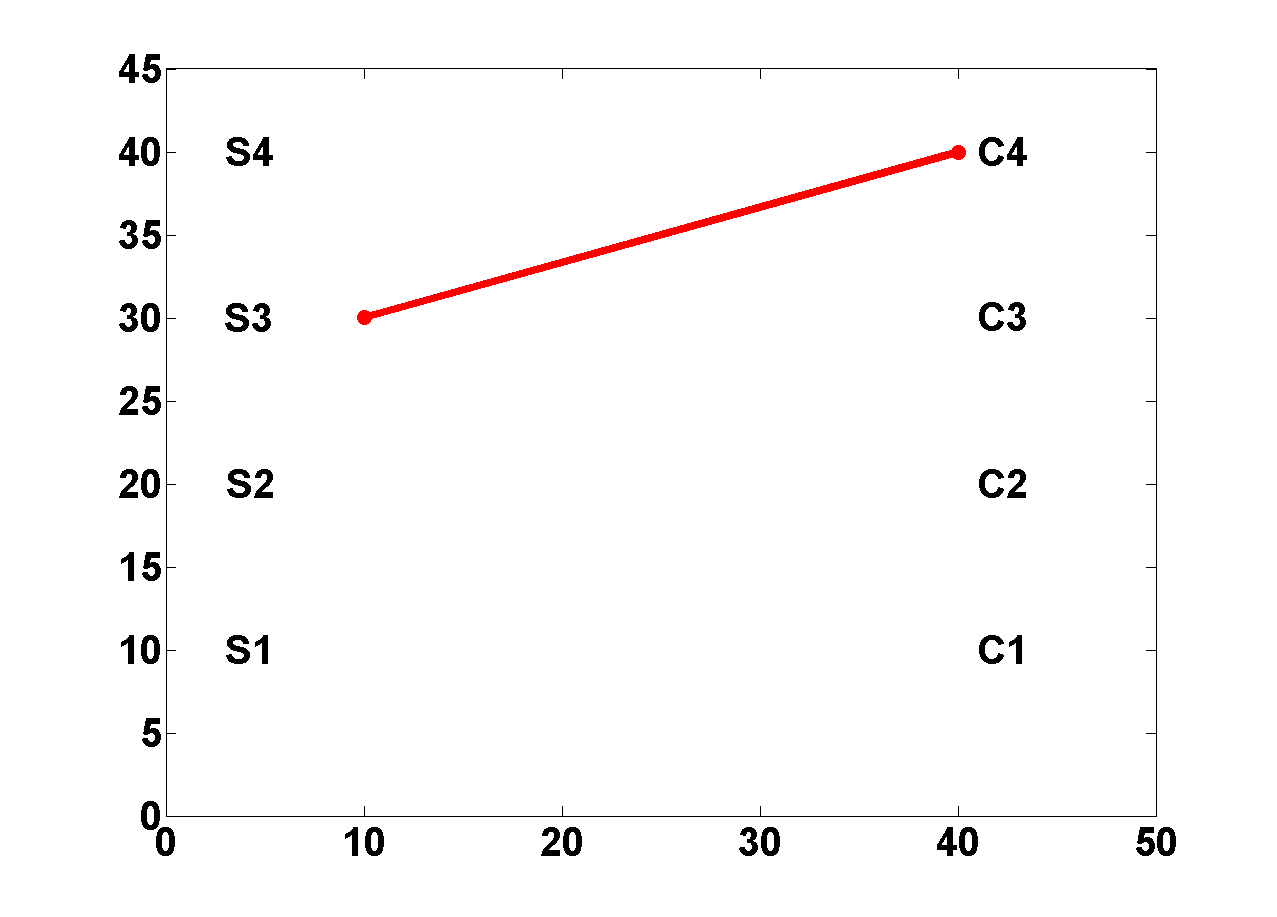}
\caption{DG-4 connection-2}
\label{fig:2_2}
\end{subfigure}
\caption{Connections available from DGs 1 and 4}
\label{fig:connection1}
\end{figure}

\begin{figure}[H]
\centering
\begin{subfigure}{0.48\textwidth}
\includegraphics[width=\linewidth, height=5cm]{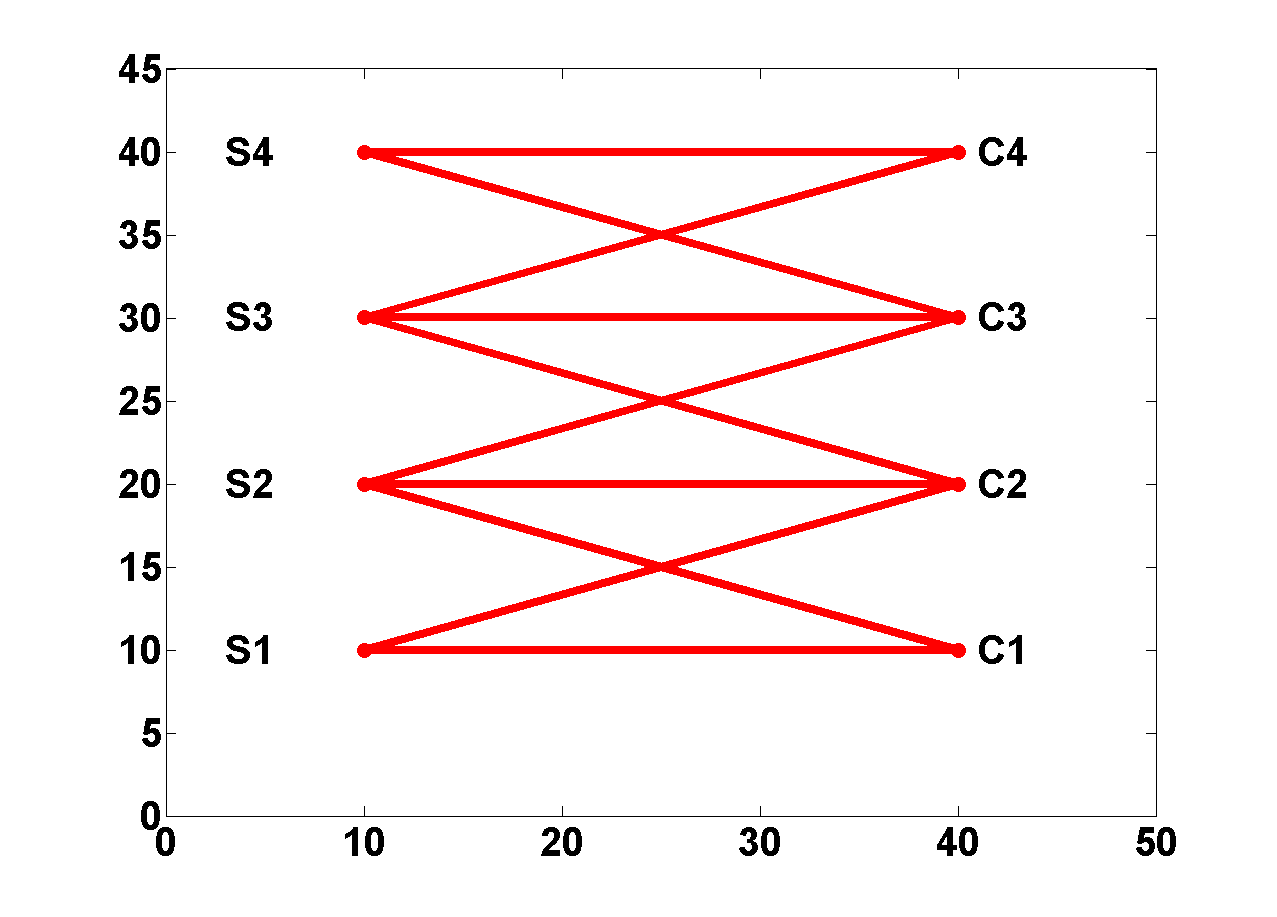}
\caption{Set of all possible network connections}
\label{fig:commconn1}
\end{subfigure}
 \begin{subfigure}{0.48\textwidth}
\includegraphics[width=\linewidth, height=5cm]{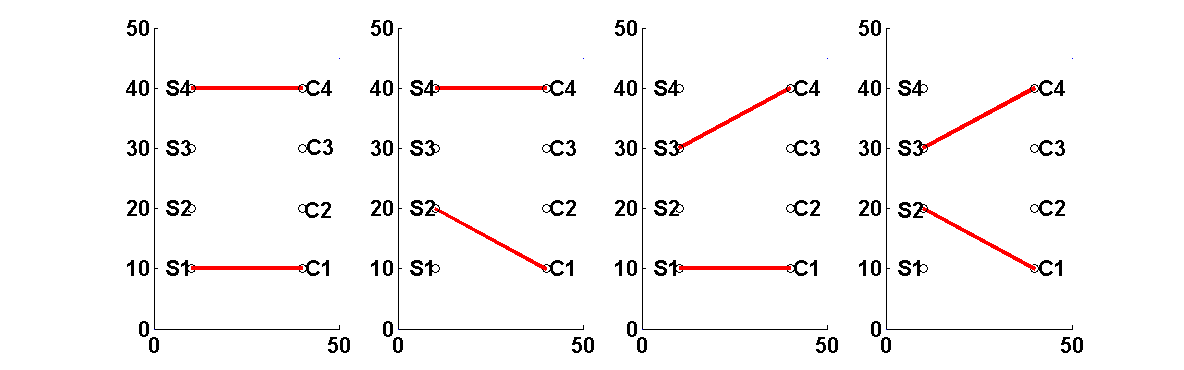}
\caption{Set $R$ after Step-12}
\label{fig:3merge}
\end{subfigure}\\
\begin{subfigure}{0.48\textwidth}
\includegraphics[width=\linewidth, height=5cm]{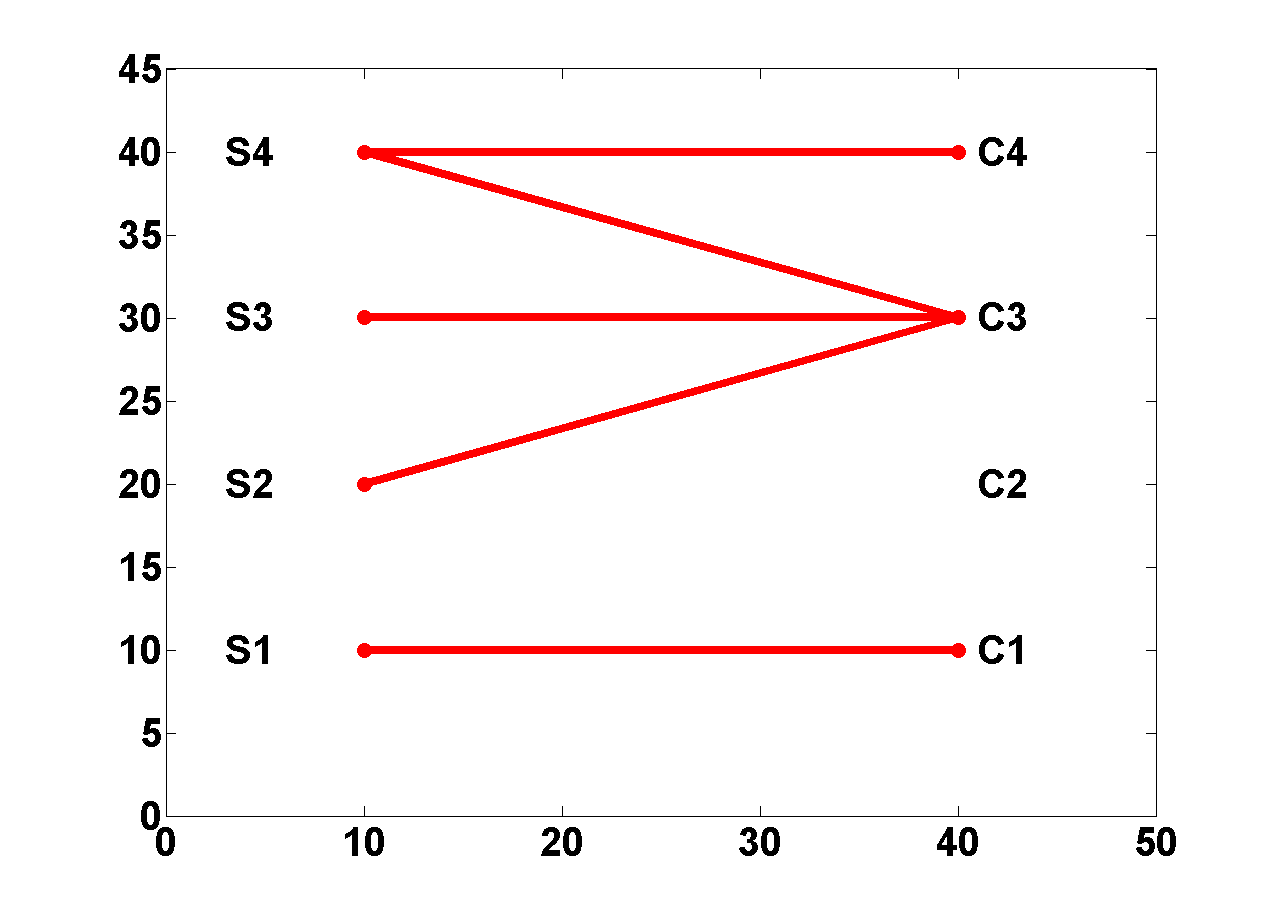}
\caption{Set $R$ after Step-15}
\label{fig:step15}
\end{subfigure}
 \begin{subfigure}{0.48\textwidth}
\includegraphics[width=\linewidth, height=5cm]{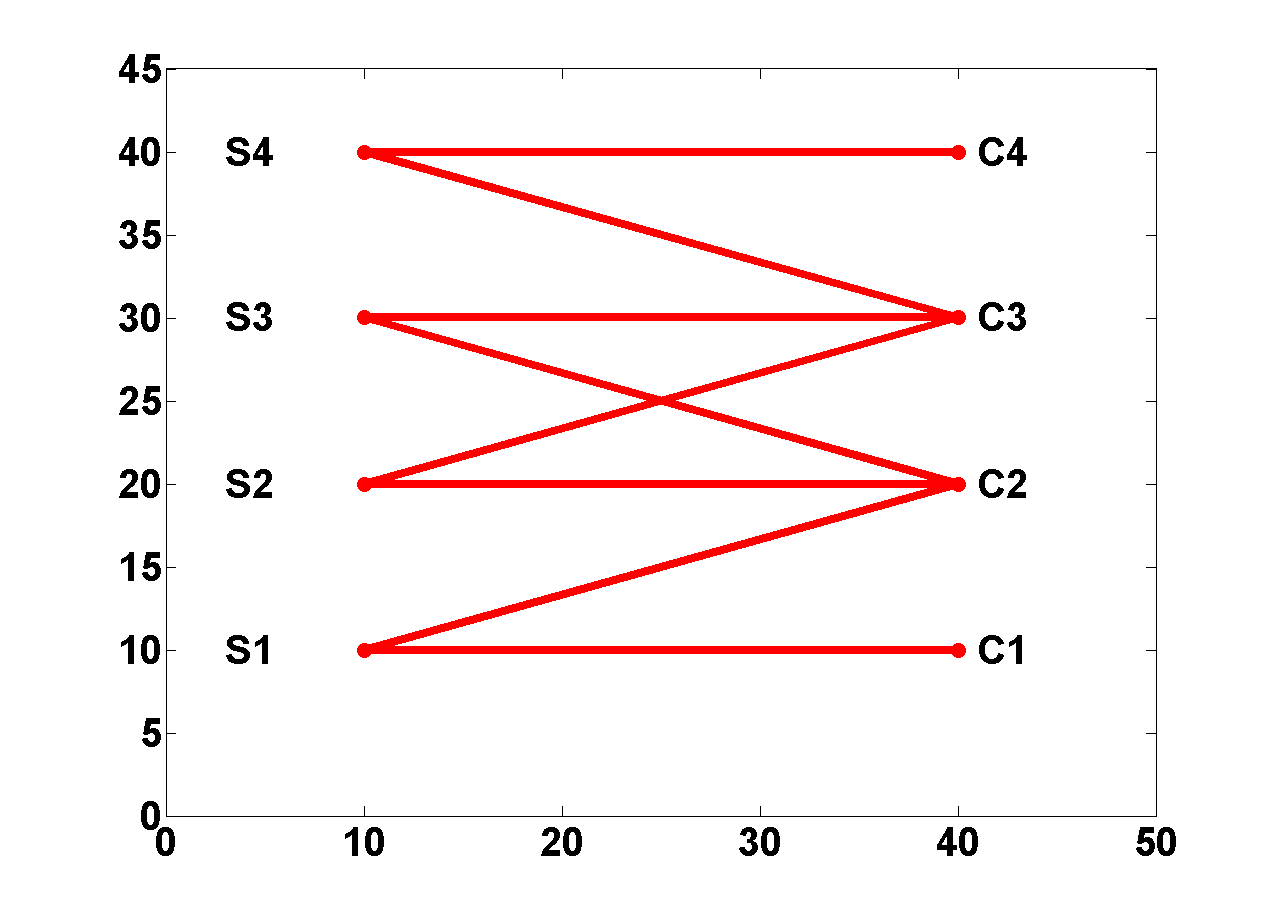}
\caption{Set $R$ after Step-16}
\label{fig:step16}
\end{subfigure}
\caption{Outputs after different steps in the algorithm}
\label{fig:connection2}
\end{figure}

The different variables used in the algorithm have been listed in Table \ref{tab:table20}.
It can be seen that assigning $ns=4$, $nc=4$, $bwc=2$ and $cc=1$, the possible connections for this configuration are shown in Fig. \ref{fig:commconn1}. Further selecting $prc=1$ and $cnc=2$ results in $cost=1331$, $cost-sortup={1 1 3 3}$ and $count=4$. Figures \ref{fig:connection1} and \ref{fig:connection2} show various stages of development of communication connections described in Step-8 to Step-16 of the algorithm. 

\begin{table}[!ht]
\begin{center}
\begin{tabular}{@{}|l|l|@{}}\hline 
Variable & Significance \\
\hline 
~~~$ns$ & number of sensors \\
~~~$nc$ & number of controllers \\
~~~$bwc$ & bandwidth constraint  \\
~~~$cc$ & connection constraint  \\
~~~$cnc$ & central cost constraint  \\
~~~$prc$ & peripheral cost constraint  \\
~~~$Cost$ & set of all possible DCSSMG costs  \\
~~~$cost$-$sortup$ & DCSSMGs sorted in ascending order of costs \\
~~~$v$ & internal variable  \\
~~~$R_1$ & internal variable  \\
~~~$count$ & internal variable  \\
~~~$R$ & set of final connections  \\
\hline
\end{tabular}
\end{center}
\label{tab:table20}
\end{table}
The set $R$ contains all the final connection sets which satisfy all the constraints provided.  		

\subsection{Controller Design Procedure}
This controller manipulates the communication constraints to handle all sorts of variation in parameters. The following procedure utilizes the connection finding algorithm and Lyapunov based optimization formulation to arrive at the value of connection parameter that would suit a particular range of parameter variation we wish to address at a particular point of time.
\begin{enumerate}
\item Make an eigen value chart with maximum eigen value on the y axis and the cyber/physical parameter on the x axis.
\item Dismember the chart into uniform smaller zones on the basis of maximum eigen value.
\item Find the least stable working system matrix for each of the $n$ individual time zones $\mathbf{A}_1,\mathbf{A}_2,\cdots,\mathbf{A}_n$.  
\item Choose a working system matrix close to the current operating parameters.
\item List all the possible constraint combinations and represent them using the variable $connection-parameter$.The connection parameter lists the set of constraints on a specific system. In short, $connection\hspace{0.1cm} parameter\hspace{0.1cm} = \hspace{0.1cm} [bwc\hspace{0.1cm} \hspace{0.1cm}  cc \hspace{0.1cm} cnc\hspace{0.1cm}  prc]$.   
\item Find the possible sensor controller combinations for all the connection parameter combinations as prescribed by the connection finding algorithm described in the previous subsection. 
\item All the different communication connections listed in  Step-5 should be explored and tested for the maximum eigen and $\gamma$ values.
\item Define the $\gamma$ tolerance value $\epsilon$ for the chosen system matrix.
\item Find the minimum set of constraints that can satisfy the $\gamma$ tolerance condition.
\begin{itemize}
\item List the maximum $\gamma$ for all bandwidths and find the highest one among all values.
\item Find the lowest $bwc$ whose maximum $\gamma$ lies within the range of tolerance value $\epsilon$.
\item Within this bandwidth, search for the lowest connection constraint value $cc$ that contains its maximum $\gamma$ within the  $\epsilon$ range.
\item Within the selected $bwc$ and $cc$, list all the connection parameters within the $\epsilon$ range and select the connection parameter that has the least sum of its digits.
\item Select the connection topology with the lowest maximum eigenvalue in the selected connection parameter.
\end{itemize}
\item List the controller $\mathbf{K}$ matrix for the obtained connection. 
\end{enumerate}

\section{Application to Voltage Control in ACSSMG}\label{Sec:descon}
This section describes the application of the general system model and control design previously developed for the purpose of controlling the bus voltages in the ACSSMG. 
\subsection{Description of the ACSSMG System}

The $n$-bus system can be represented as shown in Fig. \ref{Fig:physical grid1}. 
\begin{figure}[!ht]
\centering
\includegraphics[width=0.8\textwidth,  keepaspectratio]{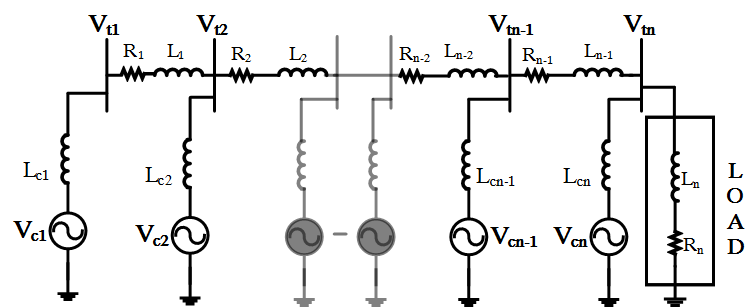}
\caption{Single line diagram of the n-bus system connected to multiple DGs.}
\label{Fig:physical grid1}
\end{figure}
The bus voltages $\mathbf{V}_t= [V_{t_1},...,V_{t_n}]'$ are controlled with the help of DCSSMGs which are modeled as voltage sources $\mathbf{V}_c= [V_{c_1},...,V_{c_n}]'$. The DCSSMG is connected to the power system with the help of an inductor. The control response of DCSSMGs is much faster than conventional sources due to low inertia and fast acting power electronic controllers. 

Upon applying Kirchoff's current law at each bus in Laplace domain and converting them to time domain it is possible to arrive at the state space representation of the system represented in \eqref{Eq:sys}.
Applying nodal analysis at bus-1:
\begin{align}
\frac{V_{t_1}-V_{c_1}}{sL_{c_1}}+\frac{V_{t_1}-V_{t_2}}{R_1+sL_1}&=0 \label{Eq:SysDer10}\\
\Big(\frac{L_1}{L_{c_1}}+1\Big) sV_{t_1}- sV_{t_2}+& \frac{R_1}{L_{c_1}}V_{t_1} -\frac{R_1}{L_{c_1}}V_{c_1} - \frac{L_1}{L_{c_1}} sV_{c_1} =0  
\end{align}
Upon converting this to time domain:
\begin{align}
\Big(\frac{L_1}{L_{c_1}}+1\Big) \dot V_{t_1}-\dot V_{t_2}+ \frac{R_1}{L_{c_1}}V_{t_1}  -\frac{R_1}{L_{c_1}}V_{c_1} - \frac{L_1}{L_{c_1}} \dot V_{c_1} =0 \label{Eq:SysDer1} 
\end{align}
In a similar way, the nodal analysis is carried out at all the four buses and the final time domain equations can be written as follows:
\begin{equation}\label{Eq:sys2}
\Delta \dot{\mathbf{V}}_t{(t)}= \mathbf{A'} \Delta \mathbf{V}_t(t) + \mathbf{B'} \Delta \mathbf{V}_c(t)+ \mathbf{M'} \Delta \dot{\mathbf{V}_c}(t)
\end{equation}
where $\Delta \mathbf{V}_t(t)=\mathbf{V}_t-\mathbf{V}_{ref}$ and $\Delta \mathbf{V}_c$ represents the change in DG voltages required for bus voltages to approach $\mathbf{V}_{ref}$. For the purpose of voltage control, $\Delta \mathbf{V}_c$ is the control input $u$ and the following state is chosen, 
\begin{equation} 
\mathbf{x}(t)=\Delta{\mathbf{V}_t(t)}-\mathbf{M'} \Delta{\mathbf{V}_c(t)}   
\end{equation}  
and applied onto \eqref{Eq:sys2} to arrive at the state space representation of the system represented in \eqref{Eq:sys}. Since the control input also holds a linear relation with the state as per \eqref{Eq:controller}, the bus voltages also approach their reference values if the state vector goes to zero.  Thus every local controller obtains the information of all the bus voltages to generate a voltage magnitude command that is to be maintained by the DG so as to maintain the bus voltage magnitudes at $\mathbf{V}_{ref}$.   

\subsection{Test System Configuration}
A sample 4-bus system is used for demonstrating voltage control whose per unit description has been given in Table-\ref{tab:table2}. The value of $R_4$ and $L_4$ depend on the amount of load resistance and reactance at a particular point of time. For all the cases $L_4$ has been taken to be $0.0148$. Also, the value of $L_{c_n}=0.001$ where $n=1,2,3,4$.

\begin{table}[H]
    \centering
    \begin{tabular}{|c c|c c |}
    \hline
Parameter & Value & Parameter & Value \\
\hline
   ~~~~~ $R_1$ &  0.175  & ~~~~~$L_1$ & 0.0005  \\
   ~~~~~ $R_2$ &  0.1667 & ~~~~~$L_2$ & 0.0004  \\
   ~~~~~ $R_3$ &  0.2187 & ~~~~~$L_3$ & 0.0006  \\  
\hline
    \end{tabular}
    \caption{Grid Parameters}
    \label{tab:table2}
\end{table}
The different system matrices used for different loading conditions $R_4= 0.001$ and $0.35$ are given as follows: 

\hspace{-0.65cm}
\begin{minipage}{0.5\textwidth}
\centering
\begin{align*}
\mathbf{A}_1 =
\begin{bmatrix}
 -150.43 & -63.00 & -21.47  & -0.000 \\
  -50.65  &-94.50 & -32.21 &  -0.000 \\
   35.60 &   9.19 & -53.69 &  -0.001 \\
  155.01 & 138.91 & 100.57 &  -0.003 \\
\end{bmatrix}
\end{align*}
\end{minipage}
\begin{minipage}{0.5\textwidth}
\centering
\begin{align}\label{Eq:A2}
\mathbf{A}_2=
\begin{bmatrix}
  -152.61 & -65.17 & -23.65 &  -2.17 \\
  -53.91 & -97.76 & -35.47  & -3.26 \\
   30.16 &   3.76 & -59.12 &  -5.43 \\
  143.04 & 126.95 &  88.61 & -11.96 \\
\end{bmatrix}
\end{align}
\end{minipage}
\begin{minipage}{0.5\textwidth}
\centering
\begin{align*}
\mathbf{B}_1=
\begin{bmatrix}
   56.13 &  -3.22 & -23.58 & -27.468 \\
   -3.22 &  39.00 &  -8.97 & -25.105 \\
  -23.58 &  -8.97 &  40.89 &  -7.801 \\
  -27.46 & -25.10 &  -7.80 &  56.555 \\
\end{bmatrix} 
\end{align*}
\end{minipage}
\begin{minipage}{0.5\textwidth}
\centering
\begin{align}\label{Eq:B2}
\mathbf{B}_2=
\begin{bmatrix}
   56.15 &  -3.20 & -23.55 & -27.394 \\
   -3.20 &  39.03 &  -8.92 & -24.994 \\
  -23.55 &  -8.92 &  40.97 &  -7.616 \\
  -27.39 &  -24.99&   -7.61&   56.964\\
\end{bmatrix} \hspace{1cm}
\end{align}
\end{minipage}

$\mathbf{C}$ has been assumed to be an identity matrix.

\subsection{Controller Design Example} 
The following example has been provided to elucidate the process of controller design for the case of varying load resistance. 

\begin{enumerate}
\item Plot the eigenvalues of the system matrix for the known range of output resistance variation.
\item Segregate the range of eigenvalues into individual smaller zones  as in Fig. \ref{Fig:maxeigvsr41}.
\item Find the worst-case working system matrix for all $n$ individual time zones $\mathbf{A}_1,\mathbf{A}_2,\cdots,\mathbf{A}_n$. Details available for two zones in the previous subsection.
\item Select a particular zone of operation depending on the time of the day. Let it be $\mathbf{A}_1$. 
\item List all the possible constraint combinations and represent them using the variable $connection-parameter$.The connection parameter (CP) of a connection gives an idea of various constraints imposed on the system. In short, $connection~ parameter~ = \hspace{0.1cm} [bwc\hspace{0.1cm} \hspace{0.1cm}  cc \hspace{0.1cm} cnc\hspace{0.1cm}  prc]$.   
\item Find the possible sensor controller combinations for all the connection parameter combinations as prescribed by the algorithm in Section \ref{Sec: GSCD}.
\item Find the $\gamma$ values and the maximum eigenvalues for all the combinations found in step-5 using the optimization formulation mentioned in Section \ref{Sec:LF}. Table-\ref{tab:table1} reflects the output of this step.

\begin{table}[!ht]
    \centering
    \begin{tabular}{|c c c|c c c|}
    \hline
C P & Max eigenvalue & $\gamma$ & C P & Max eigenvalue & $\gamma$ \\
\hline
    3202 & ~~~~~~~-2.15 & 2.04 & 3332 & ~~~~~~~-3.64 & 3.574 \\
    3212 & ~~~~~~~-2.15 & 2.04 & 3342 & ~~~~~~~-3.64 & 3.574\\    
    3222 & ~~~~~~~-2.15 & 2.04 & 3303 & ~~~~~~~-4.44 & 4.302 \\
    3232 & ~~~~~~~-2.15 & 2.04 & 3313 & ~~~~~~~-4.44 & 4.302\\
    3242 & ~~~~~~~-2.15 & 2.04 & 3323 & ~~~~~~~-4.44 & 4.302\\
    3203 & ~~~~~~~-2.838& 2.7  & 3333 & ~~~~~~~-4.44 & 4.302\\
    3213 & ~~~~~~~-2.838& 2.7  & 3343 & ~~~~~~~-4.2  & 4.01\\
    3223 & ~~~~~~~-2.838& 2.7  & 3304 & ~~~~~~~-4.44 & 4.302\\
    3233 & ~~~~~~~-2.838& 2.7  & 3314 & ~~~~~~~-4.44 & 4.302\\
    3243 & ~~~~~~~-2.15 & 2.04 & 3324 & ~~~~~~~-4.44 & 4.302\\
    3204 & ~~~~~~~-2.838& 2.7  & 3334 & ~~~~~~~-4.44 & 4.302\\
    3214 & ~~~~~~~-2.838& 2.7  & 3344 & ~~~~~~~-4.2  & 4.01\\
    3224 & ~~~~~~~-2.838& 2.7  & 4301 & ~~~~~~~-4.381& 4.304  \\
    3234 & ~~~~~~~-2.838& 2.7  & 4314 & ~~~~~~~-4.381& 4.304  \\
    3244 & ~~~~~~~-2.15 & 2.04 & 4324 & ~~~~~~~-4.381& 4.304  \\
    3302 & ~~~~~~~-3.64 & 3.574& 4334 & ~~~~~~~-4.381& 4.304 \\
    3312 & ~~~~~~~-3.64 & 3.574& 4344 & ~~~~~~~-4.381& 4.304 \\
    3322 & ~~~~~~~-3.64 & 3.574&      &       & \\    
\hline
    \end{tabular}
    \caption{Step-7 result}
    \label{tab:table1}
\end{table}
\item Define the $\gamma$ tolerance for the chosen zone. Since $\mathbf{A}_1$ is near the verge of instability, the $\epsilon_1$ will be chosen to be low like $0.1$.
\item Find the least value of constraints that can satisfy the tolerances. This process has been explained in Fig. \ref{Fig:ccc}. 
Find the maximum $\gamma$ in a particular bandwidth. Here only $bwc=3$ and $bwc=4$ connections exist with respective maximum $\gamma$ values $4.302$ and $4.304$. The lower $bwc=3$ gets selected. Further $cc=3$ is chosen which supports this maximum $\gamma$ and within the purview of these two constraints the constraint configuration 3303 is chosen since the sum of its individual digits is least. 

\begin{figure}[!ht]
\centering
\includegraphics[width=\linewidth, height=8cm]{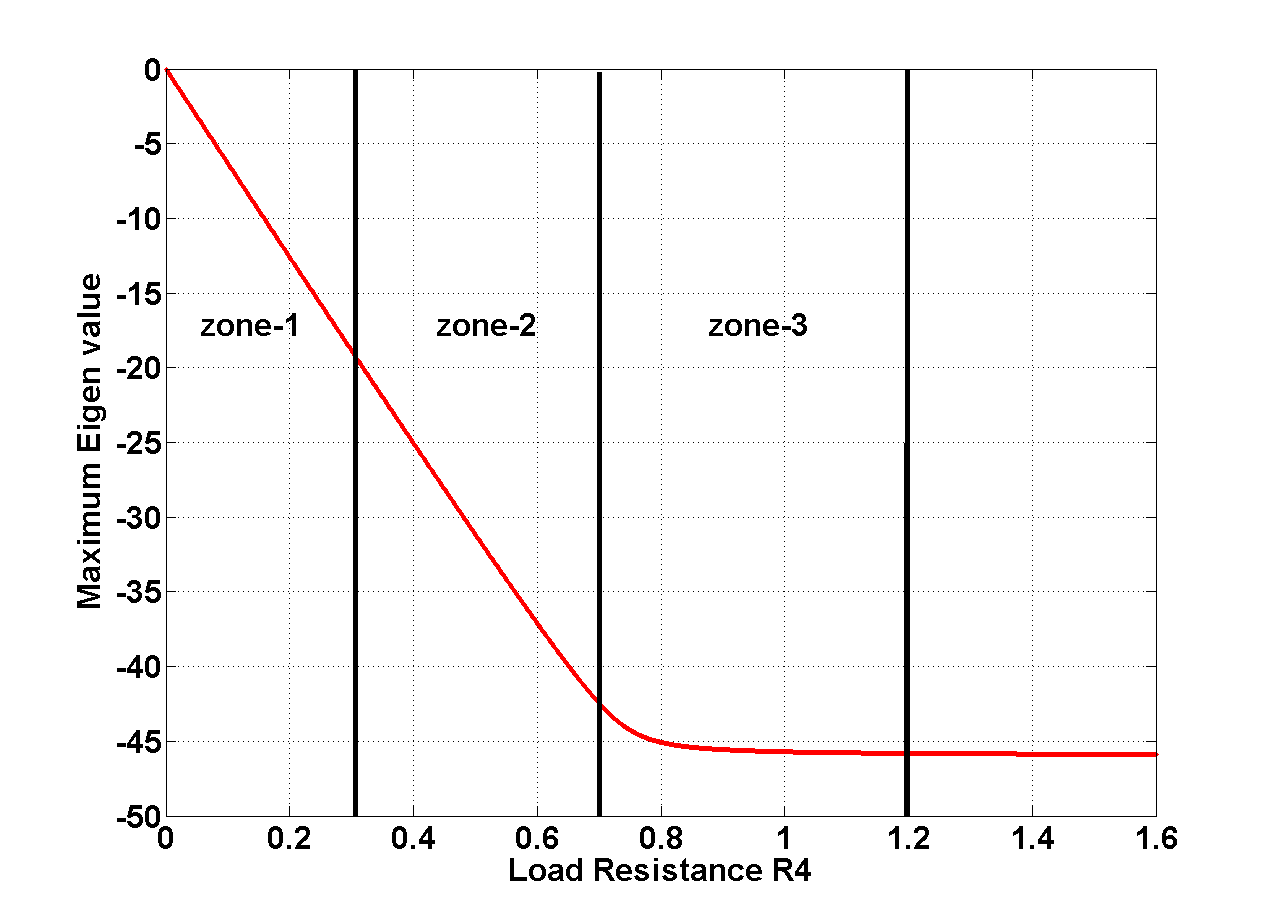}
\caption{Segregated zones}
\label{Fig:maxeigvsr41}
\end{figure}
\begin{figure}[!ht]
\centering
\includegraphics[width=\linewidth, height=7cm]{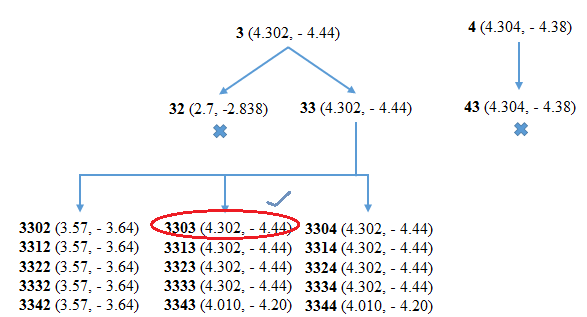}
\caption{Step-8 result}
\label{Fig:ccc}
\end{figure}

\item Find the $\mathbf{K}$ matrix for this connection.
\begin{equation*}
\mathbf{K} =
\begin{bmatrix}
0 &   -2.1345 &   -5.3123 &  0.7135 \\
-1.8397 &  -1.3090 &   -4.1146 &  0 \\
2.0861  &  0  & -3.2176  &  -9.001 \\ 
-7.5319 &   -5.5769  &   0  & -1.0024 \\
\end{bmatrix} \hspace{1cm}
\end{equation*} 
\end{enumerate}

\section{Results} \label{Sec:results}
 The performance of the controllers designed using CBSCD methodology for voltage control in a 4-bus ACSSMG are demonstrated in this section for various cases ranging from load variation to delay variation and communication failure. The controller design has been numerically executed using the YALMIP toolbox \cite{1393890}. These controllers have been designed after finding the least value of constraints for stabilizing the ACSSMG for a particular set of physical and communication setup . The maximum value of these communication constraints is generally specified by the operator based on experience and the algorithm finds the least value within these bounds that can provide maximal stability.

\subsection{Load Variation}
\begin{figure}[H]
\centering
\includegraphics[width=0.9\linewidth, height=5cm]{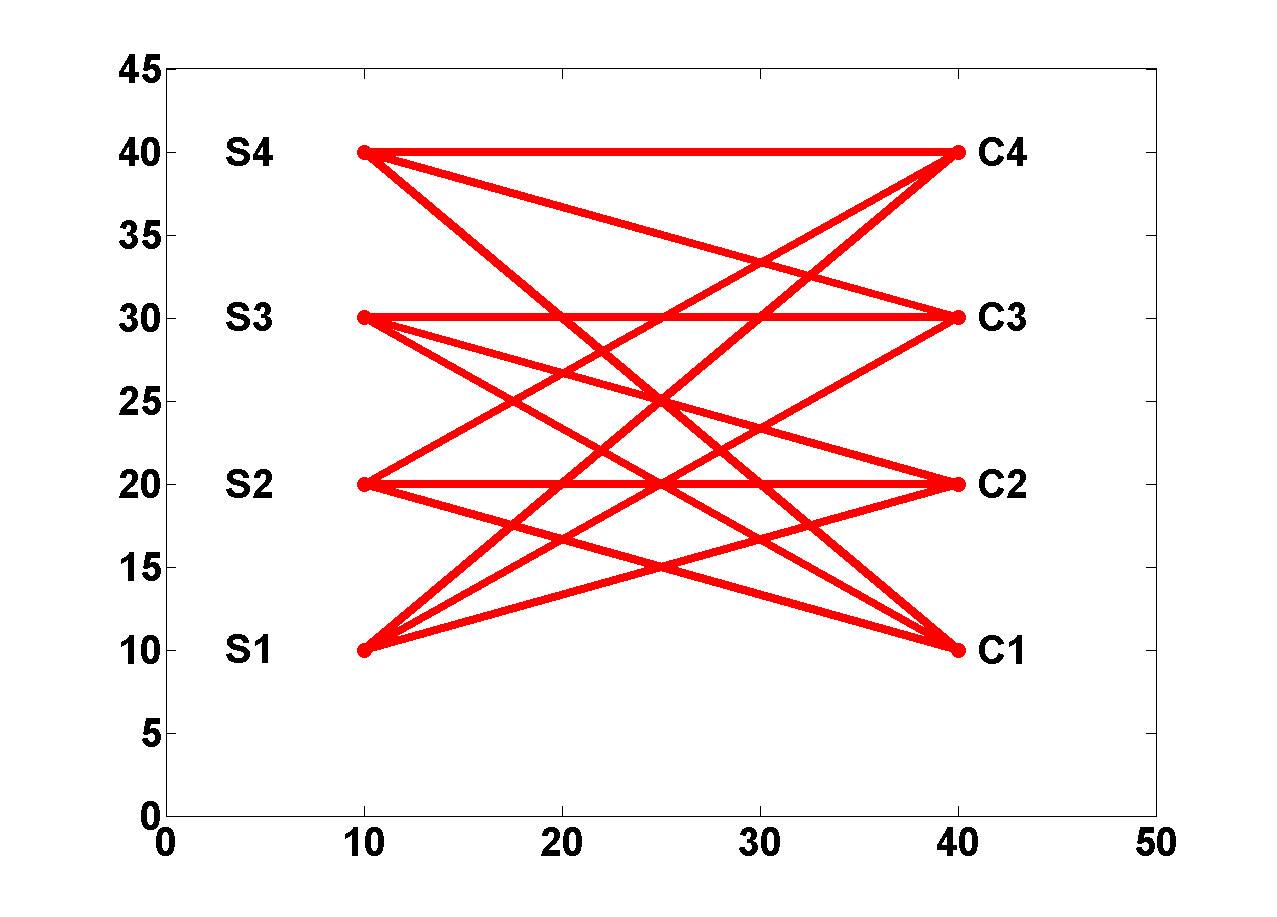}
\caption{Topology of a controller from controller set-1.}
\label{Fig:A11sw}
\end{figure}

The load on the power system is modeled as a resistance and reactance. As described previously, two zones of operation $\mathbf{A}_1$ and $\mathbf{A}_2$ with different load resistances have been selected and respective controller set-1 has been designed, which can comfortably take care of voltage control within respective zones. The values of different $\mathbf{K}$ matrices are stored at the individual DGs and get updated according to the zone of operation. \textcolor{black}{The topology of a $\mathbf{K}$ matrix is shown in Fig. \ref{Fig:A11sw}}. Table-\ref{tab:table3} and Fig. \ref{fig:result1} summarizes the results obtained for two scenarios of loading. 

\begin{table}[H]
    \centering
    \begin{tabular}{|c c c c|}
    \hline
System & Max-eigenvalue(open) & Max-eigenvalue(closed) & CP \\
\hline
~~~$A_1$ & ~~~~~~~~~-0.0063 & ~~~~~~~~~-4.4547  & 3303 \\
~~~$A_2$ & ~~~~~~~~-21.9580 & ~~~~~~~~-26.7771 & 3203  \\
\hline
    \end{tabular}
    \caption{Results-Load Variation}
    \label{tab:table3}
\end{table}

   \begin{figure}[H]
\includegraphics[width=0.9\linewidth, height=5cm]{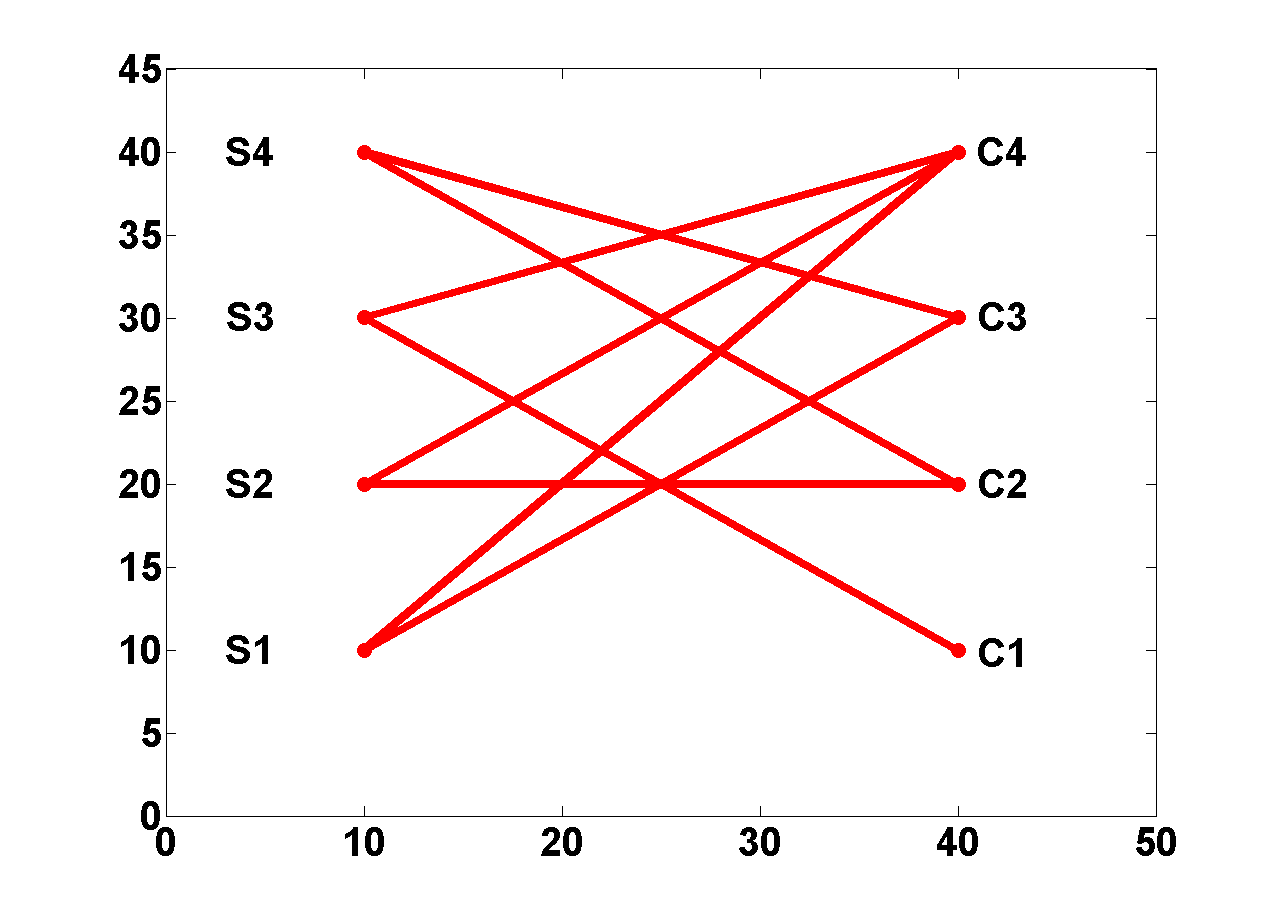}
\caption{Topology of a controller from controller set-2.}
\label{Fig:A21sw}
\end{figure}
\begin{figure}[H]
\includegraphics[width=0.9\linewidth, height=5cm]{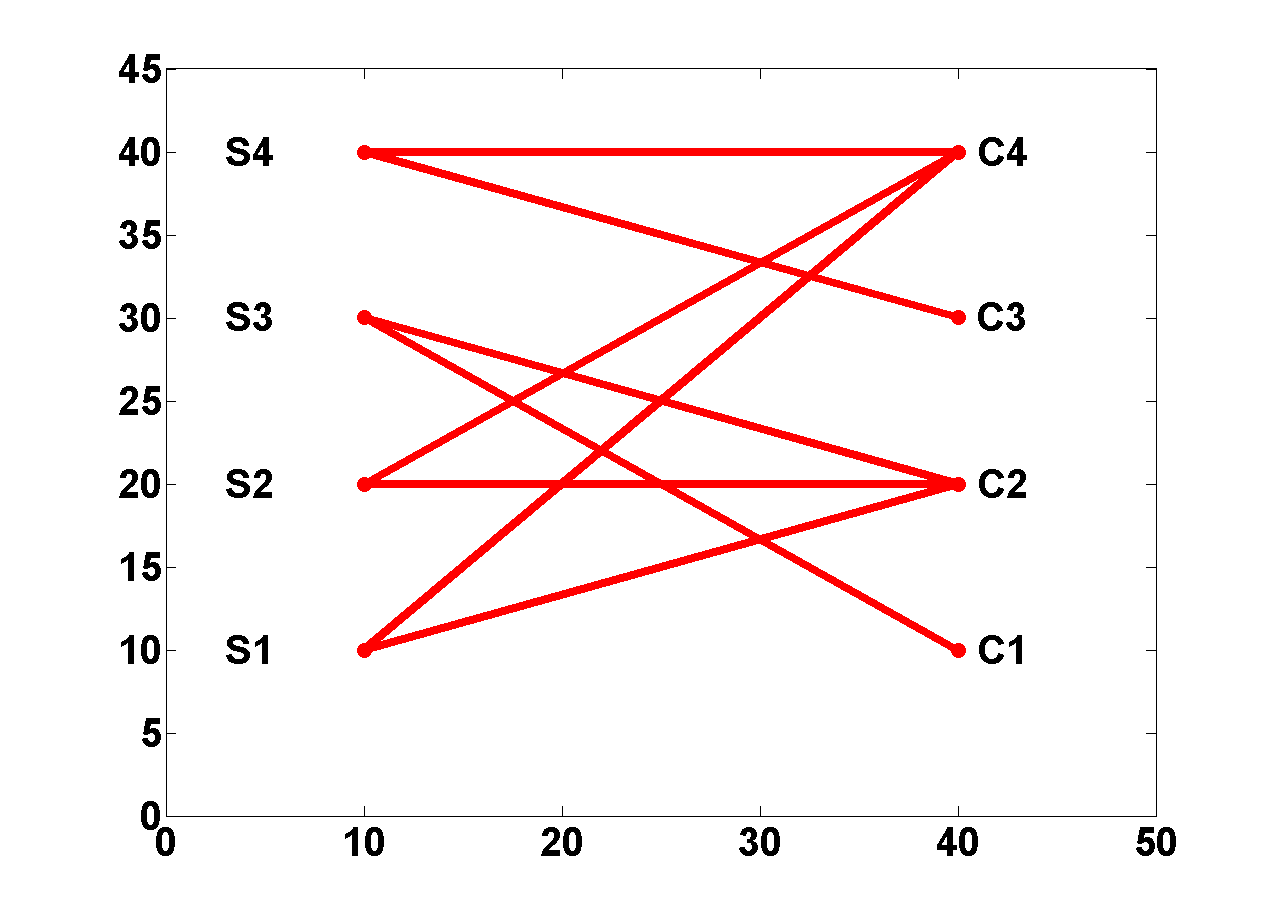}
\caption{Topology of a controller from controller set-3.}
\label{Fig:A31sw}
\end{figure}
The following is a controller from controller set-1 . 
\begin{equation}\label{Eq:resultbA11}
\textcolor{black}{\mathbf{K} =
\begin{bmatrix}
0 &   -2.1344 &   -5.3129 &  0.7135 \\
-1.8397 &  -1.3094 &   -4.1146 &  0 \\
2.0862  &  0  & -3.2174  &  -9.0016 \\ 
-7.5314 &   -5.5765  &   0  & -1.0026 \\
\end{bmatrix}} \hspace{1cm}
\end{equation}

\subsection{Variation of delay}
It is possible to map an approximate variation of the maximum delay value throughout the day based on forecast data. Thus, the controllers can be designed for the maximum delay predicted during different periods of the day. Along with this, the load variation is also an inevitable change. The controller set-2 designed using optimization formulation for delay case has demonstrated that it can stabilize the bus voltages for simultaneous change in load and delay profile.

The results in Table-\ref{table4} and Fig. \ref{fig:result2} summarize the outputs achieved when system moves from a region of no delay with load of $R_4=0.5$ to a region containing maximum delay of $1~ms$ with load of $R_4=0.001$. The $\mathbf{K}$ matrix given in \eqref{Eq:resultbA21} shows a design from controller-set 2 intended to work for operating region $\mathbf{A}_1$ and maximum delay of 1 ms. The topology for the same has been presented in Fig. \ref{Fig:A21sw}.


\begin{table}[H]
    \centering
    \begin{tabular}{|c c c c|}
    \hline
 Delay (ms) & Max-eigenvalue (open) & Max-eigenvalue (closed) & CP \\ \hline
 ~~~~~~1 &  ~~~~~~~~~-21.9580 & ~~~~~~~~~-25.9729 & 2303  \\ 
  ~~~~~   2 & ~~~~~~~~~-21.9580 & ~~~~~~~~~-24.0441 & 2302  \\
 \hline
 \end{tabular}
\caption{Results- Load variation with Delay Variation}
\label{table4}
\end{table}

\begin{equation}\label{Eq:resultbA21}
\textcolor{black}{\mathbf{K} =
\begin{bmatrix}
0	& 0 &	0.5498 &	0  \\
0	&-3.7826 &	0 &	4.0082 \\
-3.3611 &	0 &	0	& 1.2577 \\
-4.1047&	-6.7700 &	-3.3168 &	0 \\
\end{bmatrix}}
\end{equation}

\begin{figure}[H]
\centering
\includegraphics[width=0.9 \linewidth, height=8cm]{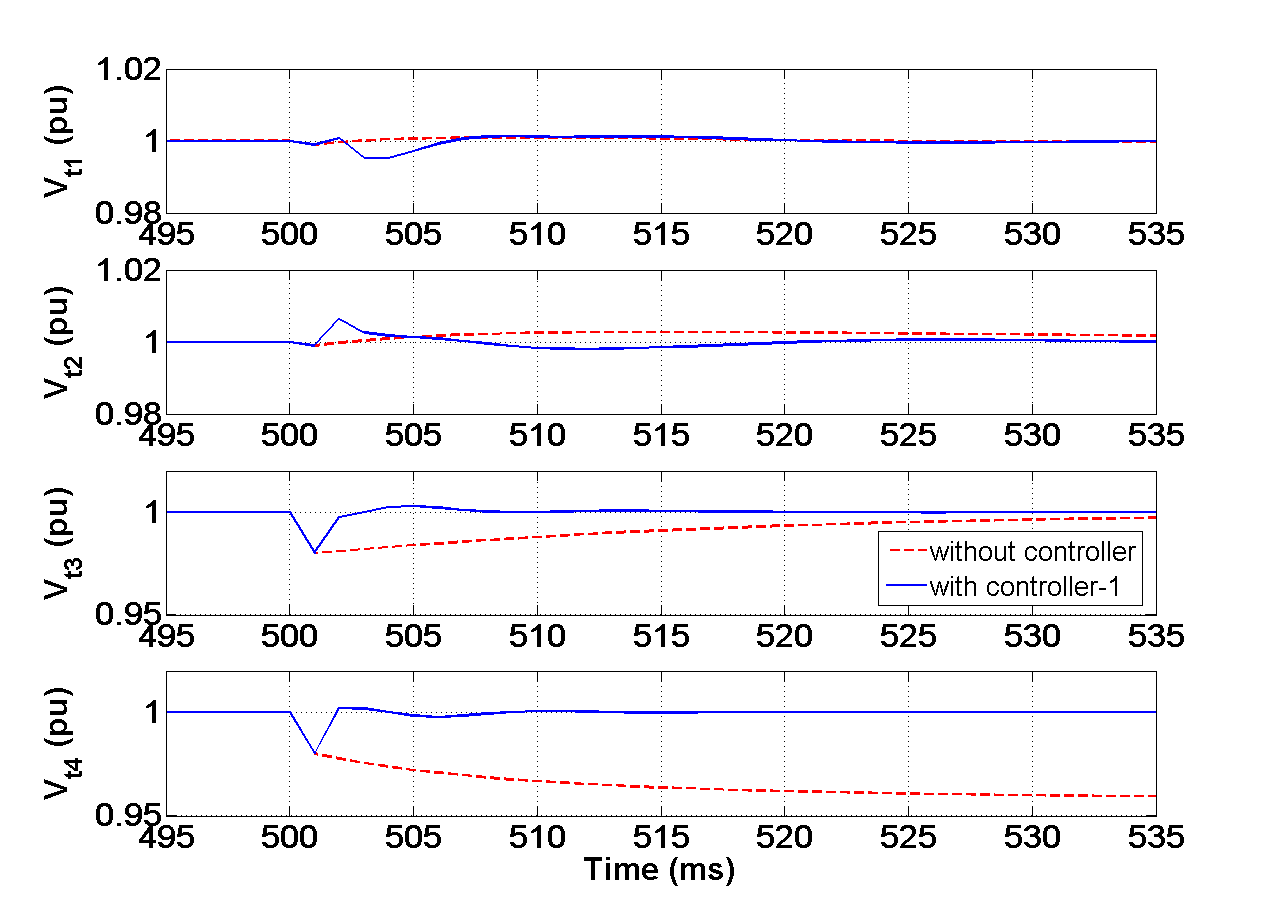}
\caption{Comparison of bus voltages between grid with controller-set 1 and uncompensated grid when load is changed at t=500 ms.}
\label{fig:result1}
\end{figure}

\begin{figure}[H]
\centering
\includegraphics[width=0.9 \linewidth, height=8cm]{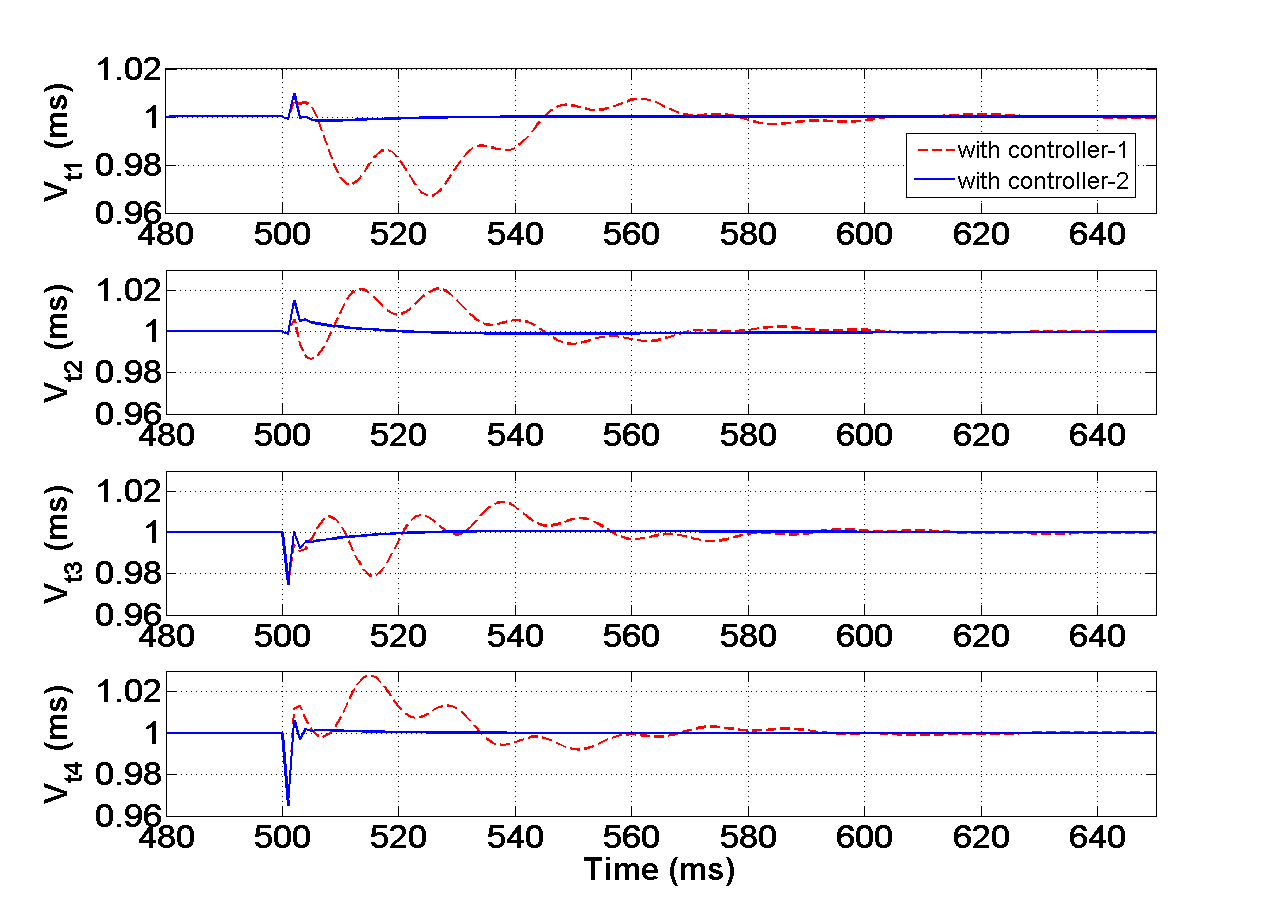}
\caption{Comparison of bus voltages between the situation with controller-set 1 designed only for load variation and the situation with controller-set 2 designed for both load and delay variation when load and delay are changed at t=500 ms.}
\label{fig:result2}
\end{figure}

\begin{figure}[H]
\centering
\includegraphics[width=0.9 \linewidth, height=8cm]{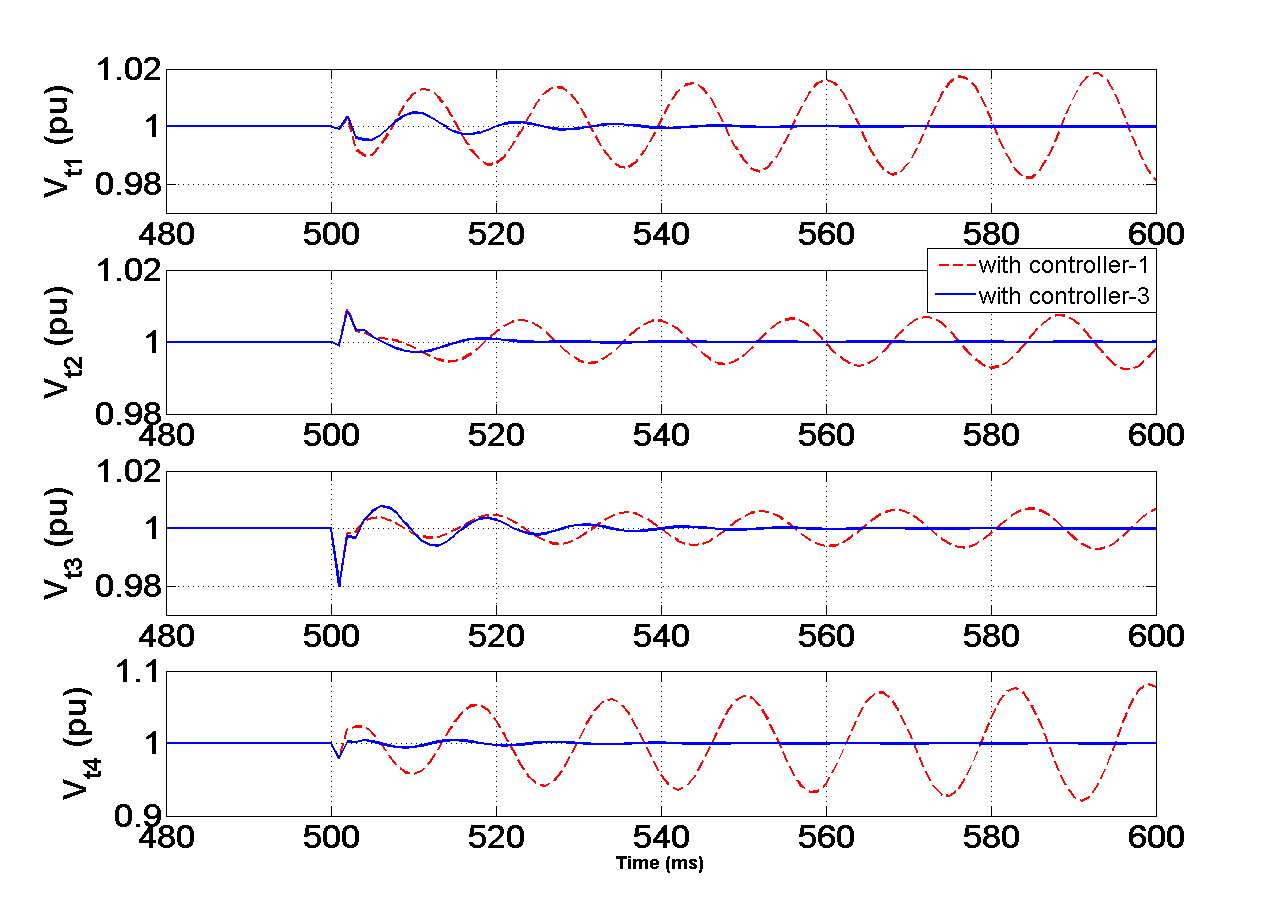}
\caption{Comparison of bus voltages between the grid with controller-set 1 designed only for load variation and grid with controller-set 3 designed for both load variation and communication link loss when link 3-3 is lost and load is changed at t=500 ms.}
\label{fig:result3}
\end{figure}

\subsection{Failure of Communication Link}
The communication signals in the ACSSMG get attenuated due to both reflection from objects along the path of the wave and shadowing, which happens due to wave obstruction. When the signal attenuation exceeds a certain limit, the particular communication node needs to be shutdown for repair. Before shutting down the communication node, the controller $\mathbf{K}$ will be calculated and put into action which ensures smooth transition without compromise in stability between the system in which all the $\mathbf{K}$ values are present and the system in which a particular $\mathbf{K}$ value is absent due to the removed connection. Table-\ref{table5} and Fig. \ref{fig:result3} summarize the results obtained for various scenarios of load when there is a communication loss in the link between controller-3 and sensor-3. The $\mathbf{K}$ matrix in \eqref{Eq:A31} shows a design from controller set-3 for a case where there has been communication failure between sensor-3 and DG-3. Moreover, the system parameters have also been changed due to change in load resistance from $\mathbf{A_2}$ to $\mathbf{A_1}$.
Fig. \ref{Fig:A31sw} shows the output topology obtained in the presence of communication failure when operating the region $\mathbf{A_1}$ with $R_4=0.0001$. The $\mathbf{K}$ matrix further confirms the alternate route chosen by the algorithm in the absence of the link between sensor-3 and controller-3.

\begin{table}[!ht]
\begin{center}
\begin{tabular}{@{}|llll|@{}}\hline 
System & Max-eigenvalue(open) & Max-eigenvalue(closed) & CP \\
\hline
~~~$A_1$ &  ~~~~~~~~~-0.0063 & ~~~~~~~~~-2.4845 &2303  \\
~~~$A_2$ & ~~~~~~~~~-21.9580 & ~~~~~~~~~-25.2524 & 2303  \\
\hline
\end{tabular}
\caption{Results- Load Variation with communication node failure}
 \label{table5}
\end{center}
\end{table}

\begin{equation}\label{Eq:A31}
\textcolor{black}{\mathbf{K} =
\begin{bmatrix}
0 & 0 &-1.6398 & 0 \\
-2.2646 & 0.5854 & -0.3357 & 0 \\
0 & 0 & 0 & -3.9057 \\
-5.3527 &-2.9395 & 0 & -0.5783 \\
\end{bmatrix}}
\end{equation} 
\textcolor{black}{Controller-1 is a controller designed without considering the possibility of communication failure between the local link between sensor-3 and controller-3. Due to this, when the communication link failure occurs due to a cyber attack, oscillations are produced in the PCC voltages leading to system instability. However, controller 3 is designed after applying the communication failure constraint in the optimization formulation. Due to this, the stability of the system is ensured through redundant communication even though the local link at PCC-3 fails. }            

A salient conclusion evident in this simulation is that reasonable stability is achieved with communication designs that require less communication cost as compared to densely connected cases.
\textcolor{black}{The summary of performance comparison of various controllers in different cases is tabulated in Table \ref{table57}. The terms WoC refer to without controller and WC-1,2,3 refer to with controller-1,2,3 and so on. Similarly, the terms Max Var refer to maximum variation/overshoot and Acc Time refer to accommodation time in Table \ref{table57}.} 

\begin{table}[!ht]
\begin{center}
\begin{tabular}{@{}|l|ll|ll|ll|ll|@{}}\hline  
Bus No. & \multicolumn{2}{c}{Peak Time} & \multicolumn{2}{c}{Max Var} & \multicolumn{2}{c}{SS Error} & \multicolumn{2}{c}{Accm Time} \\
\hline
Case-1 & WoC & WC-1 & WoC & WC-1 & WoC & WC-1& WoC & WC-1 \\
\hline
~~~1  & 1.0ms & 3.5ms & 0.1\% & 0.5\% & 0 & 0 & 7ms & 7ms \\
~~~2  & 1.0ms & 2.0ms & 0.1\% & 0.6\% & 0.3\% & 0 & 10ms & 5ms \\
~~~3  & 1.0ms & 1.0ms & 2.0\% & 2.0\% & 1.0\% & 0 & 20ms & 2ms \\
~~~4  & $-$ & 1.0ms & $-$ & 2.0\% & 4.1\% & 0 & 20ms & 2ms \\
\hline
Case-2 & WC-1 & WC-2 & WC-1 & WC-2 & WC-1 & WC-2& WC-1 & WC-2 \\
\hline
~~~1  & 25ms & 2.0ms & 3.5\% & 1.0\% & 0 & 0 & 45ms & 4ms \\
~~~2  & 14ms & 2.0ms & 2.0\% & 1.5\% & 0 & 0 & 80ms & 8ms \\
~~~3  & 16ms & 1.0ms & 2.2\% & 2.5\% & 0 & 0 & 80ms & 12ms \\
~~~4  & 15ms & 1.0ms & 3.0\% & 3.4\% & 0 & 0 & 80ms & 6ms \\
\hline
Case-3 & WC-1 & WC-3 & WC-1 & WC-3 & WC-1 & WC-3& WC-1 & WC-3 \\
\hline
~~~1  & unstable & 10ms & unstable & 0.5\% & unstable & 0 & unstable & 20ms \\
~~~2  & unstable & 2ms & unstable & 0.8\%  & unstable & 0 & unstable & 15ms \\
~~~3  & unstable & 1ms & unstable & 2.0\%  & unstable & 0 & unstable & 25ms \\
~~~4  & unstable & 1ms & unstable & 2.0\%  & unstable & 0 & unstable & 10ms \\
\hline
\end{tabular}
\caption{Performance comparison of various controllers in different cases}
\label{table57}
\end{center}
\end{table}

\section{Summary} \label{Sec:cfw}
 
This work presents a framework for designing cyber physical voltage control of an ACSSMG using variations in communication topologies in the presence of changing electrical and communication parameters. The voltage control problem in ACSSMG has been modeled using a distributed control framework. Optimization frameworks using the theories of LMIs and Lyapunov stability analysis have been used to develop robust controllers both in the presence and absence of delays. The proposed CBSCD methodology for controller design finds the set of communication links using minimal communication resources to obtain controllers providing maximal stability. Various cyber-physical conditions like load variation, link failure and propagation delay are investigated and the proposed technique showed excellent results.

\chapter{Conclusion and Future Scope}\label{sec:conclusion}
\section{Conclusion}
The overall idea of this thesis is to develop models that can better capture the interdependence of cyber and physical parameters in the smart standalone microgrids and develop distributed controllers which can make the system more stable in the presence of disturbances in both the cyber and physical parameters. Different levels of control in the hybrid AC-DC microgrid structure are fraught with different types of cyber and physical challenges. To handle these varied challenges, the hybrid microgrid has been bifurcated into DCSSMG and ACSSMG and the following objectives were formulated:
\begin{enumerate}
\item	To develop cyber-physical frameworks that can jointly model the effect of both physical and communication parameters in SSMGs.
\item	To design controllers which can counter the effect of reduced inertia in DCSSMGs with renewable rich power sources in the presence of atmospheric and load changes. 
\item	To design distributed controllers which can work with least number of sensors in the DCSSMG so as to make it robust to sensor failures and reduce overall cost.
\item	To design controllers which can achieve higher voltage stability at points of common coupling  of the ACSSMG in the presence of disturbances from changes in atmospheric conditions, load and other communication constraints like bandwidth, etc.
\end{enumerate}
While objective 1 is generic to both the AC and DC SSMGs, the rest of the objectives cover the specific issues related to the cyber and physical domains either in the ACSSMG or the DCSSMG.
Accordingly, the thesis proposes four solutions which address specific cyber physical objectives of the DCSSMG or the ACSSMG while keeping the general objective-1 in the background. 

For instance, the second and third chapters provide detailed modeling of typical DCSSMG systems with photovoltaic array and energy storage devices like battery and supercapacitor. It has been identified that these systems have low inertia due to absence of conventional power plants due to which they are highly vulnerable to both cyber and physical disturbances. Since the DCSSMG is inundated with a plethora of sensors, its stability becomes dependent on proper functioning of all the sensors. In case of failure of the sensors, the various control operations in the DCSSMG like the maximum power point tracking and DC voltage control are essentially compromised. Moreover, increasing the number of sensors in the DCSSMG also increases the cost of the system. Apart from the cyber issues, the DCSSMG is easily affected by the change in atmospheric conditions like temperature and irradiance and also by the sudden changes in load during its operation. Hence, it is very much essential to provide control techniques that can ensure stability of the DCSSMG when simultaneously subjected to sensor failure and change in atmospheric conditions/load. 

The work in chapter 2 devised an adaptive observer based backstepping control technique. The system model is conveniently broken down into smaller subsystems and back-stepping controllers are designed based on Lyapunov stability. The observers are used to estimate the disturbances in real-time which obviates the need for placing expensive sensors for measuring irradiance, temperature and other modeling parameters like battery and supercapacitor voltages. The MPP voltage of the PV array and the DCSSMG voltage are seen to stabilize between 70$ms$ to 150$ms$ without sensors as opposed to the state of the art backstepping controller which takes around 60$ms$ to 140$ms$ with extra sensors. It is seen that the developed technique performs comparably with the state-of-the-art while providing additional facility to reduce the sensors or to act as a cushion in times of sensor failure. Moreover, it is seen that the disturbance observers can be initialized with any random value obviating the need for any initial sensor measurements. 

The adaptive neural controller developed in chapter 3 goes completely model-free as opposed to the adaptive observer technique developed in chapter 2 which needs a detailed dynamic model of the DCSSMG. This technique is a black-box type of control technique which automatically captures all the system dynamics without any apriori knowledge of the system parameters including the disturbance measurements for which expensive sensors are required. Also, it has been observed that this technique cuts the settling time heavily to a mere 30$ms$ to 80$ms$ range while considerably reducing the overshoot. This technique has also clearly demonstrated its superiority over state of the art techniques in the presence of multiple changes in parameters and atmospheric conditions in the absence of extra sensor information. This can be attributed to accurate model developed by the neural networks which additionally facilitates the backstepping process to provide better control.    

Following this, the fourth and fifth chapters shift the focus on to the voltage stability in ACSSMG configuration with a capability to interconnect many DCSSMGs together to run as a single isolated power unit. Due to larger distances between different MGs, the communication of important voltage information may be subject to delays and worse, communication failure. Additionally, the ACSSMG may suddenly experience sudden changes in AC loads. Hence, hybrid and cyber-physical modeling techniques are developed in these two works to assist the necessary distributed controller design.  

The fourth chapter oversees the development of a hybrid modeling framework for capturing the various operating conditions of a densely connected ACSSMG with four buses in course of daily operation. The provision of redundant voltage information from multiple PCCs facilitate the designed feedback controllers to extend the range of operation of a single controller both in the physical and the cyber domains like change in load, change in delay and link loss while also catering well to simultaneous changes in multiple parameters.  Although the classical function of the tertiary controllers are reduced by adopting a distributed secondary structure, additional responsibilities are given to the tertiary controller to monitor the ACSSMG system on a regular basis and update the secondary controllers based on online predictions of load, delay and communication loss. Such a possibility has been explored for the first time in this work.

The fifth chapter uses a similar four bus model of the ACSSMG, but the communication network is considered to 
be of low bandwidth. It is shown that the communication structure of the distributed control framework plays a very crucial role in determining the voltage stability of the microgrid. It is also shown how flexible communication structures can aid voltage control and a communication design algorithm CBSCD is proposed to compute the communication structure in an online manner with the help of predictions by the tertiary controllers. Many cyber and physical constraints in operating the microgrid like the bandwidth, connection constraints and cost constraints are considered for developing this framework which is also a unique contribution of this work. The flexible communication structures with adaptive controllers can be used to handle multiple changes in parameters like load, communication delay and communication loss.  

The cyber physical models and novel techniques developed in this thesis will continue to gain more relevance as we proceed towards more complicated technological developments in smart grids. As computation becomes much cheaper and more accessible to the general populace, these techniques can be used widely not only for smart grids but unto a varied range of complex systems like swarm robotics, medical cyber physical systems, space applications and military.

\section{Future Scope}
The work done in this thesis presents novel directions towards designing cyber physical controllers in SSMGs. Future possibilities of this work can further establish these contributions on a generic level. The following suggestions are presented to explore such possibilities,   

\begin{itemize}
\item In this thesis, both for the ACSSMGs and the DCSSMGs, only resistive loads were used for demonstrating proof of concept. A detailed load model consisting of constant power load, motor load and zip load models can be used to make the system more realistic.
\item Only PV source has been considered in the thesis. Many other renewable sources like wind, tidal and others can be considered. Moreover, during the standalone operation of the MGs, diesel generator can also be used as a standby option.
\item Although a hybrid microgrid structure was proposed, an integrated model for the overall AC-DC microgrid still needs to be developed for effectively embedding the dynamics of different levels of control into the controller development process.
\item Direct load scheduling can be added to the microgrid to add more control in the standalone operation.  
\item Event triggered secondary control can be considered for optimizing the resources of the microgrid.
\item Grid connected modeling should be studied and using the MGs for providing ancillary grid services can be studied. 
\item The developed techniques should can also be extended to frequency control in the ACSSMG operation.
\end{itemize}

\textcolor{black}{One of the major requirements for the new age smart microgrids is its reliability. Reliability refers to the availability of dependable power for carrying out daily life businesses and critical services with an ability to withstand the effect of many disturbances in its operation. To be reliable in the short term, the microgrids must have enough regular power production and reserve power to provide services and keep system voltage and frequency in balance. To be reliable over the longer term, the microgrid must adapt to keep pace with changing consumer demand for electric energy, and the addition of new resources and technologies. With massive strides in artificial intelligence, it is possible to equip the microgrids with a high degree of reliability. Some of the directions for future work in this domain are as follows:}
\begin{itemize}
 \item \textcolor{black}{Generation forecast has become a challenging problem in the microgrids due to the highly intermittent nature of renewable generation. For instance, accurate prediction of wind speed or solar irradiance helps to reduce generation uncertainty and unexpected power fluctuation. It is seen that these climatic parameters at one generator have spatial and temporal correlations with the parameters at other generating locations. Moreover, the prediction is also very much sensitive to the error in forecasting models. It is possible to employ graph theory and deep neural networks to model these correlations to develop accurate real time predictions for power generation.}
 \item \textcolor{black}{The residential load profile in the microgrid is also equally intermittent as opposed to commercial and industrial loads. Effective load predictions coupled with economic theories such as Nash equilibrium may be used for improved pricing decisions for demand side management.}
 \item \textcolor{black}{It is to be noted that conventional current-based fault detection methods do not perform well in the electronic device-dominated microgrids because renewable generators do not yield enough fault currents to trigger protective relays. The voltage and current data available at different nodes in the microgrid may be analysed using deep neural networks to predict, detect and locate faults in the microgrids.}
 \item \textcolor{black}{Energy storage devices and electric vehicles are an integral part of microgrids and their charging and discharging patterns need to be handled in a coordinated fashion with loads and generators so as to ensure the maximum utilization of renewable power as well as to secure enough reserve to respond to sudden disturbances in the microgrid apart from reducing the operational cost of the microgrids. This can be achieved through formulation of optimal control strategies based on Adaptive Critic or Q learning techniques with neural networks for solving the optimization problem and generating the appropriate control signal to cater to multiple energy management objectives of the microgrids}.
 \item \textcolor{black}{As the microgrids continue to grow in size and modularity, it is necessary to have controllers which are adaptive and model-free. AI techniques such as multi-task learning\cite{Mishra2020TowardsQF, mccann2018natural, sanh2021multitask}, instruction-tuning\cite{mishra2021cross, mishra2021reframing, wei2021finetuned} may be used to initially train the system model and perform real-time system identification to develop adaptive intelligent controllers for the same.}
\item \textcolor{black}{It is well known that cyberphysical systems are highly prone to cyber attacks like the False Data Injection (FDI) and Denial of Service (DoS), stealth attack and distributed attack. The various DNN architectures can be used to model these attacks and also identify them so that the microgrid operator be alerted to take necessary action.}
\item \textcolor{black}{In order to enable machine learning systems in our cyber physical setup, data quality\cite{Mishra2020DQIMD, le2020adversarial, Sambasivan2021EveryoneWT} needs to monitored\cite{arunkumarreal, swayamdipta2020dataset} and kept in check\cite{mishra-sachdeva-2020-need, mishra2020dqi, mishra2021robust, Mishra2020OurEM} since it's data that guides or misguides systems in machine learning.}
\end{itemize}

\textcolor{black}{We have already established a DCSSMG prototype in the Intelligent Systems and Control Lab at IIT Kanpur. We procured an APLAB solar simulator which can provide upto 1.2kW for the DCSSMG. The BESS system contains two AMARON (12V) lead acid batteries connected in series. The PV and BESS are connected in parallel to a 100 ohm resistive load. We also designed and developed the boost and bidirectional converters with the help of PCB facility in IIT Kanpur. With the help of necessary voltage and current sensors from LEM and TMS320F28335 DSP controller from Texas Instruments, we implemented the closed loop PI control of the DCSSMG.}

\textcolor{black}{We are now working on implementing the sensor-free and adaptive algorithms developed in Chapters-2 and 3 on the PV battery based DCSSMG. We also plan to use a SEMIKRON inverter to convert the DC into AC and connect with the AC loads. We can then replicate the DCSSMGs and join them through an AC connection. My teammates in the lab have also designed WiFi communication node prototypes for placing at various nodes and collecting data from distance. All these can be used to validate the cyber physical frameworks developed in Chapters 4 and 5.} 

\chapter*{\centering List of Publications}
\thispagestyle{empty}
\addcontentsline{toc}{chapter}{List of Publications}

\section*{Journal Papers}
\begin{enumerate}
\item Meher Preetam Korukonda, Ravi Prakash, Suvendu Samanta, and Laxmidhar Behera, \enquote{Adaptive Neural Controller for Unknown Standalone Photo voltaic Distributed Generation Systems with Unknown Disturbances}, \emph{IEEE Transactions on Sustainable Energy}, Date of Accepted: 19 September 2021. 
\item Amir Hussain, Man Mohan Garg, Meher Preetam Korukonda, Shamim Hasan, and Laxmidhar Behera, \enquote{A Parameter Estimation Based MPPT method for a PV System Using Lyapunov Control Scheme}, \emph{IEEE Transactions on Sustainable}, vol. 10, no. 4, pp. 2123-2132, 2018.
\item Meher Preetam Korukonda, Swaroop Ranjan Mishra, Ketan Rajawat, Laxmidhar Behera, \enquote{Hybrid Adaptive Framework for Coordinated Control of Distributed Generators in Cyber Physical Energy Systems}, \emph{IET Cyber-Physical Systems: Theory } $\&$ \emph{Applications}, vol. 3, no. 1, pp 54-62, Dec. 2017.
\item Meher Preetam Korukonda, Swaroop Ranjan Mishra, Anupam Shukla, Laxmidhar Behera, \enquote{Handling Multi-Parametric Variations in Distributed Control of Cyber Physical Energy Systems through Optimal Communication Design}, \emph{IET Cyber-Physical Systems: Theory} $\&$ \emph{Applications}, vol. 2, no. 2, pp 90-100, June 2017.
\item Swaroop Ranjan Mishra,  Meher Preetam Korukonda, Laxmidhar Behera, and Anupam Shukla, \enquote{Enabling cyber-physical demand response in smart grids via conjoint communication and controller design}, \emph{IET Cyber-Phys. Syst.: Theory} $\&$ \emph{Appl.} vol. 4, no. 4, pp. 291-303, 2019.
\end{enumerate}

\section*{Conferences}
\begin{enumerate}
\item  Meher Preetam Korukonda, Man Mohan Garg, Amir Hussain, and Laxmidhar Behera, \enquote{Disturbance Observer based Controller Design to Reduce Sensor Count in Standalone PVDG Systems}, \emph{In IECON 2020 The 46th Annual Conference of the IEEE Industrial Electronics Society}, pp. 2975-2980. IEEE, 2020.
\item Manoranjan Satapathy, Meher Preetam Korukonda, Amir Hussain, and Laxmidhar Behera, \enquote{A Direct Perturbation Based Sensor-Free MPPT with DC Bus Voltage Control for a Standalone DC Microgrid}, \emph{In 2019 IEEE PES Innovative Smart Grid Technologies Europe (ISGT-Europe)}, pp. 1-5. IEEE, 2019.
\item Meher Preetam K, Swaroop Ranjan Mishra, Anupam Shukla,  Laxmidhar Behera, \enquote{Improving Microgrid Voltage Stability through Cyber-Physical Control}, \emph{IEEE International Conference on National Power Systems Conference (NPSC)}, pp. 1-6, December 2016.
\item Santhoshkumar Battula, Man Mohan Garg, Anup Kumar Panda, Meher Preetam Korukonda, and Laxmidhar Behera, \enquote{Analysis and Dual-loop PI Control of Bidirectional Quasi Z-Source DC-DC Converter}, \emph{In IECON 2020 The 46th Annual Conference of the IEEE Industrial Electronics Society}, pp. 2939-2944. IEEE, 2020.
\item Swaroop Ranjan Mishra, Venkata Srinath N, Meher Preetam K, Laxmidhar Behera, \enquote{A Generalized Novel Framework for Optimal Sensor-Controller Connection Design to Guarantee a Stable Cyber Physical Smart Grid}, \emph{IEEE International Conference on Industrial Informatics (INDIN)}, July 2015.
\item Anuj Nandanwar, Meher Preetam Korukonda, Laxmidhar Behera, \enquote{A Routing Scheme for Voltage Stabilization in Cyber Physical Energy Systems Advances in Control and Optimization of Dynamical Systems}, \emph{International Conference on Advances in Control and Optimization of Dynamical Systems},  Vol. 3, Part 1, PP 812-818, 2014.
\end{enumerate}

\addcontentsline{toc}{chapter}{Bibliography}
\bibliographystyle{IEEEbib}
\bibliography{refs}

\begin{thebibliography}{100}

\bibitem{iovinetase17}
A.~{Iovine}, S.~B. {Siad}, G.~{Damm}, E.~{De Santis}, and M.~D. {Di Benedetto},
\newblock ``Nonlinear control of a dc microgrid for the integration of
  photovoltaic panels,''
\newblock {\em IEEE Transactions on Automation Science and Engineering}, vol.
  14, no. 2, pp. 524--535, 2017.

\bibitem{overviewtkr2}
T.~K. {Roy}, M.~A. {Mahmud}, A.~M.~T. {Oo}, M.~E. {Haque}, K.~M. {Muttaqi}, and
  N.~{Mendis},
\newblock ``Nonlinear adaptive backstepping controller design for islanded dc
  microgrids,''
\newblock {\em IEEE Transactions on Industry Applications}, vol. 54, no. 3, pp.
  2857--2873, 2018.

\bibitem{powerindia19}
Sneha Alexander,
\newblock ``The curious case of electrification in india amid discom
  blackouts,'' Mar 2019.

\bibitem{mgmountain}
Joanna Goodrich,
\newblock ``How 14 microgrids set off a chain reaction in a himalayan
  village,'' Oct 2021.

\bibitem{mgislands}
Quan Sui, Fanrong Wei, Chuantao Wu, Xiangning Lin, and Zhengtian Li,
\newblock ``Day-ahead energy management for pelagic island microgrid groups
  considering non-integer-hour energy transmission,''
\newblock {\em IEEE Transactions on Smart Grid}, vol. 11, no. 6, pp.
  5249--5259, 2020.

\bibitem{mgdesert}
Leila Ghomri, Mounir Khiat, and Sid~Ahmed Khiat,
\newblock ``Modeling and real time simulation of microgrids in algerian sahara
  area,''
\newblock in {\em 2018 IEEE International Energy Conference (ENERGYCON)}, 2018,
  pp. 1--5.

\bibitem{mgshipboard}
Mohammad-Hassan Khooban, Tomislav Dragicevic, Frede Blaabjerg, and Marko
  Delimar,
\newblock ``Shipboard microgrids: A novel approach to load frequency control,''
\newblock {\em IEEE Transactions on Sustainable Energy}, vol. 9, no. 2, pp.
  843--852, 2018.

\bibitem{mgmilitary}
Spencer~C. Shabshab, Peter~A. Lindahl, J.~Kendall Nowocin, John Donnal, David
  Blum, Les Norford, and Steven~B. Leeb,
\newblock ``Demand smoothing in military microgrids through coordinated direct
  load control,''
\newblock {\em IEEE Transactions on Smart Grid}, vol. 11, no. 3, pp.
  1917--1927, 2020.

\bibitem{overview3}
Mostafa Farrokhabadi, Claudio~A. Cañizares, John~W. Simpson-Porco, Ehsan Nasr,
  Lingling Fan, Patricio~A. Mendoza-Araya, Reinaldo Tonkoski, Ujjwol Tamrakar,
  Nikos Hatziargyriou, Dimitris Lagos, Richard~W. Wies, Mario Paolone, Marco
  Liserre, Lasantha Meegahapola, Mahmoud Kabalan, Amir~H. Hajimiragha, Dario
  Peralta, Marcelo~A. Elizondo, Kevin~P. Schneider, Francis~K. Tuffner, and Jim
  Reilly,
\newblock ``Microgrid stability definitions, analysis, and examples,''
\newblock {\em IEEE Transactions on Power Systems}, vol. 35, no. 1, pp. 13--29,
  2020.

\bibitem{lee2008cyber}
Edward~A Lee,
\newblock ``Cyber physical systems: Design challenges,''
\newblock in {\em Object Oriented Real-Time Distributed Computing (ISORC), 2008
  11th IEEE International Symposium on}. IEEE, 2008, pp. 363--369.

\bibitem{cpsiet}
M.~Bessani, R.~Z. Fanucchi, A.~C.~C. Delbem, and C.~D. Maciel,
\newblock ``Impact of operators' performance in the reliability of
  cyber-physical power distribution systems,''
\newblock {\em IET Generation, Transmission Distribution}, vol. 10, no. 11, pp.
  2640--2646, 2016.

\bibitem{7017600}
S.~Xin, Q.~Guo, H.~Sun, B.~Zhang, J.~Wang, and C.~Chen,
\newblock ``Cyber-physical modeling and cyber-contingency assessment of
  hierarchical control systems,''
\newblock {\em IEEE Transactions on Smart Grid}, vol. 6, no. 5, pp. 2375--2385,
  Sept 2015.

\bibitem{fink2012robust}
Jonathan Fink, Alejandro Ribeiro, and Vijay Kumar,
\newblock ``Robust control for mobility and wireless communication in
  cyber--physical systems with application to robot teams,''
\newblock {\em Proceedings of the IEEE}, vol. 100, no. 1, pp. 164--178, 2012.

\bibitem{lee2012challenges}
Insup Lee, Oleg Sokolsky, Sanjian Chen, John Hatcliff, Eunkyoung Jee, BaekGyu
  Kim, Andrew King, Margaret Mullen-Fortino, Soojin Park, Alexander Roederer,
  et~al.,
\newblock ``Challenges and research directions in medical cyber--physical
  systems,''
\newblock {\em Proceedings of the IEEE}, vol. 100, no. 1, pp. 75--90, 2012.

\bibitem{parolini2010cyber}
Luca Parolini, Niraj Tolia, Bruno Sinopoli, and Bruce~H Krogh,
\newblock ``A cyber-physical systems approach to energy management in data
  centers,''
\newblock in {\em Proceedings of the 1st ACM/IEEE International Conference on
  Cyber-Physical Systems}. ACM, 2010, pp. 168--177.

\bibitem{khaitan2015design}
Siddhartha~Kumar Khaitan and James~D McCalley,
\newblock ``Design techniques and applications of cyberphysical systems: A
  survey,''
\newblock {\em Systems Journal, IEEE}, vol. 9, no. 2, pp. 350--365, 2015.

\bibitem{aalborg}
Lexuan Meng, Adriana Luna, Enrique~Rodríguez Díaz, Bo~Sun, Tomislav
  Dragicevic, Mehdi Savaghebi, Juan~C. Vasquez, Josep~M. Guerrero, Moisès
  Graells, and Fabio Andrade,
\newblock ``Flexible system integration and advanced hierarchical control
  architectures in the microgrid research laboratory of aalborg university,''
\newblock {\em IEEE Transactions on Industry Applications}, vol. 52, no. 2, pp.
  1736--1749, 2016.

\bibitem{hcmghil2}
Sri~Raghavan Kothandaraman, Ahmadreza Malekpour, Maigha Maigha, Aleksi Paaso,
  Amin Zamani, Farid Katiraei, and Muhidin Lelic,
\newblock ``Utility scale microgrid controller power hardware-in-the-loop
  testing,''
\newblock in {\em 2018 IEEE Electronic Power Grid (eGrid)}, 2018, pp. 1--6.

\bibitem{kollimalla2}
S.~K. {Kollimalla}, M.~K. {Mishra}, A.~{Ukil}, and H.~B. {Gooi},
\newblock ``Dc grid voltage regulation using new hess control strategy,''
\newblock {\em IEEE Transactions on Sustainable Energy}, vol. 8, no. 2, pp.
  772--781, 2017.

\bibitem{kabalan2016large}
Mahmoud Kabalan, Pritpal Singh, and Dagmar Niebur,
\newblock ``Large signal lyapunov-based stability studies in microgrids: A
  review,''
\newblock {\em IEEE Transactions on Smart Grid}, vol. 8, no. 5, pp. 2287--2295,
  2016.

\bibitem{wang2019continuous}
Zuo Wang, Shihua Li, and Qi~Li,
\newblock ``Continuous nonsingular terminal sliding mode control of dc--dc
  boost converters subject to time-varying disturbances,''
\newblock {\em IEEE Transactions on Circuits and Systems II: Express Briefs},
  vol. 67, no. 11, pp. 2552--2556, 2019.

\bibitem{iovine2017nonlinear}
Alessio Iovine, Sabah~Benamane Siad, Gilney Damm, Elena De~Santis, and
  Maria~Domenica Di~Benedetto,
\newblock ``Nonlinear control of a dc microgrid for the integration of
  photovoltaic panels,''
\newblock {\em IEEE Transactions on Automation Science and Engineering}, vol.
  14, no. 2, pp. 524--535, 2017.

\bibitem{mahmud2020robust}
Rasel Mahmud, Md~Alamgir Hossain, and Hemanshu Pota,
\newblock ``Robust nonlinear controller design for islanded photovoltaic system
  with battery energy storage,''
\newblock in {\em 2020 IEEE International Conference on Power Electronics,
  Smart Grid and Renewable Energy (PESGRE2020)}. IEEE, 2020, pp. 1--6.

\bibitem{bambang2014energy}
Riyanto~Trilaksono Bambang, Arief~Syaichu Rohman, Cees~Jan Dronkers, Romeo
  Ortega, Arif Sasongko, et~al.,
\newblock ``Energy management of fuel cell/battery/supercapacitor hybrid power
  sources using model predictive control,''
\newblock {\em IEEE Transactions on Industrial Informatics}, vol. 10, no. 4,
  pp. 1992--2002, 2014.

\bibitem{mane2017improving}
Sneha Mane, Manas Mejari, Faruk Kazi, and Navdeep Singh,
\newblock ``Improving lifetime of fuel cell in hybrid energy management system
  by lure--lyapunov-based control formulation,''
\newblock {\em IEEE Transactions on Industrial Electronics}, vol. 64, no. 8,
  pp. 6671--6679, 2017.

\bibitem{kong2019hierarchical}
Xiaobing Kong, Xiangjie Liu, Lele Ma, and Kwang~Y Lee,
\newblock ``Hierarchical distributed model predictive control of standalone
  wind/solar/battery power system,''
\newblock {\em IEEE Transactions on Systems, Man, and Cybernetics: Systems},
  vol. 49, no. 8, pp. 1570--1581, 2019.

\bibitem{a9}
N.~{Femia}, D.~{Granozio}, G.~{Petrone}, G.~{Spagnuolo}, and M.~{Vitelli},
\newblock ``Predictive adaptive mppt perturb and observe method,''
\newblock {\em IEEE Transactions on Aerospace and Electronic Systems}, vol. 43,
  no. 3, pp. 934--950, 2007.

\bibitem{a10}
N.~{Femia}, G.~{Petrone}, G.~{Spagnuolo}, and M.~{Vitelli},
\newblock ``Optimization of perturb and observe maximum power point tracking
  method,''
\newblock {\em IEEE Transactions on Power Electronics}, vol. 20, no. 4, pp.
  963--973, 2005.

\bibitem{a11}
M.~A. {Elgendy}, B.~{Zahawi}, and D.~J. {Atkinson},
\newblock ``Assessment of the incremental conductance maximum power point
  tracking algorithm,''
\newblock {\em IEEE Transactions on Sustainable Energy}, vol. 4, no. 1, pp.
  108--117, 2013.

\bibitem{a12}
Nahla~E. Zakzouk,
\newblock ``Improved performance low-cost incremental conductance pv mppt
  technique,''
\newblock {\em IET Renewable Power Generation}, vol. 10, pp. 561--574(13),
  April 2016.

\bibitem{a13}
M.~{Adly}, H.~{El-Sherif}, and M.~{Ibrahim},
\newblock ``Maximum power point tracker for a pv cell using a fuzzy agent
  adapted by the fractional open circuit voltage technique,''
\newblock in {\em 2011 IEEE International Conference on Fuzzy Systems
  (FUZZ-IEEE 2011)}, 2011, pp. 1918--1922.

\bibitem{a14}
A.~{Sandali}, T.~{Oukhoya}, and A.~{Cheriti},
\newblock ``Modeling and design of pv grid connected system using a modified
  fractional short-circuit current mppt,''
\newblock in {\em 2014 International Renewable and Sustainable Energy
  Conference (IRSEC)}, 2014, pp. 224--229.

\bibitem{a15}
T.~{Esram}, J.~W. {Kimball}, P.~T. {Krein}, P.~L. {Chapman}, and P.~{Midya},
\newblock ``Dynamic maximum power point tracking of photovoltaic arrays using
  ripple correlation control,''
\newblock {\em IEEE Transactions on Power Electronics}, vol. 21, no. 5, pp.
  1282--1291, 2006.

\bibitem{a16}
M.~{Rakhshan}, N.~{Vafamand}, M.~{Khooban}, and F.~{Blaabjerg},
\newblock ``Maximum power point tracking control of photovoltaic systems: A
  polynomial fuzzy model-based approach,''
\newblock {\em IEEE Journal of Emerging and Selected Topics in Power
  Electronics}, vol. 6, no. 1, pp. 292--299, 2018.

\bibitem{a17}
Muhammad Ammirrul Atiqi~Mohd Zainuri,
\newblock ``Development of adaptive perturb and observe-fuzzy control maximum
  power point tracking for photovoltaic boost dcâ€“dc converter,''
\newblock {\em IET Renewable Power Generation}, vol. 8, pp. 183--194(11), March
  2014.

\bibitem{a18}
K.~L. {Lian}, J.~H. {Jhang}, and I.~S. {Tian},
\newblock ``A maximum power point tracking method based on perturb-and-observe
  combined with particle swarm optimization,''
\newblock {\em IEEE Journal of Photovoltaics}, vol. 4, no. 2, pp. 626--633,
  2014.

\bibitem{a19}
K.~{Ishaque}, Z.~{Salam}, M.~{Amjad}, and S.~{Mekhilef},
\newblock ``An improved particle swarm optimization (pso)–based mppt for pv
  with reduced steady-state oscillation,''
\newblock {\em IEEE Transactions on Power Electronics}, vol. 27, no. 8, pp.
  3627--3638, 2012.

\bibitem{a20}
L.~M. {Elobaid}, A.~K. {Abdelsalam}, and E.~E. {Zakzouk},
\newblock ``Artificial neural network-based photovoltaic maximum power point
  tracking techniques: a survey,''
\newblock {\em IET Renewable Power Generation}, vol. 9, no. 8, pp. 1043--1063,
  2015.

\bibitem{a21}
Yousra Shaiek, Mouna~[Ben Smida], Anis Sakly, and Mohamed~Faouzi Mimouni,
\newblock ``Comparison between conventional methods and ga approach for maximum
  power point tracking of shaded solar pv generators,''
\newblock {\em Solar Energy}, vol. 90, pp. 107 -- 122, 2013.

\bibitem{gridhessref}
U.~{Manandhar}, A.~{Ukil}, H.~B. {Gooi}, N.~R. {Tummuru}, S.~K. {Kollimalla},
  B.~{Wang}, and K.~{Chaudhari},
\newblock ``Energy management and control for grid connected hybrid energy
  storage system under different operating modes,''
\newblock {\em IEEE Transactions on Smart Grid}, vol. 10, no. 2, pp.
  1626--1636, 2019.

\bibitem{microgridhessref}
G.~{Oriti}, N.~{Anglani}, and A.~L. {Julian},
\newblock ``Hybrid energy storage control in a remote military microgrid with
  improved supercapacitor utilization and sensitivity analysis,''
\newblock {\em IEEE Transactions on Industry Applications}, vol. 55, no. 5, pp.
  5099--5108, 2019.

\bibitem{olivares2014trends}
Daniel~E Olivares, Ali Mehrizi-Sani, Amir~H Etemadi, Claudio~A Ca{\~n}izares,
  Reza Iravani, Mehrdad Kazerani, Amir~H Hajimiragha, Oriol Gomis-Bellmunt,
  Maryam Saeedifard, Rodrigo Palma-Behnke, et~al.,
\newblock ``Trends in microgrid control,''
\newblock {\em IEEE Transactions on smart grid}, vol. 5, no. 4, pp. 1905--1919,
  2014.

\bibitem{rezaei2015robust}
Mohammad~Mahdi Rezaei and Jafar Soltani,
\newblock ``Robust control of an islanded multi-bus microgrid based on
  input--output feedback linearisation and sliding mode control,''
\newblock {\em IET Generation, Transmission \& Distribution}, vol. 9, no. 15,
  pp. 2447--2454, 2015.

\bibitem{mehrizi2012constrained}
Ali Mehrizi-Sani and Reza Iravani,
\newblock ``Constrained potential function—based control of microgrids for
  improved dynamic performance,''
\newblock {\em Smart Grid, IEEE Transactions on}, vol. 3, no. 4, pp.
  1885--1892, 2012.

\bibitem{zhou2016consensus}
Jianguo Zhou, Sunghyok Kim, Huaguang Zhang, Qiuye Sun, and Renke Han,
\newblock ``Consensus-based distributed control for accurate reactive,
  harmonic, and imbalance power sharing in microgrids,''
\newblock {\em IEEE Transactions on Smart Grid}, vol. 9, no. 4, pp. 2453--2467,
  2016.

\bibitem{sun2015multiagent}
Qiuye Sun, Renke Han, Huaguang Zhang, Jianguo Zhou, and Josep~M Guerrero,
\newblock ``A multiagent-based consensus algorithm for distributed coordinated
  control of distributed generators in the energy internet,''
\newblock {\em IEEE transactions on smart grid}, vol. 6, no. 6, pp. 3006--3019,
  2015.

\bibitem{bidram2013secondary}
Ali Bidram, Ali Davoudi, Frank~L Lewis, and Zhihua Qu,
\newblock ``Secondary control of microgrids based on distributed cooperative
  control of multi-agent systems,''
\newblock {\em Generation, Transmission \& Distribution, IET}, vol. 7, no. 8,
  pp. 822--831, 2013.

\bibitem{xin2011cooperative}
H~Xin, Z~Lu, Z~Qu, D~Gan, and D~Qi,
\newblock ``Cooperative control strategy for multiple photovoltaic generators
  in distribution networks,''
\newblock {\em IET control theory \& applications}, vol. 5, no. 14, pp.
  1617--1629, 2011.

\bibitem{shafiee2013distributed}
Qobad Shafiee, Josep~M Guerrero, and Juan~C Vasquez,
\newblock ``Distributed secondary control for islanded microgrids—a novel
  approach,''
\newblock {\em IEEE Transactions on power electronics}, vol. 29, no. 2, pp.
  1018--1031, 2013.

\bibitem{liang2013stability}
Hao Liang, Bong~Jun Choi, Weihua Zhuang, and Xuemin Shen,
\newblock ``Stability enhancement of decentralized inverter control through
  wireless communications in microgrids,''
\newblock {\em Smart Grid, IEEE Transactions on}, vol. 4, no. 1, pp. 321--331,
  2013.

\bibitem{majumder2012power}
Ritwik Majumder, Gargi Bag, and Ki-Hyung Kim,
\newblock ``Power sharing and control in distributed generation with wireless
  sensor networks,''
\newblock {\em Smart Grid, IEEE Transactions on}, vol. 3, no. 2, pp. 618--634,
  2012.

\bibitem{tertiary1}
Tung-Lam Nguyen, Yu~Wang, Quoc-Tuan Tran, Raphael Caire, Yan Xu, and Catalin
  Gavriluta,
\newblock ``A distributed hierarchical control framework in islanded microgrids
  and its agent-based design for cyber–physical implementations,''
\newblock {\em IEEE Transactions on Industrial Electronics}, vol. 68, no. 10,
  pp. 9685--9695, 2021.

\bibitem{tertiary3}
Ali Arefi and Farhad Shahnia,
\newblock ``Tertiary controller-based optimal voltage and frequency management
  technique for multi-microgrid systems of large remote towns,''
\newblock {\em IEEE Transactions on Smart Grid}, vol. 9, no. 6, pp. 5962--5974,
  2018.

\bibitem{tertiary4}
Lexuan Meng, Fen Tang, Mehdi Savaghebi, Juan~C. Vasquez, and Josep~M. Guerrero,
\newblock ``Tertiary control of voltage unbalance compensation for optimal
  power quality in islanded microgrids,''
\newblock {\em IEEE Transactions on Energy Conversion}, vol. 29, no. 4, pp.
  802--815, 2014.

\bibitem{tertiary5}
Seyedali Moayedi and Ali Davoudi,
\newblock ``Distributed tertiary control of dc microgrid clusters,''
\newblock {\em IEEE Transactions on Power Electronics}, vol. 31, no. 2, pp.
  1717--1733, 2016.

\bibitem{tertiary21}
Saroja~Kanti Sahoo, Avinash~Kumar Sinha, and N.~K. Kishore,
\newblock ``Control techniques in ac, dc, and hybrid ac–dc microgrid: A
  review,''
\newblock {\em IEEE Journal of Emerging and Selected Topics in Power
  Electronics}, vol. 6, no. 2, pp. 738--759, 2018.

\bibitem{tertiary22}
Josep~M. Guerrero, Mukul Chandorkar, Tzung-Lin Lee, and Poh~Chiang Loh,
\newblock ``Advanced control architectures for intelligent microgrids—part i:
  Decentralized and hierarchical control,''
\newblock {\em IEEE Transactions on Industrial Electronics}, vol. 60, no. 4,
  pp. 1254--1262, 2013.

\bibitem{tertiary20}
Sayed Mohamed, Mostafa~F. Shaaban, Muhammad Ismail, Erchin Serpedin, and
  Khalid~A. Qaraqe,
\newblock ``An efficient planning algorithm for hybrid remote microgrids,''
\newblock {\em IEEE Transactions on Sustainable Energy}, vol. 10, no. 1, pp.
  257--267, 2019.

\bibitem{trainsref}
C.~G. {da Silva Moraes}, S.~L. {Brockveld Junior}, M.~L. {Heldwein}, A.~S.
  {Franca}, A.~S. {Vaccari}, and G.~{Waltrich},
\newblock ``Power conversion technologies for a hybrid energy storage system in
  diesel-electric locomotives,''
\newblock {\em IEEE Transactions on Industrial Electronics}, pp. 1--1, 2020.

\bibitem{shipsref}
S.~{Faddel}, A.~A. {Saad}, M.~E. {Hariri}, and O.~A. {Mohammed},
\newblock ``Coordination of hybrid energy storage for ship power systems with
  pulsed loads,''
\newblock {\em IEEE Transactions on Industry Applications}, vol. 56, no. 2, pp.
  1136--1145, 2020.

\bibitem{aircraftsref}
S.~{Njoya Motapon}, L.~{Dessaint}, and K.~{Al-Haddad},
\newblock ``A comparative study of energy management schemes for a fuel-cell
  hybrid emergency power system of more-electric aircraft,''
\newblock {\em IEEE Transactions on Industrial Electronics}, vol. 61, no. 3,
  pp. 1320--1334, 2014.

\bibitem{residenceref}
D.~{Zhu}, S.~{Yue}, N.~{Chang}, and M.~{Pedram},
\newblock ``Toward a profitable grid-connected hybrid electrical energy storage
  system for residential use,''
\newblock {\em IEEE Transactions on Computer-Aided Design of Integrated
  Circuits and Systems}, vol. 35, no. 7, pp. 1151--1164, 2016.

\bibitem{datacenterref}
W.~{Zheng}, K.~{Ma}, and X.~{Wang},
\newblock ``Hybrid energy storage with supercapacitor for cost-efficient data
  center power shaving and capping,''
\newblock {\em IEEE Transactions on Parallel and Distributed Systems}, vol. 28,
  no. 4, pp. 1105--1118, 2017.

\bibitem{systemsizingref1}
H.~{Yu}, F.~{Castelli-Dezza}, F.~{Cheli}, X.~{Tang}, X.~{Hu}, and X.~{Lin},
\newblock ``Dimensioning and power management of hybrid energy storage systems
  for electric vehicles with multiple optimization criteria,''
\newblock {\em IEEE Transactions on Power Electronics}, vol. 36, no. 5, pp.
  5545--5556, 2021.

\bibitem{systemsizingref2}
D.~B.~W. {Abeywardana}, B.~{Hredzak}, V.~G. {Agelidis}, and G.~D.
  {Demetriades},
\newblock ``Supercapacitor sizing method for energy-controlled filter-based
  hybrid energy storage systems,''
\newblock {\em IEEE Transactions on Power Electronics}, vol. 32, no. 2, pp.
  1626--1637, 2017.

\bibitem{deadbeathesstse19}
B.~{Wang}, U.~{Manandhar}, X.~{Zhang}, H.~B. {Gooi}, and A.~{Ukil},
\newblock ``Deadbeat control for hybrid energy storage systems in dc
  microgrids,''
\newblock {\em IEEE Transactions on Sustainable Energy}, vol. 10, no. 4, pp.
  1867--1877, 2019.

\bibitem{decentralizedhesstpec15}
Y.~{Gu}, W.~{Li}, and X.~{He},
\newblock ``Frequency-coordinating virtual impedance for autonomous power
  management of dc microgrid,''
\newblock {\em IEEE Transactions on Power Electronics}, vol. 30, no. 4, pp.
  2328--2337, 2015.

\bibitem{distributedhesstse18}
F.~{Guo}, Q.~{Xu}, C.~{Wen}, L.~{Wang}, and P.~{Wang},
\newblock ``Distributed secondary control for power allocation and voltage
  restoration in islanded dc microgrids,''
\newblock {\em IEEE Transactions on Sustainable Energy}, vol. 9, no. 4, pp.
  1857--1869, 2018.

\bibitem{self2}
M.~P. {Korukonda}, S.~R. {Mishra}, K.~{Rajawat}, and L.~{Behera},
\newblock ``Hybrid adaptive framework for coordinated control of distributed
  generators in cyber-physical energy systems,''
\newblock {\em IET Cyber-Physical Systems: Theory Applications}, vol. 3, no. 1,
  pp. 54--62, 2018.

\bibitem{self3}
M.~P. {Korukonda}, S.~R. {Mishra}, A.~{Shukla}, and L.~{Behera},
\newblock ``Handling multi-parametric variations in distributed control of
  cyber-physical energy systems through optimal communication design,''
\newblock {\em IET Cyber-Physical Systems: Theory Applications}, vol. 2, no. 2,
  pp. 90--100, 2017.

\bibitem{selfnpsc16}
Meher~Preetam Korukonda, Swaroop~Ranjan Mishra, Anupam Shukla, and Laxmidhar
  Behera,
\newblock ``Improving microgrid voltage stability through cyber-physical
  control,''
\newblock in {\em 2016 National Power Systems Conference (NPSC)}. IEEE, 2016,
  pp. 1--6.

\bibitem{battula2020analysis}
Santhoshkumar Battula, Man~Mohan Garg, Anup~Kumar Panda, Meher~Preetam
  Korukonda, and Laxmidhar Behera,
\newblock ``Analysis and dual-loop pi control of bidirectional quasi z-source
  dc-dc converter,''
\newblock in {\em IECON 2020 The 46th Annual Conference of the IEEE Industrial
  Electronics Society}. IEEE, 2020, pp. 2939--2944.

\bibitem{lyapunovmgtsg17}
M.~{Kabalan}, P.~{Singh}, and D.~{Niebur},
\newblock ``Large signal lyapunov-based stability studies in microgrids: A
  review,''
\newblock {\em IEEE Transactions on Smart Grid}, vol. 8, no. 5, pp. 2287--2295,
  2017.

\bibitem{korukonda2021modular}
Meher~Preetam Korukonda, Man~Mohan Garg, and Laxmidhar Behera,
\newblock ``Modular design of nonlinear controllers for photovoltaic
  distributed generation systems,''
\newblock {\em Active Electrical Distribution Network: A Smart Approach},
  p.~12, 2021.

\bibitem{selfdobsiecon20}
M.~P. {Korukonda}, M.~{Mohan Garg}, A.~{Hussain}, and L.~{Behera},
\newblock ``Disturbance observer based controller design to reduce sensor count
  in standalone pvdg systems,''
\newblock in {\em IECON 2020 The 46th Annual Conference of the IEEE Industrial
  Electronics Society}, 2020, pp. 2975--2980.

\bibitem{self1}
A.~{Hussain}, M.~M. {Garg}, M.~P. {Korukonda}, S.~{Hasan}, and L.~{Behera},
\newblock ``A parameter estimation based mppt method for a pv system using
  lyapunov control scheme,''
\newblock {\em IEEE Transactions on Sustainable Energy}, vol. 10, no. 4, pp.
  2123--2132, 2019.

\bibitem{scref}
Lei Zhang, Zhenpo Wang, Xiaosong Hu, Fengchun Sun, and David~G. Dorrell,
\newblock ``A comparative study of equivalent circuit models of ultracapacitors
  for electric vehicles,''
\newblock {\em Journal of Power Sources}, vol. 274, pp. 899--906, 2015.

\bibitem{purebstep1}
Z.~{Shi} and Z.~{Zuo},
\newblock ``Backstepping control for gear transmission servo systems with
  backlash nonlinearity,''
\newblock {\em IEEE Transactions on Automation Science and Engineering}, vol.
  12, no. 2, pp. 752--757, 2015.

\bibitem{isgtpaper}
M.~{Satapathy}, M.~P. {Korukonda}, A.~{Hussain}, and L.~{Behera},
\newblock ``A direct perturbation based sensor-free mppt with dc bus voltage
  control for a standalone dc microgrid,''
\newblock in {\em 2019 IEEE PES Innovative Smart Grid Technologies Europe
  (ISGT-Europe)}, 2019, pp. 1--5.

\bibitem{puretsteobs1}
Benfei Wang, Ujjal Manandhar, Xinan Zhang, Hoay~Beng Gooi, and Abhisek Ukil,
\newblock ``Deadbeat control for hybrid energy storage systems in dc
  microgrids,''
\newblock {\em IEEE Transactions on Sustainable Energy}, vol. 10, no. 4, pp.
  1867--1877, 2019.

\bibitem{puretsteadaptive1}
Josh Davidson, Romain Genest, and John~V. Ringwood,
\newblock ``Adaptive control of a wave energy converter,''
\newblock {\em IEEE Transactions on Sustainable Energy}, vol. 9, no. 4, pp.
  1588--1595, 2018.

\bibitem{nnref11}
Z.~{Liang}, C.~{Kang}, C.~{Shengbin}, L.~{Ling}, M.~{Huanhuan}, and
  W.~{Kaiwen},
\newblock ``The bidirectional dc/dc converter operation mode control algorithm
  based on rbf neural network,''
\newblock in {\em 2019 IEEE Innovative Smart Grid Technologies - Asia (ISGT
  Asia)}, 2019, pp. 2138--2143.

\bibitem{nnref12}
Z.~{Farooq}, T.~{Zaman}, M.~A. {Khan}, {Nasimullah}, S.~M. {Muyeen}, and
  A.~{Ibeas},
\newblock ``Artificial neural network based adaptive control of single phase
  dual active bridge with finite time disturbance compensation,''
\newblock {\em IEEE Access}, vol. 7, pp. 112229--112239, 2019.

\bibitem{nnref14hybridmg}
N.~{Chettibi}, A.~{Mellit}, G.~{Sulligoi}, and A.~{Massi Pavan},
\newblock ``Adaptive neural network-based control of a hybrid ac/dc
  microgrid,''
\newblock {\em IEEE Transactions on Smart Grid}, vol. 9, no. 3, pp. 1667--1679,
  2018.

\bibitem{puretstennadaptive1}
Xuanrui Huang, Kai Sun, and Xi~Xiao,
\newblock ``A neural network-based power control method for direct-drive wave
  energy converters in irregular waves,''
\newblock {\em IEEE Transactions on Sustainable Energy}, vol. 11, no. 4, pp.
  2962--2971, 2020.

\bibitem{korukonda2021adaptive}
Meher~Preetam Korukonda, Ravi Prakash, Suvendu Samanta, and Laxmidhar Behera,
\newblock ``Adaptive neural controller for unknown standalone photovoltaic
  distributed generation systems with unknown disturbances,''
\newblock {\em IEEE Transactions on Sustainable Energy}, 2021.

\bibitem{7073775}
V.~Miñambres-Marcos, M.~Á. Guerrero-Martínez, E.~Romero-Cadaval, and
  P.~González-Castrillo,
\newblock ``Grid-connected photovoltaic power plants for helping node voltage
  regulation,''
\newblock {\em IET Renewable Power Generation}, vol. 9, no. 3, pp. 236--244,
  2015.

\bibitem{lasseter2011smart}
Robert~H Lasseter,
\newblock ``Smart distribution: Coupled microgrids,''
\newblock {\em Proceedings of the IEEE}, vol. 99, no. 6, pp. 1074--1082, 2011.

\bibitem{selfself2}
S.~R. {Mishra}, M.~P. {Korukonda}, L.~{Behera}, and A.~{Shukla},
\newblock ``Enabling cyber-physical demand response in smart grids via conjoint
  communication and controller design,''
\newblock {\em IET Cyber-Physical Systems: Theory Applications}, vol. 4, no. 4,
  pp. 291--303, 2019.

\bibitem{li2012multicast}
Husheng Li, Lifeng Lai, and H~Vincent Poor,
\newblock ``Multicast routing for decentralized control of cyber physical
  systems with an application in smart grid,''
\newblock {\em Selected Areas in Communications, IEEE Journal on}, vol. 30, no.
  6, pp. 1097--1107, 2012.

\bibitem{mishra2015generalized}
Swaroop~Ranjan Mishra, N~Venkata Srinath, K~Meher Preetam, and Laxmidhar
  Behera,
\newblock ``A generalized novel framework for optimal sensor-controller
  connection design to guarantee a stable cyber physical smart grid,''
\newblock in {\em Industrial Informatics (INDIN), 2015 IEEE 13th International
  Conference on}. IEEE, 2015, pp. 424--429.

\bibitem{7564533}
X.~Lu, N.~Chen, Y.~Wang, L.~Qu, and J.~Lai,
\newblock ``Distributed impulsive control for islanded microgrids with variable
  communication delays,''
\newblock {\em IET Control Theory Applications}, vol. 10, no. 14, pp.
  1732--1739, 2016.

\bibitem{parameterref}
Hamidreza Nazaripouya and Shahab Mehraeen,
\newblock ``Modeling and nonlinear optimal control of weak/islanded grids using
  facts device in a game theoretic approach,''
\newblock {\em IEEE Transactions on Control Systems Technology}, vol. 24, no.
  1, pp. 158--171, 2016.

\bibitem{lunze2009handbook}
Jan Lunze and Fran{\c{c}}oise Lamnabhi-Lagarrigue,
\newblock {\em Handbook of hybrid systems control: theory, tools,
  applications},
\newblock Cambridge University Press, 2009.

\bibitem{4782010}
H.~Lin and P.~J. Antsaklis,
\newblock ``Stability and stabilizability of switched linear systems: A survey
  of recent results,''
\newblock {\em IEEE Transactions on Automatic Control}, vol. 54, no. 2, pp.
  308--322, Feb 2009.

\bibitem{mahmoud2010switched}
Magdi~S Mahmoud,
\newblock ``Switched time-delay systems,''
\newblock in {\em Switched Time-Delay Systems}, pp. 109--130. Springer, 2010.

\bibitem{anjos2012handbook}
Miguel~F Anjos and Jean~B Lasserre,
\newblock ``Handbook on semidefinite, conic and polynomial optimization,
  international series in operations research \& management science, vol.
  166,'' 2012.

\bibitem{cvx1}
Michael Grant and Stephen Boyd,
\newblock ``{CVX}: Matlab software for disciplined convex programming, version
  2.1,'' \url{http://cvxr.com/cvx}, Mar. 2014.

\bibitem{cvx2}
Michael Grant and Stephen Boyd,
\newblock ``Graph implementations for nonsmooth convex programs,''
\newblock in {\em Recent Advances in Learning and Control}, V.~Blondel,
  S.~Boyd, and H.~Kimura, Eds., Lecture Notes in Control and Information
  Sciences, pp. 95--110. Springer-Verlag Limited, 2008,
\newblock \url{http://stanford.edu/~boyd/graph_dcp.html}.

\bibitem{7458197}
J.~Lai, H.~Zhou, X.~Lu, X.~Yu, and W.~Hu,
\newblock ``Droop-based distributed cooperative control for microgrids with
  time-varying delays,''
\newblock {\em IEEE Transactions on Smart Grid}, vol. 7, no. 4, pp. 1775--1789,
  July 2016.

\bibitem{selfself4}
Anuj Nandanwar, Meher~Preetam Korukonda, and Laxmidhar Behera,
\newblock ``A routing scheme for voltage stabilization in cyber physical energy
  systems,''
\newblock {\em IFAC Proceedings Volumes}, vol. 47, no. 1, pp. 812 -- 818, 2014,
\newblock 3rd International Conference on Advances in Control and Optimization
  of Dynamical Systems (2014).

\bibitem{nguyen2010distributed}
Phuong~H Nguyen, Wil~L Kling, Giorgos Georgiadis, Marina Papatriantafilou,
  Le~Anh Tuan, and Lina Bertling,
\newblock ``Distributed routing algorithms to manage power flow in agent-based
  active distribution network,''
\newblock in {\em Innovative Smart Grid Technologies Conference Europe (ISGT
  Europe), 2010 IEEE PES}. IEEE, 2010, pp. 1--7.

\bibitem{6840974}
S.~F. Bush,
\newblock ``Network theory and smart grid distribution automation,''
\newblock {\em IEEE Journal on Selected Areas in Communications}, vol. 32, no.
  7, pp. 1451--1459, July 2014.

\bibitem{7128386}
J.~Wei and D.~Kundur,
\newblock ``Goalie: Goal-seeking obstacle and collision evasion for resilient
  multicast routing in smart grid,''
\newblock {\em IEEE Transactions on Smart Grid}, vol. 7, no. 2, pp. 567--579,
  March 2016.

\bibitem{kompella1993multicast}
Vachaspathi~P Kompella, Joseph~C Pasquale, and George~C Polyzos,
\newblock ``Multicast routing for multimedia communication,''
\newblock {\em IEEE/ACM Transactions on Networking (TON)}, vol. 1, no. 3, pp.
  286--292, 1993.

\bibitem{1393890}
J.~Lofberg,
\newblock ``Yalmip : a toolbox for modeling and optimization in matlab,''
\newblock in {\em 2004 IEEE International Conference on Robotics and Automation
  (IEEE Cat. No.04CH37508)}, Sept 2004, pp. 284--289.

\bibitem{Mishra2020TowardsQF}
Swaroop Mishra, A.~Mitra, Neeraj Varshney, Bhavdeep~Singh Sachdeva, and Chitta
  Baral,
\newblock ``Towards question format independent numerical reasoning: A set of
  prerequisite tasks,''
\newblock {\em ArXiv}, vol. abs/2005.08516, 2020.

\bibitem{mccann2018natural}
Bryan McCann, Nitish~Shirish Keskar, Caiming Xiong, and Richard Socher,
\newblock ``The natural language decathlon: Multitask learning as question
  answering,''
\newblock {\em arXiv preprint arXiv:1806.08730}, 2018.

\bibitem{sanh2021multitask}
Victor Sanh, Albert Webson, Colin Raffel, Stephen~H Bach, Lintang Sutawika,
  Zaid Alyafeai, Antoine Chaffin, Arnaud Stiegler, Teven~Le Scao, Arun Raja,
  et~al.,
\newblock ``Multitask prompted training enables zero-shot task
  generalization,''
\newblock {\em arXiv preprint arXiv:2110.08207}, 2021.

\bibitem{mishra2021cross}
Swaroop Mishra, Daniel Khashabi, Chitta Baral, and Hannaneh Hajishirzi,
\newblock ``Cross-task generalization via natural language crowdsourcing
  instructions,''
\newblock {\em arXiv preprint arXiv:2104.08773}, 2021.

\bibitem{mishra2021reframing}
Swaroop Mishra, Daniel Khashabi, Chitta Baral, Yejin Choi, and Hannaneh
  Hajishirzi,
\newblock ``Reframing instructional prompts to gptk's language,''
\newblock {\em arXiv preprint arXiv:2109.07830}, 2021.

\bibitem{wei2021finetuned}
Jason Wei, Maarten Bosma, Vincent~Y Zhao, Kelvin Guu, Adams~Wei Yu, Brian
  Lester, Nan Du, Andrew~M Dai, and Quoc~V Le,
\newblock ``Finetuned language models are zero-shot learners,''
\newblock {\em arXiv preprint arXiv:2109.01652}, 2021.

\bibitem{Mishra2020DQIMD}
Swaroop Mishra, A.~Arunkumar, Bhavdeep~Singh Sachdeva, Chris Bryan, and Chitta
  Baral,
\newblock ``Dqi: Measuring data quality in nlp,''
\newblock {\em ArXiv}, vol. abs/2005.00816, 2020.

\bibitem{le2020adversarial}
Ronan Le~Bras, Swabha Swayamdipta, Chandra Bhagavatula, Rowan Zellers, Matthew
  Peters, Ashish Sabharwal, and Yejin Choi,
\newblock ``Adversarial filters of dataset biases,''
\newblock in {\em International Conference on Machine Learning}. PMLR, 2020,
  pp. 1078--1088.

\bibitem{Sambasivan2021EveryoneWT}
N.~Sambasivan, Shivani Kapania, H.~Highfill, Diana Akrong, Praveen~K. Paritosh,
  and Lora Aroyo,
\newblock ``“everyone wants to do the model work, not the data work”: Data
  cascades in high-stakes ai,''
\newblock {\em Proceedings of the 2021 CHI Conference on Human Factors in
  Computing Systems}, 2021.

\bibitem{arunkumarreal}
Anjana Arunkumar, Swaroop Mishra, Bhavdeep Sachdeva, Chitta Baral, and Chris
  Bryan,
\newblock ``Real-time visual feedback for educative benchmark creation: A
  human-and-metric-in-the-loop workflow,''
\newblock 2020.

\bibitem{swayamdipta2020dataset}
Swabha Swayamdipta, Roy Schwartz, Nicholas Lourie, Yizhong Wang, Hannaneh
  Hajishirzi, Noah~A Smith, and Yejin Choi,
\newblock ``Dataset cartography: Mapping and diagnosing datasets with training
  dynamics,''
\newblock {\em arXiv preprint arXiv:2009.10795}, 2020.

\bibitem{mishra-sachdeva-2020-need}
Swaroop Mishra and Bhavdeep~Singh Sachdeva,
\newblock ``Do we need to create big datasets to learn a task?,''
\newblock in {\em Proceedings of SustaiNLP: Workshop on Simple and Efficient
  Natural Language Processing}, Online, Nov. 2020, pp. 169--173, Association
  for Computational Linguistics.

\bibitem{mishra2020dqi}
Swaroop Mishra, Anjana Arunkumar, Bhavdeep Sachdeva, Chris Bryan, and Chitta
  Baral,
\newblock ``Dqi: A guide to benchmark evaluation,''
\newblock {\em arXiv preprint arXiv:2008.03964}, 2020.

\bibitem{mishra2021robust}
Swaroop Mishra and Anjana Arunkumar,
\newblock ``How robust are model rankings: A leaderboard customization approach
  for equitable evaluation,''
\newblock in {\em Proceedings of the AAAI Conference on Artificial
  Intelligence}, 2021, vol.~35, pp. 13561--13569.

\bibitem{Mishra2020OurEM}
Swaroop Mishra, A.~Arunkumar, Chris Bryan, and Chitta Baral,
\newblock ``Our evaluation metric needs an update to encourage
  generalization,''
\newblock {\em ArXiv}, vol. abs/2007.06898, 2020.

\end{thebibliography}
\clearpage
\phantomsection

\end{document}